# The Effects of Cultural dimensions and Demographic Characteristics on E-learning acceptance

**A thesis submitted for the degree of Doctor of Philosophy**

**By**

**Ali Tarhini**

**Department of Information Systems and Computing,**

**Brunel University**

**July 2013**

# ABSTRACT

Due to the rapid growth of internet technology, universities and higher educational institutions around the world are investing heavily in web-based learning systems to support their traditional teaching and to improve their students' learning experience and performance. However, the success of an e-learning system depends on the understanding of certain antecedent factors that influence the students' acceptance and usage of such e-learning systems. Previous research indicates that technology acceptance models and theories may not be applicable to all cultures as most of them have been developed in the context of developed countries and particularly in the U.S. So far little research has investigated the important role that social, cultural, organizational and individual factors may play in the use and adoption of the e-learning systems in the context of developing countries and more specifically there is almost absence of this type of research in Lebanon.

This study aims to fill this gap by developing and testing an amalgamated conceptual framework based on Technology Acceptance Model (TAM) and other models from social psychology, such as theory of reasoned action and TAM2 that captures the salient factors influencing the user adoption and acceptance of web-based learning systems. This framework has been applied to the study of higher educational institutions in the context of developing as well as developed countries (e.g. Lebanon and UK). Additionally, the framework investigates the moderating effect of Hofstede's four cultural dimensions at the individual level and a set of individual differences (age, gender, experience and educational level) on the key determinants that affect the behavioural intention to use e-learning.

A total of 1197 questionnaires were received from students who were using web-based learning systems at higher educational institutions in Lebanon and the UK with opposite scores on cultural dimensions. Confirmatory Factor Analysis (CFA) was used to perform reliability and validity checks, and Structural Equation Modeling (SEM) in conjunction with multi-group analysis method was used to test the hypothesized conceptual model.

As hypothesized, the findings of this study revealed that perceived usefulness (PU), perceived ease of use (PEOU), social norms (SN), perceived quality of work Life (QWL), self-efficacy (SE) and facilitating conditions (FC) to be significant determinants of behavioural intentions and usage of e-learning system for the Lebanese and British students. QWL; the newly added variable; was found the most important factor in explaining the causal process in the model for both samples. A t-test analysis proved that there are differences between Lebanese and British students in terms of PEOU, SE, SN, QWL, FC and AU; however no differences were detected in terms of PU and BI. The results of the MGA show that cultural dimensions as well as demographic factors had a partially moderated effect on user acceptance of e-learning. Overall, the proposed model achieves acceptable fit and explains for 68% of the British sample and 57% of the Lebanese sample of its variance which is higher than that of the original TAM.

Our findings suggest that individual, social, cultural and organisational factors are important to consider in explaining students' behavioural intention and usage of e-learning environments. The findings of this research contribute to the literature by validating and supporting the applicability of our extended TAM in the Lebanese and British contexts and provide several prominent implications to both theory and practice on the individual, organizational and societal levels.

# DEDICATION

This doctoral research effort would not have been completed without the only Grace of Al-Mighty Allah (swt), is dedicated to the memory of my beloved father Hussein Ahmad Tarhini (May your soul rest in heaven) who inoculated me in discipline, perseverance and strength that forever will be part of my life.

This research is also dedicated to my lovely mum, my five fabulous sisters, my precious nephews and nieces, and my parents-in-laws who always believed in me and to whom I am greatly indebted. Without your support and constant encouragement this would not have been possible.

I am particularly indebted to my beautiful wife who makes my life happy and complete. Takwa, you are my source of inspiration, I thank you for always reminding me of what life is always about. I owe you so much, I promise to make up the long time we spent away from each other. My work is finally finished and thanks to Alla (swt) it's time to be together forever.

Thank you all for giving me your moral support and unconditional love; you were my motivation and constant inspiration during all the process of my PhD and in my most difficult times. I really appreciate you, even though the value of my appreciation cannot compare with everything you have done for me. Finally, I made it.

# ACKNOWLEDGEMENTS

In the name of God, the Most Gracious, the Most Merciful. All praise be to Allah, the Creator and Master of the Universe.

The completion of this research has been assisted by a generous help of many people of whom I owe a great deal of thanks and appreciation.

First and foremost, my sincere gratitude and deepest thanks go to my supervisor professor Xiaohui Liu for his patience, sage advice, invaluable guidance, and continuous support and encouragement at every stage of this dissertation. It is a complete pleasure to work under your auspicious supervision. Thank you for always being there for me at time of need.

Equally, I wish to express my appreciation to my second supervisor Dr. Kate Hone for her constructive feedback, generous time, kind support, and invaluable advice through this research and the publications associated with it. Thank you Kate for making my learning experience became less stressful and even more fascinating!!

My heartfelt thanks and greatest gratitude go to Mahir Arzoky who spent his valuable time proofreading the thesis. Great thanks go also to Ms Ela Heaney (PhD Administrator) for her effective administrative assistance, for which I am especially thankful.

I also wish to express my appreciation to my colleagues, for interesting discussions and for keeping the office fun and exciting to work in: Valaria, Neda, Fadra, Cici, Mohsina, Djibril, Stefio, Liang and Miqing.

I am sincerely thankful to my friends and housemates. Their company and friendship have made the last four years a memorable period of my life. A special mention goes to Hadi, Bassel, Ghaytih, Bachar, Mahir, Mohammad, Hussein, Saeed, Ghorbani, Mike, Chido and Nalin.

Last but not the least; I am indebted to the 'Altajir Trust' for funding my PhD.

# DECLARATION

The following papers have been published (or submitted for publication) as a direct result of the research discussed in this thesis:

**Tarhini, A.**, Hone, K., and Liu, X. (2013), User Acceptance Towards Web-based Learning Systems: Investigating the role of Social, Organizational and Individual factors in European Higher Education. Procedia Computer Science 17, 189-197

**Tarhini, A.**, Hone, K., and Liu, X. (2013), "Factors Affecting Students' Acceptance of E-learning Environments in Developing Countries: A Structural Equation Modelling Approach", Proceedings of the 6th International Business and Social Sciences Research Conference, Dubai, UAE, 03-04 January 2013

**Tarhini, A.**, Hone, K., and Liu, X. (2013), Factors Affecting Students' Acceptance of E-learning Environments in Developing Countries: A Structural Equation Modelling Approach, (Revised version of the Proceedings of 6th International Business and Social Sciences Research Conference), International Journal of Information and Education Technology Vol 3. no. 1, pp. 54-2013

**Tarhini, A.**, Hone, K., and Liu, X. (2013), Extending TAM to Empirically Investigate the Students' Behavioural Intention to Use E-Learning in Developing Countries, Science and Information Conference (IEEE), United Kingdom, 07-09 Oct 2013

**Tarhini, A.**, Hone, K., and Liu, X. (2013), The Effects of Individual-level Culture on E-learning Acceptance, European Journal of Information Systems (under review)

**Tarhini, A.**, Hone, K., and Liu, X. (2013), A cross-cultural examination of the impact of social, organizational and individual factors on Technology Acceptance between British and Lebanese university students, British Journal of Educational Technology (under review)

**Tarhini, A.**, Hone, K., and Liu, X. (2013), The Effects of Individual Differences on e-learning users' behaviour in Developing Countries: A Structural Equation Model, Computers & Education (under review)

**Tarhini, A.**, Hone, K., and Liu, X. (2013), Measuring the moderating effect of gender and age on e-learning acceptance in England: A structural Equation Modelling Approach of an extended Technology Acceptance Model, Australasian Journal of Educational Technology (under review)

### *Papers Presented (Not Published)*

"Understanding the Behaviour of Students and Lecturers in web-based learning environments", Doctoral Consortium, May, British Computer Society (BCS), London, (2009)

"E-learning as innovation: The change of the face of Education!!", 1st NewRoute PhD Conference, Brunei Gallery London, (2009)

"Is e-Learning simply a technological solution or a process of Social, Behavioural, Individual and Cultural factors?", 3rd NewRoute PhD Conference, Brunei Gallery London, (2011)

### *Posters:*

**Tarhini, A.**, Liu,X. (2011, March). "To what extent How Cultural Differences Affects Learners' perceptions Towards Using E-learning Tools at the UK universities?" Poster presented at the Research Student Poster Conference, Brunel University, UK

# ABBREVIATIONS

| | |
|---|---|
| **AGFI** | Adjusted Goodness-of-Fit Index |
| **AMOS** | Analysis of Moment Structures |
| **AVE** | Average Variance Extracted |
| **CFA** | Confirmatory Factor Analysis |
| **CFI** | Comparative Fit Index |
| **CMS** | Course management System |
| **CR** | Critical Ratio |
| **Df** | Degree of Freedom |
| **DTPB** | Decomposed Theory of Planned Behaviour |
| **D²** | Mahalanobis Distance |
| **EFA** | Explanatory Factor Analysis |
| **e-learning** | electronic learning |
| **GFI** | Goodness-of-Fit Index |
| **HEIs** | Higher Education Institutions |
| **HCI** | Human Computer Interaction |
| **ICT** | Information and Communication Technology |
| **IDT** | Innovations Diffusion Theory |
| **IFI** | Incremental Fit Index |
| **IS** | Information Systems |
| **IT** | Information Technology |
| **KMO** | Kaiser-Mayer-Olkin |
| **LMSs** | Learning management system |
| **MI** | Modification Index |
| **NNFI** | Non-Normed Fit Index |
| **NFI** | Normed Fit Index |
| **PBC** | Perceived Behaviour Control |
| **PLS** | Partial least squares |
| **R²** | Coefficient of Determination |
| **RMSEA** | Root Mean Square Error of Approximation |
| **SD** | Standard Deviation |
| **SEM** | Structure Equation Modeling |
| **SRMR** | Standard Root Mean Square Residual |
| **SCT** | Social Cognitive Theory |
| **SMC** | Squared Multiple Correlations |
| **SPSS** | Statistical Package for Social Science |
| **TAM** | Technology Acceptance Model |
| **TPB** | Theory of Planned Behaviour |
| **TRA** | Theory of Reasoned Action |
| **TLI** | Tucker-Lewis *Index* |
| **UK** | United Kingdom |
| **UTAUT** | Unified Theory of Acceptance and Use of Technology |
| **VIF** | Variance Inflation Factor |
| **WebCT** | Web Course Tools |
| **WWW** | World Wide Web |
| **X²** | Chi Square |
| **X²/df** | Normed Chi-Square |

# TABLE OF CONTENTS

# LIST OF TABLES

# LIST OF FIGURES

# Chapter 1:  Introduction

*"We can no longer think in terms of imposing a "universal" product; culture has won this battle. A key question is how can we develop products for multiple and non-familiar cultures." (De Souza and Dejean, Cultural Influence on Design, 2000)*

## 1.1   Introduction

The current chapter introduces the PhD thesis entitled "The Effects of Individual-level Culture and Demographic Characteristics on E-learning Acceptance in Lebanon and England: A Structural Equation Modeling Approach". The following section provides an overview of the theoretical background and research problem. Section 1.3 will define the research aim and objectives. Section 1.4 presents the scope of the research. Sections 1.5 and 1.6, respectively presents the significance and contributions of the study. The research methodology employed to investigate the research questions are then introduced in Section 1.7. It is followed by the context of the research in Section 1.8. To familiarise the readers with the remainder of this dissertation, a brief overview of the contents of each chapter is provided in Section 1.9. Finally, Section 1.10 concludes this chapter.

## 1.2   Theoretical Background and Research problem

During the last two decades, with the widespread use of the World Wide Web (WWW), universities and other educational institutions have been investing in web-based information systems (such as Moodle, Blackboard and WebCT) to support both face-to-face and remote course delivery (Fletcher, 2005; Ngai *et al.,* 2007). ICT has the potential to greatly help students in their education, they reduce the cost of provision and therefore increase revenues for academic institutions (Ho and Dzeng, 2010). They also afford students with more study flexibility and improve their learning experience and performance (Christie and Ferdos, 2004).

Despite the enormous growth of e-learning in education and its perceived benefits, research indicates that failures exist (Sun *et al.*, 2008; Arbaugh and Duray, 2002; Wu *et al.*, 2006). The efficiency of such tools will not be fully utilised if the users are inclined to not accept and use the system. Therefore, the successful implementations of e-learning tools depend on whether or not the students are willing to adopt and accept the technology. Thus, it has become imperative for practitioners and policy makers to understand the factors affecting the user acceptance of web-based learning systems in order to enhance the students' learning experience (Liaw and Huang, 2011). Within this context, a number of recent studies have shown that e-learning implementation is not simply a technological solution, but also a process of many different factors such as social factors (Schepers and Wetzels, 2007), organizational factors (Sun and Zhang, 2006), individual factors (Liaw, 2008; Venkatesh and Morris, 2000), in addition to behavioural and cultural factors (Srite and Karahanna, 2006; Straub *et al.*, 1997).

In the technology acceptance and adoption literature, a considerable number of models have been applied (e.g., the theory of reasoned action (TRA), the theory of planned behaviour (TPB) and the technology acceptance model (TAM), unified theory of acceptance and use of technology (UTAUT)) to investigate and explore the determinants of user's behaviour towards adoption and using information technology. Among these models, the Technology Acceptance Model (TAM) (Davis, 1989) is the most frequently cited and influential model for explaining technology acceptance and adoption. Since it has been developed, TAM has been extensively used, tested, and extended to explain technology adoption and success in a number of application areas e.g. see (Bagozzi, 2007; Yousafzai *et al.*, 2007a;

Venkatesh and Bala, 2008) including examples in: e-government e.g., (Phang *et al.*, 2006; Walker and Johnson, 2008), e-health e.g.(Lanseng and Andreassen, 2007) and e-learning (Zhang *et al.*, 2008; Park, 2009; Saeed and Abdinnour-Helm, 2008; Yi-Cheng *et al.*, 2007; Teo, 2011).

However, the explanatory power of TAM is still questionable as it ignores the effect of social, individual and cultural influence on the acceptance of technology (Struab et al., 1997; Bagozzi, 2007). More specifically, the focus of the majority of recent studies has been within the context of developed countries such as North America (Teo *et al.*, 2008). While the internet is a global tool, the efficiency of particular applications should also be measured locally since users usually work in local/national contexts (Li and Kirkup, 2007).

Developing countries, such as Lebanon, are particularly under-researched in relation to their acceptance of e-learning applications. Such countries typically support traditional styles of pedagogy in education, due to a lack of financial resources and appropriately trained staff (Nasser, 2000; UNDP, 2002; Baroud and Abouchedid, 2010), so it is especially important to understand the factors that may encourage take up of e-learning within these developing contexts. TAM has been criticised for its cultural bias especially when tested in non-Western cultures (e.g. see McCoy et al., 2007). Some support for TAM has been shown in the Arab world in general (e.g. Rose and Straub, 1998) and for e-learning acceptance in Jordan (Abbad et al., 2009). However, in relation to e-learning, questions remain since Abbad et al. (2009) did not seek to define their sample in terms of their specific cultural characteristics.

It has also been suggested that specific cultural differences may affect the strength of some relationships within the TAM model and may help to explain some contradictory findings within the literature e.g. (Sánchez-Franco et al., 2009). This explanation has been explored explicitly in a limited number of studies through the examination of the effects of cultural variables as moderators within TAM. The most widely applied conception of culture used has been that of Hofstede (1980) which categorises countries along the following dimensions:

- Collectivist / individualistic
- Uncertainty avoidance (high / low)
- Masculinity / femininity
- Power distance (high / low)

While there is evidence that these cultural differences may explain some variations in TAM results, much of the work in this tradition is limited, as it uses nationality as a surrogate for culture which may mean that some of the specific cultural variables are confounded. Hofstede's own measurement instruments were designed to be used at macro (country) level, providing limitations even for studies which have included direct cultural measures in their methodology. More recently Srite and Karahanna (2006) have overcome this issue by using measures for Hofstede's cultural dimensions that are reliable when used at the individual level. Within e-learning relatively little attention has been given to the effect of cultural variables as potential moderators, and studies which explicitly measure culture at the individual level are particularly scarce. In the current study we therefore extend TAM to include an examination of individual level cultural variables as moderators within the model. TAM has also been criticised in some contexts for lacking explanatory power e.g. (Sánchez-Franco et al., 2009).

Furthermore, while TAM has generally been found to have acceptable explanatory power, the inclusion of moderators could improve this further (Sun and Zhang, 2006). For example, when including gender and experience in TAM2, the explanatory power increased from 35 % to 53 % (Venkatesh et al., 2003). Within this context, a number of researchers have recommended the need to incorporate a set of moderators which remain largely untested such as Experience e.g. (Venkatesh and Bala, 2008), age e.g. (Venkatesh et al., 2003) and cultural background e.g. (Qingfei et al., 2009). Therefore, it is expected that after the inclusion of moderating variables such as individual differences and culture within TAM, the predictive validity of the model will be increased and a better explanation of the inconsistencies in previous studies (Chin *et al.*, 2003; Venkatesh *et al.*, 2003; Johns *et al.*, 2003; Sørnes *et al.*, 2004).

While the inclusion of cultural moderators may address the limitation of TAM to some extent, a complementary approach to this problem is the inclusion of additional predictor variables within the model such as social, organisational, individual and cultural factors. Here we include social norms, quality of working life, facilitating technology and self-efficacy in order to examine whether the explanatory power of TAM is improved in our research context through the introduction of these additional predictor variables. Social Norms have been examined in a number of previous studies where they have been shown to be an important determinant of acceptance (e.g. see Venkatesh et al., 2003). Furthermore, Srite and Karahanna (2006) found that the impact of Social Norms was sensitive to cultural differences, providing a key rationale for the inclusion of this variable here. Quality of Working Life has also been proposed as a variable that may be a useful, culturally sensitive addition to TAM e.g. (Srite and Karahanna, 2006; Zakour, 2004). However, neither its applicability to the e-learning context, nor the influence that cultural differences have on its effects have been previously examined.

Clearly there are a number of gaps that this thesis aims to tackle. As mentioned above, there is a lack of research focusing on the individual, social, organisational and cultural factors that affect the acceptance and adoption of e-learning technologies. In addition, although there are many studies that consider the cultural values at the national level, there are very few studies that consider the individual-level culture values. The latter is critical due to the fact that although the national culture is a macro-level, however it is argued that the acceptance and adoption of technology by end-users is a micro-level concern (Srite and Karahanna, 2006). Finally, there have been little empirical studies that consider the individual acceptance of e-learning technologies within the context of developing and developed countries. This thesis aims to fill this gap.

Therefore, to address the aforementioned issues, this research aims to add new variables; namely social norms (SN), quality of work Life (QWL), computer self-efficacy (SE) and facilitating conditions (FC), Self-efficacy (SE) as a direct predictors in addition to individual-level cultural and other demographic characteristics as moderators; to the TAM research model to investigate the extent to which these variables affect students' willingness to adopt and use e-learning systems and investigate whether there are differences among these factors between developing and developed world, specifically Lebanon as developing world and England as developed world. Extending the TAM model to include social, organisational and individual factors in two cultures allows us to explore the generalizability and applicability of the proposed model in the context of e-learning in two cultures and also allows exploration of where differences may lie between the cultures involved. This will also help policy makers and practitioners to gain a deeper understanding of the students' acceptance of e-learning technology.

## 1.3   Research Aims and Objectives

The main aim of this study is to develop and test an amalgamated conceptual model of technology acceptance that explains how individual, social, cultural and organisational factors affect the students' acceptance and usage behaviour of the e-learning systems in Lebanon and England. This research aims to contribute to the stream of literature on e-learning, technology acceptance and culture. Further, it is hoped that this research will help the policy makers to establish a better understanding of the reasons for accepting or rejecting the e-learning systems across cultures. Given this context, this research aims to answer the following two questions:

- To what extent do individual, social, organisational and cultural factors affect the students' behaviour to adopt and use the web-based learning system in Lebanon and the UK?

- To what extent do individual-level cultural dimensions (power distance, masculinity\femininity, uncertainty avoidance, and individualism-collectivism) and other individual differences (age, gender, experience and educational level) — as a two sets of moderatos—impact the relationship between the main predictors and behavioural intention and usage of e-learning systems?

The meeting of the following objectives will help to meet the overall aim of the research and answer the above stated research questions:

1) Understand background and current situation

(a) To determine the current usage of Web-based learning systems in Lebanese and British universities.

(b) To review the literature related to the diffusion of innovations and technology acceptance models and theories, e-learning, and culture.

2) Develop and test a conceptual framework that captures the salient factors influencing the user adoption and acceptance of web-based learning system including:

(a) Behavioural belief
(b) QWL
(c) Social factors
(d) Role of internal and external support
(e) Individual factors
(f) Cultural factors

3) Examine the effect of two sets of moderators in the model

(a) Hofstede's cultural dimensions at the individual-level (PD, M/F, I/C, UA)
(b) Individual differences (age, gender, educational level, experience)

4) To empirically validate the model in the context of a western/developed (UK's perspective) and non-western/developing (i.e., Lebanon's perspective) countries, and examine the similarities and differences between the two settings, this will help the researcher to examine the external validity of western developed theories in non-western countries.

5) Provide recommendations that emerge from the research for practice and policy as to how adoption problems could be addressed.

## 1.4   Research Scope

It is crucial to define the scope of this study while taking into consideration the main aims and objectives of this research and the availability of resources such as time and money. This study investigates the most salient factors that affect the user's behaviour towards using the e-learning systems in the Lebanese and British context. The scope of this research can be summarised as follows:

- The area of application under investigation is the usage of learning management systems, particularly Blackboard. However, the results of this research could be generalised to other e-learning systems that share the same nature as Blackboard such as WebCT and Moodle.

- This study is focused only on students who use e-learning technologies in their studies in higher educational institutions. Although other end-users such as instructors, system administrators and university management are considered to be important in promoting the benefits of such systems, however the proposed conceptual model (Chapter 3) will only consider factors and studies that are relevant to the acceptance of e-learning systems from the students' perspective.

- The investigation for the acceptance behaviour is limited to the geographical area of Lebanon as a developing country and England as a developed country, and therefore considered representative to the areas that only share the same cultural characteristics to those two countries. Therefore, the applicability and the generalizability of the proposed conceptual model will become a questionable issue when applied in a different context or geographical area.

- This study is also limited by applying only Hofstede's (1991) cultural dimensions at the individual level as its main concern is not the cultural models by itself. However other dimensions and cultural models will be briefly discussed in the literature review (Chapter 2) but they will not be part of our research plan.

## 1.5   Significance of the study

This research provides useful insightful information not only about the effectiveness of a particular e-learning environment with a focus on online collaboration, but also its impact on students in different cultural settings. More specifically, this research helps to understand the differences and similarities between Lebanese and British students regarding their e-learning experiences. The findings can help educators and researchers from the concerned cultural contexts to be better armed with knowledge about the specific cultural-related variables. Student variables, such as behaviours and attitudes, cultural backgrounds and other demographic characteristics are important variables that influence student learning, especially in a collaborative e-learning environment. Understanding these variables is now helpful for instructors to design meaningful educational activities to promote student knowledge construction and make learning more effective and appealing.

In particular, this research helps to better understand the characteristics of students in Lebanon and England respectively, which can help policy makers, educators and experts to understand *what* the students expect from the learning management systems. This can help the management achieve the most effective deployment of such system and also helps them improve their strategic decision making about technology in the future, they can decide on the best approach that fit their students before implementing any new technology. For system developers, the findings of this study will help them understand *how* they could improve their learning management systems in the concerned cultural contexts and overcome problems that may occur during cross-cultural educational cooperation and e-learning implementation. We hope that through a better understanding cross-cultural similarities and differences, this type of research can help to overcome the potential (negative) effects of cultural differences on student's behaviour when using e-learning environments, and hopefully to create productive e-learning tools that consider different cultures. Similarly, the users (students) can understand what motivations and factors drive them into accepting the technology, and exactly know the impact of using technology on their working life and that using the technology is usually related to their social, attitudinal, cultural and individual differences.

From the academic prospective, this research developed an integrative model that combines both technology acceptance theories and cultural theories at the micro-level (individual level) rather than national level within different cultural contexts who apparently exhibited unique psychological and personal characteristics. To our knowledge, no other research has measured cultural factors at the individual level in Lebanon and England. Therefore, this study is considered a useful guide for other researchers to understand whether the acceptance of technology is mainly affected by individuals' cultural background (moderation effect) or whether the acceptance is mainly based on the key determinants of technology itself (without an indirect effect of moderation).

## 1.6   Contributions of the study

The current research will primarily contribute to the body of literature on technology acceptance and cross-cultural studies in general, and in the Arab world and England specifically. This research set out to make theoretical and practical contributions to knowledge as follows:

1) It provides a critical analysis of the literature related to the diffusion of innovations and technology acceptance models and theories, in addition, to cultural theories in order to enhance knowledge of technology acceptance and adoption from the student's perspective.

2) It empirically confirms and validates an extended Technology Acceptance Model within the context of e-learning in a western (UK's perspective) and non-western (i.e., Lebanon's perspective) countries. To the best of the author's knowledge, this research is one of the first studies that empirically and theoretically develops and test such an integrative theoretical framework. Therefore, the present research offers a deeper understanding of the interplay between student characteristics and the usability and interactivity of e-learning environment from a cross-cultural perspective.

3) This study is one of the few studies that combine technology acceptance theories and cultural theories at the micro-level (individual level) within different cultural contexts who apparently exhibited unique psychological and personal characteristics. To our knowledge, no other research has measured cultural factors at the individual level in Lebanon and England. Therefore, this study is considered a useful guide for other researchers to understand whether the acceptance of technology is mainly affected by individuals' cultural background (moderation effect) or whether the acceptance is mainly based on the key determinants of technology itself (without an indirect effect of moderation).

4) It provides an overall picture of the current usage of Web-based learning systems in Lebanese and British universities. The findings will help to understand whether additional research is needed to address the technological needs of students in efforts to close the technological gap that potentially exists between students from various socioeconomic backgrounds at the higher education institutions.

5) It develops and validates a research instrument in order to collect a valuable data from Lebanon and England respectively.

6) It integrates two sets of moderators namely; Hofstede's cultural dimensions and other individual characteristics (age, gender, experience and educational level); within the extended model and demonstrate their impact on the relationship between the main predictors and usage behavioural and behaviour intention of web-based learning tools.

7) The conceptual framework also integrates and examines the social, individual, organisational and cultural factors that impact the users' beliefs and behaviour towards using technology web-based learning system.

8) This research contributes to the trends of studies and literature in social science and IS that uses Structural Equation Modeling (SEM) technique to analyse the data. Using SEM enables a better understanding of technology adoption and acceptance in a cross-cultural context.

9) It investigates to what extent using the Web-based learning systems helps to improve the users' quality of working life.

10) It examines and understands the students' opinions of internal and external supports on influencing their usage and adoption of web-based learning systems.

11) It develops an inclusive categorisation of the similarities and differences between British and Lebanese students on the acceptance and usage of web-based learning systems. This will help in identifying and understanding any differences between the cultures of these two countries.

# 1.7   Research Design and Methodology

The nature and context of this research make it most suitable to apply the positivist approach with a quantitative strategy of analysis to answer the research questions. This research aims to test hypothesized relationships within the context of technology acceptance in an objective manner where the researcher is isolated from the aim of the study. Furthermore, the constructs and their relationships used within the well-defined conceptual model were developed and validated thoroughly in the theories and models about the adoption and technology acceptance (see Chapter 2).Therefore, a cross-sectional survey was found the most appropriate technique to collect the data. Using the survey approach, the data can be collected from a large number of participants simultaneously in a quick, easy, efficient and economical way compared with other methods such as interviews (Zikmund, 2009; Bryman, 2008; Sekaran and Bougie, 2011).Structural Equation Modeling (SEM) technique will be used to test hypotheses and moderators and to perform a number of tests such as group comparisons which require a large number of participants. Therefore, using the survey as a data collection method is appropriate from the ontological, epistemological and methodological point of view. Conversely collecting data from two different countries using other approaches such interviews was not feasible in this PhD research in terms of timing and limited financial resources.

Furthermore, this research employed a self-administrated questionnaire as a data collection method for the following reasons: it is easily designed and administrated; higher privacy of respondents because issues such as anonymity and confidentiality were dealt with in the cover letter; collecting the questionnaires immediately after being completed will assure a higher response rate; respondents can seek clarity and therefore could understand the concepts on any question they are answering which in turns minimise the outliers in the study (Sekaran and Bougie, 2011).A pre-test of the questionnaire with expert in the fields and a pilot study with potential participants were conducted to ensure reliability and validity of the questionnaires' items and scales.

The current study, as the majority of empirical research in technology acceptance (Venkatesh *et al.*, 2003; Venkatesh and Bala, 2008) and e-learning (Zhang *et al.*, 2008; Teo *et al.*, 2008), has used a non-probability convenience sampling technique as it enables the researcher to collect data from the participants based on their availability. It also helps the researcher to improvise with the resource available for the research especially when there is lack of time and financial resources. It is worth noting that a homogeneous large sample of students from Lebanon and UK was used to allow the generalisation of the results to the student populations in those two countries. In particular, a total of 2000 questionnaires were distributed to 1000 students from the UK and 1000 from Lebanon respectively, of which 1197 were returned indicating a 59.7% response rate. After screening for missing data and duplicated responses, we retained 1168 questionnaires for data analysis. These included 566 Lebanese participants and 602 British participants.

The descriptive analysis of the collected data were analysed using Statistical Package for Social Science (SPSS) version 18.0 such as data screening, frequencies and percentages, reliability analysis, explanatory factor analysis and t-test. Analysis of Moment Structures (AMOS) software version 18.0 was used to perform the Structural Equation Modeling (SEM) analysis. This research follows Hair's (2010) recommendations about evaluating the structural model using a two-steps approach (first the measurement model and then the structural model). Additionally, Multiple Group Analysis (MGA) technique is used to measure the impact of moderators on the conceptual model.

## 1.8   Context of the research

England, as a representative of developed countries, and Lebanon, as a representative of the developing world, were chosen for this study because they represent nearly reverse positions on all Hofstede's (2005) cultural dimensions, as shown in Table 1.1. England is high on individualism and masculinity, low in power distance and uncertainty avoidance. On the other hand, Lebanon is low on individualism, moderate on masculinity, high on power distance and uncertainty avoidance. Furthermore, Lebanon is largely different from England in terms of language, religion and customs.

| Country | Power Distance | Masculinity | Individualism | Uncertainty avoidance |
|---|---|---|---|---|
| **Lebanon (as part of Arab World)** | 80 | 53 | 38 | 68 |
| **England** | 35 | 66 | 89 | 35 |

**Table 1-1: Cultural differences between England and Lebanon on Hofstede's cultural dimensions (values adopted from Hofstede, 2005)**

In addition, compared to England, Lebanon remains relatively unexplored in terms of technology acceptance and the investment in technology in general and in the educational system in particular is still immature compared to western countries since universities and higher education institutions support traditional styles of pedagogy in education due to the lack of financial resources or trained staff (Nasser, 2000; UNDP, 2002; Baroud and Abouchedid, 2010), which in turn may affect the acceptance of technology within such countries. Additionally, the limitations that emerges from TAM especially on holding equally well across cultures and the inconsistency in previous studies' results (Gefen and Straub, 1997; McCoy *et al.*, 2005a; Straub *et al.*, 1997; Zakour, 2004; Srite and Karahanna, 2006) direct our research to consider these two distinctive cultures. Therefore, the applicability and thus the validity and robustness of the extended TAM will be examined in both western and non-western cultural settings.

## 1.9   Dissertation Outline

This section provides a brief overview of the eight main chapters of this thesis and the steps undertaken to fulfil the research aim and objectives.

**Chapter 1: *Introduction*** provides the 'roadmap' of the entire thesis. It first introduces the reader to the research problem along with the motivation behind conducting this research and its scope. Then it highlights the research aims and objectives. It is then followed by the research methodology and methods adopted and then followed by context of the study. Finally, the structure and organisation of the thesis are outlined.

**Chapter 2:** ***E-learning, Technology Acceptance & Cultural models*** provides a comprehensive literature review about the three main research areas that forms the basis for this research namely; technology adoption theories and models, e-learning technologies and culture. This chapter starts with a review of the nine most influential theories and models related to technology adoption including IDT, SCT, TRA, TPB, DTPB, TAM, TAM2, ATAM, and UTAUT, and also discusses their external factors which directly or indirectly are useful in developing the conceptual framework for this study. This chapter also discusses the different e-learning tools being used by higher educational institutions. Finally, this chapter will highlight the importance of cultural dimensions and in particular Hofstede's dimensions at the macro and micro level.

**Chapter 3:** ***Theoretical basis and conceptual framework*** aims to discuss the development of the proposed conceptual model to study e-learning acceptance. For this purpose, it justifies the use of Technology Acceptance Model as a theoretical basis. This chapter also provides a further justification for extending the TAM to include personal, social, and situational factors as key determinants, in addition, to the integration of individual characteristics (age, gender, educational level, experience) and Hofstede's cultural dimensions (power distance, individualism\collectivism, masculinity\femininity, uncertainty avoidance) as moderators within the model to study e-learning adoption and acceptance in Lebanon and England. Moreover, research hypotheses will be formulated and operational definitions for each constructs will be presented. The results of this chapter along with the detailed literature review in Chapter 2 helps to achieve the following objective of this research, which is *"To develop a conceptual framework that captures the salient factors influencing the user adoption and acceptance of web-based learning system"*.

**Chapter 4: *Research Design & Methodology*** describes the philosophical approach, strategy of inquiry, methods and techniques used in Information System and Social Science in general. This chapter also explain the rationale behind the chosen approach and techniques that is essential in order to empirically test the proposed conceptual model and thus to achieve the main research objectives and answer the research questions. This chapter will also discuss the sampling technique and the design of the survey and the steps taken to collect the data, instrument scale measurement, explains the data analysis and statistical techniques and finally the ethical considerations of the research study are provided.

**Chapter 5: *Preliminary data analysis*** presents the results of the pilot study to ensure the validity and reliability of the measuring instruments to be used in testing the hypotheses and including other tests such as normality, linearity, multicollinearity and detecting outliers. This chapter will then present the preliminary data analysis of the data obtained from the respondents. The Statistical Package for the Social Science (SPSS) version 18.0 will be employed for preliminary data analysis like, data screening, frequencies and percentages, reliability analysis, explanatory factor analysis and t-test. The results from the data analysis in this chapter will focus on the cross-cultural differences between Lebanon and England and also investigate the different tasks that students perform using the e-learning systems.

**Chapter 6: *Model Testing*** presents the results of the model testing phase. This chapter presents an in-depth analysis of the relationships among the constructs within the proposed research model. A two-step approach will be used during the data analysis process. The first step will assess the constructs' validity and test the model fit. The next step will test the direct relationships among the independent and dependent variables for both models as well as the moderating impact of individual-level culture and other demographic characteristics.

**Chapter 7: *Discussion*** discusses and reflects upon the main findings presented in Chapter 5 and 6 through an in-depth interpretation of the demonstrated results and findings. This chapter will help in understanding the important role that behavioural beliefs, social and organisational, individual and cultural factors plays in affecting the student's beliefs towards adoption and acceptance of e-learning technology in Lebanon and England and will also discuss the similarities and dissimilarities between the two countries at the national level.

**Chapter 8:** ***Conclusion and Further Research*** is devoted to highlighting and discussing the major methodological, theoretical and practical implications drawn from the research study. This chapter also delineates and discusses the potential limitations and finally propose directions for future research.

# 1.10 Summary

This chapter presented the foundation for the research by covering and illustrating its background and purpose. Furthermore, this chapter covered the research aim and objectives, scope, contributions, significance and context of the study. Additionally, the research approach and methodology are also presented. Finally, an outline and brief description of the thesis was discussed.

The following chapter will discuss and review the most used technology acceptance theories and models, as well as, cultural theories in the IT/IS literature, which will form the basis of the proposed research model.

# Chapter 2: E-learning, Technology Acceptance & Cultural Models

*"We seldom realize, for example that our most private thoughts and emotions are not actually our own. For we think in terms of languages and images which we did not invent, but which were given to us by our society." (Alan Watts)*

## 2.1 Introduction

This chapter reviews the literature related to e-learning environments, technology acceptance theories and models as well as cultural models and theories. Specifically, the chapter will begin with a brief discussion of the current e-learning environments including their characteristics, limitations, advantages and the major factors that affect the acceptance of such technologies. In the second section, a critical review of the models and theories used in the IS research is presented. Finally, this chapter provides a review of the different cultural theories and models with an emphasis on Hofstede's cultural theory and the importance of measuring the culture at the individual level.

## 2.2 E-learning

E-Learning is the use of Information and Communication Technology (ICT) to deliver information for education where instructors and learners are separated by distance, time, or both in order to enhance the learner's learning experience and performance (Keller et al., 2007; Horton, 2011). Govindasamy (2002) defines e-learning as a set of instructions delivered via all electronic media such as the internet, intranets, and extranets. Thus, by eliminating the barriers of time and distance, individuals can now take charge of their own lifelong learning (Bouhnik and Marcus, 2006; Fletcher, 2005). E-learning environments reduce the cost of provision and therefore increase revenues for academic institutions (Ho and Dzeng, 2010).

The universities must decide during or before the implementation phase on the best approach to deliver education, such as online learning, face to face, or apply blended approach. For the purpose of this study, e-learning with a particular focus on higher education institutions applies to the use of web-based learning systems to support face-to-face education. According to Wagner et al. (2008), this approach is the most successful learning approach compared to solely online and only face to face contact.

Learning Management Systems (LMSs) refer to the web-based delivery applications or technologies that are adopted by universities and other higher education institutions to deliver courses' contents, provide distance learning and to manage the education process (Freire *et al.*, 2012). LMS creates a variety of ways to deliver instruction and provide electronic resources for student learning. Some methods, such as using Web pages to deliver text in much the same way as hard bound texts, are very familiar to academic staff. However, a big advantage is that the Internet also supports the delivery and use of multimedia elements, such as sound, video, and interactive hypermedia (McNeil *et al.*, 2000). Different Web-based learning systems have been developed for higher education to facilitate learning in a web-based learning setting; these include Moodle, Web Course Tools (WebCT), LAMS and SAKAI, Blackboard Learn (BBL). The later will be discussed in detail in the next subsection.

## 2.2.1  Blackboard learning system

Blackboard is considered one of the most popular web-based learning systems tools in higher education today as it provides a framework for course delivery in addition to its ease of use by learners (Iskander, 2008). According to Blackboard Inc. (2012) , it is defined as "the comprehensive technology platform for teaching and learning, community building, content management and sharing, and measuring learning outcomes and consists of integrated modules, with a core set of capabilities that work together." It is been used by more than 39,000 instructors at over 1,350 colleges and universities to deliver over 147,000 courses to more than 10 million student accounts in 80 countries. It integrates communication tools, including a bulletin board, chat room and private e-mail. In addition graphics, video, and audio files can be included into a Blackboard site. Blackboard also provides instructional tools to support course content such as a glossary, references, self-test, and quiz module. Students, too, can place assignments and other materials in Blackboard for courses in which they are enrolled. Furthermore, Blackboard also gives academic staff course management tools for grading, tracking student interaction, and monitoring class progress (Iskander, 2008). Such features can facilitate interaction between academic staff and students (Iskander, 2008). These tools are available only to the students and instructor of the course, the system requires a user name and a password in order to gain access (Tella, 2012), thus protecting the intellectual property of the instructor, the privacy of the student, and the course content from external parties.

## 2.2.2  Characteristics of Web-based learning

Web-based learning does not require extensive computer skills, although familiarity with computers and software (especially Web browsers) helps to reduce the acceptance barriers (Steven, 2001). Web-based learning generally fits into one of three major categories:

*Self-paced independent study*: Students determine the schedule and study at their own pace. They can review the material for as long as necessary. Feedback from online quizzes takes the form of pre-programmed responses. Unfortunately, there is no one to whom the student can direct questions. This form of study requires the most self-motivation (Tsang *et al.*, 2007).

*Asynchronous interactive*: The students participate with an instructor and other students, although not at the same time. They attend classes whenever they need or until the course material is completed. This approach offers support and feedback from the instructor and classmates. It is usually not as self-paced as independent study. It also allows time for considered responses and so critical thinking skills are enhanced (McCombs, 2011). This can improve in-depth investigation of a topic. Moreover, it can also provide social support and encouragements for individuals and increase the total effort put forth by group members (Benbunan-Fich *et al.*, 2005). This approach will shift the attention from the instructor-centered to learners-centered (McCombs, 2011). This will produces a more egalitarian, democratic environment in which the instructor becomes a guide for knowledge (McCombs, 2011).

*Synchronous learning:* Students attend live lectures via computer and ask questions by e-mail or in real-time live chat. This format is the most interactive of the three and feels the most like a traditional classroom. Flexibility is restricted by the previously determined lecture schedule. There are limited course offerings in this format due to high delivery costs (Weimer, 2013).

For the purpose of this study, e-learning with a particular focus on higher education institutions applies to the use of web-based learning systems to support face-to-face education.

## 2.2.3  Advantages of Web-based learning

Callan et al. (2010) and Garrison  (2011) identified many advantages for e-learning technologies including:

- Less expensive to deliver, affordable and saves time
- Flexibility in terms of availability- anytime anywhere. In other words, e-learning enables the student to access the materials from anywhere at any time.

- Access to global resources and materials that meet students' level of knowledge and interest.

- Self-pacing for slow or quick learners reduces stress and increases satisfaction and retention.

- E-learning allows more affective interaction between the learners and their instructors through the use of emails, discussion boards and chat room.

- Learners have the ability to track their progress.

- Learners can also learn through a variety of activities that apply to many different learning styles that learners have.

- It helps the learners develop knowledge of using the latest technologies and the Internet.

- The e-learning could improve the quality of teaching and learning as it supports the face-to-face teaching approaches.

## 2.2.4 Disadvantages of Web-based learning

While Web-based courses have advantages, it is equally important to note that there are disadvantages. These might include little or no "in-person" contact with the faculty member, feelings of isolations, a difficult learning curve in how to navigate within the system, problems with the technology, the need for the student to be actively involved in learning, and increased lead-time required for feedback regarding assignments (Holmes and Gardner, 2006). There are also different aspects, especially in the developing countries, such as providing the required funds to purchase new technology, lack of adequate e-learning strategies, training for staff members and most importantly the student resistance to use the e-learning systems (Wagner, 2008).

Bouhnik and Marcus (2006) stated that learners' dissatisfaction in using e-learning included the following:

- Lack of a firm framework to encourage students to learn.

- A high level of self-discipline or self-direct is required, learners with low motivation or bad study habits may fall behind.
- Absence of a learning atmosphere in e-learning systems.
- The distance-learning format minimizes the level of contact, e-learning lacks interpersonal and direct interaction among students and teachers.
- When compared to the face-to-face learning, the learning process is less efficient.

## 2.2.5 Factors Affecting E-learning acceptance

Despite the enormous growth of e-learning in education and its perceived benefits, the efficiency of such tools will not be fully utilized if the users inclined to not accept and use the system. Therefore, the successful implementation of e-learning tools depends on whether or not the students are willing to adopt and accept the technology. Thus, it has become imperative for practitioners and policy makers to understand the factors affecting the user acceptance of web-based learning systems in order to enhance the students' learning experience (Liaw and Huang, 2011). However, recent studies have shown that e-learning implementation is not simply a technological solution, but also a process of many different factors such as social factors (Schepers and Wetzels, 2007), and individual factors (Liaw, 2008), organizational such as facilitating conditions (Sun and Zhang, 2006) in addition to behavioural and cultural factors (Masoumi, 2010). Such major factors play an important role in how an information technology is developed and used (Kim and Moore, 2005; Jones and Jones, 2005).

Within the Information Systems literature, various theoretical models have been used (e.g., the theory of reasoned action, the theory of planned behaviour, innovation diffusion theory, unified theory of acceptance and use technology, the technology acceptance model) to investigate and explore the determinants of user's behaviour towards adoption and using information technology. The next section will provide a detailed explanation about nine of the most influential models in literature.

## 2.3   Technology Acceptance Theories and Models

This section will review nine of the most influential technology acceptance theories and models that have been used in the area of Information System and Information Technology research. Based on the critical review, a selection of different factors will be identified. These factors will provide the basis for the proposed conceptual model presented in the following chapter.

### 2.3.1  Diffusion of Innovation theory (DOI)

Diffusion of Innovation theory (DOI) (Rogers, 1995) is considered as a base to formalise technology acceptance behaviour on which other technology acceptance models can rely. The proposal of DOI was based on S-shaped diffusion curve theory proposed by Gabriel Tarde (1903) by which DOI succeeded to set clear definitions to 'diffusion', 'innovations' and the process of 'communications'. To begin with, the 'diffusion' can be considered as a procedure to enable a set of innovations to communicate for a certain time within different social systems. 'Innovations' as a term refers to any set of new ideas, concepts or applications that need to be recognised and shared by individuals.

The process of innovations' diffusion by which technologies can be adopted or rejected includes five stages: knowledge of an innovation, attitude toward the innovation, making an adoption decision, implementation of the new idea and decision confirmation. To begin with, any individual involved in the innovation adoption decision need to be aware of an innovation's existence and able to access all knowledge and information available about it. The next stage involves obtaining reliable information about the innovation from peers and users, to form decision border line. The decision to adopt or reject the innovation is thus ready to be made. In spite of having separate stage to make the decision, innovation rejection decision can be made at any stage of the innovation adoption process. If a decision to adopt the innovation is made, the innovation has to be practically used and tested. The innovation implementation stage involves assessing the innovation complexity, ease of use and learning. Such information will have a large effect on either confirming or rejecting the decision that has been already made.

The innovation diffusion process is also dependent on the innovations factors that are divided into five categories according to Rogers (1995) (see Figure 2.1). First, Relative Advantages (RA) is known as 'degree to which an innovation is perceived as being better than its precursor' (Moore and Benbasat 1991, p.195). Using RA relies on the importance of identifying the innovation efficiency and satisfaction. Second, Compatibility (COMP) is known as 'the degree to which an innovation is perceived as being consistent with the existing values, needs and past experiences of potential adopters' (Moore and Benbasat 1991, p.195). Considering COMP in the innovation adoption process is essential because the adoption decision might differ from one group to another, according to their beliefs and cultural backgrounds. Third, Complexity (COLX) is known as 'the degree to which an innovation is perceived as relatively difficult to understand and use' (Thompson et al., 1991, p.128). Since ease of use and learning has an impact factor on adopting an innovation or not, COLX cannot be ignored. Fourth, Trialability (TRI) measures how easy using the innovation in practice is to help making the right decision about adopting/rejecting the innovation (Rogers, 2003). Finally, Observability (OBS) relies on inspecting how the innovation under study has been efficiently used within an organisation (Moore and Benbasat, 1991). In other words, OBS determines how easily the innovation is understood.

Innovators have a great impact on the process of innovation diffusion. Innovators can be classified into five groups: Innovators, early adopters, early majority, late majority, laggards (Rogers, 2003). First, *innovators* (known as system gatekeepers) are capable of understanding, dealing and coping with high amount of innovations' ideas and information as well as handling innovation uncertainties. Second, *early adopters* (known as change agents) lead the innovation adoption by modelling the role of adopters. Third, *early majority* (known as deliberation) is going for an early adoption compared to the majority of system individuals. Fourth, opposite to early adopters, *late majority* is going for a late adoption compared to the majority of system individuals. Fifth, *laggards* is the last group of the adoption process.

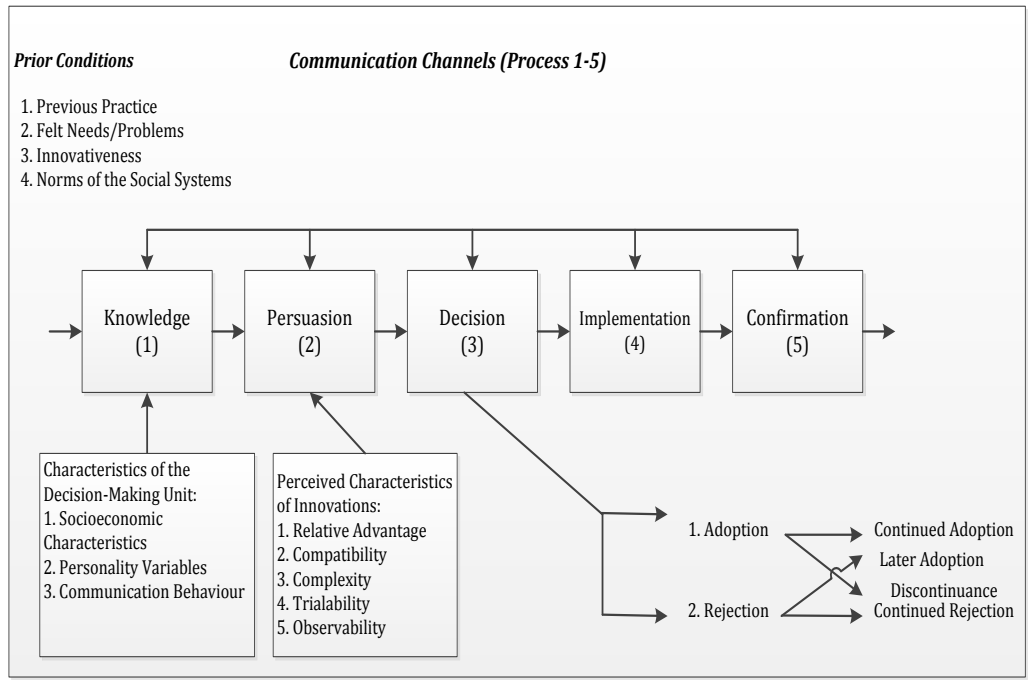

**Figure 2-1: Innovation-Decision Process (Rogers 1995)**

Several studies have relied on the DOI for the decision of the innovation adoptions (Venkatesh *et al.*, 2003). As a result, DOI itself has been under a continuous refinement and extension process. To begin with, DOI was extended by introducing two variables; *voluntariness* and *image* (Moore and Benbasat, 1991). Moreover, Agarwal and Prasad (1998) also extended the DOI by taking personal innovation into account. Introducing the concept of Personal Innovations of Information Technology (PIIT), they were able to distinguish between global and local innovations in order to minimise the risk the early adopters take. Furthermore, merging DOI with attitudinal theories has a great impact on highlighting the adopters' beliefs. The study findings revealed the impact of personal and social interest on the adoption process (Karahanna *et al.*, 1999).

In spite of all its advantages, DOI suffers from several limitations affecting its wide applicability. Some attitudes defined within DOI lack sound justifications. In other words, it is not known why certain attitudes lead to innovation adoption or rejection decision. Moreover, DOI fails to link between the innovation properties and a proper expected attitude (Karahanna *et al.*, 1999; Chen *et al.*, 2002). One solution to remedy such a problem is to propose theories taking into account the process of developing attitude such as Social Cognitive Theory (SCT).

## 2.3.2 Social Cognitive Theory (SCT)

Different from DOI, SCT considers actions and reactions of individuals (i.e., human agencies) in the process of innovation developments (Bandura, 1986). Individuals are different in their behaviours. In turn, different behaviour has different effects on the surrounding environment which affects the following behaviour. In other words, SCT relies on feedback, vicarious learning and identification. In terms of reciprocal determinism, SCT comprises three sets of factors: environmental factors, personal factors and behaviours as depicted in Figure 2.2 (Bandura, 1986). Considering these factors, the focus of SCT was successfully changed from social learning to social 'cognitive' in order to highlight the importance of cognition in people understandings and reactions.

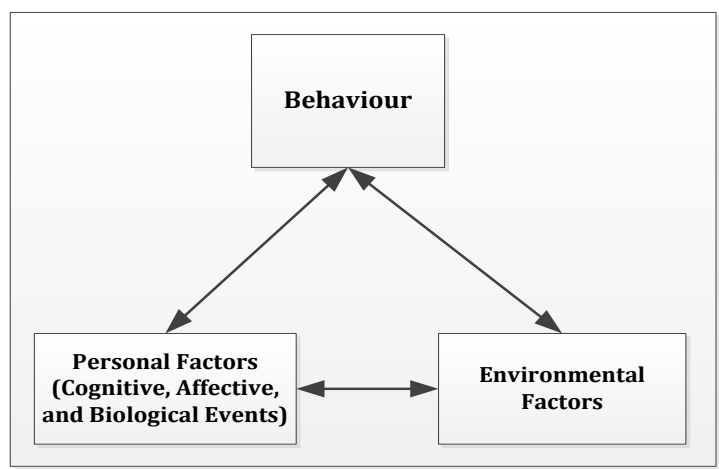

**Figure 2-2: Social Cognitive Theory (Bandura 1986)**

In order to show adoption behaviour through a thorough understanding of innovation attributes and network structures, Bandura (2006, p.119) stated that: 'Social cognitive theory distinguishes among three separable components in the social diffusion of innovation. The triadic model includes the determinants and mechanisms governing the acquisition of knowledge and skills concerning the innovation; adoption of that innovation in practice; and the social network by which innovations are promulgated and supported'.

### 2.3.3  Theory of Reasoned Action (TRA)

Forming the basis for other theories, the theory of reasoned action (TRA) focuses more on and gives insightful concepts about behaviour (Fishbein and Ajzen, 1975; Ajzen and Fishbein, 1980). TRA principles rely on a hypothesis that humans usually think about their action implications before making any decision or behaviour (Ajzen and Fishbein, 1980).

TRA comprises three main constructs as depicted in Figure 2.3: Behavioural intention (BI), Attitude (A) and Subjective norms (SN). Firstly, BI can be considered as an immediate precedent part of behaviour. BI is different from behaviour which is seen as a set of observable actions, BI forms the intension or plan towards that upcoming behaviour. In other words, BI will be totally dependent on an individual's approach in acting or reacting to certain behaviour (Fishbein and Ajzen, 1975; Ajzen and Fishbein, 1980; Sheppard *et al.*, 1988). Secondly, Attitude (A) can be considered as a valid representative to human actions and can be defined as 'individual's positive or negative evaluation of performing the behaviour' (Fishbein and Azjen, 1975, p.216). Previous experience has a paramount impact on A to be positive or negative. For instance, if the previous experience was not good, A will have a negative influence on BI. Finally, SN can be considered as 'the person's perception that most people who are important to him or her think he should or should not perform the behaviour in question' (Azjen and Fishbein, 1975, p.302). SN highlights the influence of an individual society in his/her thoughts which soon become a normative behaviour (Schepers and Wetzels, 2007). TRA can thus be explained based on its constructs as any individual intentional BI is based on the both SN and A.

TRA has received a wide research attention especially in IS area such as (Ajzen, 1985; Bagozzi, 1981). However, TRA suffers from several limitations. For instance, TRA made an assumption that any individual belief is reliant on his/her intention on how he/she behaves (Ajzen, 1985). In other words, TRA will not be able to predict individual behaviour if his/her intention to use it is not known in the first place.

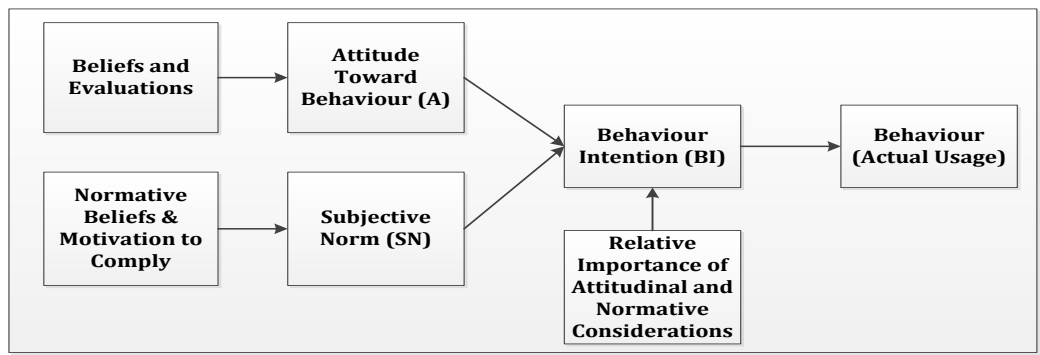

**Figure 2-3: Theory of Reasoned Action (TRA) (Ajzen and Fishbein 1980)**

## 2.3.4 Theory of Planned Behaviour (TPB)

Accroding to Ajzen (1988), the Theory of Planned Behaviour (TPB) was developed to overcome the inability of TRA in predicting behaviour in the case of which people have incomplete volitional controls. Sheppard et al., (1988, p.325) defined volitional control as 'behavioural intention will predict the performance of any voluntary act, unless intent changes prior to performance or unless the intention measure does not correspond to the behavioural criterion in terms of action, target, context, time-frame and/or specificity'.

Considering the volitional control effect, TPB thus extends TRA by adding the construct of Perceived Behaviour Control (PBC), to accommodate cases where individuals plan to show some behaviour without any success is due to the absence of confidence. In general, PBC can be defined as the 'perceived ease or difficulty of performing the behaviour' (Ajzen, 1991, p.188) or the 'perception of internal and external constraints on behaviour' (Taylor and Todd, 1995a, p.149), in the information system context.

The relations between TPB constructs are similar to those of TRA in which the BI is determined by Attitude (A), SN and PBC (Ajzen, 1991). For instance, if A affects BI positively and so does SN, PBC intends to be high and individual BI as to show behaviour is strong. Moreover, TPB also studies how these constructs can be affected by human beliefs which have been classified into behavioural beliefs, normative beliefs and control beliefs (Figure 2.4). To begin with, behavioural beliefs consider the beliefs about behaviour outcomes which affect the attitude toward that behaviour. Moreover, normative beliefs can be known as the consequences of other individuals' aspects on behaviour. They also present an expectation on what other individuals might believe in. Lastly, control beliefs correspond to any type of factor that might influence behaviour either in a supportive or unsupportive way (Ajzen, 2005).

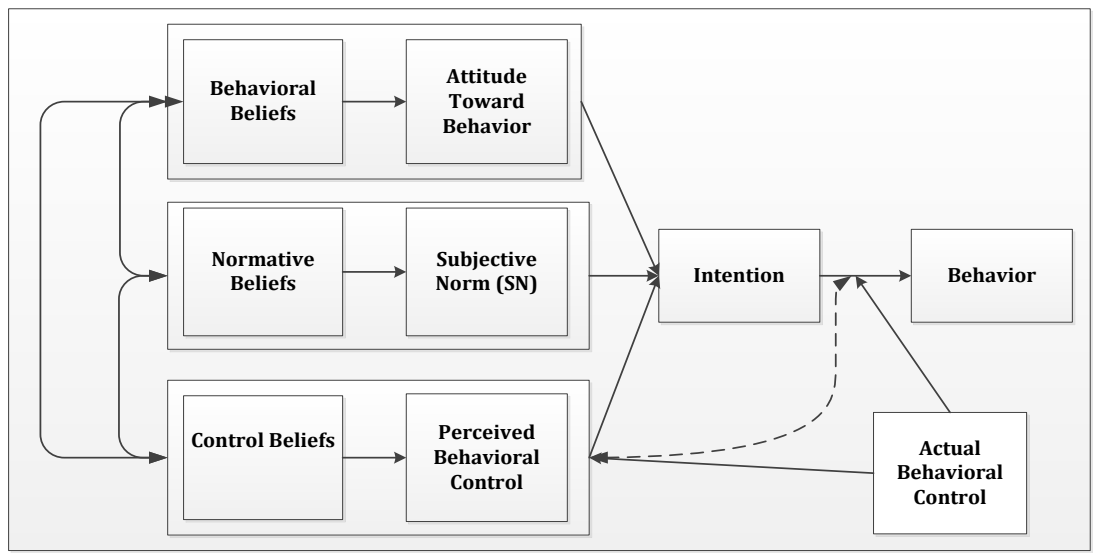

**Figure 2-4: Theory of Planned Behaviour (TPB) (Ajzen, 1991)**

TPB was also adopted in many research areas such as healthcare (Chau and Hu, 2002; Norman and Conner, 1996), and information system (Mathieson, 1991;

Taylor and Todd, 1995b; Yousafzai *et al.*, 2010) . Moreover, TPB has undergone several modifications to enhance its performance such as the work of (Armitage and Conner, 2001; Conner and Armitage, 1998; Elliott *et al.*, 2003). Most of them focus on a suitable modification of PBC construct to provide more generality to the TPB.

In spite of its generality, TPB also suffers from several limitations. For instance, there are more factors that might have an impact on behaviour but are not considered in TPB such as habit, moral obligation and self-identity (Eagly and Chaiken, 1993). Moreover, the exact prediction relation connecting BI with its correspondent behaviour has not been clearly specified yet (Foxall, 1997). The inability of TPB to move beliefs from context-specific to generalised form requires for measurement amendment every time the context or populations are different (Ajzen, 1991). Although PBC impacts BI in a sense that it predicts any individual who can perform behaviour if they think that they are able to do it, TPB fails to explain the mechanism by which the individual will perform that behaviour  (Taylor and Todd, 1995a). Finally, combining all hidden or unspecified factors affecting behaviour in PBC construct might affect the prediction accuracy (Taylor and Todd, 1995c).

## 2.3.5  Decomposed Theory of Planned Behaviour (DTPB)

Decomposed Theory of Planned Behaviour (DTPB) was proposed by Taylor and Todd (1995) to enhance the TPB model and solve the shortcomings of the operationalization of PBC by restructuring TPB in a decomposition model. Different from TPB which presented beliefs as one dimension, DTPB investigates and clarifies the multi-dimensional nature of beliefs by which the *attitudinal*, *normative* and *control beliefs* of TPB are decomposed, as depicted in Figure 2.5 (Taylor & Todd, 1995a).

First, attitudinal beliefs are known as an acquired individual mood toward a specific behaviour such as ease of use, efficiency or benefits (Fishbein and Ajzen, 1975). TRA and TPB ignored the embedded sub-beliefs when presenting attitudinal beliefs as one of their model constructs (Berger, 1993). To clarify the hidden beliefs, DTPB decomposes attitudinal belief into perceived usefulness (PU), perceived ease of use (PEOU) and compatibility. Such factors were found to be reliable with respect to IT usage (Taylor and Todd, 1995a).

Second, normative beliefs cover the opinions that have been affected by others' (Fishbein and Ajzen, 1975). The decomposition of normative beliefs into different groups highlights the difference between beliefs and its impact on the individual behaviour itself (Taylor and Todd, 1995a). For instance, to study the impact of peers on normative beliefs within DTPB, a case study was conducted among academic referents divided into three groups: peers (i.e., students), superiors (i.e., professors), and subordinates. The referents' opinions are only considered if they are different and should be ignored in the case of their similarity (Taylor and Todd, 1995a). Studying the possibilities for adopting internet services, a similar case study was conducted based on DPTP to test the decomposition of SN into two sub-categories *interpersonal* and *external beliefs.* Considering the source of influence, interpersonal beliefs represent the surrounding society effect on an individual belief where external beliefs denote the impact of non-human (i.e., machines) on an individual belief (Bhattacherjee, 2000). The experiment showed a minimum impact of external beliefs on adopting internet services (Hsu and Chiu, 2004).

Third, control beliefs consider the factors that have significant impact on encouraging (or discouraging) certain behaviour (Ajzen, 1985). In DTPB, control beliefs are decomposed into Self Efficacy (SE) and Facilitating Conditions (FC) (Taylor and Todd, 1995a). SE represented the internal beliefs of an individual that controls the decision process of performing a specific action (Bandura and McClelland, 1977). Achieving high SE would lead to a progressive influence on usage (Compeau and Higgins, 1995). On the other hand, FC is decomposed into *resource facilitating conditions* such as time and money and *technology facilitating conditions* that may put some restrictions on usage. The absence of FC may restrict usage and intention but the existence of FC may not encourage the usage (Taylor and Todd, 1995a).

In general DTPB adds more insights on belief factors because of its decomposing process. Moreover, DTPB also contributes to research in IT usage suggesting that different beliefs can affect managers targeting a specific IT usage for instance. However, the decomposition process of DTPB could lead to more complex models (than those of TRA and TPB for instance), which negatively contribute to its analysis.

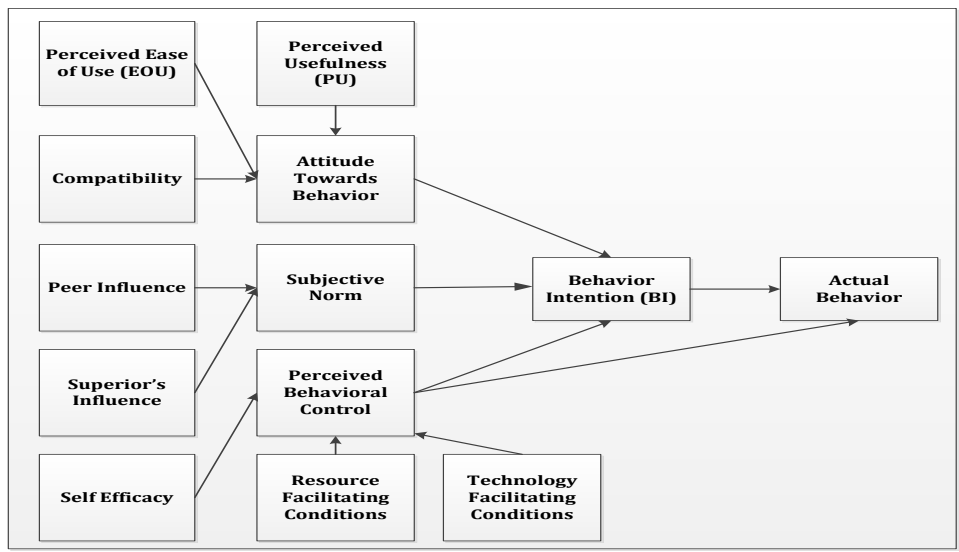

**Figure 2-5: Decomposed Theory of Planned Behaviour (DTPB) (Taylor & Todd, 1995a)**

## 2.3.6 Technology Acceptance Model (TAM)

Considering the Pros and Cons of TRA, Davis (1989) proposed a widely accepted theory for representing the technology acceptance behaviour in IT domain. This theory called Technology Acceptance Model (TAM) is considered as TRA variant in which A and BI are considered but SN is ignored due to uncertainty in theories behind it and empirical difficulties in applying it (Fishbein and Ajzen, 1975; Davis *et al.*, 1989). Moreover, different from TRA, TAM relies on Perceived Usefulness (PU) and Perceived Ease of Use (PEOU) for predicting an individual belief and attitude towards accepting certain technology. In TAM, PU tends to have a direct and an indirect impact on BI as depicted in Figure 2.6. In general, TAM comprises the following elements: Attitude (A), Behavioural Intention (BI), Perceived Usefulness (PU), Perceived Ease of Use (PEOU) and External variables.

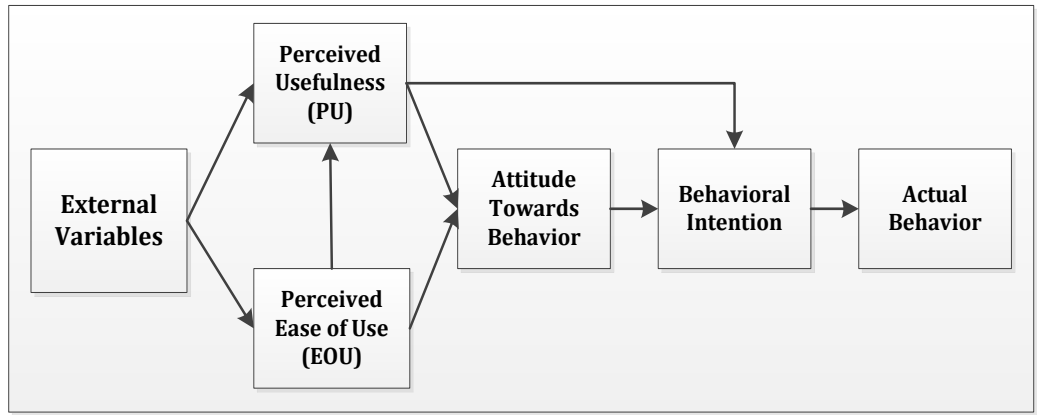

**Figure 2-6: Technology Acceptance Model (TAM) (Davis et al., 1989)**

First, A (as mentioned in TRA) considers human feelings towards acting (or not acting) a certain behaviour (Fishbein and Ajzen, 1975). Second, BI (as mentioned in TRA) measures the strength of intention to do certain behaviour. Third, PU presents the degree to which a person believes that using a particular system would enhance his/her job performance (Davis et al., 1989). PU was previously used in DTPB to evaluate the performance of technologies (Taylor and Todd, 1995a). Fourth, PEOU considers the users' expectation towards the ease of using certain technologies (Davis et al., 1989). Compared to Complexity (COLX) in DOI, which measures the difficulties in using technologies (Taylor & Todd, 1995a), PEOU can be considered as an opposite concept.

Finally, external variables represent 'the explicitly included factors in the model that has an expected impact on BI and BU through the mediation of PU and PEOU' (Davis et al., 1989, p.987). The external variables can comprise training, documentation, decision-making properties and system design. The external variables are not limited to certain ones. However, more variables can be introduced whenever they are necessary such as enjoyment, computing support, etc. (Koufaris, 2002; Moon and Kim, 2001; Agarwal and Karahanna, 2000).

In general, TAM can identify whether a technology is to be accepted by an individual according to three factors: (1) the functionality of the technology (Davis et al., 1989), (2) its ease of use (Davis et al., 1989) and (3) the benefits that might come out of using it (Dishaw and Strong, 1999). The ease of using a certain technology can influence the user's belief as a result of the increase in A and BI or the existence of external variables (Davis et al., 1989; Taylor and Todd 1995a).

TAM was developed via several stages. In its first version, the process of TAM comprises two steps. The first one includes collecting the PU and PEOU from technology's users. The next step includes assigning weights to each belief in order to anticipate behaviour strength (Davis et al., 1989). This process was not fully convincing especially in the case where positive behaviour exists. It could be better assessing the usability of technology not the technology itself by measuring the user's beliefs (Bagozzi, 1984). For instance, technologies might have a positive impact on people in spite of their unwillingness to use them. One solution to this contradiction can be provided by studying the impact of external variables on beliefs.

The process of weighting beliefs was not considered in the next version of TAM. As a result, the obtained results were mixed and misleading. For instance, users start giving contradicting evaluation outcomes for technologies under study. On the other hand, the omission of beliefs' weighting positively enables the differentiation between (A, BI, PU and PEOU), and the impact of external variables. The modification had a positive effect on TAM prediction as it appeared from the outcomes of the study conducted by Davis et al., (1989).

The ability of quick evolving can be considered as one of the most important factors that makes TAM popular. To clarify, TAM was developed to understand and explain an individual's intention toward accepting a technology on a voluntary-based (Davis et al., 1989). However, TAM has evolved and been used in different context including example, graphics (Adams et al., 1992; Karahanna et al., 1999), in e-government e.g. (Phang *et al.*, 2006; Walker and Johnson, 2008) and e-health e.g. (Lanseng and Andreassen, 2007).

TAM has been extensively studied and evaluated, Lee et al., (2003) classified TAM evolving phases as the following: (1) introduction, (2) validation, (3) extension and (4) elaboration. To begin with, the first phase (i.e., introduction) focused on the TAM elaboration process by applying it on different applications. The research during this phase also ran several comparison studies with previous theories such as TRA (Taylor and Todds, 1995a). The second phase (i.e., validation) included the studies interested in examining the concept of TAM within different contexts (Chin and Todd, 1995; Davis *et al.*, 1989). The third phase (i.e., extension) comprised the studies focusing on extending TAM with new constructs in order to enhance its performance (Agarwal and Prasad, 1999; Venkatesh *et al.*, 2003). Finally, the elaboration phase focused on introducing newer versions of TAM such as the one that includes external variables (Venkatesh and Davis, 2000).

The large number of studies applying TAM enabled the detection of TAM limitations. To begin with, self-reported usage (Davis, 1993) can be considered as the most frequent reported limitation due to the bias it adds to TAM process (Agarwal and Karahanna, 2000). Moreover, the lack of real full measures that prove TAM validity can be considered as another limitation. In other words, the existing measures were only based on beliefs, attitude and intentions not on usage behaviour (Taylor and Todd, 1995a; Mathieson, 1991). Furthermore, TAM fails to explain the reasons of some results. The explanatory power of TAM is still questionable as it ignores the effect of social, individual and cultural influence on the acceptance of technology (Straub *et al.*, 1997; Teo *et al.*, 2008; Straub and Burton-Jones, 2007; Bagozzi, 2007; Benbasat and Barki, 2007). More specifically, nowadays, the focus of the majority of studies has been within the context of developed countries such as North America. Finally, TAM also suffers from the weakness in illuminating the design process behind the acceptance behaviour (Venkatesh and Davis, 1996; Venkatesh and Davis, 2000). In other words, no feedback is available for improvement concepts (Taylor and Todd, 1995a; Venkatesh et al., 2003).

### 2.3.7 Revised Technology Acceptance Model (TAM 2)

The continuous evolution of TAM leads to the inclusion of other additional constructs related to the social influence (Venkatesh & Davis, 2000). In details, TAM 2 mainly added two main factors: (1) social influence processing factors and (2) cognitive instrumental processing factors, as depicted in Figure 2.7 (Venkatesh and Davis, 2000).

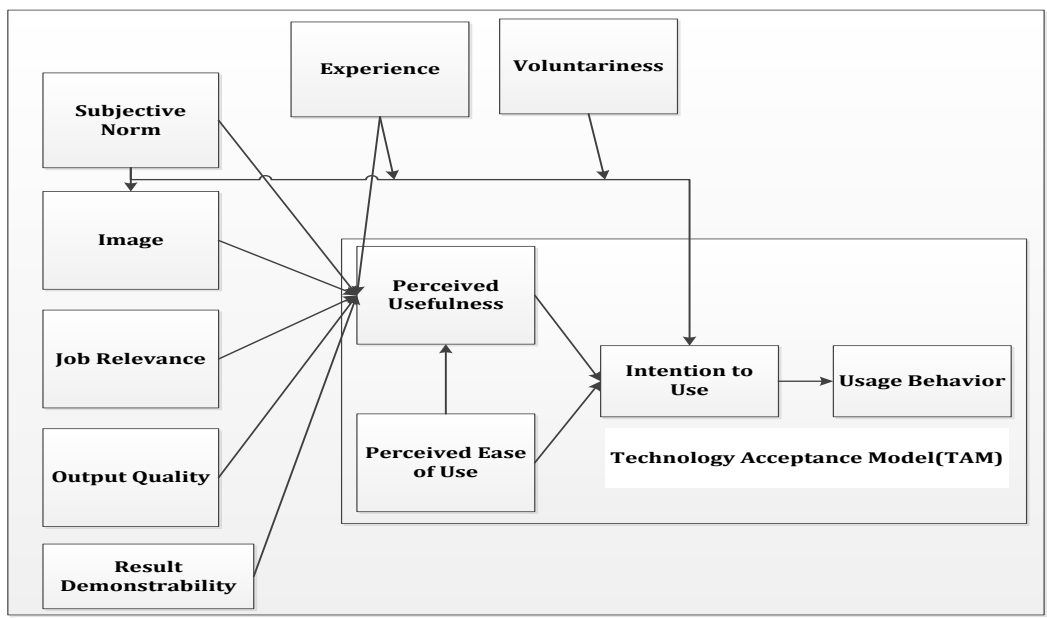

**Figure 2-7: TAM2 (Venkatesh & Davis 2000)**

The *social influence processing* factors include any factors which have a direct impact on innovation adoption decision. Several variables can be categorised under this group: Subjective Norms (SN), voluntariness (VOL), experience (EXP) and image (IMG).

First, SN represents the implication of other people on an individual's adoption decision (Fishbein and Ajzen, 1975). Second, IMG which extends DOI can be known as 'the degree to which use of an innovation is perceived to enhance one's ... status in one's social system' (Moore and Banbasat, 1991, p.195). IMG might be influenced by SN since combining the intension of doing certain behaviour with actions increases the individual standing within the group (Venkatesh and Davis, 2000). Third, VOL is defined as an 'extent to which potential adopters perceive the adoption decision to be non-mandatory' (Venkatesh & Davis, 2000, p.188). VOL tends to decrease SN's impact compared with mandatory settings (Barki and Hartwick, 1994; Moore and Benbasat, 1991). Finally, EXP can point to the degree of familiarity towards a particular system (Venkatesh and Davis, 2000). The larger EXP, the lower the SN influence on PU and BI (e.g., Harwick & Barki, 1994; Agrawal & Prasad, 1997).

On the other hand, *cognitive process* includes factors related to decision process between a system capability and system usefulness in terms of the intended job such as Job Relevance (JR), Output Quality (OQ), Result Demonstrability (RD) and PEOU. First, similar to COMP factor in DOI, JR can be defined as an individual's perception regarding the degree to which the target system is applicable to his or her job (Venkatesh & Davis, 2000, p.191). Second, OQ considers the performance of a system. To clarify, OQ assess whether a system can achieve its designed objectives or not. The higher the OQ, the higher the effect is on PU (Venkatesh & Davis, 2000). Third, RD can be defined as 'tangibility of the results of using innovations' (Moore and Banbasat, 1991, p.203).

To validate the model of TAM2, different studies were conducted within different domains such as examining Website usage (Wu *et al.*, 2008), e-learning (Van Raaij and Schepers, 2008). The results showed that the capability of TAM2 in explaining variance was considerably better than TAM. The significant impact of SN on BI in TAM2 was also noted (Venkatesh and Davis, 2000).

TAM2 was proposed to overcome TAM limitations. However, TAM2 still suffers from some inherited TAM limitations. To begin with, there might still be a bias in results especially after considering self-reported usage as a valid measure.

Moreover, TAM2 presumed that any action needs to be made on an absolutely voluntarily basis; that would not be possible in practice for all factors such as limited skill, limited time, environmental limits and unconscious habits (Wilkins *et al.*, 2009). The ability of TAM2 to explain the impact of external variables only on PU can be considered another limitation.

## 2.3.8 Augmented version of TAM (A-TAM)

The limitations of TAM to measure the social and other control factors directs Taylor and Todd (1995b) to propose the augmented version of TAM (A-TAM) as a hybrid model of TAM and TPB. Figure 2.8 depicts the A-TAM constructs; PU and PEOU, SN, PBC, BI and BU.

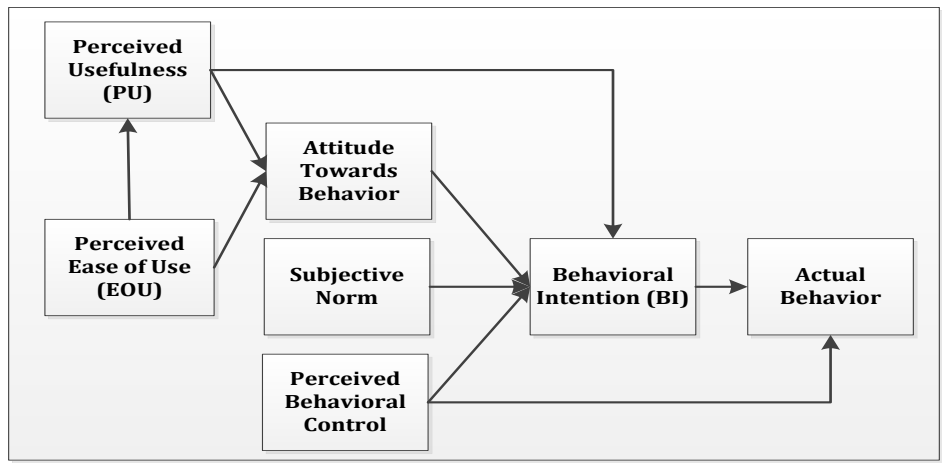

**Figure 2-8: Augmented TAM (Taylor and Todd 1995)**

A-TAM gives a higher priority to the social factors (e.g., SN and PCB) especially in IT (Compeau and Higgins, 1991; Mathieson, 1991; Moore and Benbasat, 1991). Considering new and experienced users within the model can be considered as another reason for adopting A-TAM as shown in the Taylor and Todd (1995) study conducted among experienced and inexperienced students. The result of the study confirms the capability of A-TAM in predicting the usage behaviour according to the users' experience level.

On the other hand, A-TAM also poses some limitations. First, the A-TAM model misses other important variables (e.g., age and gender) that are related to the user experience. Second, the study results could not be generalised as it was conducted with homogenous technology.

## 2.3.9 Unified Theory of Acceptance and Use of Technology (UTAUT)

The Unified Theory of Acceptance and Use of Technology (UTAUT) were developed by Venkatesh, Morris, Davis and Davis (2003) in order to fill the gap of other models. UTAUT consists of four main constructs namely (see Figure 2.9); 1) performance expectancy (PE) which is similar to PU in TAM, effort expectancy (EE) which is similar to PEOU in TAM, social influence (SI) which is similar to SN in TRA, and facilitating conditions (FC) which is similar to PBC in TPB/DTPB. Those factors may affect the BI and usage of technology. UTAUT also included a set of four moderators (age, gender, voluntariness of use and experience) that may influence the relationships between the main determinants and behavioural intention and usage (Venkatesh et al. 2003).

While the UTAUT was received little attention compared to other models, it has been used in different context such as health/hospital IT e.g. (Kijsanayotin et al., 2009), online-banking e.g. (Abu-Shanab and Pearson, 2009; YenYuen and Yeow, 2009), and E-government e.g. (Carter and Weerakkody, 2008).

Although the UTAUT proved to be a good model for measuring the intention and usage behaviour, however there are some criticisms concerning its parsimony and explanatory power e.g. (Williams *et al.*, 2011). A detailed comparison of UTAUT with other models will be discussed in the next section.

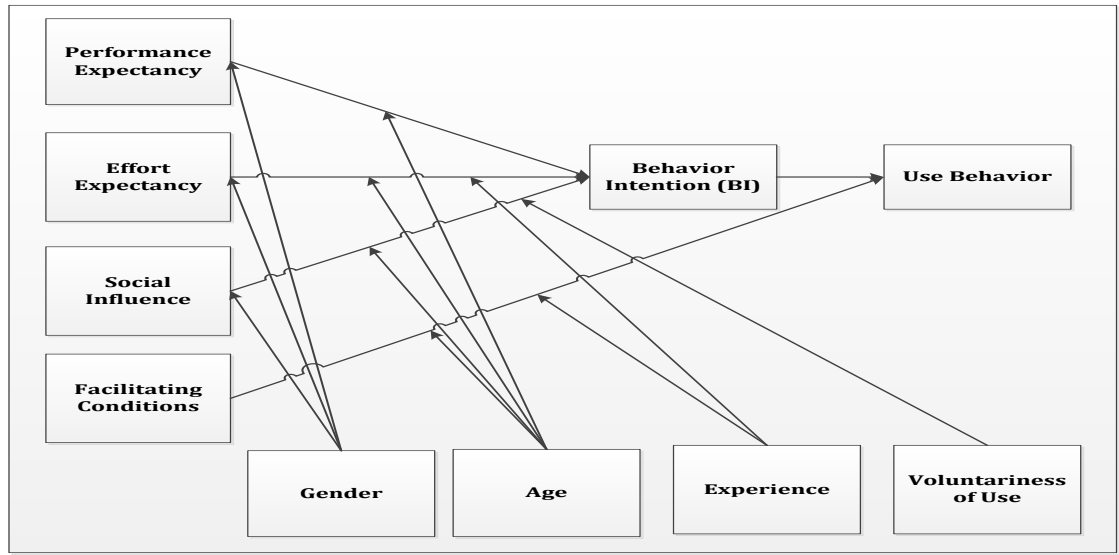

**Figure 2-9: Unified Theory of Acceptance and Use of Technology (UTAUT) (Venkatesh et al. 2003)**

## 2.3.10 Comparison of Models

The existence of different user acceptance models in the literature gives a sort of flexibility. On the other hand, choosing the best model that can fulfil the case study requirements is considered a difficult decision to make. As a result, several comparative studies to highlight the pros and cons of each user acceptance model have been conducted. The following subsections highlight some of the comparison studies.

### 2.3.10.1 TAM and TRA

Different studies compared the performance of TAM and TRA such as that of Davis, Bagozzi and Warshaw (1989). In this study, the convergence of TAM and TRA led to have stronger structure which is based on behaviour intention (BI), perceived usefulness (PU) and perceived ease of use (PEOU). In spite of its importance as a determinant of behavioural intention, Social Norms (SN) does not have that much attention in the study. SN can be considered the main construct that differentiates TRA from TAM. Where TAM does not consider SN as constructer influencing BI, TRA has theorised SN to be a main construct. However, the importance of SN is still controversial. Due to its poor scaling, Davis, Bagozzi and Warshaw (1989) discussed that SN may not affect BI especially when SN scales up and the systems usage is personal on a voluntary basis.

In summary, comparing TAM with TRA highlights the simplicity and applicability of TAM over TRA with a variety of applications in spite of the possibility of losing rich information because of not considering SN (Han 2003).

## 2.3.10.2 TAM, TPB and DTPB

Mathieson (1991) highlighted several points in the comparison between TAM and TPB. First, user intentions were successfully represented by both TAM and TPB. Second, TPB might be more useful than TAM during the system development cycle and post-implementation evaluation. This is due to the capability of TPB to provide specific information, reasons and justifications on the users' intention to use a technology. Third, compared with TPB, TAM was easier to follow and much quicker in collecting information about users' perception of a technology.

Another study of Taylor and Todd (1995b) compared TAM with TPB and a decomposed version of TPB (DTPB). With regards to the usage of IT, DTPB outperformed TAM due to several factors. First, similar to TAM, DTPB can classify the beliefs that may affect the usage of IT such as perceived usefulness and perceived ease of use. Second, different from TAM, DTPB includes additional influencing factors on behaviour (i.e., Normative beliefs, self-efficacy, and facilitating conditions) (Ajzen 1991). Considering normative beliefs highlights the significance of the user's communications and participations to measure the IT usage. Self-efficiency emphasises the importance of training which influences the user acceptance. Facilitating conditions would provide an informative guidance during the implementation phase. Including such factors increases the capability of DTPB to integrate the studies of IT usage with the influence of organisation decisions on it.

Another study was conducted by Chau and Hu (2001)to compare TAM, TPB and DTPB. This study used those three models to understand a physician's usage of telemedicine technology. TAM, TPB and DTPB were capable to illustrate 40%, 32% and 42% of the physicians' acceptance respectively.

In summary, TAM, TPB and DTPB can be considered useful in finding the IT usage behaviour and users' intention towards the usage. However, TPB and DTPB outperformed TAM by providing more explanatory power when users' intension behaviour is included. To clarify, TAM is capable of predicting IT usage behaviour but DTPB gives more insights on intension behaviour by taking normative and control beliefs into account. As a result, DTPB can supervise the implementation process of an IT system (Taylor & Todd 1995b).

### 2.3.10.3 UTAUT and Other Theories

According to Bagozzi (1992), the model which have good parsimonious, in other words the one which can have best predictive power as well as fewest constructs. On the other hand, Venkatesh et al. (2003) mentioned that detailed understanding of the phenomena under investigation is more important than parsimony. While Taylor and Todd (1995b) argued that the balance between parsimony and their contribution to understanding should be thought of when evaluating the models.

In this regards, Venkatesh et al. (2003) compared eight models based on their percentage of explained variance. They indicate that the variance explained in user intention for these 8 models varied between 17% and 53%. However, after including the core determinates as well as the four moderators in the UTAUT. They found that the variance explained 69% of behavioural intention and usage behaviour. With this rationale, this study will adopt some of UTAUT's constructs in addition to the four moderators.

## 2.4   Culture

In this section relevant literature related to the culture is reviewed. First a definition of culture is presented. This is followed by a brief explanation of the different cultural models including their scope and criticisms. We then proceed with a brief explanation about Hofstede's cultural dimensions. Finally, we provide a justification for measuring the cultural dimensions at the individual level.

### 2.4.1  Definition of Culture

The study of culture has been always important to explain behaviour and differences within national, organizational and individual culture. However, culture is one of the most difficult and complex terms to define (Williams, 1985) and it is also, therefore, difficult to operationalise and measure. Traditionally, the topic of culture had been addressed by anthropologists to describe a group of people who have common aspects of life.

Cultural models usually compare the similarities and differences of two or more cultures by using cultural variables (Hoft, 1996). This research follows the definition and broad characterisation of culture provided by Hofstede (1991) since it is generally well accepted within the IS literature and provides a clear framework for defining cultural dichotomies. Hofstede (1991) describes culture as the software of the mind which distinguishes the member of one group or category of people from another. He mentioned that culture is learned and not inherited, and people usually acquire patterns of thinking, feeling and actions which remain with them until later stages of their lives. He also argues that culture is affected by the social environment where people are interacting. Hofstede (1991) argues that a person may carry many layers of culture and this is highly dependent on his/her situational states. The different layers of culture include: "(1) national or country level; (2) a regional and/or ethnic and/or religious and/or linguistic affiliation level; (3) a gender level; (4) a social class level; (5) an organizational level; (6) an individual level" (Hofstede, 1991). This research aims to provide an understanding on how individual-level of culture might affect students' behaviour when using Web-based learning tools in developing and developed countries. The cultural models and variables in addition to the reasons behind choosing the individual level culture are described in more detail in the next sections.

## 2.4.2  Cultural Models

In general, cultural models usually compare the similarities and differences of two or more cultures by using cultural variables in which there are categories that organise cultural data (Hoft, 1996). There are many models that study the culture and each of those models uses its own scope and variables to identify culture characteristics. The following section will discuss some of most important cultural models.

## 2.4.3   Schwartz's Cultural Model

Schwartz (1994) is another cultural framework to calculate, compare and study the cultural differences between nations or sub-groups. These cultural variables were derived from both individual level values and cultural level values. In his framework, Schwartz suggested seven cultural domains based on universal human value types where were defined and labelled by Schwartz (1994):

1. *Conservatism*: in such societies, people emphasize close relations among other group's members, and usually avoids actions that disturb traditional order.

2. *Intellectual autonomy*: in such kind of societies, individuals are recognized as autonomous entities that are entitled to follow their own intellectual interests and desires.

3. *Affective autonomy*: Again in such societies, individuals are recognized as autonomous entities that are entitled to follow their stimulation and hedonism interests and desires.

4. *Hierarchy*: A society that emphasizes the legitimacy of hierarchical roles and resource allocation.

5. *Mastery*: in such kind of societies, individuals can get ahead of other people by emphasizing active mastery of the social environment.

6. *Egalitarian Commitment*: A society that emphasizes the transcendence of selfless interests.

7. *Harmony*: A society that emphasizes harmony with nature.

Schwartz's (1999) framework originally focussed on teachers and students with 35000 respondents, its cultural values or variables were derived theoretically with 57 questions within 49 different nations that included 41 cultural groups, collected between 1988 and 1992. He concluded that the seven value types "efficiently captures the relations among national cultures" (Schwartz, 1999, p. 38).

## 2.4.4  Hall's Cultural Model

For Hall (1973), culture is not something to simply study, but a "critical site of social action and intervention, where power relations are both established and potentially unsettled." (p.13). In other words, culture is a program of behaviour. Hall developed a method for the analysis of culture through defining the basic units of culture. These units or variables are *Context*, *Space* and *Time*. As for *Time*, Hall differentiates between *Monochronic* and *Polychronic* time orientation; he argues for example that people with *Monochronic* time orientation deal with time in a sequential way, while *Polychronic*-oriented people deal with time in simultaneous way. As for *Space*, this was another area that reveals important cultural differences among societies. As an example, Hoft (1996) cited that Latin countries such as Spain or Italy and Arabs seem to prefer a closer relation and are close to half the body distance one would find acceptable in the U.K or United States. As for *Context*, it refers to the amount of information that a person can comfortably manage, *context* can vary from a high context culture where background information is implicit to low context culture where much of the background information must be made explicit in an interaction, for example by the use of language. In other words, in high-context cultures which value relationships and information more than schedules, the information tends to be very fast and free, while in low-context cultures, where people usually follow procedures, the information flow tends to be slow (Hoft, 1996).

## 2.4.5 Trompenaars' Cultural Model

Trompenaars (1993) defines culture as the way in which a group of people solves problems. He identifies his cultural model using a three model layers; he mentioned that "Culture comes in layers, like an onion. To understand it, you have to unpeel it layer by layer". The first layer is *Outer Layer* which is made up of explicit, external, and observable products and behaviors and contains all aspects of life. The second layer is *Middle Layer* and it reflects deeper layers of culture and concerns with the norms and values of an individual groups. Trompenaars (1993) defined norms as "the mutual sense of a group of what is 'right' and 'wrong.'", and so it gives us a feeling of 'this is how I normally should behave,' While values determine the definition of 'good and bad', and closely related to ideals shared by a group, so values give us a feeling of 'this is how I aspire or desire to behave.'. The third layer is the *Core*, which is the assumption about existence; it can be defined as "Groups of people organize themselves in such a way that they increase the effectiveness of their problem-solving processes." (Trompenaars,1993 p.23). Hoft (1996) reported that Trompenaars collected his data from a multinational survey that contains 16 questions across 30 companies and 50 countries by which the respondents were 15,000 managers.

The model includes seven dimensions and these dimensions are useful in understanding different interactions between people from different national cultures. One of the main variables of culture in this model is *Universalism* versus *particularism*, by which the *Universalism* defines the morality and ethics and it is the belief that ideas and practices can be applied everywhere in the world without modification. While, *Particularism* is the belief that circumstances can influence how ideas and practices should be applied and some things cannot be done the same way everywhere, people tend to focus on relationships, working things out to suit those involved, for example Chinese people tend to find the solutions to the problem on the relationship with each other's. Another main variable is *Neutral* versus *Emotional*, where *Neutral* culture refers to culture where people try not to show their feelings to others. While, *Emotional* culture refers to a culture in which emotions are expressed openly and naturally, for example, people smile, may talk loudly, greet each other with enthusiasm, show happiness or unhappiness. *Achievement* versus *Ascription* is another variable in Trompenaar's model, while in *Achievement* culture people are accorded status based on how well they perform their work and what they have accomplished, while in Ascription culture, people are attributed based on who or what a person is, for example, status may be accorded on the basis of age, gender, family and ethnic group. The last four variables are *Sequential* versus *Synchronic*, *Internal* versus *External* control, *Specific* versus *Diffuse* and *Individualism* versus *collectivism*.

## 2.4.6  Hofstede's cultural model

During the late 1960s and early 1970s, while working for IBM, Hofstede initiated a survey of the company's employees across the world. The original aim of the work was not specifically to study cross-cultural differences. However, in analysing the results Hofstede began to identify systematic variations in the way certain groups of questions were answered in different parts of the world. Subsequent follow-up work focused more closely on identifying those questions which showed variation between countries. As a result of his research Hofstede identified four[1] dimensions of national culture difference:

---

[1] In his later work Hofstede also included a fifth dimension, long term vs. short term orientation, which was identified through collaborative research between Hofstede and Michael Bond Hofstede, G. & Bond, M. H. (1988). The Confucius connection: From cultural roots to economic growth. *Organizational dynamics* **16,** 4-21.. The respective poles of this dimension contrast an orientation towards the future with an orientation towards the present and past. This dimension is not considered within the work reported here.

- Power distance – this refers to the extent to which individuals expect and accept differences in power between different people, in other words it reflects the attitudes to authority and power.

- Individualism-collectivism – this refers to the extent to which individuals are integrated into groups. In other words, *Individualism* is defined as a situation in which people are supposed to look after themselves and their immediate family only. While in contrast, Collectivism can be defined as a situation where people who belong to the same group should look after each other's for loyalty. This can be considered as reflecting attitudes to group membership.

- Masculinity-femininity – this refers to the extent to which traditional gender roles are differentiated. In general, *Masculinity* refers to a situation where the dominant values in society are success, money and other things, while in contrast, *Femininity* refers to a situation which is a preference for relationships, caring for the weak, and quality of life

- Uncertainty avoidance – this refers to the extent to which ambiguities and uncertainties are tolerated.

Hofstede's work has been very widely cited (for example Hofstede is among the 100 most cited authors in the Social Science Citation Index; Hofstede and Hofstede, 2001). However, while Hofstede's work has been highly influential in a number of disciplines, it has also been heavily criticized. Some challenge Hofstede's ideas on the fundamental issue of whether culture is a construct that can be dimensionalized and measured empirically. Criticisms from within the positivist tradition focus on the specifics of Hofstede's data gathering methodology and on the statistical properties of his measurement scales. For example, Schwartz (1999) argues that the survey as a data collection is not the appropriate instrument to measure cultural differences especially when the variable being measured is a value which culturally sensitive and subjective. Hofstede's defence of his work in such cases is typically robust e.g see (Hofstede, 2002). Another criticism is the number of years that has passed since Hofstede's work took place between 1968 and 1972, some researchers questions the validity of such model in the evaluation of technologies as what may be applicable at that time may not be applicable in the present (McCoy, 2003). Problems in the literature are also argued to have arisen from the inappropriate use of Hofstede's scales to measure cultural variables at the individual level rather than at the country (group) level (Eckhardt and Houston, 2002).

Despite the criticism on Hofstede's model, Hofstede's research is considered one of the most widely used pieces of research among scholars and social science research (Rose and Straub, 1998; Søndergaard, 1994; Nakata, 2009). In addition, Hofstede's model was cited in more than 9000 articles in peer-reviewed journals between 1981 and 2011 which reinforce the value of his work. Furthermore, Hofstede's research framework was based on systematic data collection and coherent theory and it was supported by many empirical studies; and this is what the scholars and researchers had been asking for (Søndergaard, 1994). Nevertheless, in his book *Culture's Consequences: Comparing Values, behaviours, institutions, and organizations across nations*, Hofstede (2001) noted that most of the research that has validated his framework reported that these variables are reliable and valid.

Despite finding many examples of the use of Hofstede's framework in the Information Systems literature, Ford et al. (2003) found relatively little theory development work in this area. They therefore make several recommendations for developing a theoretical basis for the integration between IS and culture. They recommend that researchers should have a theoretical basis for including Hofstede's dimensions in any study. Ford et al (2003) also suggest some ways in which Hofstede's dimensions could be incorporated into the design of research. One suggestion is to hypothesize a moderating role for the dimensions, particularly in relation to existing theories used within IS such as the technology acceptance model (Davis, 1989). Our research approach adheres to these suggestions.

## 2.4.7 Cultural Variables

As previously mentioned, there are many models that study the culture and each of those models uses its own scope and variables to identify culture characteristics. A summary of these cultural variables or dimensions and its definitions will be summarized in Table 2.10. This section helps in identifying the variables that may affect the individual's cognitive behaviour.

| Cultural Variables | Researcher | Interpretation |
|---|---|---|
| Power Distance (high vs. low power distance) | Hofstede | The extent to which the less powerful members of institutions and organisations within a country expect and accept that power is distributed unequally.<br>High Power Distance / Low Power Distance<br>- centralized power / - decentralized power<br>- tall hierarchies / - flat hierarchies<br>- superior/subordinates unequal / - equal |
| Uncertainty Avoidance (high vs. Low uncertainty avoidance) | Hofstede | The extent to which the members of a culture feel threatened by uncertain or unknown situations.<br>High Uncertainty Avoidance / Low Uncertainty Avoidance<br>- emotions to be shown / - emotions not to be shown<br>- expressive people / - quiet/ controlled people<br>- what is different is dangerous / - what is different is curious |
| Individualism vs. collectivism | Hofstede | The extent to which individuals are integrated within groups.<br>Individualism / Collectivism<br>- right to privacy / - group invade private life<br>- individual decisions / - group decisions<br>- laws and rights same for all / - laws and rights per group<br>- everyone looks after himself / - group protect individuals |
| Masculinity vs. Femininity | Hofstede | The extent to which roles of women versus men are different in the society.<br>Masculinity / Femininity<br>- focus on work goals / - focus on personal goals<br>- assertiveness/ competitive / - modesty<br>- concern for material success / - concern for quality of life |
| Confucian Dynamism (long-term vs. Short term) | Hofstede | The extent to which long-term and short-terms gratification of needs is traded-off.<br>Short Term / Long Term<br>- traditions respected / - traditions modernized<br>- unlimited social obligations / - limited social obligations<br>- quick results expected / - persistence for slow results<br>- concern with 'face' / - concern with purpose |
| Universalism vs. Particularism | Trompenaars | The extent to which, in a problem, people base their solution on rules versus relationship with others.<br>Particularist / Universalist<br>- relationship based / - rule based<br>- break rules if necessary / - strictly apply rules |
| Specific vs. Diffuse | Trompenaars | The extent to which public and private life and public and private personal spaces are separated.<br>Diffuse / Specific<br>- public/ private life diffused / - public/ private life separated<br>- personal nature for business / - business/ friendship separate<br>- sympathy reactions / - judgementally reactions |
| Achievement vs. Ascription | Trompenaars | The extent to which achieving versus being values are stressed.<br>Being culture / Doing culture |

| | | - emphasis social relations<br>- emotional oriented<br>- words for social effect | - emphasis accomplishments<br>- activity oriented<br>- words match actions |
|---|---|---|---|
| Low-context vs. High-context | Hall | The extent to which meaning is found in the context versus in the code.<br>High Context<br>- meaning in context<br>- implicit<br>- direct and obvious | Low Context<br>- meaning in message<br>- explicit<br>- indirect and non-obvious |
| Time Perception (Polychronic vs. Monochronic time perception) | Hall | The extent to which time variable is perceived.<br>Polychronic<br>- many things at once<br>- simultaneous/ concurrent<br>- interruption accepted<br>- time duty objective<br>- committed human relations<br>- change plans easily<br>- life time relationship | Monochronic<br>- one thing at a time<br>- sequential/ linear<br>- interruption refused<br>- time duty critical<br>- committed to the job<br>- strict to plans<br>- short term relationship |
| Hierarchy vs. Egalitarian | Schwartz | Degree to which people in a country believe in freedom and equality and a concern for others (Egalitarian), vs. emphasis the legitimacy of fixed roles and resources (Hierarchy) | |
| Harmony vs. Mastery | Schwartz | Degree to which people in a country concerned with overcoming obstacles in the social environment (Mastery) vs. concern beliefs about unity with nature and fitting harmoniously into the environment. | |
| Conservatism vs. Affective/intellectual autonomy | Schwartz | Degree to which people in a country emphasis maintenance of status quo (Conservatism), or emphasis creativity or affective autonomy emphasis the desire for pleasure and an exciting life. | |

**Table 2-1: Culture Dimensions (Adopted from Al Said, 2005 and Ali and Alshawi, 2005)**

## 2.4.8 Cultural dimensions at the country vs. individual level

Although few would disagree that cultural factors are important in theory, there is surprisingly little published literature concerning the effect of national cultural aspects of online learning and teaching (Elenurm, 2008; Ya-Wen Teng, 2009; Hannon and D'Netto, 2007; Sánchez-Franco *et al.*, 2009). Most of the literature about cultural effects in IS research is based on the national or organizational level. A typical approach is to use nationality as a surrogate for culture, comparing similar samples of participants from two or more countries and attributing any differences to the assumed cultural dichotomies between the respective countries. This approach is problematic for several reasons. First, researchers often rely on historical findings regarding the cultural characteristics of particular countries or regions (dating from Hofstede's original findings). Research by McCoy et al (2005a) suggests that shifts may have occurred over the last 30 years and that assumptions based on Hofstede's work may therefore no longer be valid. This finding argues for the importance of directly measuring participant cultural values within any new research study. A second problem is that as there are several cultural dichotomies within Hofstede's model and these will covary between different countries. It can therefore be difficult to infer which cultural factor is responsible for differences between samples from different countries (for example if two countries vary in both uncertainty avoidance and collectivism, it may be unclear which of these variables might be having an influence in any observed results). As a result, it can be a challenge to find samples that usefully isolate the cultural variables that may be of interest. A third problem is that within the same country, individuals will vary on cultural dimensions. While national culture is a macro-level phenomenon, the acceptance of technology by end-users is an individual level phenomenon. Individual behaviour cannot be measured or predicted using the national measurement score since there are no means to generalise cultural characteristics about individuals within the same country especially for measuring actual behaviour in the adoption and acceptance of technology (Ford *et al.*, 2003; McCoy *et al.*, 2005a; Straub *et al.*, 2002). Hofstede himself mentioned that his country-level analysis was not able to predict the individual behaviour (Hofstede, 1984). This means that it is problematic to include national culture constructs within individual level models such as TAM (McCoy *et al.*, 2005b). Therefore, to avoid the limitations of this research conceptual model at the micro level of analysis and to avoid over generalization of cultural typology, the four cultural dimensions were measured at the individual level.

McCoy et al. (2005a) recommend that individual level versions of Hofstede's instrument (such as that developed by Dorfman and Howell (1988) should be used with individual level research models. Srite and Karahanna (2006) followed this approach in two studies of the general acceptance of computing technology (PCs in the first study and PDAs in the second study). They argue that the effect of culture on individuals depends on the degree the individual is willing to get involved and engage with the values of his/her own culture. They therefore used scales derived from the work of Hofstede (1980) and Dorfman and Howell (1988) to measure cultural values at the individual level. The scales were found to have adequate psychometric properties and Srite and Karahanna (2006) were able to successfully integrate them with a model derived from TAM. The research in this study therefore follows this approach and measures culture at the individual level, enabling the moderating effects of culture within the TAM model to be meaningfully explored. The direct measurement of cultural values within a contemporary Lebanese and British samples also allow exploration of whether the shift in cultural values observed over the last 30 years by McCoy et al (2005a) applies to this group.

## 2.5   Summary

This chapter has presented and discussed available literature related to e-learning, technology acceptance and cultural models. These three essential parts will provide the theoretical background of this research. In particular, an overview of the e-learning environment was presented in the first section. In this regards, the advantages, disadvantages as well as the factors that affect the acceptance of such technologies were presented. Secondly, a review of the nine most influential models that has been used to study human behaviour were critically reviewed and compared. From the preceding critical literature, we can note that some of the models have good parsimony (e.g., TAM) but lack the comprehensive cover of many major factors, whereas other models include more complex factors but compromise on the parsimony of the model. However, compared to other models, the TAM was found to have an acceptable explanatory power and also have good parsimony. In this regards, TAM has received extensive empirical support in the IS implementation area. However, there have been some criticisms concerning the theoretical contributions of the model, specifically its ability to fully explain technology adoption and usage. Additionally, the existing parameters of the TAM neglected investigating other essential predictors and factors that may affect the adoption and acceptance of technology such as social, individual and cultural factors. Taking into consideration the above limitations, this research will extend the TAM to include other factors in order to increase its predictive power.

Finally, this chapter presented and discussed the different cultural models that have been used in the literature with an emphasis on Hofstede's cultural theory (1991). Although the acceptance of technology by end-users is an individual level phenomenon, surprisingly it was found that most of the literature about cultural effects in IS research is based on the national (macro-level) or organizational level. Therefore, in an attempt to overcome this gap, this research will examine the influence of culture at the individual level within the context of developing (Lebanese) and developed (UK) countries. The following chapter will discuss the theoretical framework and hypotheses of this study.

# Chapter 3:  Theoretical basis and Conceptual Framework

*"Data and facts are not like pebbles on a beach, waiting to be picked up and collected. They can only be perceived and measured through an underlying theoretical and conceptual framework, which defines relevant facts, and distinguishes them from background noise." (Wolfson M., 1994, p.309)*

## 3.1   Introduction

The literature review chapter discussed the various theories and models that are related to technology acceptance with its components and external factors which directly or indirectly are useful in developing the conceptual framework for this study. Therefore drawing on previous chapter, the aim of this chapter is to explain and discuss the development of the proposed conceptual model to study e-learning acceptance. This chapter also provides a further justification for including the personal, social, and situational factors as key determinants, in addition to the integration of individual characteristics and Hofstede's cultural dimensions as moderators within the model to study e-learning adoption and acceptance. Moreover, research hypotheses are drawn and operational definitions are presented. The proposed model will form the basis for the empirical data collection and analysis.

The chapter is divided into 6 sections. Section 3.2 provides a review of the theoretical background of the proposed conceptual framework. Section 3.3 briefly discusses and justifies the inclusion of the key determinants for the research study; these include PEOU, PU, SN, FC, SE, QWL, BI and AU. In sections 3.4 and 3.5, two sets of moderators are introduced, more specifically, Hofstede's cultural dimensions at the individual level are presented in sections (3.4.1, 3.4.2, 3.4.3, and 3.4. 4), while demographics characteristics are presented in sections (3.5.1, 3.5.2, 3.5.3 and 3.5.4). Finally, Section 3.6 concludes the chapter.

## 3.2   Theoretical Framework

Chapter two has presented a thorough review of the models and theories frequently used in explaining behaviours related to acceptance and adoption of IT/IS. In addition, this research identified a set of factors that fall in different domains such as personal, social, behavioural, cultural and technological that might affect the use and adoption of e-learning systems. As discussed in Chapter 2, these constructs have been established in the literature as salient predictors of technology acceptance.

Therefore, in this chapter, we discuss the development of the proposed conceptual model for this research by explaining the different factors in greater details. It is worth noting that the proposed model is trying to obtain a complete understanding of a phenomenon under investigation which requires some sacrifice in the degree of parsimony (Taylor and Todd, 1995c). The development of the proposed model will show the influence of independent variables on the value of dependent variables (Mandell, 1987) and will help the researcher to hypothesise and test the relationships between the identified constructs in order to check if the theorised model is valid or not (Sekaran and Bougie, 2011). However, drawing on the fact that prior models related to acceptance and adoptions of IT have some limitations (see Chapter 2), therefore, the most appropriate approach was to select the relevant constructs related to context of this study from the various models.

In particular, this study includes 3 categories of variables to be tested, the first category include the key determinants (independent variables) that may have an effect on BI and AU of the e-learning system. These constructs are PEOU, PU, SN, FC, SE and QWL. The second category includes two dependent variables which are BI and AU. In this study BI is expected to influence the AU of the e-learning system. While the third category integrates two sets of moderators that may have an impact on the relationships between the key determinants and BI, the first group integrate Hofstede's cultural dimensions (MF, IC, UA, and PD) at the individual level and the second is demographic variables (age, gender, educational level and experience). Figure 3.1 depicts the proposed research model and a detailed explanation of each category is presented in the next sections of this chapter.

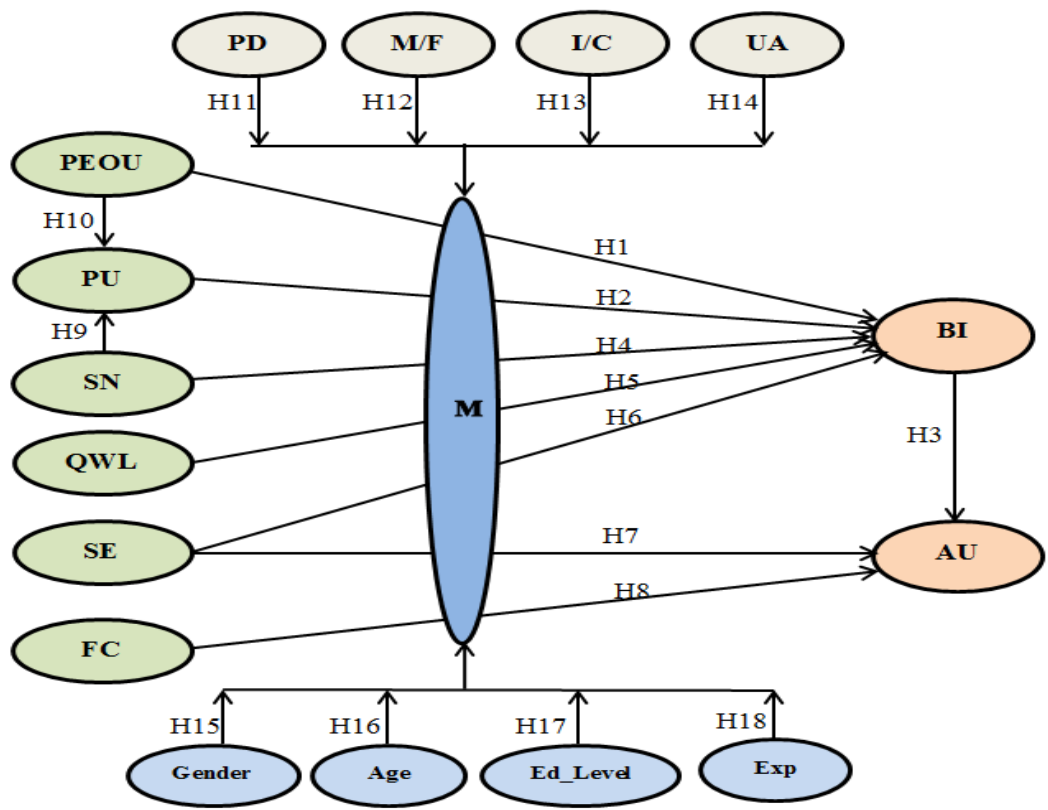

**Figure 3-1: The proposed Conceptual Model**

## 3.3   Direct Determinants

In the previous section, the proposed research model comprised the three groups of variables that may affect the acceptance and adoption of e-learning systems. In this section, a detailed explanation about the key determinants that may have an influence on BI and AU will be discussed with a justification for including each of these constructs within the proposed model.

### 3.3.1 Perceived ease of use

Perceived ease of use (PEOU) is defined as 'the degree to which a person believes that using a particular system would be free of effort' (Davis et al., 1989 p.320) and is similar to *effort expectancy* in UTAUT (Venkatesh et al., 2003). In the TAM, TAM2 and DTPB, PEOU was theorised as a direct determinant of BI. In addition, a number of researchers found a support to the indirect relationship of PEOU on BI through PU (Sun and Zhang, 2006; Yousafzai *et al.*, 2007b). Strong evidence supported the important role that PEOU play in predicting the BI (Davis, 1989; Igbaria *et al.*, 1997; Venkatesh and Morris, 2000; Chan and Lu, 2004; Reid and Levy, 2008).

Reviewing the literature, the majority of the subsequent studies about student perceptions on using technologies support the important role that PEOU plays in predicting the BI (Saadé and Galloway, 2005; Liu *et al.*, 2010; Chang and Tung, 2008; Liu *et al.*, 2005; Teo and Noyes, 2011). However, the degree of significance was different between the findings in the literature. The difference in the findings was based on the field of study, sample size, or techniques used for analysing. For example, Peng et al. (2009) found that PEOU was the strongest determinant on the intention to use the system, which supported the findings of Chao and Tung's (2008) study. Furthermore, Saeed and Abdinnour (2008) found that PEOU have a direct and significant influence on BI. However, it was not the strongest predictor on the BI to use to the system. In contrast, Chesney (2006) concluded that perceived ease of use did not have a direct and significant influence on the intention to use the system.

In the context of this study, the inclusion of PEOU was to investigate students' beliefs of whether the system is free of effort and to predict their behavioural intention to use the Blackboard system. It is expected that if the students find the Blackboard system easy to use, then they are more likely to adopt and use the system. Therefore, based on many models and previous research that consider the direct relationship of PEOU on BI and indirectly through PU, we propose the following hypothesis:

*H1a,b: Perceived Ease of Use will have a direct positive influence on the intention to use web-based learning System in the British and Lebanese context.*

*H1c: Students' mean Perceived Ease of Use will be significantly higher in the United Kingdom, compared to Lebanon.*

*H10a.b: Perceived Ease of Use will have a direct positive influence on Perceived Usefulness of web-based learning system in the British and Lebanese context.*

## 3.3.2 Perceived Usefulness (PU)

Perceived usefulness (PU) is defined as "the degree to which a person believes that using a particular system would enhance his/her job performance" (Davis, 1989, p. 453). PU is similar to *relative advantage* in the model DOI and *performance expectancy* in UTAUT (Venkatesh et al., 2003). In other words, it is the extent to which benefits are seen as outweighing costs. In the TAM, TAM2 and Augmented TAM, PU was theorised as a direct determinant of BI. In addition, many researchers have provided evidence of its direct determinant on AU (Gefen and Straub, 1997; Igbaria *et al.*, 1996; Lederer *et al.*, 2000; Szajna, 1994). Compared to the other behavioural belief construct (PEOU), PU was found to have a significantly greater correlation with BI than did perceived ease of use (Davis, 1989) and the same result has been found in e-learning studies (Liu *et al.*, 2010; Chang and Tung, 2008). Davis (1989) concluded that users are mostly driven to adopt and use the system primarily because of the functions it performs for them.

In the present context of the study, PU was used to investigate the students' beliefs about the potential benefits in using the Blackboard system. Many research studies have highlighted the important rule that PU plays on BI to use Web-based learning tools (Amoako-Gyampah, 2007; Chesney, 2006; Chang and Tung, 2008; Saadé and Galloway, 2005; Liu *et al.*, 2005; Landry *et al.*, 2006; Liu *et al.*, 2010; Rodriguez and Lozano, 2011; Šumak *et al.*, 2011). For example, Liu et al. (2010) applied an extended TAM to explore the factors that affect the intention to use an online learning community. They found that PU was the most influential variable in predicting the intention to use the web-based learning system. In contrast, Saeed (2008) found that PU has an influence on the intention to use but was not the most influential factor. In this cross-sectional study and in accordance with the TAM, TRA and TPB studies, it is expected that if students think that the Blackboard system is useful and will add value to their education then they are more likely to adopt and use the system. In contrast, students may resist educational technologies if they are sceptical of their educational value. Therefore, it is hypothesised that PU will have a positive significant influence on the intention to use the Blackboard system. Therefore, the researcher hypothesised:

*H2a,b: Perceived Usefulness (PU) will have a direct positive influence on the intention to use web-based learning system in the British and Lebanese context.*

*H2c: Students' mean perceived usefulness will be significantly higher in the United Kingdom, compared to Lebanon.*

### 3.3.3 Social Norm

Social norm also known as Social Influence, is defined as 'the person's perception that most people who are important to him or her think he or she should or should not perform the behaviour in question' (Ajzen and Fishbein, 1980; Ajzen, 1991). In other words, SN refers to the social pressure coming from external environment which surrounds the individuals and may affect their perceptions and behaviours of engaging in a certain action (Ajzen, 1991 p.188). SN was included in many theories such as TRA, TPB, DTPB and TAM2 and is similar to social influence in UTAUT; and image in IDT (see Chapter 2).

SN was studied in some research as an antecedent of BI and in other studies as an antecedent PU. However, as mentioned by Venkatesh et al. (2003) the influence of SN is very complex. SN was found to be an important determinant of behaviour in TRA and TPB, and directly and significantly related to the behavioural intention (Ventaktesh et al., 2003., Venkatesh and Morris, 2000, Taylor and Todd, 1995). The direct effect of SN on BI is justified from the fact that people may be influenced by the opinion of others and thus involved in certain behaviour even if they don't want to. Venkatesh and Davis (2000) argue that the effect of SN occurs only in mandatory environments and has less influence in a voluntary environment. Therefore, following the guidelines of TPB and since the use of VLE is mandatory (i.e. students must use the e-learning system in order to complete their course), this research will study the direct effect of SN on behavioural intention as well as on PU.

According to Taylor and Todd (1995a), SN is decomposed into two groups and usually determined by peer and superior influences. In the context of e-learning technologies, student's decision to adopt and accept such technologies is usually influenced by other colleagues/students and superiors/lecturers pressures (Grandon *et al.*, 2005; Ndubisi, 2006).

However, there are inconsistencies in the findings when studying the direct impact of SN on BI. For example, while some scholars found a significant influence of SN on BI such as (Park, 2009; Venkatesh and Morris, 2000; Venkatesh and Davis, 2000; Grandon *et al.*, 2005; Van Raaij and Schepers, 2008) others failed to find any influence (Lewis *et al.*, 2003; Chau and Hu, 2002; Ndubisi, 2006; Davis, 1989). Davis (1989) omitted the SN construct from the original TAM due to theoretical and measurement problems, however SN was added later in TAM2 due to its importance in explaining the external influence of others on the behaviour of an individual.

This research assumes that students will be influenced by their colleagues and instructors in the Lebanese and British contexts, whereas the effect of SN on BI is highly considered in the Lebanese context. The rationale is based on the cultural index which is proposed by Hofstede (1980). He indicated that power distance (PD) and Masculinity (M/F) are high and Individualism (I/C) is low in Lebanon, while PD and Masculinity (M/F) are low and Individualism (I/C) is high in England. Therefore, based on the inconsistencies of the above findings and the importance of SN in establishing behavioural intention towards adoption and acceptance of a technology and its impact on PU, and in an attempt to overcome the limitation of TAM in measuring the influence of social environments (Venkatesh and Davis, 2000), it is hypothesised:

*H4a,b: Social Norm will have a positive influence on student's behavioural intention to use and accept the e-learning technology in the British and Lebanese context.*

*H4c: Students' mean Social Norm will be significantly higher in Lebanon, compared to the United Kingdom.*

*H9a,b: SN will have a positive influence on perceived usefulness of web-based learning system in the British and Lebanese context.*

## 3.3.4  Quality of work life (QWL)

The Quality of Work Life (QWL) seeks to achieve integration among technological, human, and societal demands (Cascio and McEvoy, 2003). Reviewing the literature, the term "Quality of Work Life" has appeared in Research Journals in USA only in 1970's and since then it regained an interest by scholars and researchers. Quality of Life (QWL) was included based on a number of suggestions in the IS literature that this extension may improve the TAM model (Zakour, 2004; Kripanont, 2007; Srite and karahanna, 2000).

QWL has not previously been considered within an educational context and the current study therefore explores whether it plays a role within this context. In this research, QWL is defined in terms of students' perception and belief that using the e-learning system will improve their quality of work life such as saving expenses when downloading e-journals, or in communication when using email to communicate with their instructors and colleagues. Generally speaking, a mismatch between students and the impact of technology on their lives can be disadvantageous for both students and institutions and which in turn affect their behavioural intention to use the e-learning systems. Thus, the emphasis is given to QWL construct due to the economic gains and increases in opportunities for advancement in students' lives and it is expected that the higher the QWL the better the acceptance of the technology.

Srite and Karahnana (1999) found a moderating effect of a set of individual differences including gender and culture on the relationship between QWL and behavioural intention. They found that people with high masculinity will focus more on materiality and might be concerned with usefulness and work goals rather than perceived ease of use and quality of work life that is dominated in the feminine cultures. Therefore, it is expected that introducing the QWL construct will enable a better capturing of cultural influence on the acceptance of e-learning systems especially the impact of masculinity/femininity on intention to use the system. According to Zakour (2004), the include of QWL in TAM will help in better understanding the technology acceptance by users and conclude that future research should highly consider this construct due to its importance.

Thus, we propose that understanding the relationship between QWL and BI is an important goal, in order to satisfy the various needs of the students and in return eliciting favourable behavioural intention. It is worth noting that QWL can be considered the operationalize construct of PU. We therefore expect the relationship between those two constructs to be correlated. Based on previous discussions, we hypothesise the followings:

*H5a,b: QWL will have a positive influence on student's behavioural intention to use the web-based learning system in the British and Lebanese context.*

*H5c: Students' mean QWL will be significantly higher in the United Kingdom, compared to Lebanon.*

### 3.3.5 E-learning self-efficacy

Self-efficacy (SE) - as an internal individual factor - has been defined as the belief "in one's capabilities to organise and execute the courses of action required to produce given attainments" (Bandura, 1997, pp.3). In the Social Cognitive Theory (SCT), SE is a type of self-assessment that helps the understanding of human behaviour and performance in a certain tasks e.g., (Bandura, 1997; Bandura, 1995). In the context of IT, self-efficacy has been defined as "an individual's perceptions of his or her ability to use computers in the accomplishment of a task rather than reflecting simple component skills" (Compeau and Higgins, 1995, p.192).

According to Marakas et al. (1998), SE is categorized into two types, the first is related to general use of computers and is known as 'general Computer self-efficacy', whereas the second is related to a specific task on the computer and is known as 'task-specific computer self-efficacy'. Several studies have found SE to be an important determinant that directly influences the user's behavioural intention and actual usage of IT e.g. (Downey, 2006; Shih and Fang, 2004; Guo and Barnes, 2007; Yi and Hwang, 2003; Hernandez *et al.*, 2009) and e-learning acceptance (Chang and Tung, 2008; Yuen and Ma, 2008; Park, 2009; Vijayasarathy, 2004; Chatzoglou et al., 2009; Roca et al., 2006). On the contrary, Venkatesh et al.(2003) did not find a casual direct relationship between SE and BI.

In the context of this study, e-learning self-efficacy is defined as a student's self-confidence in his or her ability to perform certain learning tasks using the e-learning system. In general, it is expected that e-learning users with higher level of self-efficacy are more likely to be more willing to adopt and use the system than those with lower self-efficacy. Therefore, consistent with previous research that integrated self-efficacy as a direct predictor that has effects on behavioural intention and actual usage of the system, we propose the following hypotheses:

*H6a,b: Computer self-efficacy will have a positive influence on student's behavioural intention to use the web-based learning system in the British and Lebanese context.*

*H7a,b: Computer self-efficacy will have a positive influence on the actual usage of the web-based learning system in the British and Lebanese context.*

*H7c: Students' mean Computer self-efficacy will be significantly higher in the United Kingdom, compared to Lebanon.*

### 3.3.6 Facilitating conditions

The facilitating condition (FC) has been defined as "the degree to which an individual believes that an organizational and technical infrastructure exists to support use of the system" (Venkatesh et al. 2003, p. 453). More specifically, it comprises the availability of external resources (time, money and effort), which is known in the literature as Resource *Facilitating Conditions "RFC"* (Fu *et al.*, 2006; Lin, 2006; Guo and Barnes, 2007; Ajjan and Hartshorne, 2008) and also the availability of the technological resources (PCs, broadband, accessible network, network security etc...), which is referred to as the *Technology Facilitating Conditions "TFC"* (Fu *et al.*, 2006; Taylor and Todd, 1995c). In other words, it is providing the external resources that are needed to facilitate the performance of a particular behaviour (Ajzen, 1985). In the context of this study, FC will be measured by the perception of students of whether they are able to access the required resources and the necessary support to use the e-learning services.

Reviewing the literature, FC construct is considered an important antecedent of the UTAUT model and is similar to *perceived behavioural control* (PBC) from TPB, C-TAM/TPB and *compatibility* from IDT (Ajzen, 1985; 1991; Taylor and Todd, 1995; Venkatesh et al., 2003). FC was included as a direct determinant of BI and AU in many theories and by many researchers in the field of technology studies. For example, the relationship between FC and BI was found to be significant in several studies e.g. (Shih and Fang, 2004; Yi *et al.*, 2006). On the contrary, other researchers found FC non-significant in predicting BI but significant in determining usage (Chang *et al.*, 2007; Limayem and Hirt, 2000).

Thus, the importance of the external influence of facilitating conditions on the decision-making process is a crucial antecedent of human behavioural roles within information system studies (Tan and Teo, 2000; Shih and Fang, 2004) and within the e-learning context (Teo, 2009b; Ngai *et al.*, 2007; Maldonado *et al.*, 2009). Therefore, it is very important to investigate whether FC has a direct influence on the actual usage of the e-learning system, as the absences of facilitating resources may represent barriers to usage (Taylor & Todd, 1995a, p.153). Hence, it is expected that these external resources will lead the students to adopt the learning management systems. Based on the above discussion, the researcher proposes the following hypotheses:

*H8a,b: Facilitating conditions will have a positive influence on actual usage of web-based learning system in the British and Lebanese context.*

*H8c: Students' mean facilitating conditions will be significantly higher in the United Kingdom, compared to Lebanon.*

### 3.3.7  Behavioural Intention

The presence of behavioural intention (BI) in TAM is one of the major differences with TRA. BI is considered to be an immediate antecedent of usage behaviour and gives an indication about an individuals' readiness to perform a specific behaviour.

Ajzen (1981) claims that as a general rule, "the stronger the intention to engage in a behavior, the more likely should be its performance". In TAM, both PU and PEOU influence an individual's intention to use the technology, which in turns influence the usage behaviour (Davis, 1989).

There is considerable support in literature for the relationship between BI and usage behaviour in general (Davis *et al.*, 1989; Taylor and Todd, 1995b; Taylor and Todd, 1995c; Venkatesh and Davis, 2000; Venkatesh *et al.*, 2003). This has recently been extended to the e-learning context (Zhang *et al.*, 2008; Yi-Cheng *et al.*, 2007; Chang and Tung, 2008; Park, 2009; Saeed and Abdinnour-Helm, 2008; McCarthy, 2006; Liu *et al.*, 2010; Walker and Johnson, 2008; Teo *et al.*, 2011). In addition, the path from BI to AU is significant in the TAM, DTPB, and TPB and models. BI has a large influence on AU. However, it is worth mentioning that when individuals have prior experience with using the technology, the effect of BI is more predictive on AU (Taylor and Todd 1995b).

In the context of information system research, system usage were studied as a dependent variable and is often measured by only BI (Agarwal and Karahanna, 2000; Venkatesh and Morris, 2000; Gefen and Straub, 2000), or by only AU (Szjna, 1994; Davis, 1989), or even both BI and AU (Venkatesh and Davis 2000; Venkatesh et al., 2003, Taylor & Todd 1995a).

In the context of this research and similar to previous studies, this research considered both BI and AU as dependent variables in the theoretical framework. It is expected that BI will have a direct influence in predicting the usage behaviour of students to accept and use the Blackboard system in the future (Self-reported usage measures). Therefore, the researcher proposes the following hypotheses:

*H3a,b: Student's BI will have a positive effect on his or her actual use of web-based learning system in the British and Lebanese context.*

*H3c: Students' mean intention to use e-learning systems will be significantly higher in the United Kingdom, compared to Lebanon.*

## 3.4   Hofstede's Cultural dimensions

This section will discuss the first category of factors which is related to Hofstede's cultural dimensions (MF, ID, PD, UA) that may impact the relationships between the core constructs and behavioural intention to use the system. The rationale behind the inclusions of moderators within our conceptual model is to increase the predictive power of the model.

## 3.4.1 Power distance (PD)

PD determines the extent to which individuals expect and accept differences in power between different people (Hofstede, 1993; Hofstede and Peterson, 2000). A number of authors have suggested that Power Distance might be expected to moderate the relationship between Social Norms and Behavioural Intention e.g. (McCoy *et al.*, 2005a; Srite and Karahanna, 2006; Dinev *et al.*, 2009; Li *et al.*, 2009). The general prediction is that users with PD values would be more likely to be dependent on referent power in decision making, i.e. they would be more influenced by the views of others, particularly superiors, in deciding whether to adopt technologies. While this argument appears logical, evidence in support of it has been ambiguous at best. Dinev et al. (2009) compared samples from South Korea and the US in the context of adoption of protective (e.g. anti-virus) software. They found that the relationship between SN and BI was significant for the South Korean sample (a high PD culture) but not for the US sample (low PD). While this finding is in line with the discussion above, several other cultural factors co-varied between the two sample groups and Dinev et al. (2009) attribute the result to a cumulative effect of individualism, masculinity, power distance and uncertainty avoidance. McCoy et al (2005a) compared email users in Uruguay and the USA, predicting that the relationship between SN and BI would be stronger for the Uruguay sample based on a number of cultural differences including Power Distance; however, they found no significant effect of SN on BI in either sample. Li et al. (2009) compared China and the US in the context of adoption of a web portal. While they measured culture at the individual level, they found no moderating effects. Srite and Karahanna (2006) also measured culture at the individual level. In one part of their study they found that PD was a significant moderator of the relationship between SN and BI, this was in the opposite direction to that predicted. The second part of their study found no significant effect of PD. The effect of PD in the context of educational technology adoption does not appear to have been explored directly in past research. Therefore, it is expected that the relationship between SN and BI will be stronger for high PD users. Applying the concept of PD on e-learning implementation, students with low PD cultural values will implement many ways of learning which include using the technology, and the education would be student-centred.

McCoy et al. (2005b) additionally predicted that PD would moderate the relationship between Perceived Usefulness and Behavioural Intention, such that the relationship would be stronger for low PD samples. While the justification for this is not clearly articulated, it could be hypothesised that this would be the case because in low PD cultures users might feel free to use their own intention judgements based on usefulness, rather than rely heavily on the views of those with higher perceived power. McCoy et al. (2005a) were not able to demonstrate the predicted effect in their work, with both the US and Uruguay samples showing similar strength PU->BI relationships. However, since the samples co-varied on a number of other cultural factors it is difficult to interpret the result. Our study provides the opportunity to explore this in more detail with individual level cultural values data. Furthermore, individuals with high PD are characterized with lower rates of innovation and acceptance (Zmud, 1982) and thus the effect of PD will be weaker on the relationship between behavioural belief (PEOU,PU) and BI (Harris, 1997). This argument is consistent with psychological studies which suggests that individuals are less innovative when they have less autonomy and freedom (Mumford and Licuanan, 2004). In contrast, the freedom in working environment could increase the behavioural intention to adopt and use the new technological system (Straub et al., 1997) and individuals will use their own skills (i.e SE) and accept the technology due to its importance to perform the required job (i.e. PU) rather than under the pressure from their superior. Moreover, it is expected that individuals with low PD cultural values will mainly use the technology if it helps improve their quality of work life (QWL). Generally speaking, a mismatch between students and the impact of technology on their lives can be disadvantageous for both students and institutions and which in turn affect their behavioural intention to use the e-learning systems. There are evidence of a worldwide decrease in PD in recent years especially in the Arab world, however as previously mentioned Lebanon, as an Arab country, differs socially and culturally from England based on Hofstede's PD dimension. Lebanon historically has a significantly higher level of PD (value index = 80) than England (value index = 35).

With respect to power distance and based on previous discussion, it is therefore expected that the relationships between PEOU, PU, SE and QWL will be more related to students with low power distance, while students with higher power distance are expected to be influenced by SN, FC. Therefore, we propose the following hypotheses:

*H11a1,a2,a3,a4,a5,a6: The relationship between (PEOU, PU, SN, QWL, SE, FC) and Behavioural Intention and actual usage of the e-learning system is moderated by the* **Power Distance** *value in the British context.*

*H11b1,b2,b3,b4,b5,b6: The relationship between (PEOU, PU, SN, QWL, SE, FC) and Behavioural Intention and actual usage of the e-learning system is moderated by the* **Power Distance** *value in the Lebanese context.*

*H11c: Power Distance mean will be significantly higher in Lebanon, compared to the United Kingdom.*

## 3.4.2  Masculinity/Femininity

According to Hofstede's (1980, 1984, 1991, 2001) definition of the masculinity/femininity cultural dimension, a high masculinity culture (low femininity) will emphasise work goals, such as earning and promotions. In general, individuals high on masculinity usually value competitiveness, assertiveness, ambition and focus on performance and material possessions. On the other hand, low masculinity individuals (high on femininity) are encouraged to follow more traditional, tender and modest roles.

Those holding high femininity cultural values are characterised as being more people-oriented than those with high masculinity values. For this reason they would be expected to be more influenced by interpersonal contact and a number of authors have therefore predicted a moderating effect of social norms on behavioural intention, such that the relationship will be stronger for more feminine samples (Li et al., 2009; Dinev et al., 2009; Srite and Karahanna, 2006). Srite and Karahanna (2006) found support for this hypothesis in one study using individual measures of culture, but a non-significant effect in their second study. Dinev et al. (2009) showed the predicted effect in a cross-country sample, with the relationship between SN and BI stronger for South Korea (more feminine culture) than the USA (more masculine culture), though a number of other cultural differences between the samples could also contribute to this result. Other studies were unable to show the predicted effect e.g. (McCoy *et al.*, 2005a; Li *et al.*, 2009). Here we predict that in an educational context the relationship between SN and BI will be moderated by Masculinity-Femininity, such that it will be stronger for those espousing more feminine values.

Conversely, those high in masculinity would be expected to focus more on instrumental values and so would be expected to be more influenced by features that enhance the achievement of work goals compared to those high on femininity. PU is usually determined through the cognitive instrumental process by which individuals use a mental representation for considering the match between the potential goals and the consequences of performing an act of using the technology (Venkatesh, 2000). A moderating relationship between PU and BI would therefore be predicted, with the relationship expected to be stronger for those with more masculine values. Srite and Karahanna (2006) hypothesised this effect but failed to find a significant result. Srite (2006) found an effect in the opposite direct to that expected, with a significant effect of PU on BI for a US sample (more feminine) but no significant effect for a Chinese sample (more masculine). The relationship has not been explicitly explored in relation to e-learning technology acceptance; we predict that in this context the relationship will be stronger for high masculinity individuals.

In contrast to PU (an instrumental variable), PEOU captures the hedonic experience of using a technology. Authors such as Srite and Karahanna (2006) argue that such experiences will be more important for users who espouse feminine values since feminine cultures tend to emphasise the creation of more pleasant work environments. A number of authors found evidence to suggest that the relationship between Perceived Ease of Use and Behavioural Intention is stronger for those with more feminine values e.g. (Srite, 2006; Srite and Karahanna, 2006; McCoy *et al.*, 2007; Qingfei *et al.*, 2009). We therefore predict the same effect in an educational context.

Quality of Working Life (QWL) has been suggested by several authors as a potentially relevant factor when considering the impact of masculinity-femininity on technology acceptance e.g. (Zakour, 2004; Srite and Karahanna, 2006; Srite, 2006); however, none of these studies directly considered QWL. Quality of Working Life is generally valued more within masculine cultures QWL is closely related to the obtained benefits of the technology (Zakour, 2004; Kripanont, 2007; Srite and karahanna, 2000), so would be expected to be a more important predictor of technology acceptance for users with more masculine values. In an educational context we hypothesise that QWL will be stronger predictor of BI for those expressing more masculine cultural values. Similarly, individuals with high masculine values perceive analytical and competitive approaches to solving problems (Venkatesh et al., 2004; Venkatesh and Morris, 2000) which will lead to higher score on Self-efficacy. Such individuals are also expected to rate a higher performance towards FCS with respect to service aspects and the working environment.

It is worth noting that England differs socially and has a significantly higher score of masculinity (value index= 66) than Lebanon (value index= 53). At the national level Lebanon is considered as moderate in masculine index compared to 66 in England which indicates a very masculine society.

Therefore, based on previous discussions it is expected that students with high masculine values will be more concerned about the usefulness of the e-learning system and QWL, while students with feminine values perceive a higher importance of PEOU, SN, SE and FC. Thus, we hypothesise the followings:

*H12a1,a2,a3,a4,a5,a6: The relationship between (PEOU, PU, SN, QWL, SE, FC) and Behavioural Intention and actual usage of the e-learning system is moderated by the **masculinity/femininity** value in the British context.*

*H12b1,b2,b3,b4,b5,b6: The relationship between (PEOU, PU, SN, QWL, SE, FC) and Behavioural Intention and actual usage of the e-learning system is moderated by the **masculinity/femininity** value in the Lebanese context.*

*H12c: **masculinity** mean will be significantly higher in to the United Kingdom, compared to Lebanon.*

### 3.4.3  Individualism/Collectivism

According to Hofstede (1980), the terms individualism/collectivism refer to the extent to which individuals are integrated into groups. In individualistic societies, individuals focus on their own achievements and personal goals rather than on the group they belong to, such people seem to innovative, and value personal time, challenges and freedom, while in collectivistic societies people prefer loyalty and group success on their individual gain (Kagitcibasi, 1997). The level of individualism/collectivism of an individual strongly affects the relationship between a person and the group of which one is a member (Benbasat and Weber, 1996; Hofstede, 1991).

Individualism/collectivism was found to have a moderating impact on the relationship between PU, PEOU, SE, FC, SN, QWL and behavioural intention (BI) to use a specific technology, such that PU and BI will be highly considered with individualistic users, and PEOU, SE, FC, QWL, SN and BI will be highly considered with collectivistic users.

A number of authors have hypothesised that the relationship between SN and BI would be stronger in more collectivist cultures due to the views of others in-group members being considered as more important within such cultures e.g. (Zakour, 2004; McCoy *et al.*, 2005a; Srite and Karahanna, 2006; Li *et al.*, 2009; Dinev *et al.*, 2009). Srite and Karahanna (2006) for instance, argue that normative influences may be a more important determinant of intended behaviour for those who espouse collectivist values. Unfortunately their empirical data did not support the predicted relationship. Other researchers who predicted this moderating effect of Individualism/Collectivism but failed to find support from their data include McCoy et al. (2005a) using a cross-country comparison and Li et al. (2009) using individual level cultural data. On the other hand Srite (2006) found that while SN was a significant predictor of BI in a Chinese sample (collectivist culture), there was no significant effect of SN on BI in the USA (individualistic culture), a result they attribute as potentially due to the moderating effect of individualism. Dinev et al's. (2009) argued that the relationship was significant for South Koreans (collectivist) but not for a US sample, but this particular cultural comparison confounded a number of cultural differences and the authors attribute the result to a cumulative effect of these differences. Therefore evidence is limited to support the moderating role of individualism on the effect of SN in technology acceptance and does not appear to have been investigated explicitly in relation to e-learning adoption. We predict that those espousing more collectivist values will be more likely to be guided by SN in their decision to adopt e-learning technology.

Some authors have also suggested that Individualism/Collectivism may play a role in other TAM relationships. Lee et al. (2007) predicted and found that Individualism has a direct positive effect on both PU and PEOU. Other authors have hypothesised a moderating role of Individualism/Collectivism on the relationship between PU and Behavioural Intention e.g. (McCoy *et al.*, 2005a; Sánchez-Franco *et al.*, 2009). Individualistic cultures are characterised by an emphasis on the achievement of individual goals, so PU would appear to be a highly relevant factor for technology adoption in such settings, relating as it does to technology as a means for the achievement of specific goals. In an educational context Sanchez-Franco et al. (2009) predicted that the relationship between PU and BI would be higher for individualistic users. Their results support this in that they showed that Nordic (individualist culture) users' intentions were more influenced by PU than those of Mediterranean users (collectivist culture). On the other hand McCoy et al. (2005a) failed to find the predicted difference in a comparison of Uruguay and US samples. We re-examine this potential moderating effect in an educational setting, using individual level measures of Individualism/Collectivism, predicting a stronger relationship for those expressing individualist values. Fewer authors have considered whether the effect of PEOU might be moderated by Individualism/Collectivism. McCoy et al. (2005b) explicitly state that they expect no influence of IC here. However, McCoy et al. (2007) found that the path from PEOU and BI was impaired in collectivist settings and speculate that people within these settings may be more willing to endure poor usability so long as they are achieving goals that are valued by the wider group. We therefore also consider in the current work whether this may be the case. Furthermore and as TAM demonstrates, perceived ease of use affects individual attitudes through two mechanisms: instrumentality and self-efficacy and this also can be improved with the facilitation conditions (Csikszentmihalyi, 2000; Davis, 1989; Bandura, 1997; Thatcher et al., 2003). The rationale could be that users with high collectivistic cultural values will need training, physical conditions and use of skills which is associated collectivism.

It is worth noting that Lebanon s' cultural score is significantly lower on individualism (value index= 38) compared to England (value index = 89) which is considered very high on individualism. Therefore, based on previous discussions, it is expected students with high individualistic values will be more concerned about the usefulness of the e-learning system and QWL, while students with feminine values perceive a higher importance of PEOU, SE, and SN. Hence, we postulate the following hypotheses:

*H13a1,a2,a3,a4,a5,a6: The relationship between (PEOU, PU, SN, QWL, SE, FC) and Behavioural Intention and actual usage of the e-learning system is moderated by the* **individualism /collectivism** *value in the British context.*

*H13b1,b2,b3,b4,b5,b6: The relationship between (PEOU, PU, SN, QWL, SE, FC) and Behavioural Intention and actual usage of the e-learning system is moderated by the **individualism /collectivism** value in the Lebanese context.*

*H13c: **individualism** mean will be significantly higher in the United Kingdom, compared to Lebanon.*

### 3.4.4 Uncertainty avoidance (UA)

According to Hofstede (1980), UA refers to the extent to which ambiguities and uncertainties are tolerated. Actually, the level of stress and anxiety for individuals with high UA increases more when uncertain situation occurs compared to individuals with low UA (Marcus and Gould, 2000). In other words, individuals with high UA cultural values will establish formal rules and might reject deviant ideas and behaviours since it has been associated with anxiety and the need for security  (Hofstede, 1984). Conversely, individuals with low UA cultural values might have a greater willingness to take risks (Hofstede, 1984; Pfeil *et al.*, 2006) and will feel less anxiety with unfamiliar situations and problems (Dorfman and Howell, 1988).

Several authors propose a moderating effect of UA in the relationship between SN and Behavioural Intention e.g. (Zakour, 2004; Kim, 2008; Sanchez-France et al., 2009; Dinev et al., 2009). The prediction is that SN will be more important in a high UA context because the opinions of referent groups provide a useful means for people to reduce the uncertainty associated with the uptake of new technology. Support for this hypothesis comes from the work of Srite and Karahanna (2006) and Dinev (2009), while Li et al. (2009) were unable to show a moderating effect of UA in their study. In an educational context we predict the relationship between SN and BI will be higher for those espousing high UA values.

UA has also been hypothesised to play a moderating role in a number of TAM relationships and these have been explored with mixed results as described below. In an educational context, Sanchez-Franco et al. (2009) predicted that UA would have a moderating effect on the relationships between both PU and PEOU on Behavioural Intention, arguing that these factors would help to resolve unclear situations and that this information would have a relatively greater influence on the behaviour of high UA samples. They conducted a comparison of samples of educators from Nordic (high individualism, low UA) and Mediterranean (low individualism, high UA) cultural settings. The results supported a moderating effect of PEOU on BI, with PEOU more likely to encourage uptake among the Mediterranean e-learning system users (where UA was higher). However, the results with respect to PU were counter-intuitive, with PU having a bigger effect on BI for Nordic users (the lower UA group). This may be due to the confounding impact of individualism (which was expected, and found, to influence the effect of PU, such that it played a greater role in the more individualist Nordic culture). This study again illustrates the difficulties of inference from studies that examine culture at the group, rather than individual level. McCoy et al. (2007) found that the PU->BI and PEOU->BI paths in TAM were only significant in high UA settings, not in lower UA settings, supporting the moderating relationship of these two variables. However, in an earlier study they failed to find the predicted difference between Uruguay (high UA) and US (low UA) samples for either PU or PEOU (McCoy et al., 2005a). The current study re-examines the moderating impact of UA on PEOU effects in an e-learning context with culture measured at the individual level, predicting that this factor will play a bigger role for those espousing higher UA values.

On the other hand, the relationship between PU and BI will be stronger for individuals with low UA cultural values (Parboteeah et al., 2005). Previous studies suggest that people with an orientation low on UA prefer situations that are free and not bound by pre-defined rules and regulations (McCoy, 2002). In such situations, individuals will relatively accept the technology as they are less likely to be cautious towards technology and therefore perceive the system to be more useful than those in high UA culture (McCoy et al., 2007; Straub et al., 1997). In contrast, individuals rated high on UA will feel uncomfortable and thus will resist to change easily especially in situations in which technology is dominant and therefore will not perceive the usefulness of technology (Zakour, 2004; Garfield and Watson, 1997).

Furthermore, the relationship between SE, FC, QWL and BI will be stronger for individuals with high UA cultural values. This is due to the fact that individuals with high UA tends to hold lower perceptions of SE and be more concerned about the risks associated with technology. Adoption and acceptance of technology involve taking risks and doing something new (Stoneman, 2002). Bandura (1986) argued that the relationship between SE and PEOU are reciprocal to each other. This means that higher anxieties will decrease the SE and eventually decreasing the overall performance. In such situations and to reduce uncertainty and anxiety and improve performance, individuals will rely more on FC from the social environment (Venkatesh and Morris, 2000; Hwang, 2005). Similarly, individuals with low UA cultural values are associated with a strong motivation to achieve and more ambition and therefore will perceive the importance of the technology on their life.

It is worth noting that Lebanon has higher degree of UA (index value = 68) than England (index value = 35). Based on the above discussion, we posit the following hypotheses in line with this discussion:

*H14a1,a2,a3,a4,a5,a6: The relationship between (PEOU, PU, SN, QWL, SE, FC) and Behavioural Intention and actual usage of the e-learning system is moderated by the **Uncertainty Avoidance** value in the British context.*

*H14b1,b2,b3,b4,b5,b6: The relationship between (PEOU, PU, SN, QWL, SE, FC) and Behavioural Intention and actual usage of the e-learning system is moderated by the **Uncertainty Avoidance** value in the Lebanese context.*

*H14c: **Uncertainty Avoidance** mean will be significantly higher in Lebanon, compared to the United Kingdom.*

# 3.5   Individual characteristics

Beside the first category which comprised of Hofstede's cultural dimensions, there are also another category of factors which is individual characteristics (gender, age, educational level and experience) that may have moderating effects on the relationships between the core constructs and BI and AU. These variables will be discussed in the following section sequentially.

## 3.5.1  Gender

Gender is defined as a hierarchical separation between women and men embedded in both social institution and social practices (Jackson and Scott, 2001). The consideration of gender in models of behaviour was introduced in gender schema theory (Bem, 1981) and other technology acceptance models (e.g. TAM 2 and TPB). Previous studies has shown that men and woman are different in decision-making processes and usually use different socially constructed cognitive structures (Venkatesh and Morris, 2000).

Previous research has suggested that gender plays an important role in predicting usage behaviour in the domain of IS research e.g. (Venkatesh and Morris, 2000; Gefen and Straub, 1997; Porter and Donthu, 2006; Venkatesh et al., 2003; He and Freeman, 2010; Wang et al., 2009; Morris and Venkatesh, 2000). For example, Venkatesh et al. (2003) found that the explanatory power of TAM significantly increased to 52% after the inclusion of gender as a moderator. More specifically, gender was found to have a moderating impact on the relationships between PU, PEOU, SE, SN, QWL and BI as well as AU.

Venkatesh et al (2003) found gender to influence the relationship between performance expectancy (similar to PU) and BI, with the relationship significantly stronger for men compared to women. Their findings are consistent with literature in social psychology, which emphasizes that men are more pragmatic compared to women and highly task-oriented (Minton et al., 1980). It is also argued that men usually have a greater emphasis on earnings and motivated by achievement needs (Hoffmann, 1980; Hofstede and Hofstede, 2005) which is directly related to usefulness perceptions. This suggests that men place a higher importance on the usefulness of the system. Their argument is also supported by other researchers e.g. (Srite and Karahanna, 2006; Venkatesh and Morris, 2000; Terzis and Economides, 2011). In contrast, Wang et al (2009) did not find any moderating effect of gender on the relationship between performance expectancy (similar to PU), effort expectancy (similar to PEOU) and BI. It is expected that gender will also affect the relationship between QWL and BI since it focuses on the benefits of the technology and this is considered a more salient issue for males than females (Kripanont, 2007).

In terms of the moderating impact of gender on the relationship between PEOU, SE, FC, SN and BI, it is expected to be stronger for women compared to men. Venkatesh et al. (2003) reported that the intention to adopt and use a system is more highly affected by effort expectancy for women than men. Their results is consistent with gender role studies (Lynott and McCandless, 2000; Schumacher and Morahan-Martin, 2001). The reason could be that women compared to men generally have higher computer anxiety and lower computer self-efficacy (SE). The difference is based on the correlational relationship which is closely related to PEOU, so that higher computer self-efficacy will lead to lowering of the importance of ease of use perception (Venkatesh and Morris, 2000). This is also supported in previous research in psychology e.g.(Cooper and Weaver, 2003; Roca et al., 2006) which suggests that men  perceive analytical and competitive approaches to solving problems which will lead to higher score on Self-efficacy (Venkatesh et al., 2004). Furthermore, it is expected that gender will have an impact on facilitating conditions (FC) and that the relationship will be stronger for women compared to men. This argument is based on Hofstede's cultural theory (Hofstede and Hofstede, 2005) proposition and more specifically related to masculinity/femininity cultural dimensions, which indicates that women compared with men rated a higher importance towards FCs with respect to service aspects and the working environment. Additionally, it has been found that gender affects the relationship between SN and BI such that the effect is stronger for women (Venkatesh and Morris, 2000; Venkatesh et al., 2003; Kripanont, 2007; Huang et al., 2012). Women are found to rely more than men on others' opinion (Venkatesh and Morris, 2000; Hofstede and Hofstede, 2005) as they have a greater awareness of others' feelings compared to men and therefore more easily motivated by social pressure and affiliation needs than men. Therefore it is expected that the relationship between SN and BI will be stronger for women than for men (Wang *et al.*, 2009).

In line with previous discussion, it is expected that the relationship between PU, QWL and BI will be stronger for male students, whereas the relationship between PEOU, SE, FC, SN and BI will be stronger for female students. Thus we propose the following hypotheses:

*H15a1,a2,a3,a4,a5,a6: The relationship between (PEOU, PU, SN, QWL, SE, FC) and Behavioural Intention and actual usage of the e-learning system will be moderated by the **gender** in the British context.*

*H15b1,b2,b3,b4,b5,b6: The relationship between (PEOU, PU, SN, QWL, SE, FC) and Behavioural Intention and actual usage of the e-learning system will be moderated by the **gender** in the Lebanese context.*

## 3.5.2 Age

Research has shown that age is an important demographic variable that has direct and moderating effects on behavioural intention, adoption and acceptance of technology e.g. (Chung *et al.*, 2010; Venkatesh *et al.*, 2003; Wang *et al.*, 2009; McCoy *et al.*, 2005a; Yousafzai *et al.*, 2007b; King and He, 2006; Walker and Johnson, 2008; Sun and Zhang, 2006; Akhter, 2003; Porter and Donthu, 2006). Venkatesh et al (2003) reported that age was an important moderator within his UTAUT model. They found that within an organizational context, the relationships between performance expectancy (similar to PU), FC and BI was stronger for younger employees, while the relationship between effort expectancy (similar to PEOU) and SN was stronger for older employees in accepting and using the technology (Venkatesh *et al.*, 2003). They concluded that ''increased age has been shown to be associated with difficulty in processing complex stimuli and allocating attention to information on the job'' (Venkatesh et al., 2003, p. 450). They also found that age moderate the relationship between facilitating conditions and behavioural intention. Similarly, Morris and Venkatesh (2000) found the same moderating effects of age. It could be that age increased the positive effect of SN due to greater need of affiliation e.g. (Morris and Venkatesh, 2000; Burton-Jones and Hubona, 2006).

In contrast with this, Chung et al (2010) did not find any moderating effect of age on the relationship between PEOU, PU and BI in online communities. In sharp contrast, Wang et al (2009) found that age differences moderate the relationship between effort expectancy (similar to PU) and BI and was stronger for older adults but did not find any moderating effect of age on effort expectancy (similar to PEOU) and BI. Sun and Zhang (2006) found that the relationship between PU and BI was stronger for younger adults in the adoption decision. Correspondingly and since QWL may be correlated with PU since its perceived the importance of technology on user's quality of work life, it is expected that the relationship between QWL and BI will be stronger for younger users. Additionally, with respect to social and psychological influence on the adoption decision, Jones et al (2009) found the relationship between SN and BI to be stronger for older adults. Similarly, Wang et al (2009) found that age moderates the relationship between SN and BI, and the effect was stronger for older adults on using m-learning technology.

In terms of computer and internet self-efficacy, it was found that older people have low self-efficacy in use of technology (Czaja et al., 2006). The rationale could be that older adults often think that they are too old to learn a new technology (Turner et al., 2007). Previous research also found that age differences influence the perceived difficulty of learning a new software application (Morris *et al.*, 2005; Morris and Venkatesh, 2000). There is a clear evident that younger adults have lower levels of computer anxiety than their older counterparts (Chaffin and Harlow, 2005; Saunders, 2004) and that lower levels of computer anxiety are associated with lesser reluctance to engage in opportunities to learn new Internet skills (Jung et al., 2010).

Despite the inconsistencies that have been found in previous research about the direct or moderating effect of Age on the influence of various determinants on behavioural intention, many researchers support the important role that age plays in the context of technology acceptance. Therefore, in the context of this study, it is expected that the effect of age on the relationship between PEOU, SE, SN and BI will be stronger for older students, while the influence of PU, QWL on BI will be stronger for younger students. Therefore, we propose the following hypotheses:

*H16a1,a2,a3,a4,a5,a6: The relationship between (PEOU, PU, SN, QWL, SE, FC) and Behavioural Intention and actual usage of the e-learning system will be moderated by the **age** in the British context.*

*H16b1,b2,b3,b4,b5,b6: The relationship between (PEOU, PU, SN, QWL, SE, FC) and Behavioural Intention and actual usage of the e-learning system will be moderated by the **age** in the Lebanese context.*

## 3.5.3  Educational level

In previous studies, education level was related to knowledge and skills which in turns affect the behavioural beliefs (PU and PEOU) towards acceptance and usage of new technologies (Rogers, 2003; Agarwal and Prasad, 1999). Educational level, like other individual factors, has been studied as an antecedent of PU or PEOU (Agarwal and Prasad, 1999) and as a moderator that affects the relationship between main determinates and behavioural intention (Burton-Jones and Hubona, 2006). In particular, educational level was found to influence the relationships between PEOU, PU, SN and BI (Porter and Donthu, 2006; Rogers, 2003; Sun and Zhang, 2006; Zakaria, 2001; Mahmood et al., 2001; Burton-Jones and Hubona, 2006).

Venkatesh et al., (2000) found a positive correlation between the level of education and PU, Similarly, Burton-Jones and Hubona (2006) suggested that higher education level leads to positive association with PU and those users are less sensitive to PEOU since it will reduce the computer anxiety and improve the overall attitude. In contrast, Agarwal and Prasad (1999) found that there was no relationship between educational level and PU, but there was with PEOU. Similarly, Al-Gahtani (2008) found that educational level only moderate the influence of PEOU on BI, while no moderating impact were found on the relationship between PU and BI towards using computer applications on a voluntary basis in the context of Arab countries. Abu-Shanab (2011) found a moderating effect of educational level on the relationship between most of the key determinants of UTAUT and acceptance of internet banking in Jordan. Moreover, educational level was also found to negatively affect the social influence on behaviour when adopting new technology in an organization as both education and experience will empower the users (Burton-Jones and Hubona, 2006; Lymperopoulos and Chaniotakis, 2005).

The moderating impact of educational level on the relationship between quality of life and behavioural intention has not been investigated in literature. Nevertheless, it is expected that educational level will have an impact on the relationship between QWL and BI such that the relationship will be stronger for students with higher educational level. The rationale is that students who have higher level of education will perceive the e-learning system and value the impact of this system on their career.

Despite mixed results, however the moderating role that educational level can play on the adoption and acceptance of technology is indisputable (see meta-analysis of Mahmood et al (2001) and Sun and Zhang (2006)). Hence, in the context of this study, it is expected that the relationships between (PU, QWL) and BI will be stronger for users with higher educational level, while the relationships between (SN, PEOU) will be stronger for users with lower educational level. We thus propose the following hypotheses:

*H17a1,a2,a3,a4,a5,a6: The relationship between (PEOU, PU, SN, QWL, SE, FC) and Behavioural Intention and actual usage of the e-learning system will be moderated by **Educational Level** in the British context.*

*H17b1,b2,b3,b4,b5,b6: The relationship between (PEOU, PU, SN, QWL, SE, FC) and Behavioural Intention and actual usage of the e-learning system will be moderated by **Educational Level** in the Lebanese context.*

### 3.5.4  Experience

The concept of experience refers to the involvement of an individual in something over a period of time. In the technology context, an individuals' experience is measured by the level of experience and number of years in using a specific technology and will result in a stronger and more stable behavioural intention relationship (Venkatesh and Morris, 2000; Venkatesh *et al.*, 2003; Poon, 2007). Users may employ the knowledge that have gained from their prior experience to form their intentions (Fishbein and Ajzen, 1975). Experience was not incorporated in the original TPB and DTPB; however it was added later after follow-on studies due to its importance on the intentions (Morris and Venkatesh, 2000). Moreover, TAM2 clearly incorporated Experience as a moderator that affects the relationship between main determinants and behavioural intention (Venkatesh and Davis, 2000).

Previous research has found that a user's degree of relevant experience moderates a number of relationships within TAM e.g. (Lymperopoulos and Chaniotakis, 2005; Al-Jabri and Al-Khaldi, 1997; Venkatesh *et al.*, 2003; Venkatesh and Bala, 2008). The relationship between behavioural intention and usage was empirically confirmed to be more statistically significant for expert users compared to novice users (Taylor and Todd, 1995b; Venkatesh *et al.*, 2004; Venkatesh and Davis, 2000) and thus experience will have a positive influence on the strength of the relationship between BI and AU.

As for PU, Taylor and Todd (1995a) reported that experience significantly moderates the relationship of PU and BI such that the relationship was stronger for inexperienced users, their results were not expected. This means that experienced users tended to give less consideration on PU and based their consideration to control information in formation their intentions (Taylor & Todd 1995a). In contrast, Venkatesh et al. (2003) did not find a significant moderating effect of experience on the relationship between ''performance expectancy'' (similar to PU) on BI. This suggested that PU has a strong impact on BI for inexperienced users.

Additionally, the moderating effect of experience on the relationship between PEOU and BI is clear and stable in the literature (Venkatesh and Davis, 2000; Venkatesh and Morris, 2000). Generally speaking, when users have prior knowledge in using the technology, this will provide the users a more robust base to learn as users will relate their incoming information with what they already know (Cohen and Levinthal, 1990). In other words, experienced users will perceive PEOU as not a big issue when learning a new technology (Taylor and Todd, 1995a; Venkatesh et al., 2003). Venkatesh (2002) found that the direct influence of PEOU on BU will decrease over time due to the experience that individuals obtain during the time using the system. In contrast, inexperienced users with no prior knowledge will prefer to use the technology which is easy to use.

With respect to SN, empirical evidence has demonstrated that experience was also found to significantly moderate the relationship between SN on BI (Venkatesh and Davis, 2000; Venkatesh and Morris, 2000). Venkatesh and Davis (2000) argued that the influence of SN on BI will decrease over time. Where users already have extensive experience, the role of SN will be expected to be lower as users are more able to draw on their own past experiences to shape their perception rather than the opinions of others (Venkatesh and Davis, 2000; Venkatesh and Morris, 2000). Similarly, Karahanna, Straub & Chervany (1999) found that inexperienced users are more driven by SN than experienced users. It is expected that the relationship will be stronger for inexperienced users since they will be more sensitive to their colleagues' opinion (Venkatesh et al., 2003).

In terms of self-efficacy and facilitating conditions, it is noteworthy to mention that SE has been studied as a direct determinant and moderators on behavioural intention and usage behaviour to use the technology and is similar to "indirect" experience e.g. (Park *et al.*, 2012; Vijayasarathy, 2004; Ong and Lai, 2006; Roca *et al.*, 2006). In addition, experience was found to influence the relationship between facilitating conditions and behavioural intention (Venkatesh et al., 2003). The authors found that the relationship was stronger for experience. They suggested that when the experience increases, this will lead to user's wider options for help and support and this will lead to more usage of the system.

In the context of this study, it is expected that experience will play an important role on the relationship between main determinants and behavioural intention to use the e-learning system. It is expected that when students experience increases; they will be more aware of the benefits of the e-learning system on their education e.g. (Evanschitzky and Wunderlich, 2006; Stoel and Lee, 2003). Therefore, we propose the following hypotheses:

*H18a1,a2,a3,a4,a5,a6: The relationship between (PEOU, PU, SN, QWL, SE, FC) and Behavioural Intention and actual usage of the e-learning system will be moderated by* **Experience** *in the British context.*

*H18b1,b2,b3,b4,b5,b6: The relationship between (PEOU, PU, SN, QWL, SE, FC) and Behavioural Intention and actual usage of the e-learning system will be moderated by* **Experience** *in the Lebanese context.*

## 3.6   Conclusion

In this chapter we proposed a theoretical framework that might be helpful in understanding the various factors that are expected to influence the adoption and acceptance of e-learning systems in Lebanon and England in the context of an HEI. The research model is based on prominent well known technology acceptance models and theories that have been discussed in Chapter 2, e.g. TAM, TAM2, TRA, DTPB and UTAUT which are relevant to the context of this research.

These factors reflect personal, social, and situational factors and specifically include social norm, facilitating conditions, E-learning self-efficacy, quality of work life, perceived ease of use, perceived usefulness and behavioural intention and usage. In addition, culture and demographic characteristics were integrated as two sets of moderators in the model.

Therefore this research proposes and tests three types of hypotheses, in the first category, this study proposes 10 direct hypotheses from H1 to H10. In the second category, 4 individual characteristics and 4 cultural variables were hypothesised to have a moderating impact on the relationship between the main determinants and BI. While in the third category, 12 hypotheses were proposed to test the differences between the Lebanese and British sample at the national level. It is expected that extending the TAM to include SN, SE, QWL and FC, in addition to the two sets of moderators, may help in explaining more of the variance of behavioural intention and actual usage as well as explore reasons for why the model may hold better in some contexts than others.

The conceptual framework will be tested empirically in two countries, Lebanon and England, to achieve the objective of the research, as discussed in previous chapters. Therefore, the next chapter discusses the chosen methodology, and include a detailed explanation of the method of data collection, questionnaire development and the various tools to be used for the data analysis.

# Chapter 4:  Research Design & Methodology

*"The Formulation of the problem is often more essential than its solution" Albert Einstein*

## 4.1  Introduction

Chapter 2 reviewed the theories and concepts in technology acceptance and culture. In Chapter 3, the conceptual model was developed to examine the influence of culture and other demographic characteristics on the acceptance of technology on e-learning environment in Lebanon and England. This chapter describes and justifies the philosophical approach, methods and techniques used in this research to achieve the main research objectives and to answer the research questions.

Technically speaking, this research employed a quantitative method in order to understand and validate the conceptual framework. A survey research approach based on positivism was employed to guide the research. Questionnaire was used as a data collection technique. Additionally, Structural Equation Modelling (SEM) using Analysis of Moment Structures (AMOS) version 18.0 was employed as a data analysis technique.

This chapter is structured as follows. Section 4.2 and 4.3 examines the philosophical assumptions with a justification and focus on the reasons behind choosing the positivist epistemology and ontology which form the basis of this research.  Section 4.4 discusses two main research strategies qualitative and quantitative with a discussion focus on the justification for the selection of quantitative research strategy. Section 4.5 discusses the different research approaches available in the IS field and justify the use of survey research approach in this study. Section 4.6 describes the research design and the research framework. Section 4.7 explains the different available sampling strategies with a justification of using convenience sampling technique in this research in. Sections (4.8 and 4.9) describe the different stages in developing the questionnaire and measurement scales. Section 4.10 provides an overview of the pilot study. Sections 4.11 outlined the issues related to data analysis and techniques using SEM with AMOS. Before concluding this chapter in Section 4.13, the ethical considerations related to this research were reported in Section 4.12.

## 4.2   Underlying research Assumptions

*Webster Dictionary* defines the word 'paradigm' as "an example or pattern: small, self-contained, simplified examples that we use to illustrate procedures, processes, and theoretical points". According to Denzin and Lincoln (2005), paradigms are a broad framework of perceptions, beliefs, and feelings with which theories and practice operate. For Guba and Lincoln (1994), research philosophies are the set of feelings about how the world works (ontology) and how it should be understood (epistemology) and studied (methodology). Whereas ontology raises questions about the nature and form of reality to be known, epistemology raises questions about nature of the relationship of the Knower (researcher) and what can be known (the problem under investigation). Methodology refers to general principles which underline how we investigate the social world and how we demonstrate that the knowledge generated is valid (Blaikie, 2000; Mingers, 2003; Orlikowski and Baroudi, 1991).

According to Guba and Lincoln (1994) and Lincoln *et* al. (2011), positivism, post-positivism, critical theory and constructivism or interpretivism are the 4 schools or thoughts that underline the major paradigms that structure the social science research. The following discussion and Table 4.1, shows the major differences between these four paradigms and their methods and approaches.

| Philosophical Assumptions | Positivism | Post-positivism | Critical theory | Constructivism or Interpretivism |
|---|---|---|---|---|
| **Ontology** | *Naive realism:* real reality exists but apprehensible. Knowledge of 'the way things are' is summarised in the form of time-and-context-free generalisations, and takes the form of cause-and-effect laws. | *Critical realism:* real reality is exists but only imperfectly and probabilistically apprehensible. | *Historical realism:* virtual reality shaped by social, political, ethnic, cultural, economic, and gender values crystallized over time. | Relativism: local and specific constructed realities. There are multiple realities where the mind plays an important role by determining categories and shaping realities. In this case, there is no separation of mind and objective as the two linked together. |
| **Epistemology** | Dualist/Objectivist: Findings true, the investigator and the investigated 'object' don't affect each other's and are supposed to be independent entities; enquiry takes place in the cause-effect relationships, the observer does not influence or is influenced by the object. Replicable empirical results are 'true'. | Modified dualist/objectivist: critical tradition and community; findings probably true but always subjects to falsifications. | *Transactional/ Subjectivist:* the findings are value-mediated, and its aim is a critique to the knowledge. | *Transactional/ subjectivist:* The observer and the object to be observed are supposed to be interactively linked so that the 'findings' are created as the investigation proceeds by the investigator's interpretation. created findings |
| **Methodology** | Experimental/Manipulative: verification of hypotheses; chiefly quantitative methods | Modified Experimental/ Manipulative; critical multiplism; falsification of hypotheses; may include qualitative methods | *Dialogic/Dialectic:* A change in practice and social relationships is a results based on the dialogue between the observer and the participants in order to extract more accurate knowledge from the ignorance. | Hermeneutic/dialectical |

**Table 4-1: Basic Beliefs of Alternative Research Paradigms (Adapted from Guba and Lincpln, 1994)**

According to Orlikowski and Baroudi (1991), a research is positivist if there was evidence of formal propositions, quantifiable measures of variables, hypothesis testing, and the drawing of inferences about a phenomenon from the sample to a stated population (Orlikowski and Baroudi, 1991). For Straub *et al.* (2005), positivism was described from the statistical viewpoint. They argued that the objective of statistics is to falsify the null hypothesis. Thus, the theoretical hypothesis is supported if the null hypothesis is rejected (Struab *et al.*, 2005). In previous IS studies, the positivism approach was the dominant between the other 3 approaches with more than 75% research employing this school of thought, 17% interpretivist and only 5% critical research (Mingers, 2003). Examples of positivist include the work of (Yin, 2009; Straub *et al.*, 2004; Walsham, 1995; Galliers, 1992). This research employs positivist approach since it includes research hypothesis testing (Chapter 5 and 6) and quantifiable measures of variables (Chapter 4) towards e-learning adoption and also provides evidence of propositions (Chapter 2). A further discussion of selecting this approach is provided in the next section.

Post-positivist approach is positioned between positivism and interpretivism (Lincoln *et al.*, 2011). Post-positivists recognise that; when studying the behaviour and actions of humans; the researchers cannot be 'positive' about their claims of knowledge (Creswell, 2008). The findings that the researchers obtain from the post-positivist studies are based on observation and measurement of the objective reality that usually exists 'out there' in the world. There is no difference in kind between positivist and post-positivist approach, only a difference in degree. Both approaches conduct empirical and quantitative research (Creswell, 2008). This school of thought was not chosen since there is no lack of values and ethical questioning in our theoretical model. The constructs used in this study were heavily applied within information systems acceptance and tested within North American and Western countries. Therefore, post-positivist would need more effort, money and thus wasting a lot of time. Additionally, this approach fails to explain the unpredictable nature of human (Onwuegbuzie, 2002)

Walsham (1993) described interpretivism research in terms of "aiming at producing an understanding of the context of the information system, and the process whereby the information system influences and is influenced by the context"(p.4-5). According to Bryman and Bell (2011), in an interpretative research, the access to reality will be achieved through social actors and constructions. Contrary to the positivist research, interpretive research does not predefine dependent and independent variables (Kaplan and Maxwell, 2005). Although this school of thought produces deep insights into social phenomena since it employs qualitative data collection (Myers and Avison, 1997; Struab *et al.*, 2005). It lacks the ability to generalise the findings to larger population (Winfield, 1991) and thus considered to be less appropriate to our research compared with positivist approach.

According to Myers and Avison (1997), critical researches assume that "social reality is historically constituted and that it is produced and reproduced by people. Although people can consciously act to change their social and economic circumstances, critical researchers recognise that their ability to do so is constrained by various forms of social, ethnic and political domination" (Myers and Avison, 1997, p.7). Interview and observation is the two main methods of enquiry in critical studies (Bryman and Bell, 2011). As previously mentioned and similar to interpretivist approach, this school of thoughts was considered to be less relevant for our research compared to positivist approach. Critical research is still immature and unclear as a legitimate approach in the IS discipline since it lacks an agreed theoretical basis (Falconer, 2008; Kvasny and Richardson, 2006). Furthermore, some other researchers describe it as "a missing paradigm" in IS research due to the little research that consider this approach (Chen and Hirschheim, 2004; Richardson and Robinson, 2007).

# 4.3 Choosing the positivism Paradigm for our research

The positivist approach was selected after considering the differences between all the other three underlying approaches and the nature of the study being addressed (Hall and Howard, 2008). More specifically, choosing this approach was based on the following points:

- As previously mentioned, the positivist approach was the dominant between the other 3 approaches with more than 75% research employs this approach (Mingers, 2003) and especially in the adoption and technology acceptance research.

- The current research aim to investigate the moderating effect of culture and demographic variables on e-learning behaviour within two culturally different contexts (Lebanon and United Kingdom). Exploring the direct impact of student's acceptance beliefs on technology acceptance was also a part of the study. Thus, this research is related to social subjects where student's behaviour is measured and where the researcher is isolated from the aim of the study (Saunders *et al.*, 2009). Therefore, the positivist approach was justifiable from the ontological point of view.

- This research posits a number of hypothesised relationships to be tested and quantitatively measured within the context of technology acceptance. The positivist approach is mostly linked with Quantitative methodology, which in turn uses a deductive process (Bryman, 2008). Therefore, this research is also justified from the methodological point of view.

- The purpose of the research requires a well-defined conceptual framework where the relationships between the constructs are clearly defined and with precise measurements. In the current study, a number of constructs about the adoption and technology acceptance from many developed and validated theories and models were used and presented in chapter 2. Therefore, this research is justified from the epistemological perspective.

- This research will use Structural Equation Modelling technique in order to test hypotheses and moderators and to perform a number of tests such as group comparisons (Chapter 5 and 6). The statistical packages used describe the positivist approach (Struab *et al.*, 2005).
- Finally, the findings of this research can be replicated in a different study or in different context which is one of the major advantages of using the positivist approach (Winfield, 1991).

Having discussed the reasons behind choosing the positivist paradigm for this research, the following section describes the research method.

## 4.4   Strategy of Inquiry: Quantitative and Qualitative

The next step after underlying philosophical assumptions is the research method (which shifts the focus to research design and data collection). The choice to employ a qualitative or quantitative methodology will influence how to collect the data in later stages of the research. This section discusses the differences between the two approaches in order to justify employing the quantitative approach in this research.

According to Bryman (2008), quantitative research methods seek to collect numerical data and  verify the relationships between measureable variables in a universal cause-effect way. Furthermore, quantitative methods use a deductive approach that is associated with hypothesis testing in order to modify or support the existing theory (David and Sutton, 2004). They are linked mainly with positivist epistemology which uses scientific procedures and involves statistical methods and usually presents data numerically (Creswell, 2008). On the other hand, qualitative research methods tend to explore and discover meanings and patterns instead of numbers (Creswell, 2008). The research of a qualitative nature employs an inductive approach to derive the theories through the process of collecting and analysing the data (Punch and Punch, 2005; Creswell, 2008). The following Table (Table 4.2) summarises the differences that are agreed upon between qualitative and quantitative according to Johnson and Christensen (2010) and Lichtman (2006).

| Criteria | Qualitative Research | Quantitative Research |
|---|---|---|
| **Aim /Purpose** | To understand and interpret social interactions in order to provide a complete description. | To test hypotheses, look at cause and effect, and generalise results. |
| **Group Studied** | Smaller & not randomly selected. | Larger & randomly selected. |
| **Type of Data Collected** | Words, pictures, or objects. | Numbers and statistics. |
| **Nature of Reality and Form of Data Collected** | Multiple realities; subjectivity is expected (In-depth interviews, open-ended responses, participant observations) | Single reality; Objectivity is critical (precise measurements using validated data-collection instruments) |
| **Type of Data Analysis** | Identify patterns, features, themes | Identify statistical relationships. |
| **Role of Researcher** | Researcher may influence the participants (Subjectivity is expected) | Researcher cannot influence the participants. The characteristics of the participants are intentionally hidden from the researcher. |
| **Results** | The results are less generalisable, and the findings are particular and specialised to a certain subject. | The findings are more generalisable and can be applied to different contexts and other populations. |
| **Scientific Method** | Exploratory or bottom–up | Confirmatory or top-down |
| **Research Objectives** | Explore, discover, and construct. | Describe, explain, and predict. |

**Table 4-2: Qualitative versus Quantitative research (Source: Johnson and Christensen, 2010; Lichtman, 2006)**

The use of a quantitative research strategy which is rooted in the positivist ontology is supported by various researchers in the domain of information systems (Yin, 2009; Straub *et al.*, 2004; Walsham, 1995; Galliers, 1992). This research aims to examine and test hypothesised relationships within the context of technology acceptance in an objective manner where the researcher is isolated from the aim of the study. Additionally, the constructs and their relationships used within the conceptual model were developed and validated thoroughly in the theories and models about the adoption and technology acceptance (see chapter 2). Furthermore, data survey method was employed to collect data from a large number of participants in order to analyse the data using SEM technique and this data was generally presented in numbers and thus belong to the quantitative strategy rather than qualitative (Bryman, 2008; Creswell, 2008).

After the justification for the quantitative research approach, the next section explains the data collection methodology employed in this research.

## 4.5   Survey Research Approach

There are a number of different research approaches that have been used in literature when conducting any research, including field experiment, lab experiment, field study, phonological research, narrative research, opinion research, ethnography, grounded theory (Creswell, 2008; Guba and Lincoln, 1994; Neuman, 2006; Crotty, 1998; Orlikowski and Baroudi, 1991). For the current research, the researcher employed the survey approach in order to collect data from the participants in both Lebanon and England for the following reasons:

- Survey research approach is the most dominant approach (with at least 50% of the total number of articles) used in Information system journals (e.g, MIS Quarterly, European Journal of Information systems, Journal of Information system for example see (Mingers, 2003; Orlikowski and Baroudi, 1991; Choudrie and Dwivedi, 2005)).
- Survey approach is most widely considered within technology adoption research, for example Mingers (2003) found out that more than 74 % of the articles in the previously mentioned journals that are related to the technology adoption employed survey research, while the case study method were employed by the remaining 26% only. For examples see Venkatesh and Davis (2000), Venkatesh and Morris (2000), Srite and Karahanna (2006), Venkatesh and Bala (2008).

- The aim of this study is to examine the individuals' technology acceptance behaviour within Lebanon and England which involves collecting data from a large number of participants especially when using Structural Equation Modelling (SEM) technique in data analysis, employing another research approach will be very expensive and time-consuming (Hair *et al.*, 2010).

- A number of research hypotheses need to be empirically tested within the proposed conceptual model (see Chapter 3) which is only appropriate using the survey research approach.

- The survey approach is associated with the research using positivist-quantitative methodologies (Saunders *et al.*, 2009).

- Since a large amount of data is being collected when using the survey approach, this allows the findings to be generalised to the entire population.

- Another major factor that affects the research approach is the extent to which a researcher is involved within the context being studied, this research is related to social subjects where student's behaviour is measured and where the researcher is isolated from the aim of the study (Saunders *et al.*, 2009); therefore it is more appropriate and feasible to use survey approach than others such as ethnography and case studies.

Within the survey research approach, data is usually collected through a number of methods such as mail, telephone interview, email, and self-administrated questionnaire (Zikmund, 2009). This research employed the self-administrated questionnaire as a data collection method for the following reasons:

- Data can be collected from a large number of participants simultaneously in a quick, easy, efficient and economical way compared with other methods such as interviews (Zikmund, 2009; Bryman, 2008; Sekaran and Bougie, 2011).

- It is easily designed and administrated. For example, interviews usually require much administrative skills (Sekaran and Bougie, 2011).

- Higher privacy of respondents because issues such as anonymity and confidentiality were dealt with in the cover letter.
- Collecting the questionnaires immediately after being completed will assure a higher response rate (Sekaran and Bougie, 2011).
- Respondents can seek clarity and therefore could understand the concepts on any question they are answering which in turns minimise the outliers in the study (Aaker *et al.*, 2009).
- A questionnaire as a data collection method has been widely used in studies similar to the context of this study. For example see, Venkatesh and Morris (2000), Venkatesh and Bala (2008), Srite and Karahanna (2006).

Based on the previous discussion of the research methodologies, the next section explains the research design and summarises the overall research framework.

## 4.6   Research Design

According to Bryman and Bell (2011), a research design provides an overall guidance and framework for the data collection and analysis of the study. It is critical in terms of linking the theory and the empirical data collected in order to answer the research questions (Nachmias and Nachmias, 2008). A choice of an appropriate research design will influence the use and type of data collection, sampling techniques, and the budget (Hair *et al.*, 2010). Additionally, when designing a study the researcher should make a sequence of rational decisions regarding the purpose of the study, location of the study, the investigation type, role of the researcher, time horizon and the level of data analysis (Sekaran and Bougie, 2011).

Based on the guidelines provided by Sekaran and Bougie (2011) about the research design, the purpose of the current study was to test the hypotheses generated from the conceptual model. The relationships that exist among variables can be easily understood through Hypothesis testing, as such studies usually explain the nature of certain relationships among variables, or establish the differences among different groups. A correlational type of study (i.e., field studies) is chosen over casual type to delineate the variables that are associated with the research objectives and examine the salient relationships between the main determinants of an individual's behaviour with a set of individual differences in a non-western nation compared to a western nation. This study is conducted in a non-contrived setting which is similar to all studies that use a correlation type of investigation. Since the data collection method used in this study was based on survey, therefore there was no intervention from the researcher. Furthermore, based on the aims and objective of the research it was obvious that the unit of analysis is an individual university student within Lebanon and England. This study selected a cross-sectional design as data can be collected just once and over a fixed period of time due to the fact that using SEM requires a relatively large number of respondents and it is beyond the timeframe of this research to collect longitudinal data in a different time in order to examine the change in the dependent variables.

Figure 4.1 illustrates the research design followed in this research which is based on a sequence of interrelated step by step process (Sarantakos, 1993). The literature review (Chapter 2) was critical in the first stage to gain a deep understanding about the research problem. A conceptual model with constructs was developed in order to test the generated hypotheses (Chapter 3). In the later stage, the decisions about how to go about finding the solution to the research problem was through methodology (Chapter 4), for example collecting data using a quantitative methods, stages in the development of survey (explanatory, validity, pilot test), sampling technique and data collection process.

To test the proposed model, a descriptive analysis of the collected data was essential (chapter 5), where chapter 6 provides the results of testing the research model. The final step was the discussion (Chapter 7) and conclusion and future work in the final chapter (Chapter 8).

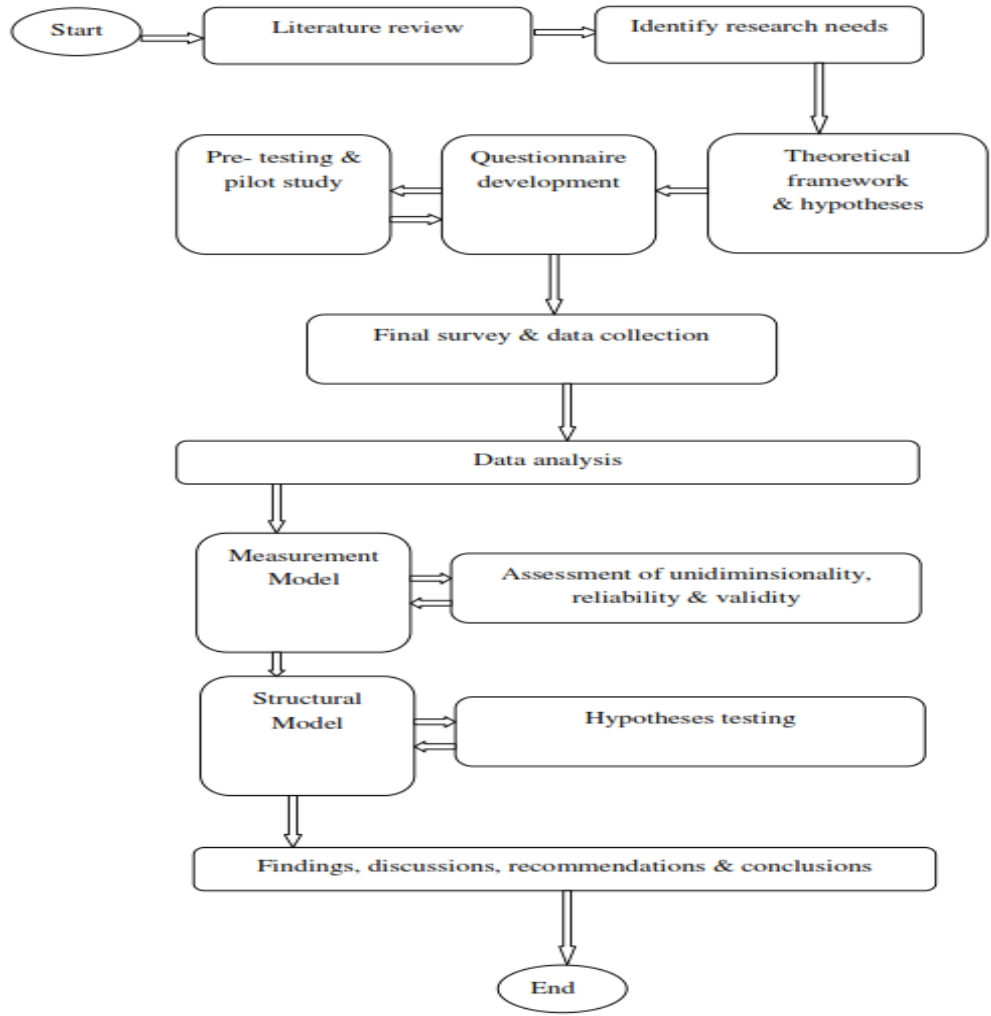

**Figure 4-1: Research Design**

## 4.7   Population and Sampling

Before proceeding to data collection process, the sampling technique is considered a critical concern to the research  in order to represent the targeted population and to eliminate the bias in the data collection methods and thus generalise the results (Bryman and Bell, 2011; Russ-Eft and Preskill, 2009). According to Fowler (2009), there are four critical issues to be considered when designing the sample as follows; (1) the choice of probability or non-probability sample technique; (2) the sample frame; (3) the size of sample; (4) the response rate.

## 4.7.1  The Sampling choice

It is common for a research to recognise the importance of collecting information from the respondents that represent the entire population due to time and financial constraints. According to Blumberg et al (2008), when designing a sample, the researcher should consider several decisions and take into account the nature of the research problem and the specific questions that evolve from the question, objectives, time and budget. Probability and non-probability are the two types of sampling technique (Krathwohl, 2004; Bryman and Bell, 2011). The following Table (Table 4.3) provide a description about each sampling methods within each sampling technique.

| A classification of Sampling technique | | |
|---|---|---|
| Sampling Technique | 1. Probability Sampling Techniques | a. Simple Random Sampling<br>b. Systematic Sampling<br>c. Stratified Sampling<br>d. Cluster Sampling<br>e. Other Probability Sampling Techniques |
| | 2. Non-probability Sampling Techniques | a. Convenience Sampling<br>b. Judgmental Sampling<br>c. Quota Sampling<br>d. Snowball Sampling |

**Table 4-3: A Classification of Sampling Techniques Source: (Groves et al., 2009)**

A random selection of the sample is the base of the concept in probability sampling. This guarantees a controlled procedure to ensure that each person within the population has a known chance of selection (Groves *et al.*, 2009; Blumberg *et al.*, 2008). The random probability technique consists of simple random, stratified, systematic and cluster sampling.

In Simple Random Sample (SRS) each element of the population has an equal probability of selection in the sample; this method requires a complete numbered list of the population where each possible selection of a given size (*n*) from probability of size (N), the probability of selection is n/N. Similar to SRS methods, systematic sampling require a complete list of the population. However, researchers select the subjects using set skip interval from starting element and add skip interval to find the next elements until *n* elements found (N/n). This method is simpler, quicker, and cheaper than SRS. In stratified random sampling, two-step process is required, in the first step the researchers tend to partition the population either proportionally or disproportionally into mutually exclusive and collectively exhaustive subgroups or strata, while in the second step researchers use SRS method sampling in order to select from each subgroups. Stratified sampling reduces sampling error by increasing precision without increasing in cost (Blumberg *et al.*, 2008; Bryman and Bell, 2011). In cluster sampling, researchers tend to reduce the data collection costs by selecting the sample for the targeted population from a number of small geographic areas. First, the researcher partitions the population into exclusive and exhaustive subgroups or clusters. They later use SRS in order to select a random sample of clusters. After that, for each selected cluster, either includes all the subjects within each subgroup in the sample are included (one-stage), or a probability sample of subjects is selected from each subgroup or cluster in the sample (two-stage). This method is more cost efficient than other random methods but less statistically efficient.

Contrary to probability sampling, the concept of non-probability sampling is based on non-random selection of sample and thus not all the elements within the population has equal or known chance of selection (Blumberg *et al.*, 2008; Groves *et al.*, 2009). The non-probability sampling technique includes judgmental, Quota, snowball and convenience sampling methods.

In judgmental; also called purposive; sampling, the selection of units from the targeted population is based on the knowledge and professional judgment of the researcher. In this method, usually the characteristics of the subjects needed are already clear for the researcher. The researcher then targets the potential sample members in order to check if they are suitable to meet the criteria of the research. Quota sampling divides the targeted population into control categories and then the selection of sample is based on convenience or judgment methods to ensure equal representation of subjects. In snowball sampling, researchers start with an initial respondent who meets the criteria of the study which is usually chosen randomly, then other potential subjects are identified and included in the research based on the recommendations of the initial subjects. This method is used when the numbers of individuals are limited. This method is time consuming and results depend on characteristics of the respondents in the sample and thus are hardly representative of the population. The convenience sampling methods allow the researcher to select the sample subjects from the targeted population based on who are willing and easily accessible to be recruited and included in the research. This method is the least expensive, least time-consuming among all other techniques. According to Stangor (2010), the convenience sampling method is the most common used method in behavioural and social science studies (p.151). The following Table (Table 4.4) explains the strength and weakness for each of the sampling methods.

| Technique | Advantages | Disadvantages |
|---|---|---|
| *Probability sampling* | | |
| Simple random sample (SRS) | Easy to implement, analysis, and interpretation, results projectable. | Require a complete list of population, expensive, time-consuming, produces high error rate. |
| Systematic | Simpler, quicker, cheaper expensive than SRS. Sampling distribution of mean or proportion is easy to be determined. | Costly, lower representative than SRS, the results and sample may be skewed due to the periodicity within the population. |
| Stratified random | The sample size in strata is controlled by researcher. Include all important subgroups, decrease sampling error | Expensive, more complex, and also researcher should make a greater effort compared to simple random; the sample size in strata must be carefully defined |
| Cluster | Cost effective, quick, good for large population, easy to do without a list of population. | Imprecise, not easy to compute results, the fact that sub-groups (clusters) are being homogenous rather heterogeneous this will lead to lower statistical efficiency. |
| *Non-probability sampling* | | |
| Convenience | The least expensive, least time-consuming and administration to ensure sufficient participants of a study, most convenient and common among other methods. | Selection bias, cautious when generalisation of findings as the sample is not a representative of the whole population. |
| Judgmental or Purposive | Low cost, not time-consuming, ensures balance of group sizes. | The subjectivity of the researcher may lead to bias and thus reliability and generalisability of the results may be questionable. |
| Quota | Low cost, and not time-consuming, the researcher select subgroups with controlled characteristics and number of participants of which is related to the study. | Results depends on the characteristics of the respondents within the sample and thus not easy to defensible the results as a representative of targeted population. |
| Snowball | It is very efficient where individuals are very rare. It is also possible to include participants even if there is no known list in advance. | Time-consuming, questionable to guarantee whether the sample is a representative of the entire population. |

**Table 4-4: Advantages and disadvantages of the sampling methods Source: (Blumberg et al., 2008; Black, 1999)**

Based on the previous discussion of the sampling techniques and methods, the next section is explaining the reasons behind employing the convenience sampling technique in this research.

## 4.7.2  The justifications behind using Convenience sampling in this research

As previously mentioned, the convenience sampling method is the most commonly used method in behavioural and social science studies. This method allows the researcher to select the sample subjects from the targeted population based on who is willing and easily accessible to be recruited in the research. This method is the least expensive and least time-consuming among all other techniques. In the present study, it was not feasible to access data to allow random sampling to take place, as well as time and budget constraints led to decision to employ the non-random approach with the potential to greatly collect the sample sizes needed for the analysis.

The targeted population are Lebanese and British students studying full time at universities and higher educational institutions who use web-based learning system. Those students share many common similarities, for example, the participants are all university students with very close age groups, and balance in representation of gender (53 % male and 47 % female). In terms of technology and internet usage, it was found that they are mostly intermediate or expert in using web-based learning system.

Although this research employed convenience sampling technique in collecting data which assumes homogeneous population and thus generalisation of results to the entire population should be done with caution, however based on the characteristics of the respondents in this study which also share many similarities with other university students demographically and in technology usage, then it could be argue that a random sampling was partially used.

Furthermore, in a statistical meta-analysis study of applying the technology acceptance model within the field of information system, King and He (2006) used 88 published studies to analyse the conditions where TAM may produce different results based on user types and usage types, they included that "*in terms of type of user and type of use demonstrated that professionals and general users produce quite different results. However, students, who are often used as convenience sample respondents in TAM studies, are not exactly like either of the other two groups*". Moreover, several of other similar studies that involve students used the convenience sampling and have contributed valuable results, for example:

a) Comparison at national culture and e-learning acceptance (Sánchez-Franco *et al.*, 2009; Li and Kirkup, 2007; Keller *et al.*, 2007).

b) Cross-cultural and perceptions of e-learning (Arenas-Gaitán *et al.*, 2011; Teo *et al.*, 2008).

c) Individual-level culture and technology (Srite and Karahanna, 2006; Min *et al.*, 2009).

d) Behaviour towards using e-learning (Zhang *et al.*, 2008; Martinez-Torres *et al.*, 2008; Liaw and Huang, 2011).

Additionally, in an attempt not to limit collecting data from one geographical area and to increase the reliability of the findings of this study, participants from 2 different universities in Lebanon and one central university in England were targeted.

Since the target population for this research is too large especially in England, this thesis adopted the convenience sampling method based on previous discussion and due to time and financial constraints.

## 4.7.3  Population

For Zikmund (2009), the target population are the entire group of subjects by which the researcher is interested to investigate to answer the research objectives.

Four categories defined for the population are inclusion, exclusion, expected effect size and feasibility (Light *et al.*, 1990).

With respect to the Lebanese population in this thesis, based on statistics retrieved from the official website of the Center of Educational Research and Development in Lebanon (http://crdp.org/CRDP/English/en_construction.asp) for the year 2006-07, there are 41 private higher educational institutions and one public. Based on the categories of private higher institutions, there are 18 universities and 23 institutes. The total number students studying full time for the academic year 2011 was 160,364, with 45% studying in the Lebanese University and 55% studying in private institutions.

With respect to the target population in England, according to the Higher Education Statistics Agency (HESA) (http://www.hesa.ac.uk/index.php/content/view/1973/239/), there are 130 universities based on different geographical areas, of which 82 universities uses VLE to support face to face education. The total number of British students is 1746060, of which 76.5 % are undergraduate students, and 68.5% are in full-time education.

In terms of inclusion criteria, this study targets all Lebanese and British full time students from those 100 universities (18 in Lebanon and 82 in England) who use web-based learning system in their education.

Regarding the feasibility criteria, the data has been collected from 2 universities (AUB, LAU) located in the capital of Lebanon (Beirut) and one central university (Brunel University) from the capital of England (London). Regarding the exclusion criteria, the UK sample excluded non-British students who studies in England the Lebanese sample excluded students studying in the only public university in Lebanon (The Lebanese University). Also, in terms of mode of study, part time students from both countries were excluded. This study also excluded students where the use of web-learning systems in the education is voluntary.

## 4.7.4  Sample Size

It is critical for the research to specify the sample size within the targeted population. According to Bryman  (2011), using a large sample within the study cannot guarantee precision and thus will waste time and money. On the contrary, using a small size especially when statistical data analysis such as SEM is required will result in lower accuracy of the results (Hair *et al.*, 2010). The targeted population within this study was very large especially in England. Therefore, the sample size was determined based on the rules of thumb for using structural equation modelling within AMOS. According to Roscoe (1975), the following rules of thumb should be considered when considering the sample size:

a) Sample size > 30 and < 500 are appropriate for most research.

b) When categorising the sample into sub-groups (e.g., older/younger, postgraduate /undergraduate), a minimum size of 30 is required within each category.

c) In multivariate research (e.g., SEM), the required sample size should exceed by several times (preferably 10 times) the number of variables within the proposed framework or study.

Similarly, Kline (2010) suggested that a sample of 200 or larger are appropriate for a complicated path model. While a sample size varies between 50 and 1000 of which 50 as very poor and 1000 as excellent (Comrey and Lee, 1992). Accordingly, Hair *et al.* (2010) recommended that a sample size should be estimated in terms of the number of respondents per estimated parameter, and therefore should consider the complexity of the model which takes into account the number of constructs and variables within the model. For example, a sample size of 400 or more is required for a model with 6 or more constructs with 3 indicators in each. Additionally, Schreiber et al. (2006) suggested that a generally agreed-on value is 10 respondents for every estimated parameter within the model.

In line with the above suggestions about SEM assumptions (Schreiber *et al.*, 2006; Hair *et al.*, 2010), and considering the complexity of the proposed model in terms of variables and ratio of respondents (estimation of approximately 48 parameters), the sample size required from each of the two countries should be at least 400.

## 4.7.5  Execution of the sampling process

For the present study, the selection of participants from Lebanon and England were based on participants' availability and the researcher self-selection. As previously discussed, this method of selection might affect the generalizability of the results to the entire population as there is no equal chance for other students to participate in the study. However, based on the literature about the effect of user types and usage types in the TAM results, students share many similarities and thus produce similar results (King and He, 2006). Therefore, the effect of this method on generalizability of the results is decreased to minimum.

## 4.7.6  Non-response bias

The sample is intended to be a representative of the entire population, and thus a relatively high response rate to acquire a large sample will increase the level of confidence and decrease the bias from the collected data (Saunders *et al.*, 2009). There are two reasons of non-response: (1) refusal to respond to individual questions, (e.g. leaving a few blank questions); and (2) refusal to respond to any questions without even giving a reason (Saunders *et al.*, 2009). When a relatively high rate of non-response occurs, there is a high risk to effect the validity of the survey.

According to (Vogt, 2005), a non-response bias is "*the kind of bias that occurs when some subjects choose not to respond to particular questions and when the non-responders are different in some way (they are a non-random group) from those who do respond*" (p. 210). The non-response bias occurs when those who respond differ in the outcome variable from those who do not respond. The type of data collection methods relatively affects the nature of bias. For example, a high non-response bias occurs when using postal survey, telephone or even interview. For the current study, taking into account that better-educated people (i.e., students) will return the questionnaire within a reasonable rate compared to those who are less educated (Fowler, 2009) and in an attempt to reduce the bias to minimum, this research used a paper-based questionnaire to collect the data.

Furthermore, based on the pre-test and pilot study results of the questionnaire (Chapter 4), a high response rate were acquired with also a high satisfaction about the length, clarity of wording and layout of the self-administrated questionnaire. Therefore, this will also help reducing the bias in the research. Additionally, a case deletion is performed with all the biased questionnaires in this study.

## 4.8   Data collection Development

In this section, the researcher presents the different stages of developing the questionnaire, instrument pre-testing and pilot study.

### 4.8.1  Questionnaire Design and Development

A questionnaire was developed to collect the data required to answer the research questions and thus achieve the main objectives of the study (Saunders *et al.*, 2009). The questionnaire items were mainly obtained from reviewing the literature about technology acceptance models, culture and e-learning outlined in Chapter 2 and more specifically based on the conceptual framework and the research hypotheses outlined in Chapter 3. This research followed Sekaran and Bougie's (2011) and Ghauriand Grønhaug's (2005) procedures to develop a questionnaire which is based on 1) conceptualisation of each construct and 2) operationalise the constructs.

In order to assure that there are neither ambiguous nor confusing questions and keeping in mind the main objective of the research,  the questionnaire design went through different stages and took over a year before it was finalised (from November 2010 till September 2011). The questionnaire consisted of 4 pages, a consent form and a cover letter. The purpose of the study was briefly explained to the respondents in the covering letter with other information, which indicate that their participation will be strictly confidential (see Appendix A). The main questionnaire consisted of 6 sections. Section one includes the moderating demographic variables such as gender, age, educational level and internet and web-based learning experience. Section B, C, D and E covered the direct determinates within the proposed conceptual framework, while section F covered the 4 moderating cultural variables. *The conceptualisation and operationalisation of the constructs and their variables used in the questionnaire are measured as follows:*

**Section A**: this section includes the demographic background of the respondents

- **Demographics characteristics:** refer to nationality, gender, age and educational level. These variables are comprised in 4 questions (Q1-Q4) all measured on nominal scale and were measured as moderators within the proposed conceptual model. These questions were critical so they were put in the first part of the questionnaire.
- **Experience:** refers to individual's experience in using web-based learning systems and general internet skills. This moderator is measured on a nominal scale and consisted of two questions (Q6-Q7).

**Section B**: this section includes the main determinants of TAM

- **Perceived usefulness (PU):** refers to the degree to which an individual believes that using the web-based learning system would enhance his or her performance (Davis, 1989). This construct is consisted of 5 questions (Q8- Q12) measured using seven-point Likert scale ranging from 1 (strongly disagree) to 7 (strongly agree).
- **Perceived ease of use (PEOU):** refers to the degree to which an individual believes that using the web-based learning system will be free of effort (Davis, 1989). This construct is based on 5 questions (Q13-Q17) and measured using seven-point Likert scale ranging from 1 (strongly disagree) to 7 (strongly agree).

**Section C**: this section includes the extended TAM determinants'

- **Self-efficacy (SE):** refers to individual's belief in his/her own competence (Ormrod, 2010) in order to understand how and why individuals perform differently at various tasks (Bandura, 1997). This construct is based on 6 questions (Q18-23) and measured using seven-point Likert scale ranging from 1 (strongly disagree) to 7 (strongly agree).

- **Facilitating conditions (FC):** refers to degree to which an individual believes that an organisational and technical infrastructure exists to support use of the system(Venkatesh *et al.*, 2003). This construct is based on 4 questions (Q24-27) and is measured using seven-point Likert scale ranging from 1 (strongly disagree) to 7 (strongly agree).

- **Subjective Norms:** refers to individual's perception that most people who are important to him or her think he or she should or should not perform the behaviour in question (Ajzen, 1991; Ajzen and Fishbein, 1980). This construct were included in many models (TRA, TAM2, TPB/DTPB And C-TAM-TP) and is based on 4 questions (Q28-Q31) and is measured using seven-point Likert scale ranging from 1 (strongly disagree) to 7 (strongly agree).

- **Perceived Quality of work life (QWL):** QWL is defined in terms of students' perception and belief that using the technology will improve their quality of work life such as saving expenses when downloading e-journals, or in communication when using email to communicate with their instructors and friends. This construct is based on 5 questions (Q32-36) adapted from the work of (Kripanont, 2007; Zakour, 2004; Srite and karahanna, 2000) and are measured using seven-point Likert scale ranging from 1 (strongly disagree) to 7 (strongly agree).

Section D: this section measures the behavioural intention to use the web-based learning system

- **Behavioural intention (BI):** refers to the degree to which an individual has formulated conscious plans to engage in a given behaviour (Davis, 1989). This construct is based on 3 questions (Q37-Q39) and are measured using seven-point Likert scale ranging from 1 (strongly disagree) to 7 (strongly agree).

**Section E**: this section measures the actual usage of the web-based learning system and also serves the profile description about what the students use the system for.

- **Actual usage:** refers to individual's actual use of web-based learning system (Davis, 1989) and based on 2 questions (Q40-Q41). The first question (Q40) measures how frequently the students use the system and is measured using 6-point Likert scale where 1= less than a month, 2= once a month, 3= a few times a month, 4= a few times a week, 5= about once a day, 6= several times a day. The second question (Q41) measures the students' average daily using of the system and is measured using 6-point Likert scale where 1= almost never, 2= less than 30 minutes, 3= from 30 minutes to one hour, 4= from one hour to two hours, 5= from two to three hours, 6= more than three hours.

This section also includes a subsection about questions related to what the students use the system for. This question (Q42) gather information about the extent to which they use the system to perform the following tasks: 1) lecture note, 2) Announcements, 3) Email, 4) Assessments, 5) course handbook, 6) discussion board, 7) browsing websites, 8) take quizzes, 9) previous exams. These questions are measured using 5-Likert scale ranging from 1 (not at all) to 5 (to a great extent).

**Section F**: this section measures the four moderating cultural factors suggested by Hofstede (1980). The reasons cultural variables were measured at the individual level and not at the national level were discussed and outlined in Chapter 3.

- **Power distance (PD):** refers to the extent to which individuals expect and accept differences in power between different people (Hofstede, 1980). This construct is based on 6 questions (Q43-Q48) adapted from the work of Dorfman and Howell's (1988) and are measured using seven-point Likert scale ranging from 1 (strongly disagree) to 7 (strongly agree).

- **Masculinity/Femininity (MF):** refers to the extent to which individuals traditional gender roles are differentiated (Hofstede, 1980). This construct is based on 6 questions (Q49-Q54) adapted from the work of Dorfman and Howell's (1988) and are measured using seven-point Likert scale ranging from 1 (strongly disagree) to 7 (strongly agree).

- **Individualism/ collectivism(IC):** refers to the extent to which individuals are integrated into groups (Hofstede, 1980). This construct is based on 6 questions (Q55-Q60) adapted from the work of Dorfman and Howell's (1988) and are measured using seven-point Likert scale ranging from 1 (strongly disagree) to 7 (strongly agree).

- **Uncertainty Avoidance (UA)**: refers to the extent to ambiguities and uncertainties are tolerated (Hofstede, 1980). This construct is based on 5 questions (Q61-Q65) adapted from the work of Dorfman and Howell's (1988) and are measured using seven-point Likert scale ranging from 1 (strongly disagree) to 7 (strongly agree).

## 4.8.2  Methods to Achieve high rates of response

According to Manfreda *et al.* (2002), there are many reasons that may affect the response rate and thus cause refusal from potential participants to help filling the questionnaire such as length of the questionnaire, asking un-interesting questions and difficult or sensitive questions.

Therefore, the following steps were followed in this research in order to enhance the response rate and eliminate non-response bias:

- The items within the questionnaire were measured either as nominal or 7-point Likert scale so participants can focus on the questions.
- The questionnaire uses easy and simple language and avoids the use of open-ended questions. For example, personal and demographic information were put at the first part of the questionnaire so students will be encouraged to take place in the study.
- In order to encourage participation and engage curiosity, an interesting covering letter explaining the purpose and impact of the study were provided to each participant prior to his/her participation. It also indicates that their personal information will remain strictly confidential.
- Keeping in mind the complexity of the proposed model (74 items), the researcher produced a short and concise questionnaire and also avoided the use of dull or uninteresting questions.

The questionnaire was distributed in Lebanon and England to a total number of 2000 students (1000 within each country) between the period of December 2010 and February 2011. Out of the 1000 questionnaires distributed in Lebanon, 640 questionnaires were filled by students.

## 4.8.3  Pre-testing the questionnaire

Pre-testing is considered to be critical and essential part of the questionnaire design in order to provide valid, reliable and unbiased results and to detect any potential problems in the questionnaire such as difficulty, time, wording and also see how it works and whether changes are necessary before the start of the actual survey (Sekaran and Bougie, 2011; Creswell, 2008).

In this study, two stages of pre-test took place before producing the final version of the questionnaire, one with the academics (Expert knowledge) and the other with potential participants.

The researcher sought the help of academic experts from Lebanon and England in order to obtain content validity. The experts that participated in the pre-testing stage in England were 5 academics who were mainly in the department of IS at Brunel university (Professor XiaoHui Liu, Professor Panos Louvieris, Dr. Juile Barnett, Dr Kate Hone) and one academic from the Business school (Dr. Maged Ali), whereas in Lebanon 2 experts were participated (Dr. Abbass Tarhini, lecturer at LAU and Dr. Anwar Tarhini, Head of Business department at Islamic University). This procedure was critical in order to overcome the different cultural terminologies and avoid collecting useless data (Saunders *et al.*, 2009). The meetings were mainly focused on refining the questionnaire items in order to delete or even generate additional ones, and also to check the relevancy and accuracy of the items in answering the purpose it was designed. A very constructive feedback and suggestions were provided by the experts which led the researcher to revise the list of questions 5 times before producing the final version of the questionnaire. After adjustments to the instrument and final approval were obtained from the experts, it was critical to pre-test the questionnaire to a sample that is expected to respond in a similar way in each cultural context (Douglas and Craig, 2006).

For the second stage of pre-testing, the questionnaires were distributed to 32 students studying at Brunel University from different disciplines (10 PhD, 8 MSc and 14 undergraduates). 13 of these students were Syrian and Lebanese students to ensure the readability and easiness of the questionnaire items in the Lebanese context. Twenty one questionnaires were returned which indicate a high response rate (65%). This stage was critical to capture potential wording ambiguities, difficulty, length and timing of the questionnaire. The results of this stage is considered a success, for example, one female student mentioned that she has a problem to answer her age in years and suggested that it is better to present age in different interval groups, another respondent highlighted some potential problem with wording one of the questions (Q18), the researcher then tried to simplify the question and make it clear and easy as possible. In general, the results indicated that the questionnaire items were clearly worded and easy to understand. Therefore, preliminary support for reliability and validity of the questionnaire was obtained before proceeding to the data collection stage. Again, after the adjustments to the instrument, copies of the questionnaire were distrusted for pilot test.

## 4.9   Instrument scale measurement

This study developed a questionnaire technique incorporating nominal and ordinal scale (see Appendix A). Nominal scales were mainly used to determine the participants' demographic characteristics such as nationality, age, gender, educational level and experience. Likert scales were used to measure the participants' beliefs and opinions towards technology acceptance in e-learning environment. This scale was first developed by Rensis Likert in 1932 (Likert, 1932) and provides a sequential point scale that varies from 1 to 10 separated by equal intervals. In order to allow participants to indicate their agreement or disagreement with a certain statements or questions related to their attitudes and beliefs, the 7-point Likert scales were used to allow varieties in the answers as participants in this study share a lot of similarities in their characteristics. In addition, this scale is widely used by many scholars in the IS and social science literature (Davis, 1989; Venkatesh and Davis, 2000; Dorfman and Howell, 1988)

.

## 4.10 Pilot Study Results

It was essential to pilot test the questionnaire prior to its use within this study in order to examine the validity and reliability of the instrument and to improve questions, format and scales (Creswell, 2008). A pilot study was conducted in England and Lebanon before the actual questionnaires were distributed. The main purpose of the pilot study was to ensure the readability, clarity, and easiness of the questionnaire items and to check if the data collected answers the investigated questions and provide face validity (Presser *et al.*, 2004). The researcher then analyses the data to discover any drawbacks or potential threats within the questionnaire items and thus decisions can be made regarding the items to be deleted, kept or even added.

The questionnaire were distributed to a convenience sample of 50 (N=100) students in each country with a duration of two weeks. Of which 74 were returned which indicate a high response rates (74%). The number of usable questionnaires was 65 representing the average targeted sample in terms of age, gender etc were analysed. The completion time of the questionnaire was 12 minutes which is relatively reasonable. Based on the suggestions of the respondents and the results of the basic statistical analysis, the researcher removed and modified some questions. The result of the pilot study is presented in Chapter 5 (Section 5.2).

## 4.11 Data Analysis

For this research, the collected data was analysed into 2 different stages. In the first stage, SPSS 18.02 version was used for the purpose of descriptive statistics about the respondents and the preliminary data analysis (see Chapter 5) such as missing value, outliers and extreme values, mean and standard deviation, multicollinearity and Skewness. While in the second stage Structural Equation Modelling (SEM) were used to test and examine the relationships among variables within the proposed conceptual model (see Chapter 6 and 7). This section briefly describes and justifies the use of SEM as the main data analysis technique used in the research.

## 4.11.1  Structural Equation Modelling (SEM)

Due to its wide and general acceptance among researchers in IS, behavioural and social science (Blunch, 2008; Janssens *et al.*, 2008; Gefen *et al.*, 2000), the structural equation modelling (SEM), also known as path analysis, covariance structure analysis, simultaneous equation models,  is used to test and examine the hypothesised relationships among variables within the proposed conceptual model. SEM; as an example of second generation of multivariate analysis, which differ greatly from first-generation techniques such as factor analysis, discriminant analysis or regression; is a statistical technique for *simultaneously* testing and estimating a set of hypothesised relationships among multiple independent and dependent variables (Gefen *et al.*, 2000). Similarly, Hair et al. (2010) define SEM as a multivariate technique, which combines features of multiple regression and factor analysis in order to estimate a multiple of networking relationships simultaneously. Thus, SEM helps the researcher to answers a set of interrelated research questions in a single, systematic, and comprehensive analysis (Gefen *et al.*, 2000). According to Tabachnick and Fidell (2007), SEM is mostly used to generate theories and concepts. One of the most important steps in structural equation modelling is assessing whether a specified model 'fits' the collected data or not (Yuan, 2005). SEM has the ability to model theoretical constructs that are hard or impossible to measure directly.

In the context of our study, the selection of SEM as the main analysis technique was based on the following reasons:

- Structural equation modelling is more appropriate than other statistical technique when one exogenous (dependent) variable becomes an endogenous (independent) variable (Tabachnick and Fidell; 2000). The Behavioural Intention (BI) latent factor will act as an endogenous variable that affect the actual usage of the e-learning system. BI will affected by the main determinants of the proposed research model and thus will act as exogenous variable. In this case, the model will be tested simultaneously. However, a large number of multiple analyses would be required when using first generation statistical tool.

- The proposed conceptual model aim to contribute to understanding the e-learning acceptance in the context of developing and developed countries which considered a complex model, and thus scarifies the parsimony. Using first generation statistical tools are not applicable to test complex modelling whereas SEM is more valuable when testing complex mathematical models (Gefen et al, 2000).

- This research will test a set of hypothesized relationships within the constructs of the proposed research model which is more suitable for SEM as it employs confirmatory modelling strategy (Tabachnick and Fidell, 2000).

According to Hair *et al*. (2010), there are 6 stages in SEM decision process (see Figure 4.2); "*1) Defining individual constructs, 2) Developing the overall measurement model, 3) Designing a study to produce empirical results, 4) Assessing measurement model validity, 5) Specifying the structural model, and 6) Assessing structural model validity*" (Hair et al. 2010, p. 654).

The first 4 stages are usually covered within the *measurement model* while the last 2 stages usually covered in the *Structural model*. The use of the 6-stages in SEM techniques is heavily discussed in Chapter 5, 6 and 7.

There are two families of SEM: 1) Covariance-based modelling using software such as LISREL, Mplus, AMOS and EQS and 2) Variance-based modelling – partial least squares (PLS) (Gefen *et al.*, 2000). The covariance-based SEM is appropriate when the main objective of the research is theory testing and confirmation, while PLS-SEM is more appropriate when main objective of the research is prediction and theory development. For the current study, Analysis of Moment Structures (AMOS version 18.0), a covariance-based SEM approach is used to examine and analyse the data within the proposed model.

As discussed in previous paragraph, this study follows Hair's (2010) recommendations about evaluating the structural model using a two-steps approach (first the measurement model and then the structural model). Additionally, Multiple Group Analysis (MGA) technique is used to measure the impact of moderators on the conceptual model. The next two Chapters provide a detailed explanation about employing SEM in this research.

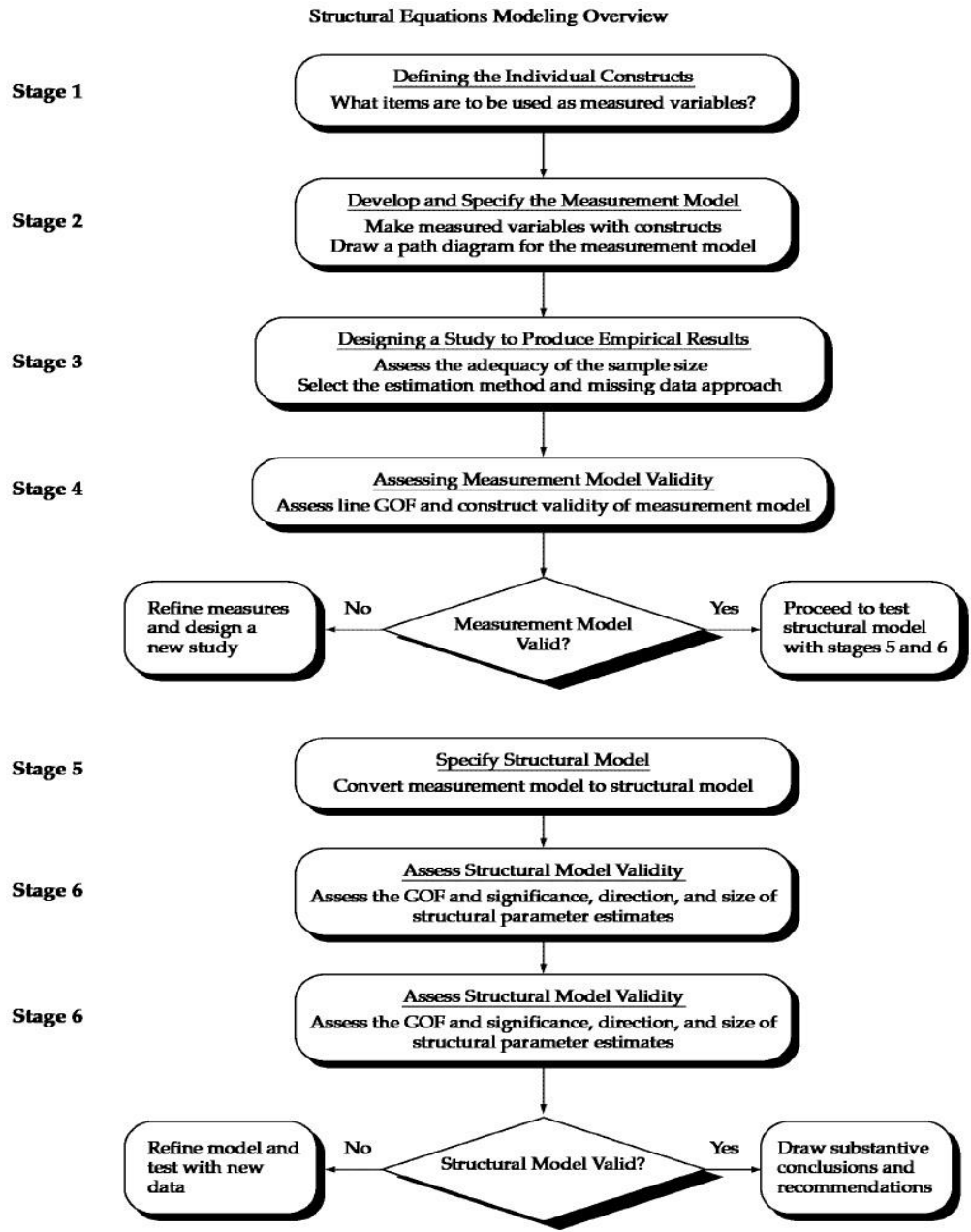

**Figure 4-2: the six-stage Process for structural equation modelling used in this research (Source: (Hair et al., 2010).**

# 4.12 Ethical considerations

The ethical consideration is highly critical in any research and more specifically in the ones that aims to study the social behaviour of participants (Hesse-Biber and Leavy, 2010). For the present study, based on Brunel university guidelines about ethical considerations when collecting the data, all participants were asked to sign a printed consent form which indicate their rights to withdraw from the study at any time if they choose to, and that their participation is voluntary, and also that confidentiality and anonymity will be guaranteed at all stages of the research. This procedure was critical for the research validity as participants can answer the questionnaire with complete honesty. This research also considers other ethical issues such as the role of the researcher after the data collection procedure especially at the data analysis stage (Sekaran and Bougie, 2011). A copy of the cover letter is included in Appendix-B.

Additionally, a cover letter was provided to all the participants. This letter included the title, purpose and impact of the study, in addition to length and time of the questionnaire, and also the contact details of the researcher and the supervisor if there are any later questions, additionally the contact details of the ethical committee at Brunel University regarding the ethical elements of the research.

An approval from the Department of Information System and Computing at Brunel University was received before the start of the data collection stage.

## 4.13 Conclusion

The main aim of this chapter was to present and justify the philosophical perspectives, approaches, methods and statistical techniques used in this research to achieve the main research objectives and to answer the research questions as well as pilot study.

This research employed a quantitative method in order to understand and validate the conceptual framework (see Chapter 3). A survey research approach, which is based on positivism to guide the research, was found to be best appropriate for the research.

A questionnaire was used as a data collection technique and a detailed account of the various stages including the development, scale, pre-test and pilot study of the questionnaire was offered. In this study, two stages of pre-test took place before producing the final version of the questionnaire, one with the academics (Expert knowledge) and the other with potential participants.

In this chapter, a detailed explanation of sampling size and different techniques were provided with a justification for the selection of convenience non-probability sampling technique.

The main data analysis technique adopted in this research is based on Structural Equation Modelling (SEM) using Analysis of Moment Structures (AMOS) version 18.0. A justification for the selection of this technique due to the complexity of the conceptual model was provided. CFA and the structural model are the two-step approach in the SEM analysis that applied in this study to examine and test the relationships among independent and dependent variables. Finally, the chapter considers the ethical issue related to this research.

The next chapter (Chapter 5) presents the preliminary data analysis including the pilot study results and sample screening.

# Chapter 5:  Preliminary Data Analysis

*"The competent analysis of research-obtained data requires a blending of art and science, of intuition and informal insight, of judgement and statistical treatment, combined with a thorough knowledge of the context of the problem being investigated" (Green, Tull and Albaum, 1988: 379)*

## 5.1  Introduction

Chapter 3 described the proposed conceptual model that is used to examine the impact of culture and demographic characteristics on the adoption and acceptance of e-learning technologies in Lebanon and England. In Chapter 4, the research method that guided the research was discussed and justified the choice of the survey research approach to test the hypotheses and thus answer the research questions.

This chapter presents the preliminary data analysis of the data obtained from the respondents.  The Statistical Package for the Social Science (SPSS) version 18.0 was employed for preliminary data analysis including data screening, frequencies and percentages, reliability analysis, exploratory factor analysis and t-tests. The presentation of the results from the data analysis in this chapter focuses on the cross-cultural differences between Lebanon and England and also investigates the different tasks that students perform using the e-learning systems.

This chapter comprises of 8 sections. The next section describes the results of the pilot study to ensure the validity and reliability of the measuring instruments to be used in testing the hypotheses. Section 5.3 then presents the results of the preliminary examination of the main study. This is followed by presenting the results of the reliability analysis for the main study in Section 5.4. Section 5.5 describes the demographic profile of the respondents. Following that, a detailed descriptive analysis of all the constructs included in the proposed research model is provided in Section 5.6 and 5.7 respectively. This is followed by results of the t-test analysis at a cross-cultural level in Section 5.8. The tasks that students perform using the e-learning systems are illustrated in section 5.9. Finally, Section 5.10 presents the summary and conclusions of this chapter.

## 5.2   Pilot Study Results

It was essential to pilot test the questionnaire prior to its use within this study in order to examine the validity and reliability of the instrument and to improve the questions, format and scales (Creswell, 2008). A pilot study was conducted in England and Lebanon before the actual questionnaires were distributed. The main purpose of the pilot study was to ensure the readability and clarity of the questionnaire items and to check if the data collected answers the investigated questions and provide face validity (Presser et al., 2004; Sekaran and Bougie, 2011; Zikmund, 2009). It is worth noting that all the items (questions) used by this research have been drawn from the literature, where they were quoted to be reliable and valid to measure constructs of the phenomena that they intend to represent. More specifically, all of the items were used as part of questionnaires in studies investigating how individual differences affect users' perception and behaviour in the context of technology acceptance. The fact that the items were developed for and tested within similar contexts to the current study supports their applicability here. The researcher seeks the help of academic experts from Lebanon and England in order to obtain content validity (refer to Chapter 4, Section 4.8.3). Based on the feedback and suggestions from the potential participants (students), very minor changes were suggested on the questions wording and the questionnaire layout by the respondents, and thus face validity was established. The researcher then analysed the data to discover any drawbacks or potential threats within the questionnaire items and thus decisions can be made regarding the items to be deleted, kept or even added.  Below are some examples of the minor changes:

1) Although some of the measurement scales for some constructs were valid, they were not relevant in the current study. For example, we removed the Voluntariness construct as most of the participants believe that it is mandatory to use the system during their learning process.

2) The term 'Web-based learning system' was used instead of 'e-learning tools'

3) "A specific person was available to provide assistance" instead of "someone was available to provide assistance"

The questionnaires were distributed to a convenience sample of 50 (N=100) students in each country within a duration of two weeks. According to Nargundkar (2003), the sample size for the pilot study should be relatively small (up to 100) but representative for the population being investigated. Of the 100 questionnaire being distributed, 74 were returned which indicates a high response rates (74%). The number of usable questionnaires was 65, which were analysed. The completion time of the questionnaire was 12 minutes which is relatively reasonable. The next step after content validity was the reliability check.

The reliability refers to the consistency of a measure used within the research (Sekaran and Bougie, 2011). A test is considered reliable if we are able to get the same results when we repeat the same research with different samples, assuming we provide the same initial conditions for the test (Last and Abramson, 2001). The reliability of the constructs in this research was checked using Cronbach's Alpha (Cronbach, 1951). Cronbach's Alpha measures how well a set of items measures a single unidirectional latent construct. Different reliability values were considered satisfactory by different researchers. For instance, it should be at least 0.7 according to (DeVellis, 2003; Robinson et al., 1991) or  0.6 is considered satisfactory while a value of 0.8 or higher is preferred according to (Nunnally, 1970). In other words, if Cronbach Alpha gets closer to 1.0, this means that the constructs have high reliability. SPSS was used to analyse the reliability tests of the pilot study which are presented in table 5.1.

| Factor | Number of Items | Cronbach Alfa | Inter-Item Correlation | Item-to-total correlation |
|--------|-----------------|---------------|------------------------|---------------------------|
| **PU** | 5 | .951 | .7 - .88 | .798 -.894 |
| **PEU** | 5 | .962 | .754 - .935 | .813 -.95 |
| **SE** | 6 | .891 | .432 - .817 | .649 -.761 |
| **FC** | 4 | .866 | .422 - .94 | .572 - .792 |
| **SN** | 4 | .855 | .514 - .661 | .652 - .758 |
| **QWL** | 5 | .926 | .591 - .825 | .74 - .931 |
| **BI** | 3 | .932 | .795 - .849 | .853 - .893 |
| **AU** | 2 | .792 | .721 - .721 | .721 - .721 |
| **PD** | 6 | .865 | .336 - .83 | .475 - .758 |
| **MF** | 6 | .922 | .54 - .8 | .662 - 861 |
| **IC** | 6 | .879 | .366 - .78 | .621 - .757 |
| **UA** | 5 | .925 | .595 - .801 | .777 - .837 |

**Table 5-1: Cronbach's Alpha, Inter-item correlation for the pilot study**

The results in Table 5.1 suggest that the constructs had adequate reliability, with a score ranging from 0.792 for AU to .962 for PU. This means that the items related to each construct used in the proposed model were positively correlated to one another.

Table 5.1 also presents the results of anther two internal consistency reliability indicators, namely inter-item correlation and item-to-total correlation. Hair et al. (2010) suggests that the values should exceed 0.5 for the item-to-total correlation and 0.3 for inter-item correlation. The results of the pilot study exceeded the cut-off value for all the constructs used in the questionnaire except for PD. However, after examining each item in PD, it was obvious that question 'It is frequently necessary for instructors to use authority and power when dealing with students' showed a lower item-to-total correlation (0.475) than cut-off value 0.5 (Hair et al., 2010). Therefore, based on the results of the basic statistical analysis, PD6 has been removed from the questionnaire. Additionally, to avoid the misleading answers and to check the seriousness of the respondents in answering the questions a Mann-Whitney-U-test were performed on the first construct (PU) and compared to the last construct (UA) of the questionnaire. The process was to run the test on groups belonging to the same category; in this case the Gender was used.

As can be shown in Table 5.2, there is enough evidence to conclude that no statistical significant difference between male and female on the PU and UA were found and the significance for all the items were above 0.5 probability value (Pallant, 2010).   Therefore, there was no need for other versions of the questionnaire. Moreover, when comparing the Z-score for all the items for PU and UA, we detected that none of construct' items are totally higher than the other (e.g, PU1 > UA1, PU2> UA2 etc…). Therefore, we assumed that the participants within the pilot study did accept the length of the questionnaire.

|  | PU1 | PU2 | PU3 | PU4 | PU5 |
|---|---|---|---|---|---|
| **Mann-Whitney U** | 313.500 | 333.000 | 254.000 | 329.500 | 297.000 |
| **Wilcoxon W** | 566.500 | 586.000 | 507.000 | 582.500 | 550.000 |
| **Z** | -.876 | -.524 | -1.913 | -.588 | -1.153 |
| **Asymp. Sig. (2-tailed)** | .381 | .600 | .056 | .557 | .249 |
|  | UA1 | UA2 | UA3 | UA4 | UA5 |
| **Mann-Whitney U** | 342.500 | 315.000 | 323.000 | 358.500 | 342.000 |
| **Wilcoxon W** | 903.500 | 568.000 | 884.000 | 919.500 | 903.000 |
| **Z** | -.360 | -.853 | -.705 | -.079 | -.368 |
| **Asymp. Sig. (2-tailed)** | .719 | .393 | .481 | .937 | .713 |

**Table 5-2: Grouping Variable: Gender**

Having discussed the results of the pilot study, the next section examines and discusses the results of the main study including data screening, normality, homogeneity and multicollinearity.

## 5.3   Preliminary Examination of the Main study

The aim of the preliminary examination of the data is to detect missing data, outliers, as well as normality and Homogeneity of the data in the two data files through SPSS statistical package and AMOS 18.0. This process is very important in preparing the collected data for final analysis later on. The next section presents the findings and treatment of the above identified tests for both samples.

## 5.3.1 Data screening

All the questionnaires used in this research were screened for any missing answers before the data entry. Although this process was simple but it was very critical to facilitate data entry. Furthermore, a check at the descriptive statistics for each item was undertaken in order to ensure the accuracy of data. In this regards, we compared the answers of questions that produced out of range values with the original questionnaires for more accuracy.

## 5.3.2 Missing Data

According to Hair et al (Hair *et al.*, 2010), missing data is considered one of the most continuing problems in data analysis that may affect the results of the research objectives. The impact of missing data is even more critical when using Structural Equation Modelling in AMOS (Arbuckle, 2009). For example, Chi-Square and other fit measures such as Goodness-of-Fit-Index and also modification indices (refer to Chapter 6) cannot be computed if there are any missing data in the sample.

Furthermore, it is important to determine the type of missing values to know whether the missing data were occurring randomly or non-randomly (Pallant, 2010). In this regards, if the missing values are non-randomly distributed within the items of the questionnaire, then such data can be ignored. However, if the missing values are non-randomly distributed, then the generalizability of the results will be affected (Tabachnick and Fidell, 2007).

Schumacker and Lomax (2004) suggests that missing data up to 5% is considered acceptable. After the initial screening in the SPSS 18.0, it was found that all missing data for both samples were distributed in a random manner and were below the 5 % threshold. The maximum percentage of missing data was 2.5 and 2.4 for the British and Lebanese samples respectively. This percentage of missing data is very low and can be considered acceptable. Therefore, the researcher applied the 'mean substitution' method to replace missing data for the categorical variables while missing data for nominal variables were excluded later during the multi-group analysis, as suggested by many scholars e.g.,(Pallant, 2010; Byrne, 2006; Arbuckle, 2009). The frequency and percentage of missing data for both samples is presented in Table1 and 2 in Appendix-C.

### 5.3.3  Outliers

According to (Hair et al, 2006, p.73), an outlier is defined as "observations with a unique combination of characteristics identifiable as distinctly different from the other observation". Therefore, detecting and treating outliers is critical since it may affect the normality of the data and can seriously distort statistical tests (Tabachnick and Fidell, 2007). In this regards, Tabachnick and Fidell (2007) suggest that extreme outliers should be deleted while keeping the mild-outliers. Hair et al (2006) identified two methods to detect the outliers: Univariate outliers and multivariate outliers.

In the current study, SPPS was used to identify the univariate outliers for the two data files by determining frequency distributions of z-score, as suggested (Kline, 2010). While there are no specific rules to identify extreme values in literature, for a large sample (more than 80) a value up to $\pm$ 3.29 can be accepted.  The decision was to delete the row that has more has two univariate outliers from the dataset. Accordingly, three rows (222, 455, and 143) were deleted from the Lebanese sample and two outliers from the British sample (903 and 1197). Table 3 and 4 in Appendix D presents the results of the univariate outliers for both samples.

The other type of outliers is known as multivariate outliers, this test involves observation and analysis of more than one statistical outcome variable at a time. In the current study, we used Mahalanobis $D^2$ measure to determine the multivariate outlier (Hair *et al.*, 2010; Kline, 2010). Mahalanobis $D^2$ measure the distance of a particular case from the centroid of the remaining cases. In this research, Mahalanobis $D^2$ was measured using AMOS version 18.0. For all records that p1 value < 0.05 would consider influential outlier and that the correlation between the variables for these responses are significantly different or abnormal comparing to the rest of the dataset (Tabachnick and Fidell, 2007). Twelve and fifteen multivariate outliers were detected in the British and Lebanese sample respectively. However, the researcher retained the outliers to the dataset because they were not found to be problematic due to their limited number compared to the whole database and so were suitable to be included in further analysis (Hair *et al.*, 2010). Results of multivariate outliers for both samples are shown in Table 5 and 6 in Appendix D.

### 5.3.4 Testing the normality assumption

According to Hair et al. (2010), testing the presence of normality is essential in multivariate analysis. In other words, if the data is not normally distributed then it may affect the validity and reliability of the results.

In the current study, we employed Jarque-Bera (skewness-Kurtosis) test to check whether the data is normally distributed or not. The skewness value indicate the symmetry of the distribution (Pallant, 2010). A negative skew indicates that the distribution is shifted to the right; whereas positive skew indicates a shift to the left. Kurtosis provide information about the height of the distribution (Pallant, 2010). The positive kurtosis value indicates a peaked distribution; whereas a negative value indicates a flatter distribution. According to Tabachnick and Fidell (2007), the normal range for skewness-kurtosis value is $\pm 2.58$. Following this recommendation, all the items in the dataset for both samples were found to be normally distributed (i.e, $< \pm 2.58$). More specifically the skewness and Kurtosis value in each case was in the range of $\pm 1$ which is considered negligible. Table 7 and 8 in Appendix E shows the means, standard deviation, skewness and kurtosis values for each variable. This confirms that there was no major issue of non-normality of the data.

## 5.3.5 Homogeneity of Variance in the Dataset

Tabachnick and Fidell (2007), state that homogeneity is "the assumption of normality related with the supposition that dependent variable(s) display an equal variance across the number of independent variable (s)". According to Hair et al (2010), it is essential to determine the presence of the homogeneity of variance within multivariate analysis as it might lead to incorrect estimations of the standard errors. In this study, Levene's test in SPSS 18.0 was used to determine the presence of homogeneity of variance in the data (see Table 5.3) using (gender) as a non-metric variable in the t-test. The results revealed that most of the constructs were non-significant (i.e. p>0.05) except PU and IC for British sample and PEOU for the Lebanese sample. The results confirmed the homogeneity of variance in the data and suggest that variance for all the variables were equal within groups for male and female.

| | England | | | | Lebanon | | | |
|---|---|---|---|---|---|---|---|---|
| | Levene Statistic | df1 | df2 | Sig. | Levene Statistic | df1 | df2 | Sig. |
| **PU** | 4.847 | 1 | 600 | .032 | 2.800 | 1 | 564 | .095 |
| **PEOU** | 1.201 | 1 | 600 | .274 | 4.512 | 1 | 564 | .034 |
| **SE** | 3.035 | 1 | 600 | .082 | .029 | 1 | 564 | .866 |
| **FC** | 3.269 | 1 | 600 | .071 | .626 | 1 | 564 | .429 |
| **SN** | 1.353 | 1 | 600 | .245 | 3.206 | 1 | 564 | .074 |
| **QWL** | 3.197 | 1 | 600 | .074 | 3.637 | 1 | 564 | .057 |
| **MF** | .926 | 1 | 600 | .336 | .132 | 1 | 564 | .716 |
| **IC** | 4.216 | 1 | 600 | .043 | .259 | 1 | 564 | .611 |
| **PD** | .155 | 1 | 600 | .694 | 1.374 | 1 | 564 | .242 |
| **UA** | .871 | 1 | 600 | .351 | .085 | 1 | 564 | .771 |

**Table 5-3: Test of Homogeneity of Variances**

### 5.3.6 Multicollinearity

According to (Pallant, 2010), multicollinearity occurs when two or more variables are highly correlated to each other. Different values were suggested to be satisfactory by different scholars. For instance, correlations up around 0.8 or 0.9 is considered highly problematic according to (Tabachnick and Fidell, 2007) while a value 0.7 or higher is considered reason for concern according to (Pallant, 2010). The presence of multicollinearity is determined by two values: tolerance and VIF (Variance Inflation Factor) (Pallant, 2010). If the value of tolerance is greater than 0.10 and VIF value less than 3.0, then there is no multicollinearity. Given all the independent constructs had VIP value less than 3.0 and tolerance value above 0.10 (see Appendix F) this suggests that the absence of multicollinearity in both samples. After completion of the data screening and testing for multivariate normality, the data was investigated further using SPSS version 18.0.

This section discussed the results of the preliminary examination of the two data files through SPSS statistical package and AMOS 18.0 including the detecting of the missing data, outliers, as well as normality, homogeneity and multicollinearity. The next section discusses the results of reliability tests of the two data sets.

## 5.4 Reliability

Similar to pilot study, the reliability of the constructs in the main study was checked by Cronbach's Alpha (Cronbach, 1951). SPSS was used to analyse the reliability tests of the main study which are presented in Table 5.4. The results suggest that the constructs had adequate reliability, with a score ranging from 0.705 for AU to .936 for UA for the British sample, whereas the lowest score was .655 for the AU and the highest was .929 for PEOU within the Lebanese sample. This means that the items related to each construct used in the proposed model were positively correlated to one another (Hair et al., 2010).

| Factor | Items | England | | | Lebanon | | |
|---|---|---|---|---|---|---|---|
| | | Cronbach Alfa | Inter-Item Correlation | Item-to-total correlation | Cronbach Alfa | Inter-Item Correlation | Item-to-total correlation |
| **PU** | 5 | .922 | .642 - .765 | .779 -.824 | .903 | .544-.744 | .681-.799 |
| **PEOU** | 5 | .923 | .647 - .786 | .736 -.837 | .929 | .625-.798 | .774-.837 |
| **SE** | 5 | .841 | .513 - .827 | .518 -.683 | .768 | .273-.831 | .487-.816 |
| **FC** | 4 | .881 | .525 - .867 | .661 -.776 | .9 | .595-.805 | .707-.818 |
| **SN** | 4 | .833 | .445 - .666 | .55 -.734 | .813 | .387-.664 | .586-.726 |
| **QWL** | 5 | .889 | .554 - .728 | .701 -.767 | .832 | .376-.656 | .519-.718 |
| **BI** | 3 | .893 | .701 - .813 | .737 -.821 | .864 | .607-.788 | .663-.797 |
| **AU** | 2 | .705 | .549 - .549 | .549 -.549 | .655 | .551-.551 | .551-.551 |
| **PD** | 5 | .896 | .521 - .693 | .694 -.792 | .899 | .460-.677 | .650-.772 |
| **MF** | 6 | .87 | .383 - .723 | .612 -.757 | .886 | .557-.687 | .678-.761 |
| **IC** | 6 | .861 | .394 - .687 | .548 -.718 | .851 | .394-.757 | .512-.732 |
| **UA** | 5 | .936 | .693 - .803 | .807 -.836 | .89 | .511-.706 | .683-.769 |

**Table 5-4: Cronbach's Alpha, Inter-item correlation for the British and Lebanese sample**

Table 5.4 also presents the results of anther two internal inconsistency reliability indicators, namely inter-item correlation and item-to-total correlation. Hair et al. (2010) suggests that the values should exceed 0.5 for the item-to-total correlation and 0.3 for inter-item correlation. The results of the main study exceeded the cut-off value for all the constructs used in the questionnaire except for SE within the Lebanese sample. After examining each item in SE construct, we found that SE5 and SE6 showed a lower item-to-total correlation (0.487) than cut-off value 0.5 (Hair et al., 2010). However, the researcher retained these items as they will be checked again during the structural equation modeling analysis stage (see Chapter 6), this will help the researcher to understand the complete picture of the lower correlation of these two items before their deletion.

## 5.5   Profile of respondents

The target sample for this survey was British and Lebanese students that use e-learning systems provided by their university. These students were full time students studying for Masters or undergraduate degrees from one university in England located in London, and two universities in Lebanon located in Beirut. A total of 2000 questionnaires were distributed to 1000 students from the UK and 1000 from Lebanon respectively, of which 1197 were returned indicating a 59.7% response rate overall. After screening for missing data and duplicated responses, we retained 1168 questionnaires for data analysis. These included 566 Lebanese participants and 602 British participants (see Table 5.5).

|  | **Frequency** | **Percent** |
|---|---|---|
| **Lebanon** | 566 | 48.5% |
| **England** | 602 | 51.5% |
| **Total** | 1168 | 100.0 |

**Table 5-5: frequency and percentage of respondents**

The demographic information for both samples is discussed next.

|  | **England** | | **Lebanon** | |
|---|---|---|---|---|
|  | Frequency | Percent | Frequency | Percent |
| **Male** | 315 | 52.3% | 305 | 53.9% |
| **Female** | 287 | 47.7% | 261 | 46.1% |

**Table 5-6: frequency and percentage of respondents in terms of their gender**

Within the British sample, there were 315 (52.5%) male respondents and 287 (47.7%) female respondents. While within the Lebanese sample there were 305 (53.9%) male and 261 (46.1%) female respondents. The proportion of male and female respondents was almost adequately distributed. The above table (5.6) presents the gender category including the frequencies and percentage of each category. Next, the respondents' age were analysed in Table 5.7 below.

|  | **England** | | **Lebanon** | |
|---|---|---|---|---|
|  | Frequency | Percent | Frequency | Percent |
| **17-22** | 370 | 61.5% | 409 | 72.3% |
| **>22** | 232 | 38.5% | 157 | 27.7% |

**Table 5-7: frequency and percentage of respondents' age group**

In terms of age, the majority of the respondents for both samples were in the age group 17-22 years old, with 370 (61.3%) and 409 (72.3%) within the British and Lebanese respectively. While respondents belonging to the older group were 232 (38.55) and 157 (27.7%) within the British and Lebanese respectively. Next, the respondents' educational levels were analysed in Table 5.8 below.

|  | **England** | | **Lebanon** | |
|---|---|---|---|---|
|  | Frequency | Percent | Frequency | Percent |
| **Undergraduate** | 347 | 57.6% | 364 | 64.3% |
| **Postgraduate** | 255 | 42.4% | 202 | 35.7% |

**Table 5-8: frequency and percentage of respondents' educational level**

From the educational level perspective, similar to the age, the majority of respondents for both samples were undergraduate students with 347 within the British sample and 364 within the Lebanese sample, while the rest were studying for their master degree. Next, the respondents' internet experience were analysed in Table 5.9 below.

|  | **England** | | **Lebanon** | |
|---|---|---|---|---|
|  | Frequency | Percent | Frequency | Percent |
| **Some Experience** | 192 | 31.9% | 225 | 39.5% |
| **Experienced** | 410 | 68.1% | 344 | 60.5% |

**Table 5-9: frequency and percentage of respondents' web experience**

The category internet experience revealed that the majority of the students within the two samples were experienced in using the internet, with 410 (68.1%) and 341 (60.2%) within the British and Lebanese sample respectively. The lowest group include the students who have some experience in using the internet with 192 British respondents and 225 Lebanese respondents. Next, the respondents' computer skills were analysed in Table 5.10 below.

| | England | | Lebanon | |
|---|---|---|---|---|
| | Frequency | Percent | Frequency | Percent |
| **Novice** | 96 | 15.9% | 88 | 15.5 % |
| **Moderate** | 279 | 46.3% | 253 | 44.7 % |
| **Expert** | 227 | 37.7% | 225 | 39.8% |

**Table 5-10: frequency and percentage of respondents' computer experience**

The results revealed that the majority of respondents were found moderate on computer skills, with 46.3% and 44.7% respondents within the British and Lebanese sample respectively. This was followed by 37.7% British respondents and 39.8% Lebanese respondents who were found to be expert in using the computer. Finally, there were only around 15% of the respondents who evaluated themselves as novice in using the computer. Next, the numbers of courses using e-learning tools that the respondents have used within the current academic year were analysed.

| | England | | Lebanon | |
|---|---|---|---|---|
| | Frequency | Percent | Frequency | Percent |
| **1-2** | 121 | 20.1% | 103 | 18.2% |
| **3-5** | 232 | 38.5% | 278 | 43.8% |
| **>5** | 249 | 41.4% | 215 | 38.0% |

**Table 5-11: frequency and percentage for number of courses delivered using e-learning tools**

The results in Table 5.11 revealed that the majority of respondents (41.4%) within the British sample were studying more than five courses using e-learning tools for the current academic year, this was followed by 38.5% that fulfil the 3-5 courses intervals, whereas 20.1% fulfil the interval 1-2 courses. While within the Lebanese sample, the majority of respondents (43.8%) fulfil the interval 3-5 courses, this is followed by 38.0% fulfil the interval >5, and the lowest one with 18.2% was the first interval (i.e., 1-2 courses).

## 5.6   Descriptive statistics of construct items

The descriptive statistics including the means and standard deviation for each independent and dependent variable used in the proposed research model are presented in the following subsections. Overall, all means were greater than 4.42 for the British sample (N=602) and 4.21 for the Lebanese sample (N=566) which indicate that the majority of participants express generally positive responses to the constructs that are measured in this study. The standard deviation (SD) values showed a narrow spread around the mean. The descriptive statistics for both samples is discussed next.

### 5.6.1  Perceived ease of use (PEOU)

The PEOU construct is conceptualised, in this thesis, to extract the information about the students' belief that using the e-learning tools is easy to use and understand. This variable is measured by 5 items adopted from the work related to TAM (Davis, 1989; Ngai *et al.*, 2007; Pituch and Lee, 2006) and was measured using a  7 point Likert scale, ranging from 1 - strongly disagree to 7- strongly agree. The results in table 5.12 shows that the mean for the items related to PEOU range between  5.34(±1.264) and  5.44(±1.321)  for  the  British  sample,  and 5.54(±1.237) and 5.8(±1.206) for the Lebanese one. The results indicated that the students found the system easy to use and understand and unexpectedly it was higher in the Lebanese sample.

| | England | | Lebanon | |
|---|---|---|---|---|
| **Item** | **Mean** | **Standard Deviation** | **Mean** | **Standard Deviation** |
| **PEOU1** | 5.41 | 1.377 | 5.80 | 1.206 |
| **PEOU2** | 5.34 | 1.264 | 5.54 | 1.237 |
| **PEOU3** | 5.36 | 1.234 | 5.59 | 1.195 |
| **PEOU4** | 5.40 | 1.257 | 5.67 | 1.173 |
| **PEOU5** | 5.44 | 1.321 | 5.74 | 1.171 |

**Table 5-12: Descriptive statistics of PEOU construct**

## 5.6.2 Perceived usefulness

The PU construct is conceptualised in this thesis to extract the information about the students' belief that using the e-learning tools will improve his/her performance and productivity. Five items were adopted from the work related to TAM (Davis, 1989); (Ngai et al., 2007),(Pituch and Lee, 2006) and was measured using a 7 point Likert scale, ranging from 1 - strongly disagree to 7- strongly agree. As can be shown in Table 5.13, the mean for each item related to PU construct range between 5.17 ($\pm$1.289) and 5.37 ($\pm$1.288) for the British sample, and between 5.02 ($\pm$1.323) and 5.41 ($\pm$1.225) which indicate that the majority of participants agrees that e-learning system is useful in their education process.

| | England | | Lebanon | |
|---|---|---|---|---|
| **Item** | **Mean** | **Standard Deviation** | **Mean** | **Standard Deviation** |
| **PU1** | 5.25 | 1.33 | 5.41 | 1.225 |
| **PU2** | 5.17 | 1.289 | 5.06 | 1.244 |
| **PU3** | 5.37 | 1.288 | 5.36 | 1.288 |
| **PU4** | 5.23 | 1.312 | 5.07 | 1.310 |
| **PU5** | 5.24 | 1.338 | 5.02 | 1.323 |

**Table 5-13: Descriptive statistics of PU construct**

### 5.6.3  Social Norm (SN)

The SN construct is conceptualised, in this thesis, to extract information related to students' perceptions related to e-learning system which usually influenced by others' opinion such as other colleagues/students and lecturers. Four items were adopted from the work of (Park, 2009), (Van Raaij and Schepers, 2008) and (Venkatesh *et al.*, 2003) and were measured using 7-point Likert scale. The results of the descriptive statistics in Table 5.14 shows that the mean for the SN items ranged from 4.62 (±1.535) and 5.14(±1.381) within the British sample, whereas ranged between 5.38(±1.521) and 5.51(±1.039) within the Lebanese sample. The results indicate that students were moderately influenced by their colleagues and instructors.

| | England | | Lebanon | |
|---|---|---|---|---|
| **Item** | **Mean** | **Standard Deviation** | **Mean** | **Standard Deviation** |
| **SN1** | 4.75 | 1.514 | 4.88 | 1.521 |
| **SN2** | 4.62 | 1.535 | 4.38 | 1.419 |
| **SN3** | 4.96 | 1.392 | 5.23 | 1.280 |
| **SN4** | 5.14 | 1.381 | 5.51 | 1.039 |

**Table 5-14: Descriptive statistics of SN construct**

### 5.6.4  Quality of work life (QWL)

The QWL construct is conceptualised, in this thesis, to extract information related to students' perception that using the technology and more specifically the e-learning system will improve their quality of work life. This construct is measured by five items adopted from the work of (Kripanont, 2007), (Zakour, 2004), (Srite and karahanna, 2000) and measured using 7 point Likert scale. The results in Table 5.15 shows that the mean ranges between 5.29(±1.347) and 5.7(±1.318) for British students, whereas ranged between 4.49(±1.309) and 5.54(±1.27) for Lebanese students. This indicates that the majority of the students in both samples agree that using the technology will help improve their quality of work life.

| | England | | Lebanon | |
|---|---|---|---|---|
| **Item** | **Mean** | **Standard Deviation** | **Mean** | **Standard Deviation** |
| **QWL1** | 5.64 | 1.342 | 5.52 | 1.291 |
| **QWL2** | 5.70 | 1.318 | 5.54 | 1.270 |
| **QWL3** | 5.29 | 1.347 | 4.94 | 1.397 |
| **QWL4** | 5.56 | 1.321 | 5.49 | 1.309 |
| **QWL5** | 5.65 | 1.301 | 5.51 | 1.142 |

**Table 5-15: Descriptive statistics of QWL construct**

## 5.6.5 Computer Self-Efficacy (SE)

The SE construct is conceptualised, in this thesis, to extract information about students' self confidence in his/her ability to perform certain learning tasks using the e-learning system. This construct is measured by 6-items using a 7 point Likert scale, ranging from 1 - strongly disagree to 7- strongly agree and was adopted from the work of (Chang and Tung, 2008), (Compeau et al., 1999), (Vijayasarathy, 2004), (Yuen and Ma, 2008). The results in Table 5.16 shows that the mean for the items that measure the SE construct is ranged between 4.99 (±1.683) and 5.07(± 1.470) for the British sample, while the mean for SE items within the Lebanese sample ranged between 4.86(±1.696) and 5.7(±1.014) which indicate that the majority of the respondents were agreeable on this construct.

| | England | | Lebanon | |
|---|---|---|---|---|
| **Item** | **Mean** | **Standard Deviation** | **Mean** | **Standard Deviation** |
| **SE1** | 5.06 | 1.596 | 5.33 | 1.476 |
| **SE2** | 5.07 | 1.470 | 5.32 | 1.407 |
| **SE3** | 4.76 | 1.561 | 4.98 | 1.446 |
| **SE4** | 4.93 | 1.563 | 5.70 | 1.014 |
| **SE5** | 5.05 | 1.459 | 4.96 | 1.440 |
| **SE6** | 4.99 | 1.683 | 4.86 | 1.696 |

**Table 5-16: Descriptive statistics of SE construct**

## 5.6.6 Facilitating Condition (FC)

The FC construct is conceptualised, in this thesis, to extract information related to students' perception of students of whether they are able to access the required resources and the necessary support to use the e-learning services that the organizational and technical infrastructure exists to support his/ her use of the e-learning system. This construct was measured by four items adopted from the work of (Teo, 2009a), (Maldonado et al., 2009) and (Venkatesh et al., 2003) using a 7 point Likert scale, ranging from 1 - strongly disagree to 7- strongly agree. As can be shown in table 5.17, the results shows that the mean range between 4.67 (±1.743) and 5.29(±1.553) within the British sample, and 5.36 (±1.352) and 5.51(±1.149) within the Lebanese sample indicating agreement on the importance of the availability of the technological resources.

| | England | | Lebanon | |
|---|---|---|---|---|
| **Item** | **Mean** | **Standard Deviation** | **Mean** | **Standard Deviation** |
| **FC1** | 5.29 | 1.553 | 5.47 | 1.355 |
| **FC2** | 5.29 | 1.525 | 5.36 | 1.352 |
| **FC3** | 4.93 | 1.565 | 5.51 | 1.149 |
| **FC4** | 4.67 | 1.743 | 5.43 | 1.159 |

**Table 5-17: Descriptive statistics of FC construct**

## 5.6.7 Behavioural Intention (BI)

The BI construct is conceptualised, as a dependent variable in this thesis, to extract information related to students' behavioural intention to use the Blackboard system in the future. Three items adopted from the work of (Moon and Kim, 2001),(Davis, 1989), (Park, 2009) and (Chang and Tung, 2008) and were measured using 7 point Likert scale ranging from 1-strongly disagree to 7-strongly agree. The results of the descriptive analysis in Table 5.18 shows that the mean ranged between 5.51(±1.346) and 5.77(±1.309) for the British sample, whereas it ranged between 5.42(±1.29) and 5.67(±1.26) for the Lebanese sample.

The results revealed that British and Lebanese students showed agreement on this variable.

| Item | England | | Lebanon | |
|---|---|---|---|---|
| | Mean | Standard Deviation | Mean | Standard Deviation |
| BI1 | 5.51 | 1.346 | 5.42 | 1.290 |
| BI2 | 5.77 | 1.258 | 5.84 | 1.219 |
| BI3 | 5.76 | 1.309 | 5.67 | 1.260 |

**Table 5-18: Descriptive statistics of BI construct**

## 5.6.8  Actual Usage (AU)

The AU construct is conceptualised, as a dependent variable in this thesis, to extract the information about students' actual use of the system. It is worth noting that actual use was measured using self-reported measures. This construct is measured by two items adopted from the work of (Davis, 1989), (Venkatesh et al., 2000), (Venkatesh et al., 2003). Usage Behaviour construct uses scales from 1 to 6 (1= less than once a month and 6 = several times a day) to assess the frequency and of using web-based learning system and (1= Almost never and 6 = more than 3 hours) to measure the average of daily usage per hour. As can be seen in Table 5.19, the mean ranged between 4.01($\pm$1.291) and 4.84($\pm$1.136) for the British sample, whereas ranged between 3.5 ($\pm$1.267) and 4.94($\pm$1.074) for the Lebanese sample. Therefore, the results suggest that there were high levels of usage in general.

| Item | England | | Lebanon | |
|---|---|---|---|---|
| | Mean | Standard Deviation | Mean | Standard Deviation |
| FreqUsage | 4.84 | 1.136 | 4.94 | 1.074 |
| DailyUsage | 4.01 | 1.291 | 3.50 | 1.267 |

**Table 5-19: Descriptive statistics of AU construct**

# 5.7   Culture variables

This section presents the result of the four individual-level cultural values for the British and Lebanese students based on the results of the exploratory factor analysis. All the items were measured by 6 items except UA and PD, which were measured by 5 items. Items were measured using a 7 point Likert scale ranging from 1-strongly disagrees to 7-strongly agree. The constructs adopted from the work of (Srite and Karahanna, 2006) who in return developed  their measures based on the work of Dorfman and Howell  (1988) and Hofstede (1980) and were modified to fit the context of the study. The following subsections provide a detailed description of the four cultural variables used in the study.

## 5.7.1  Power Distance (PD)

This construct is conceptualised to measure the students' perception about the acceptance of power from his/her instructor. The results in Table 5.20 revealed that both British and Lebanese students were low on power distance, with a mean ranged between 2.39(±1.233) and 2.65(±1.326) for the British students, whereas ranged between 3.05(±1.514) and 3.49(±1.658) for the Lebanese students. Our results is inconsistent with Hofstede's (1980) finding who indicate that Arab countries are high on power distance.

| | England | | Lebanon | |
|------|------|-----------------------|------|-----------------------|
| Item | Mean | Standard Deviation | Mean | Standard Deviation |
| PD1 | 2.65 | 1.326 | 3.12 | 1.541 |
| PD2 | 2.57 | 1.273 | 3.49 | 1.658 |
| PD3 | 2.39 | 1.233 | 3.20 | 1.568 |
| PD4 | 2.60 | 1.252 | 3.05 | 1.514 |
| PD5 | 2.56 | 1.228 | 3.18 | 1.460 |
| PD6 | 2.64 | 1.218 | 3.15 | 1.442 |

**Table 5-20: Descriptive statistics of PD construct**

## 5.7.2 Masculinity/Femininity (MF)

This construct is conceptualised, in this thesis, to observe the masculine/feminine nature of the students. A students have high masculinity (low femininity) values will emphasise on the work goals and material accomplishments, such as earnings and promotions, whereas low masculinity individuals (high on femininity) place more value on human relationships and quality of life and usually encouraged to follow more traditional, tender and modest roles. As can be shown in Table 5.21, the results indicate that both students were rated high on masculinity with the mean ranged between 2.42(±1.228) and 2.71(±1.332) for the British students, whereas ranged between 3.06(±1.593) and 3.66(±1.568) for the Lebanese students.

Although this study studied the cultural dimensions at the individual level and there is no previous literature to support our findings, However, comparing it to the national level, the results from the Lebanese sample were unexpected as at the national level Lebanon is considered moderate in masculine index compared to England which indicates a very masculine society (Hofstede, 1980).

| Item | England | | Lebanon | |
|------|---------|---|---------|---|
| | Mean | Standard Deviation | Mean | Standard Deviation |
| MF1 | 2.42 | 1.228 | 3.13 | 1.671 |
| MF2 | 2.51 | 1.229 | 3.17 | 1.644 |
| MF3 | 2.52 | 1.285 | 3.06 | 1.593 |
| MF4 | 2.43 | 1.273 | 3.21 | 1.597 |
| MF5 | 2.71 | 1.332 | 3.21 | 1.600 |
| MF6 | 2.62 | 1.328 | 3.66 | 1.568 |

**Table 5-21: Descriptive statistics of PEOU construct**

### 5.7.3 Individualism/Collectivism (IC)

This construct is conceptualised, in this thesis, to reflect the extent to which individuals are integrated into groups. In individualistic societies, individuals focus on their own achievements and personal goals rather than on the group they belong to. Table 5.22 presents the descriptive statistics of the IC cultural values. The mean average for the British students ranged between $3.36(\pm1.471)$ and $3.88(\pm1.576)$, whereas the mean ranged between $4.62(\pm1.477)$ and $5.21(\pm1.220)$ for the Lebanese students. The results indicate that British students had individualistic values whereas Lebanese students had collectivist values. Our results is consistent with Hofstede's (1980) finding at the national level.

| | England | | Lebanon | |
|---|---|---|---|---|
| **Item** | **Mean** | **Standard Deviation** | **Mean** | **Standard Deviation** |
| **IC1** | 3.36 | 1.471 | 4.74 | 1.412 |
| **IC2** | 3.88 | 1.576 | 5.21 | 1.220 |
| **IC3** | 3.71 | 1.418 | 4.99 | 1.290 |
| **IC4** | 3.65 | 1.492 | 5.00 | 1.319 |
| **IC5** | 3.53 | 1.435 | 5.04 | 1.368 |
| **IC6** | 3.50 | 1.434 | 4.62 | 1.477 |

**Table 5-22: Descriptive statistics of PEOU construct**

### 5.7.4 Uncertainty avoidance (UA)

This construct is conceptualised, in this thesis, to measure the extent to which ambiguities and uncertainties are tolerated. In other words, individuals with high UA cultural values will establish formal rules and might reject deviant ideas and behaviours since it has been established with anxiety and the need for security. Conversely, individuals with low UA cultural values might have a greater willingness to take risks. As can be shown in Table 5.23, the average for the British sample ranged between $4.08(\pm1.304)$ and $4.14(\pm1.432)$, whereas the mean ranged between $5.32(\pm1.203)$ and $5.48(\pm1.240)$ for the Lebanese students. The results indicate that British students were neutral about uncertainty avoidance compared to the Lebanese students who were found to have high uncertainty avoidance. This results is very close to Hofstede's (1980) finding at the national level.

| Item | England | | Lebanon | |
|------|---------|---------|---------|---------|
| | **Mean** | **Standard Deviation** | **Mean** | **Standard Deviation** |
| **UA1** | 4.10 | 1.346 | 5.47 | 1.168 |
| **UA2** | 4.08 | 1.304 | 5.48 | 1.240 |
| **UA3** | 4.10 | 1.366 | 5.32 | 1.203 |
| **UA4** | 4.12 | 1.335 | 5.33 | 1.173 |
| **UA5** | 4.14 | 1.432 | 5.46 | 1.179 |

**Table 5-23: Descriptive statistics of PEOU construct**

## 5.8 Cross-cultural differences between the two samples

A T-test was employed to examine the similarities and differences between the British and Lebanese students at the national level. The results of this section will help to achieve another objective of this research, which is *"to develop an inclusive categorization of the similarities and differences between British and Lebanese students on the acceptance and usage of web-based learning systems at the national level"*. This will help in identifying and understanding any differences between the cultures of these two countries.

Table 5.24 shows the mean and group differences (t-test) for each construct between the Lebanese and British students at the national-level culture.

| T-test for Equality of Means | | | | |
|---|---|---|---|---|
| | **Mean** | | **t-test** | **Sig. (2-tailed)** | **Mean Difference** |
| | **Lebanon** | **England** | | | |
| **PU** | 5.182 | 5.25 | -1.033 | .302 | -.06750 |
| **PEOU** | 5.668 | 5.390 | 4.336*** | .000*** | .27781 |
| **SE** | 5.191 | 4.977 | 3.414** | .002** | .21351 |
| **FC** | 5.442 | 5.045 | 5.464*** | .000*** | .39687 |
| **SN** | 5.001 | 4.868 | 2.010* | .044* | .13294 |
| **QWL** | 5.4 | 5.566 | -2.695** | .007** | -.16578 |
| **BI** | 5.643 | 5.681 | -.563 | .573 | -.03795 |
| **AU** | 4.222 | 4.424 | -3.325** | .002** | -.20269 |
| **PD** | 3.1603 | 2.567 | 9.377*** | .000*** | .63235 |
| **MF** | 3.22 | 2.535 | 10.401*** | .000*** | .70630 |
| **IC** | 5.0114 | 3.822 | 21.102*** | .000*** | 1.32902 |
| **UA** | 5.464 | 4.4 | 20.171*** | .000*** | 1.0130 |
| Note: *p<.05, **p<.01,***p<.001 | | | | | |

**Table 5-24: T-test for all Constructs**

Regarding the group differences at the national level, the results revealed that no significant differences in terms of perceived usefulness (t=-1.033, main difference=.06750) and BI (t=-.563, mean difference= -.03795) between Lebanese and British students. This means that both Lebanese and British students found the system useful in their education and have good behavioural intention to use the system in the future.

In terms of PEOU, the mean score is 5.668 for Lebanon and 5.390 for England; unexpectedly the results revealed that PEOU is significantly higher in Lebanon compared to England (t=4.336, p<.001). In what concerns self-efficacy, the mean score is 5.191 for Lebanon and 4.977 for England, which is unexpectedly higher for Lebanon than England (t=3.414, p<.01).

Similarly, the results in Table 5.24 revealed that facilitating conditions (FC) was significantly higher in Lebanon (t=5.464, p<.001), with a mean 5.442 for Lebanon and 5.045 for England. These results were unexpected; however this could be due to the data being collected from two private universities in Lebanon, who invest a large sum of money in their educational system.

The results also showed that there is slightly significant difference for Lebanese and British students on social norm (SN), with a mean value 5.001 for Lebanon and 4.868 for England (t=2.010, p<.05), thus indicating that Lebanese students are more willing to accept the pressure from external environments (i.e., peers and superiors) to use the e-learning system.

The results in Table 5.24 also showed that there is a significance difference for Lebanese and British students on QWL, with the mean score 5.4 for Lebanon and 5.566 for England, indicating that a higher significant mean in England (t=-2.695, p<.01). This means that both Lebanese and British students are aware of the impact of technology on their quality of life. Similarly, with respect to AU, the mean score is 4.424 for England and 4.222 for Lebanon, thus revealing that the British students are using the e-learning system more than the Lebanese students (t=-.20269, p<.01)

Regarding the differences between Lebanon and England on the individual- level cultural values, the results revealed that there is a significant difference on each of the four variables.

In terms of PD, the mean score is 3.16 for Lebanon and 2.567 for England, indicating that both British and Lebanese students were low on PD with the mean being significantly higher in Lebanon (t=9.377, p<.001). Our results support the original findings of Hofstede's (1980) in terms of the British sample which indicate that the UK is low on PD, however it deviates from the national score of the Arab countries as Hofstede's indicate that Arab countries are high on power distance.

Regarding masculinity/femininity (MF), the mean is 3.22 for Lebanon and 2.535 for England, indicating a masculine cultural values in both samples with the mean is significantly higher in Lebanon (t= 10.401, p<.001). This results is inconsistent with Hofstede's (1980) finding who indicate that Arab countries are high on femininity.

In what concerns individualism/collectivism (IC), the mean score is 5.011 for Lebanon and 3.82 for England with the mean significantly higher in Lebanon (t=21.102, p<.001). The result reveals that British students have individualistic cultural values, while Lebanese students had collectivistic cultural values. The results is consistent with Hofstede's (1980) findings about Lebanon and England.

Finally, concerning uncertainty avoidance (UA), the mean score is 5.46 for Lebanon, whereas for British students the mean is 4.4, revealing that the mean is significantly higher in Lebanon (t=20.171, p<.001). This means that Lebanese students find the ambiguity more stressful and avoid unclear situation than British students. Our results is consistent with Hofstede's (1980) finding about the Arab countries and England.

The next section will provide a more detailed explanation of each item used in the survey questionnaire.

## 5.9   The tasks the students perform using the web-based learning system

This section will help in determining the current usage of Web-based learning systems in Lebanese and British universities which help achieve the fourth objective of this research. A descriptive analysis using SPSS were used to describe the current usage of e-learning in both countries, where the T-test were employed to examine the similarities and differences between the Lebanese and British students on using the web-based learning system

Regarding the actual tasks the students perform using the web-based learning system, the participants were asked to circle each questions based on the following measurement scales, where 1= *Do not use it at all*; 2=*to a small extent*; 3= *to some extent*; 4= *to a moderate extent*; and 5= *to a great extent*. Overall, the majority of students in both samples use the e-learning system to perform specific tasks. Tables 5.25 and 5.26 explain the percentage and frequencies for each country, whereas the differences between the two groups are presented in Table 5.27.

| England | | | | | |
|---|---|---|---|---|---|
| **Item** | **Not at All** | **Small extent** | **Some extent** | **Moderate extent** | **To a great extent** |
| **Announcement** | 9.5% (57) | 15% (90) | 20.8% (125) | 28.4% (171) | 26.4% (159) |
| **Email** | 5.0% (30) | 9.8% (59) | 14% (84) | 24.3% (146) | 47% (283) |
| **Assessment** | 2.5% (15) | 4.8% (29) | 13% (78) | 24.4% (147) | 55.3% 333 |
| **Lecture note** | 2.5% (15) | 5% (30) | 11% (66) | 25.7% (155) | 55.8% (336) |
| **Course Handbook** | 9.8% (59) | 16.1% (97) | 21.6% (130) | 22.9% (138) | 29.6% (178) |
| **Past Papers** | 10% (60) | 13.5% (81) | 15.3% (92) | 23.5% (143) | 37.5% (226) |
| **Discussion Board** | 23.8% (143) | 25.2 % (152) | 18.4% (111) | 16.9% (102) | 15.6% (94) |
| **Websites** | 11.8% (71) | 17.4% (105) | 21.9% (132) | 21.4% (129) | 27.4% (165) |
| **Quizzes** | 29.2% (176) | 16.4% (99) | 15.8% (95) | 19.9% (120) | 16.6% (112) |

**Table 5-25: British sample actual usage of the system**

In terms of the British sample, the results in Table 5.25 showed that the majority of participants responded "to a great extent" on all other tasks except for "Announcement" which scored "to a moderate extent" and "Discussion Board" with the majority responded "to a small extent", whereas "Quizzes" scored the lowest mean (M=2.82) among other tasks with the majority of the participants answered "Not at all".

| Lebanon | | | | | |
|---|---|---|---|---|---|
| **Item** | **Not at All** | **Small extent** | **Some extent** | **Moderate extent** | **To a great extent** |
| **Announcement** | 3.4% (19) | 8.7% (49) | 16.8% (95) | 29.3% (166) | 41.9% (237) |
| **Email** | 10.1% (57) | 9.4% (53) | 15.2% (86) | 26.0% (147) | 39.4% (223) |
| **Assessment** | 8.5% (48) | 14.7% (83) | 20.7% (117) | 30.6% (173) | 25.6% (145) |
| **Lecture note** | 3.9% (22) | 4.4% (25) | 11.5% (65) | 24% (136) | 56.2% (318) |
| **Course Handbook** | 10.1% (57) | 12% (68) | 22.4% (127) | 29% (164) | 26.5% (150) |
| **Past Papers** | 12.9% (73) | 13.1% (74) | 18.2% (103) | 28.4% (161) | 27.4% (155) |
| **Discussion Board** | 28.6% (162) | 20.7% (117) | 19.3% (109) | 19.1% (108) | 12.4% (70) |
| **Websites** | 17.3% (98) | 17.8% (101) | 20.8% (118) | 23.1% (131) | 20.8% (118) |
| **Quizzes** | 11.5% (65) | 16.4% (93) | 24% (136) | 26.9% (152) | 21.2% (120) |

**Table 5-26: Lebanese sample actual usage of the system**

In terms of the Lebanese sample, Table 5.26 showed that the participants response was either 'to a moderate extent' or 'to a great extent', with the majority of the participants responding "to a moderate extent" for "Assessment", "Course Handbook", "Past Papers", "Websites", "Quizzes". The means, standard deviation and t-test analysis are presented in the next table.

| | Lebanon | | England | | t-test | | |
|---|---|---|---|---|---|---|---|
| **Item** | **Mean** | **Std. Deviation** | **Mean** | **Std. Deviation** | **t** | **Sig (2-talied)** | **Mean Difference** |
| **Announcement** | 3.98 | 1.112 | 3.49 | 1.299 | **6.848***** | .000*** | .484 |
| **Email** | 3.75 | 1.330 | 3.99 | 1.205 | **-3.123**** | .002** | -.232 |
| **Assessment** | 3.50 | 1.252 | 4.25 | 1.019 | **-11.2***** | .000*** | -.751 |
| **Lecture note** | 4.24 | 1.071 | 4.27 | 1.009 | -.526 | .599 | -.032 |
| **Course Handbook** | 3.50 | 1.276 | 3.46 | 1.324 | .457 | .648 | .035 |
| **Past Papers** | 3.44 | 1.354 | 3.65 | 1.359 | **-2.657**** | .008** | -.211 |
| **Discussion Board** | 2.66 | 1.387 | 2.75 | 1.393 | -1.169 | .243 | -0.095 |
| **Websites** | 3.12 | 1.387 | 3.35 | 1.355 | **-2.847**** | .004** | -.228 |
| **Quizzes** | 3.30 | 1.286 | 2.82 | 1.5 | **5.81***** | .000*** | .476 |
| Note: *p<.05, **p<.01,***p<.001 | | | | | | | |

**Table 5-27: t-test for the tasks performed using e-learning system**

Regarding the group differences on the tasks performed using e-learning system, the results in Table 5.27 revealed no significant differences in "lecture note", "course Handbook" and "Discussion Board" between Lebanon and England, where significant difference were higher in Lebanon in terms of "Announcement" and "Quizzes" and lower in terms of," Email"," Past Papers"," Assessment" and "Websites". This means that Lebanese students use the e-learning more for "Announcement" and "Quizzes" and lower for "Email", "Past Papers", "Assessment" and "Websites than British students. It is worth noting that the highest mean for both samples was observed for "lecture note", with 318 (Mean=4.24) and 336 (mean=4.27) of respondents responded 'to a great extent' in Lebanon and England respectively.

# 5.10 Summary and Conclusions

This chapter reported the findings of the preliminary data analysis using the statistical package of social science (SPSS) version 18.0. The different statistical techniques used in this chapter helped the researcher examine the preliminary results of the dataset.

This chapter first reported the results of the pilot study to ensure the validity and reliability of the measuring instruments of the survey items. Although not all the survey items; and especially the individual-level cultural variables; were previously used in Lebanon or England. However, the results confirmed that the survey items had high reliability (internal consistency) and all the constructs had a Cronbach's alpha above (0.7).

The second section involved the data screening of the dataset such as missing data, outliers, testing the normality assumption, homogeneity and multicollinearity. The results revealed that all missing data for both samples were distributed in a random manner and were below 5 %. The maximum percentage of missing data was 2.5 and 2.4 for the British and Lebanese samples, respectively. This percentage of missing data is very low and can be considered acceptable.

Therefore, the researcher applied 'mean substitution' method were undertaken to replace missing data for the categorical variables while missing data for nominal variables were excluded later during the multi-group analysis, as suggested by many scholars e.g.,(Pallant, 2010; Arbuckle, 2009; Byrne, 2006). Mahalanobis D2 revealed that twelve and fifteen multivariate outliers were detected in the British and Lebanese sample respectively. However, the researcher retained the outliers to the dataset because they did not find them to be problematic and so to be included in further analysis (Hair et al., 2010). The results of the Jarque-Bera (skewness-Kurtosis) test indicate that the value in each case was in the range of ±1, which confirms that there were no major issues of non-normality of the data. Finally, given that all the independent constructs had VIP value less than 3.0 and tolerance value above 0.10, this suggests the absence of multicollinearity in both samples.

To show the relationship of variables to factors, an explanatory factor analysis (EFA) was employed for the dataset. Based on the results of the EFA, 12 factors were extracted. EFA also suggested the deletion of SE5, SE6 and PD6 as they were highly cross-loaded on another latent construct. However, the researcher retained those variables for further investigation using the SEM.

The third section captured the respondents' demographic characteristics such as age, gender, nationality, educational level and experience in using internet and web-based learning systems. A total of 2000 questionnaires were distributed to 1000 students from the UK and 1000 from Lebanon respectively, of which 1200 were returned indicating a 60.0% response rate.

The fourth section presented the results of the descriptive statistics including the means and standard deviation for each independent and dependent variable used in the proposed research model. Overall, all means were greater than 4.42 for the British sample (N=602) and 4.21 for the Lebanese sample (N=566) which indicate that the majority of participants express generally positive responses to the constructs that are measured in this study. In terms of the cultural variables, the results indicate British students were scored low on (PD, MF, IC and UA), whereas Lebanese students were scored low on (PD, MF), and high on (IC and UA).

The results of the t-test between the Lebanese and British students are presented in Section 5. The results showed that all the constructs were significant except for perceived usefulness and behavioural intention. In addition, a summary related to the main determinants as well as Hofstede's cultural dimensions are summarised in Table 5.28 and 5.29 respectively.

The chapter ended with an elaboration on the differences between British and Lebanese students on tasks performed using the e-learning system, the results in Section 8 revealed that no significant differences in "lecture note", "course Handbook" and "Discussion Board" between Lebanon and England, where significant difference were higher in Lebanon in terms of "Announcement" and "Quizzes" and lower in terms of "Email", "Past Papers", "Assessment" and "Websites".

The next chapter (Chapter 6) will provide a further analysis using structural equation modeling (SEM) with AMOS version 18.0 to observe the construct validity and convergent validity and to test the direct relationship between predictors and behavioural intention and actual usage. These results will be discussed against hypotheses in Chapter 8.

| T-test for Equality of Means (main determinants) | Study Results |
|---|---|
| *H1c: Students' mean Perceived Ease of Use will be significantly higher in the United Kingdom, compared to Lebanon.* | (Significant but higher in Lebanon), Not Supported |
| *H2c: Students' mean perceived usefulness will be significantly higher in the United Kingdom, compared to Lebanon.* | Not Supported |
| *H4c: Students' mean Social Norm will be significantly higher in Lebanon, compared to the United Kingdom.* | **Supported** |
| *H5c: Students' mean QWL will be significantly higher in the United Kingdom, compared to Lebanon.* | **Supported** |
| *H7c: Students' mean Computer self-efficacy will be significantly higher in the United Kingdom, compared to Lebanon.* | (Significant but higher in Lebanon), Not Supported |
| *H8c: Students' mean facilitating conditions will be significantly higher in the United Kingdom, compared to Lebanon.* | Not Supported |
| *H3c: Students' mean intention to use e-learning systems will be significantly higher in the United Kingdom, compared to Lebanon.* | Not Supported |
| *H3c: Students' mean actual usage of e-learning systems will be significantly higher in the United Kingdom, compared to Lebanon.* | (Significant but higher in Lebanon), Not Supported |

**Table 5-28: Summary of results related to the difference in Means for the main predictors hypotheses**

| T-test for Equality of Means (Hofstede's cultural variables) | Study Results |
|---|---|
| *H11c: Power Distance mean will be significantly higher in Lebanon, compared to the United Kingdom.* | **Supported** |
| *H12c: **masculinity** mean will be significantly higher in to the United Kingdom, compared to Lebanon.* | **Supported** |
| *H13c: **individualism** mean will be significantly higher in the United Kingdom, compared to Lebanon.* | **Supported** |
| *H14c: **Uncertainty Avoidance** mean will be significantly higher in Lebanon, compared to the United Kingdom.* | **Supported** |

**Table 5-29: Summary of results related to the difference in Means for the Cultural dimensions hypotheses**

# Chapter 6:  Model Testing

*"The competent analysis of research-obtained data requires a blending of art and science, of intuition and informal insight, of judgement and statistical treatment, combined with a thorough knowledge of the context of the problem being investigated" (Green, Tull and Albaum, 1988: 379)*

## 6.1   Introduction

Chapter 5 presented the preliminary data analysis. This chapter presents an in-depth analysis of the relationships among the constructs within the proposed research model. Two steps were used during the data analysis process. In the first step, the confirmatory factor analysis (CFA) was employed to assess the constructs' validity and test the model fit. The next step employed the structural equation modeling (SEM) technique to test the hypothesised relationships among the independent and dependent variables. Using a two-step approach assures that only the constructs retained from the survey that have good measures (validity and reliability) will be used in the structural model (Hair et al., 2010).

Accordingly, this chapter is organised as follows: in the first section, the assessment of the measures will be discussed including the measures of the validity and reliability of the constructs. This is then followed by testing the direct relationships among the constructs of the revised measurement model in Lebanon and England respectively. Section 6.4 will discuss the criteria and methods used to measure the impact of the moderators.

This chapter will then investigate the moderating effects of Hofstede's four cultural dimensions (PD, M/F, I/C, UA) and then the demographic characteristics such as age, gender, and experience on the relationships between the exogenous (independent) and endogenous (dependent) constructs in the generated model in Section 6.5 and 6.6. Finally, Section 6.7 provides a summary of the main results. This research has employed the Analysis of Moment Structures (AMOS) version 18.0 to test the causal relationships among the variables. AMOS is considered easy-to-use and user-friendliness compared to other software such as LISREL, EQS (Blunch, 2008). AMOS also has the ability to estimate and present the model.

## 6.2   Analysis of measurement model

To examine the relationships among the different constructs within the conceptual model, this study employed a confirmatory factor analysis (CFA) based on AMOS 18.0 (Arbuckle, 2009). To assess the measurement model in CFA, the researcher first considered the measurement model fit and then evaluated the validity of the measurement model.

In the CFA, there is no need to distinguish between dependent and independent variables while it is necessary during the model testing stage. As can be shown in Figure 6.1, all the variables are linked together and the construct items (measured variables) are represented in rectangular shapes.  The covariance is usually represented by two-headed arrows, whereas a causal relationship from a construct to an indicator is represented by one-headed arrow. In the current study, the researcher worked on each sample separately in order to generate a model that best fit each sample. In total, 33 items were used in the CFA which derived from the EFA (see Chapter 5).

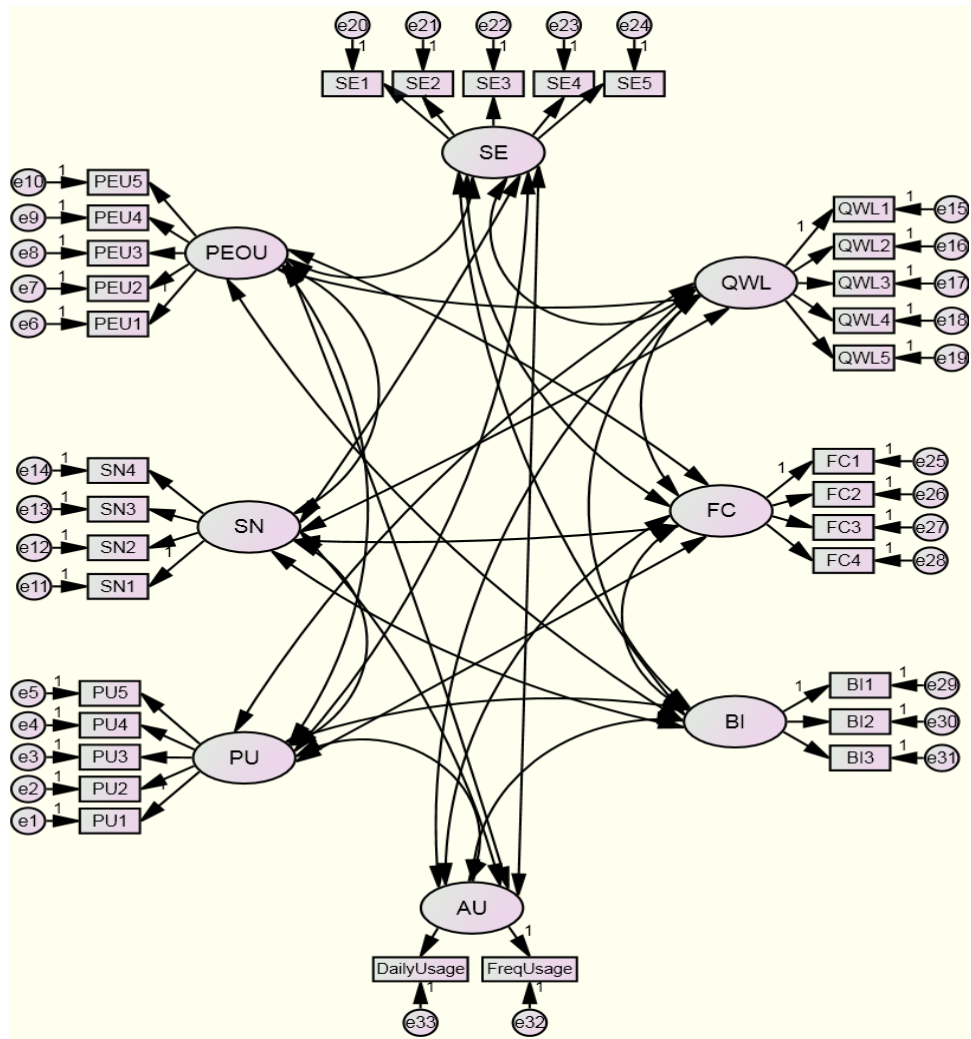

**Figure 6-1: Hypothesised CFA model derived from EFA**

## 6.2.1 Goodness of fit indices

We adopt the maximum-likelihood method to estimate the model's parameters where all analyses were conducted on variance-covariance matrices (Hair et al., 2010). There are some fit indices that should be considered in order to assess the model goodness-of-fit (Kline, 2005; Hair *et al.*, 2010). First, it was determined using the minimum fit function $\chi^2$. However, the $\chi^2$ was found to be too sensitive for our sample size (Hu and Bentler, 1999), in this case it is more likely that the model will reject something true, and also very small differences between the observed model and the good model fit might be found significant (Hair et al., 2010).

Therefore, other fit measures were used to overcome this problem. First, the ratio of the $\chi^2$ statistic to its degree of freedom ($\chi^2$/df) was used, with a value of less than 3 indicating acceptable fit (Carmines and McIver, 1981). Hair (2010) suggest the following indices to indicate acceptable fit "Goodness of Fit Index (GFI); Normed Fit Index (NFI); Parsimony Normed Fit Index (PNFI); Root Mean Square Residuals (RMSR); Comparative Fit Index (CFI); Adjusted Goodness-of-Fit Index (AGFI); the Root Mean Square Error of Approximation (RMSEA)". Table 6.1 shows the level of acceptance fit obtained with the survey data.

The first run of the model revealed the following results for the British sample [$\chi^2$ =1640.667; df= 467; $\chi^2$/df= 3.513; GFI =.853; AGFI=.823; CFI=.921; RMSR=.125; RMSEA=.065, NFI=.893; PNFI=.790] and [$\chi^2$ =1377.704; df= 467; $\chi^2$/df= 2.905; GFI =.860; AGFI=.832; CFI=.923; RMSR=.084; RMSEA=.059, NFI=.888; PNFI=.785] for the Lebanese sample. These results indicated a further room for improvement to indicate a good measurement model fit of the data. The complete results are shown in Appendix G.

| Fit Index | Recommended Value (Hair, 2006) | England Measurement Model | Lebanon Measurement Model |
|---|---|---|---|
| $\chi 2$ | Non-significant at p <0.05 | 1640.667 | 1377.704 |
| Degrees of freedom (df) | n/a | 467 | .467 |
| $\chi 2$ /df | <5 preferable <3 | 3.513 | 2.905 |
| Goodness-of-fit index (GFI) | >0.90 | .853 | .860 |
| Adjusted Goodness-of-fit index (AGFI) | >0.80 | .823 | .832 |
| Comparative fit index (CFI) | >0.90 | .921 | .923 |
| Root mean square residuals (RMSR) | <0.10 | .125 | .084 |
| Root mean square error of approximation (RMSEA) | <0.08 | .065 | .059 |
| Normed fit index (NFI) | >0.90 | .893 | .888 |
| Parsimony normed fit index (PNFI) | >0.60 | .790 | .785 |

**Table 6-1: Model fit summary for the measurement model**

The modification indices in AMOS provide information on the improvement in model fit. The following measures were applied by the researcher in order to achieve a better fit of the model.

- the standardised residual covariance should be within |2.58| (Byrne, 2006)
- Factor loading (Standardised regression weight) should be greater than 0.5 and preferably above 0.7 (Byrne, 2006)
- The Squared multiple correlations (SMC) value should be greater than 0.5 (Byrne, 2006)
- Modification indices (MI) that reveal a very high covariance and demonstrate high regression weights should be deleted (Byrne, 2006; Hair *et al.*, 2010).

In terms of the British sample, the results revealed that only two items SE4 (3.9) and SE5 (4.9) had a standardised regression weight below 0.5 (see Appendix H). Furthermore, the standardised residuals values showed that items (SE4, SE5, FC4, SN4 and FreqUsage) were not within the acceptable value |2.58|. In addition and as can be shown in Table 6.2, the MI showed that items (FC3, FC4, SE4, SE5 and SN4) had very high covariance.

| Errors | MI-covariance | Path | MI-regression weight |
|---|---|---|---|
| e27 → e28 | 179.176 | FC3 → FC4 | 104.479 |
| | | FC4 → FC4 | 77.914 |
| e23 → e24 | 185.263 | SE4 → SE5 | 155.086 |
| | | SE5 → SE4 | 137.600 |
| e14 → e31 | 35.596 | SN4 → BI1 | 38.838 |
| | | BI1 → SN4 | 27.082 |

**Table 6-2: British sample selected text output**

To ensure a good fit model and as can be shown in Table 6.2, some indicators (SE4,SE4, FC4, SN4) have to be deleted from the initial measurement model for the British sample since they were demonstrating high covariance and also had high regression weight (Byrne, 2006). The process was to delete one indicator at a time and then re-estimate the model.

Regarding the Lebanese sample and following the recommended criteria mentioned above, the output from AMOS revealed that only one items SE5 (3.41) had a standardised regression weight below 0.5 (see Appendix F-2). In addition, the standardised residuals values showed that items' value for (SE4, SE5, QWL3, FC4, PEOU2) were not within the recommended range |2.58|. Furthermore and as can be shown in Table 6.3, the MI showed that items (FC3, FC4, SE4, SE5, QWL3) had high covariance.

| Errors | MI-covariance | Path | MI-regression weight |
|--------|---------------|------|----------------------|
| e27 → e28 | 108.302 | FC3 → FC4 | 48.749 |
| | | FC4 → FC4 | 30.090 |
| e23→ e24 | 87.482 | SE4 → SE5 | 76.520 |
| | | SE5 → SE4 | 62.227 |
| e17→ e18 | 52.362 | QWL3 → QWL4 | 18.331 |
| | | QWL4 → QWL3 | 20.332 |

**Table 6-3: Lebanese sample selected text output**

In this regards, items (SE4, SE5, FC4 and QWL3) have to be deleted from the initial measurement model in order to achieve a good fit model. Table 6.4 shows the level of acceptance fit and the fit indices for the Lebanese sample after the improvement in model fit.

| Fit Index | Recommended Value (Hair, 2006) | England Final Measurement Model | Lebanon Final Measurement Model |
|-----------|-------------------------------|--------------------------------|--------------------------------|
| $\chi 2$ | Non-significant at p <0.05 | 840.368 | 866.050 |
| Degrees of freedom (df) | n/a | 349 | 349 |
| $\chi 2$ /df | <5 preferable <3 | 2.408 | 2.482 |
| Goodness-of-fit index (GFI) | >0.90 | .908 | .902 |
| Adjusted Goodness-of-fit index (AGFI) | >0.80 | .886 | .873 |
| Comparative fit index (CFI) | >0.90 | .963 | .951 |
| Root mean square residuals (RMSR) | <0.10 | .063 | .067 |
| Root mean square error of approximation (RMSEA) | <0.08 | .048 | .051 |
| Normed fit index (NFI) | >0.90 | .939 | .922 |
| Parsimony normed fit index (PNFI) | >0.60 | .807 | .792 |

**Table 6-4: England and Lebanon final measurement model fit**

After achieving the good measurement model for both samples, we can proceed to assess the validity and reliability in order to evaluate whether the psychometric properties of the measurement model are adequate.

## 6.2.2  Construct Validity and reliability

It is an essential step before proceeding to test the hypotheses in the proposed research model to examine the validity and reliability of the measures  as this may affect the results and thus the objective of the research (Hair et al., 2010). Although these two test are separate from each other, they are closely related (Bollen, 1989). According to Holmes-Smith (2011), a measure may have high reliability (consistency) but not be valid (accurate), and a measure may have high validity (accuracy) but not be reliable (consistent). Hair et al (2010) defined validity as "extent to which a set of measured variables actually represent the theoretical latent construct they are designed to measure". Construct validity can be examined by convergent validity, discriminant validity and nomological validity.

Convergent validity refers "to the extent to which measures of a specific construct should converge or share a high proportion of variance in common" (Hair *et al.*, 2010). In other words, it is the degree to which two measures of constructs that theoretically should be correlated (related), are in fact correlated (related), whereas discriminant validity; also known as divergent validity; is the extent to which a construct or concepts is not unduly related to other similar, yet distinct, constructs (Hair *et al.*, 2010). In other words, it is the extent to which a construct is divergent from other constructs within the model.

According to (Hair et al., 2010), validity and reliability can be measured using: Composite Reliability (CR), Average Variance Extracted (AVE), Maximum Shared Squared Variance (MSV), and Average Shared Squared Variance (ASV). The AVE measures the amount of variance that is captured by the construct in relation to the amount of variance due to measurement error. To establish reliability, Hair et al (2010) suggest that CR should be greater than 0.6 and preferably above 0.7. To establish convergent validity the AVE should be greater than 0.5 and CR is greater than the AVE, discriminant validity is supported if MSV is less than AVE and ASV is less than AVE. According to Tabachnick and Fidell (2007), convergent validity can be assessed using factor loading and (AVE). However, AMOS does not automatically calculate the AVE and CR for each construct. Therefore the researcher followed the following two formulas.

According to Fornell and Larcker (Fornell and Larcker, 1981) the AVE can be calculated using the following formula:

*AVE*= (summation of squared factor loadings)/(summation of squared factor loadings) (summation of error variances).

$$AVE = \frac{\sum_{i=1}^{n} \lambda_i^2}{n}$$

In the formula mentioned above $\lambda$ represents factor loadings (standardised regression weights) and $i$ represents the total number of items.

Regarding the CR, it can be calculated using the following formula (Chau and Hu, 2001):

CR = (square of summation of factor loadings)/(square of summation of factor loadings) + (summation of error variances).

$$CR = \frac{(\sum_{i=1}^{n} \lambda_i)^2}{\left(\sum_{i=1}^{n} \lambda_i\right)^2 + \left(\sum_{i=1}^{n} \delta_i\right)}$$

In the formula mentioned above $\lambda$ represents factor loadings (standardised regression weights) and $i$ represents the total number of items and $\delta_i$ represents the error variance term for each latent construct.

In terms of the British sample and as can be shown in Table 6.5, the average extracted variances were all above 0.575 and above 0.725 for CR. Additionally all the items had a standardised regression above 0.6 where the cut-off value is below 0.5 (see Table 6.7). Therefore, all factors have adequate reliability and convergent validity. Additionally, the total AVE of the average value of variables employed within the proposed model is larger than their correlation value, except of AU (MSV>AVE), thus there were discriminant validity issues. However, since AU is measured by two items only, deleting one of the variables might cause un-identification problems, we thus established discriminant validity.

| Factor Correlation Matrix with VAVE on the diagonal | | | | | | | | | | | | | |
|---|---|---|---|---|---|---|---|---|---|---|---|---|---|
| | α | CR | AVE | MSV | ASV | SE | PU | PEOU | SN | QWL | FC | AU | BI |
| SE | .903 | 0.907 | 0.765 | 0.711 | 0.370 | 0.875 | | | | | | | |
| PU | .922 | 0.922 | 0.703 | 0.507 | 0.393 | 0.567 | 0.839 | | | | | | |
| PEOU | .923 | 0.925 | 0.711 | 0.416 | 0.328 | 0.624 | 0.645 | 0.843 | | | | | |
| SN | .836 | 0.838 | 0.633 | 0.271 | 0.207 | 0.391 | 0.521 | 0.385 | 0.795 | | | | |
| QWL | .889 | 0.889 | 0.617 | 0.579 | 0.343 | 0.493 | 0.699 | 0.563 | 0.436 | 0.785 | | | |
| FC | .883 | 0.893 | 0.738 | 0.516 | 0.360 | 0.653 | 0.580 | 0.580 | 0.458 | 0.560 | 0.859 | | |
| AU | .705 | 0.725 | 0.575 | 0.711 | 0.414 | 0.843 | 0.638 | 0.529 | 0.467 | 0.518 | 0.718 | 0.758 | |
| BI | .893 | 0.898 | 0.745 | 0.579 | 0.425 | 0.586 | 0.712 | 0.637 | 0.507 | 0.761 | 0.616 | 0.707 | 0.863 |

**Table 6-5: Construct reliability, convergent validity and discriminant validity for the British sample**

Regarding the Lebanese sample and as can be shown in Table 6.6, all items loading were above 0.631 where the cut-off value is 0.5. In addition, the factor loading for all the average extracted variances were all above 0.521 and above 0.704 for CR. Therefore, all factors have adequate reliability and convergent validity. Additionally, the total AVE of the average value of variables used for the research model is larger than their correlation value; therefore we also established discriminant validity.

| Factor Correlation Matrix with √AVE on the diagonal | | | | | | | | | | | | | |
|---|---|---|---|---|---|---|---|---|---|---|---|---|---|
| | α | CR | AVE | MSV | ASV | SE | PU | PEOU | SN | QWL | FC | AU | BI |
| SE | .889 | 0.895 | 0.740 | 0.392 | 0.234 | 0.860 | | | | | | | |
| PU | .903 | 0.904 | 0.655 | 0.293 | 0.191 | 0.317 | 0.809 | | | | | | |
| PEOU | .929 | 0.929 | 0.723 | 0.281 | 0.225 | 0.522 | 0.512 | 0.850 | | | | | |
| SN | .813 | 0.824 | 0.542 | 0.188 | 0.146 | 0.383 | 0.356 | 0.343 | 0.736 | | | | |
| QWL | .831 | 0.836 | 0.563 | 0.460 | 0.255 | 0.444 | 0.541 | 0.492 | 0.425 | 0.750 | | | |
| FC | .895 | 0.897 | 0.744 | 0.392 | 0.238 | 0.626 | 0.294 | 0.502 | 0.364 | 0.475 | 0.863 | | |
| AU | .655 | 0.733 | 0.521 | 0.402 | 0.238 | 0.527 | 0.441 | 0.387 | 0.361 | 0.436 | 0.570 | 0.705 | |
| BI | .864 | 0.871 | 0.694 | 0.460 | 0.301 | 0.499 | 0.525 | 0.530 | 0.434 | 0.678 | 0.500 | 0.634 | 0.833 |

**Table 6-6: Construct reliability, convergent validity and discriminant validity for the Lebanese sample**

Consequently, the internal consistency of each construct was assessed using Cronbach's Alpha. Cronbach's Alpha measures how well a set of items measures a single unidirectional latent construct. As shown in Table 6.5 and 6.6, the value of Cronbach's Alpha for all factors in the model had a coefficient above the cut-off value of 0.7 for both samples (DeVellis, 2003; Robinson *et al.*, 1991). Therefore, the result indicates that the constructs within the two samples had adequate reliability.

Therefore, after checking the goodness of fit indices in addition to validity and reliability, the final refined model results in deleting SE4, SE5, FC4 and SN4 from the British sample. The final measurement model is depicted in Figure 6.2.

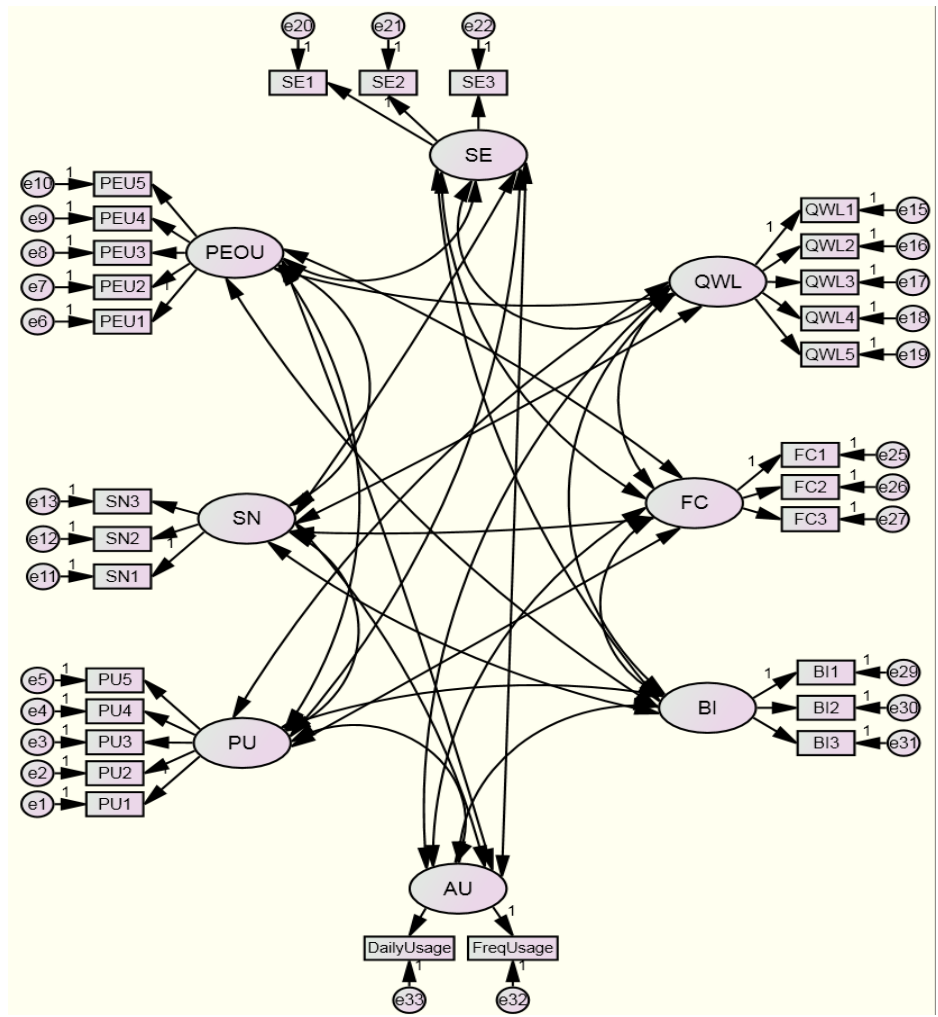

**Figure 6-2: The refined measurement model for the British sample**

The refined model results in deleting SE4, SE5, FC4 and QWL3 from the Lebanese sample. The final measurement model is depicted in Figure 6.3.

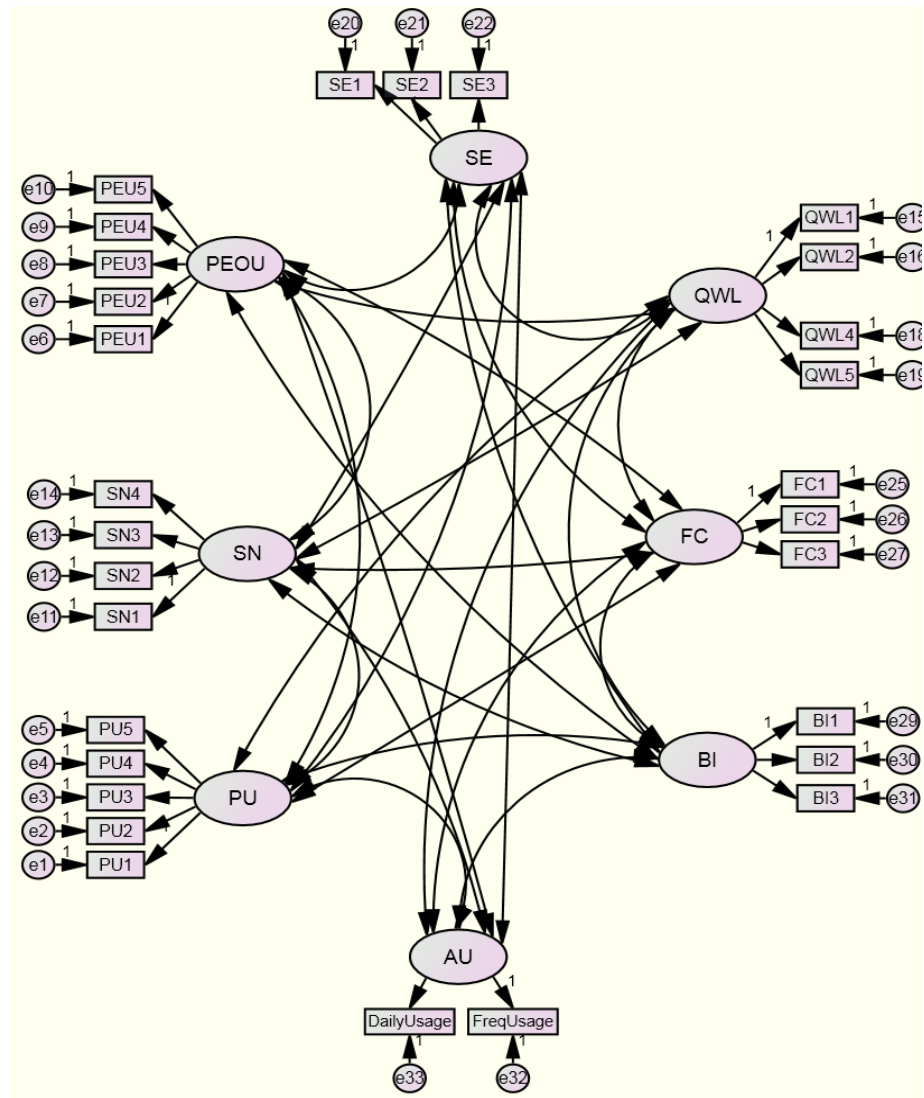

**Figure 6-3: The refined measurement model for the Lebanese sample**

## 6.3   Analysis of the structural model and Hypotheses testing

Having established reliability, convergent validity and discriminant validity for both samples, the next step is to test the relationships between the exogenous (independent) and endogenous (dependent) latent variables which can be done during the structural model stage (Arbuckle, 2009; Hair *et al.*, 2010).

Unlike the CFA, there is a need to distinguish between dependent and independent variables. SEM assumes the covariance between the independent variables, which is represented by two-headed arrows, whereas the causal relationship from an independent variable to a dependent variable is represented by one-arrow. Therefore, the relationship between constructs is specified after the transition from the measurement model to the structural model.

The following hypotheses were used to test the direct relationships between the 5 exogenous (independent) and 2 endogenous (dependent) latent variables. These relationships were identified in Chapter 3 during the model development stage. The exogenous constructs were perceived ease of use, perceived usefulness, social norm, self-efficacy, and facilitating conditions, while endogenous constructs were behavioural intention and actual usage.

*H1a,b: Perceived Ease of Use will have a direct positive influence on the intention to use web-based learning System in the British and Lebanese context.*

*H2a,b: Perceived Usefulness will have a direct positive influence on the intention to use web-based learning system in the British and Lebanese context.*

*H3a,b: Students' BI will have a positive effect on his or her actual use of web-based learning system in the British and Lebanese context.*

*H4a,b: Social Norm will have a positive influence on student's behavioural intention to use and accept the e-learning technology in the British and Lebanese context.*

*H5a,b: QWL will have a positive influence on student's behavioural intention to use the web-based learning system in the British and Lebanese context.*

*H6a,b: Computer self-efficacy will have a positive influence on student's behavioural intention to use the web-based learning system in the British and Lebanese context.*

*H7a,b: Computer self-efficacy will have a positive influence on the actual usage of the web-based learning system in the British and Lebanese context.*

*H8a,b: Facilitating conditions will have a positive influence on actual usage of web-based learning system in the British and Lebanese context.*

*H9a,b: SN will have a positive influence on perceived usefulness of web-based learning system in the British and Lebanese context.*

*H10a,b: PEOU will have a positive influence on perceived usefulness of web-based learning in the British and Lebanese context.*

The results of the analysis of the structural model will be discussed in separate subsections.

## 6.3.1 Testing the Structural Model for the British sample

The structural model for the British sample is depicted in Figure 6.4.

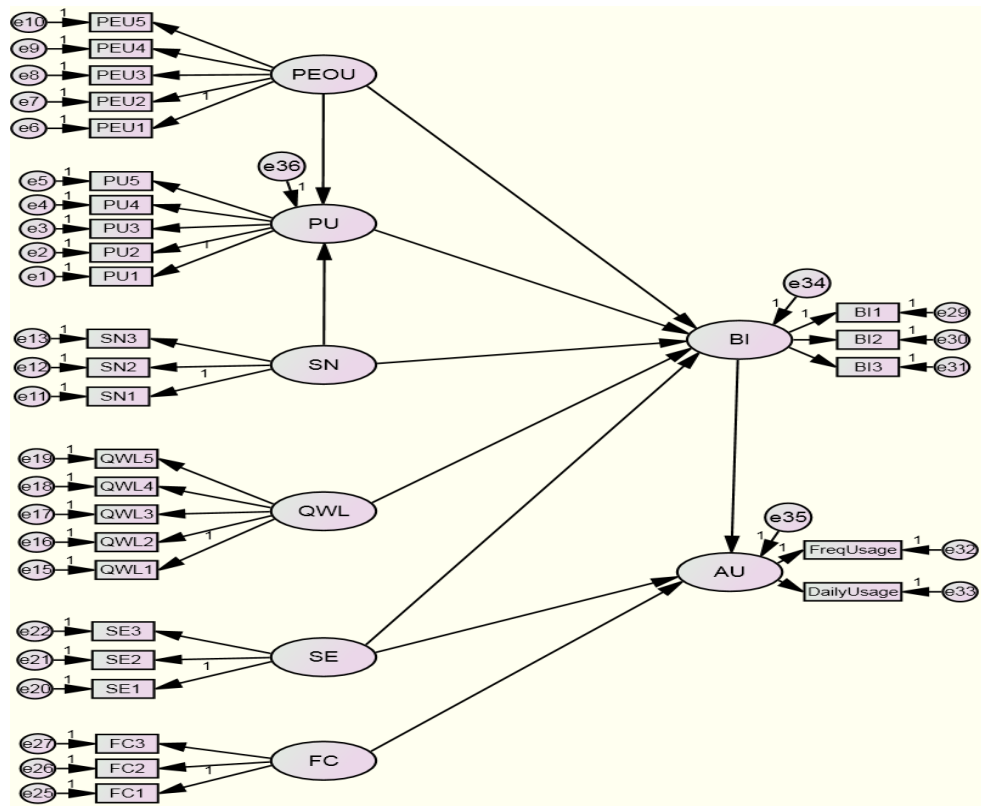

**Figure 6-4: the structural model for the British sample Note: the covariance (the two-headed arrows) between 6 exogenous variables was deleted for the purpose of presentation.**

Based on the same criteria used for measurement model to measure the goodness-of-fit for the proposed model, the results of the fit indices for the first run were as follows: [CMIN=892.592; df= 355; CMIN/DF = 2.514; GFI=.903; AGFI=.882; CFI=.960; RMSR=.069, RMSEA=.050, NFI=.935; PNFI=.818] indicating a very good fit of the model. Thus, we proceed to examine the hypothesized relationships within the model. Table 6.8 depicts the path coefficients for the hypothesised relationships within the proposed research model.

| H# | Proposed Relationship | Effects Type | Path coefficient | Study Results |
|---|---|---|---|---|
| Perceived usefulness, Intention and Actual Usage Prediction | | | | |
| H1$_a$ | PEOU (+) ⟶ BI | Direct effect | 0.120** | Supported |
| H2$_a$ | PU (+) ⟶ BI | Direct effect | 0.201*** | Supported |
| H4$_a$ | SN (+) ⟶ BI | Direct effect | 0.082* | Supported |
| H5$_a$ | QWL (+) ⟶ BI | Direct effect | 0.430*** | Supported |
| H6$_a$ | SE (+) ⟶ BI | Direct effect | 0.054 | Not supported |
| H7$_a$ | SE (+) ⟶ AU | Direct effect | 0.282*** | Supported |
| H8$_a$ | FC (+) ⟶ AU | Direct effect | 1.00** | Supported |
| H3$_a$ | BI (+) ⟶ AU | Direct effect | 0.175*** | Supported |
| H9 $_a$ | SN (+) ⟶ PU | Direct effect | 0.299*** | Supported |
| H10 $_a$ | PEOU (+) ⟶ PU | Direct effect | 0.398*** | Supported |
| Notes: * p<0.1; ** p<0.05; *** p<0.01; NS p>0.1 | | | | |

**Table 6-7: The summary of Direct Hypothesized results for the British sample**

As can be shown in Table 6.7, 9 out of 10 direct hypotheses were supported in the model. PEOU ($\gamma$=0.12**) and PU ($\gamma$=0.201***) were found to have a significance positive influence on behavioural intention to use web-based learning system, supporting H1a and H2a. The influence of colleagues and instructors on students' behavioural intention to use the system was significant, SN ($\gamma$=0.082*) supporting H4a. Moreover, BI was also influenced by the quality of work life ($\gamma$=0.430***) which supports H5a. However, the data fails to support the direct relationship between SE ($\gamma$=0.054) and BI, indicating that H6a was rejected. The results of the squared multiple correlations (SMC), which provides information about the extent to which the model explains variance in the data set, indicated that PEOU, PU, SN,SE and QWL account for 68% ($R^2 = 0.68$) of the variance of BI, with QWL contributing the most to behavioural intention compared to the other constructs. The results indicate that the higher the perceived ease of use, perceived usefulness, social norm, quality of work life, the higher behavioural intention to use the e-learning system in education. Additionally, SN ($\gamma$=0.299***) and PEOU ($\gamma$=0.398***) were found to have a significance positive influence on PU, supporting H9a and H10a. Together SN and PEOU explained 54.1% of the variance in PU with PEOU contributing the most.

AU is affected directly by BI, SE, and FC and indirectly by PEOU, PU, SN, SE and QWL. The indirect effect is the product of the paths that are linked to the dependent variable. The total indirect effect is the sum of all these paths. The results of the direct, indirect and total effects on AU are presented in Table 6.8.

| H# | Variables | Direct | Indirect | Total |
|---|---|---|---|---|
|  | PEOU | - | 0.039 | 0.039 |
|  | PU | - | 0.035 | 0.035 |
|  | SN | - | 0.27 | 0.27 |
|  | SE | - | 0.018 | 0.018 |
|  | QWL | - | 0.076 | 0.076 |
| H7$_a$ | SE | 0.282*** | - | 0.284*** |
| H8$_a$ | FC | 0.1** | - | 0.1*** |
| H3$_a$ | BI | 0.175*** | - | 0.175*** |
| Notes: * p<0.1; ** p<0.05; *** p<0.01; NS p>0.1 | | | | |

**Table 6-8: The summary of Direct, Indirect, and Total Effects in predicting actual usage for the British sample**

Table 6.8 shows that both SE and FC have a strong direct positive effect on AU ($\gamma$=0.284*** and $\gamma$=0.1**) which supports H7a and H8a respectively. AU is also positively influenced by BI (ß=0.175***), supporting H3a. The results indicate that BI, SE and FC account for 79 % ($R^2 = 0.79$) of the variance of AU, with SE contributing the most to actual usage compared to the other constructs.

In conclusion, there was no need for model refinement for the British sample as the first run indicates that all the indices were in the acceptable level. Furthermore, the results of the structural model revealed that all the hypothesised relationships were supported except the relationship between SE and BI. The next subsection reports the testing of the structural model for the Lebanese sample.

## 6.3.2  Testing the Structural Model for the Lebanese sample

The structural model for the Lebanese sample is depicted in Figure 6.5.

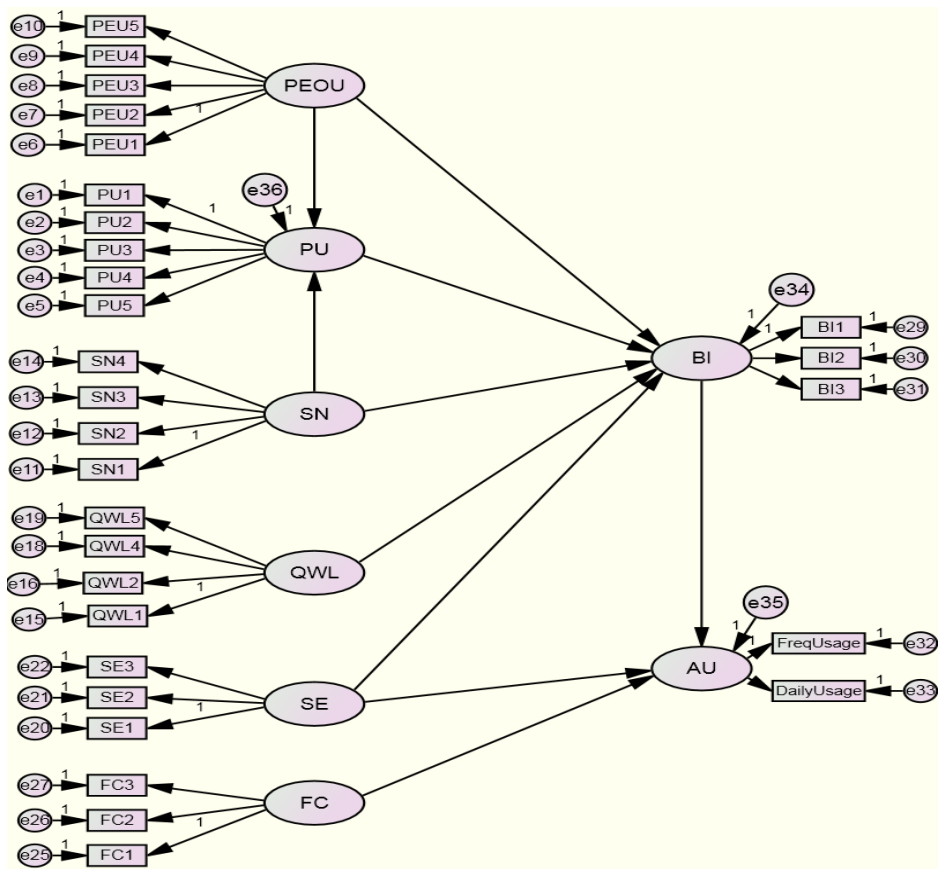

**Figure 6-5: the structural model for the Lebanese sample** *Note: the covariance (the two-headed arrows) between 6 exogenous variables was deleted for the purpose of presentation.*

As with the British sample, the results of the fit indices for the first run of the Lebanese sample suggest a good fit of the model (CMIN=898.677; df= 355; CMIN/DF = 2.531; GFI=.894; AGFI=.871; CFI=.949; RMSR=.074, RMSEA=.052, NFI=.915; PNFI=.803). Therefore, the second stage is to examine the hypothesized relationships within the proposed research model. The results of the path coefficients for the hypothesised relationships within the proposed research model are presented in Table 6.9.

| H# | Proposed Relationship | Effects Type | Path coefficient | Study Results |
|---|---|---|---|---|
| Intention and Actual Usage Prediction | | | | |
| H1$_b$ | PEOU (+) $\longrightarrow$ BI | Direct effect | 0.106* | Supported |
| H2$_b$ | PU (+) $\longrightarrow$ BI | Direct effect | 0.195*** | Supported |
| H4$_b$ | SN (+) $\longrightarrow$ BI | Direct effect | 0.095* | Supported |
| H5$_b$ | QWL (+) $\longrightarrow$ BI | Direct effect | 0.408*** | Supported |
| H6$_b$ | SE (+) $\longrightarrow$ BI | Direct effect | 0.041 | Not supported |
| H7$_b$ | SE (+) $\longrightarrow$ AU | Direct effect | 0.105* | Supported |
| H8$_b$ | FC(+) $\longrightarrow$ AU | Direct effect | 0.176*** | Supported |
| H3$_b$ | BI (+) $\longrightarrow$ AU | Direct effect | 0.374*** | Supported |
| H9$_b$ | SN(+) $\longrightarrow$ PU | Direct effect | 0.167** | Supported |
| H10$_b$ | PEOU(+) $\longrightarrow$ PU | Direct effect | 0.318*** | Supported |
| Notes: * p<0.1; ** p<0.05; *** p<0.01; NS p>0.1 | | | | |

**Table 6-9: The summary of Direct Hypothesized results for the Lebanese sample**

The factor loadings (regression weights) output in AMOS indicates that 9 out of 10 hypothesised relationships for the Lebanese sample were supported (refer to Table 6.9). More specifically, both PEOU ($\gamma$=0.106*) and PU ($\gamma$=0.195***) were found to have a strong significant positive influence on behavioural intention to use web-based learning system, hence H1b and H2b were supported. Additionally, the hypothesised path from SN ($\gamma$=0.095*) to BI was also significant, supporting H4b. Moreover, behavioural intention was also influenced by the quality of work life ($\gamma$=0.408***) which supports H5b. However, contrary to our expectation, the hypothesised relationship between SE ($\gamma$=0.041) and BI was rejected. PEOU, PU, SN and QWL account for 57% ($R^2 = 0.57$) of the variance of BI, and similar to the British sample, QWL contributed the most to behavioural intention compared to the other constructs. Furthermore, PU was found to be influenced by SN ($\gamma$=0.167**) and PEOU ($\gamma$=0.318***), supporting H9 $_b$ and H10 $_b$. Both SN and PEOU explained 32% of the PU.

The results of the direct, indirect and total effects on AU are presented in Table 6.10.

| Variables | Direct | Indirect | Total |
|-----------|--------|----------|-------|
| PEOU | - | 0.068 | 0.068 |
| PU | - | 0.073 | 0.073 |
| SE | - | 0.048 | 0.048 |
| SN | - | 0.043 | 0.043 |
| QWL | - | 0.154 | 0.154 |
| SE | 0.105* | - | 0.105* |
| FC | 0.176*** | - | 0.174*** |
| BI | 0.374*** | - | 0.374*** |
| Notes: * p<0.1; ** p<0.05; *** p<0.01; NS p>0.1 | | | |

**Table 6-10: Direct, Indirect, and Total Effects in predicting actual usage for the Lebanese sample**

As can be shown in Table 6.10, the direct path from SE ($\gamma$=0.105*) and FC ($\gamma$=0.176***) is significant, which supports H7b and H8b respectively. Furthermore, actual usage is influenced by the behavioural intention (ß=0.374***), supporting H3b. The results of the squared multiple correlation indicate that BI, SE and FC account for 52 % ($R^2 = 0.52$) of the variance of AU (indicating a reasonable explanation for AU), with BI having the highest impact on the actual usage compared to the other constructs. These results indicate that the higher the level of behavioural intention, the higher the actual usage will be.

Overall, the results indicated that there was no need to refine the model since the first run indicates that all the fit indices were within an acceptable range. The results also indicate that; except for the relationship between SE and BI; all the other hypothesised direct relationships were supported. A detailed discussion about the results will take place in Chapter 7.

In summary, the determinants PEOU, PU, SN, SE and QWL accounted for 68% and 57% of the variance of BI for the British and Lebanese sample, respectively. Moreover, the determinants SE, FC and BI accounted for 79% of the variance of AU for the British sample and 52% for the Lebanese sample.

The refined model indicated that the similarities between the Lebanese and British model are greater than the differences. More specifically, PEOU, PU, SE, FC, BI and AU were represented with the same indictors, whereas differences were detected in terms of SN (SN4 was deleted from the British sample) and QWL (QWL3 was deleted from the Lebanese sample).

Having tested the direct path for both samples, the next step is to test the moderating effect of culture and other demographic characteristics on the relationships between the 6 exogenous (PEOU, PU, SN, QWL, SE, FC) and the two endogenous (BI and AU) constructs.

## 6.4   Moderating effects

This section will discuss the moderating effect of  Hofstede's four cultural dimensions (MF, ID, PD, UA) at the individual level as well as four demographic variables (gender, age, educational level and experience) on the relationships between the exogenous (PEOU, PU, SN, QWL and FC) and the endogenous (BI and AU) latent constructs.

The impacts of these moderators were investigated through using multi-group analysis. In this approach, the data-sample is divided into subsamples and then the same structural model is run at the same time for both samples. It is then followed by pairwise comparison in path coefficients across the two groups (high vs. low), considering the critical ratio for differences among the groups in order to establish a reliability and validity. However, following Hair et al (2010) recommendation, we first examined the assessment of the measurement model (i.e. goodness of fit including discriminant and convergent in addition to reliability for each construct) before proceeding to examine the impacts of moderators on relationship between the constructs. The hypotheses related to the moderating impact of culture and other individual variables that have been identified in Chapter 3 during the model development stage are as follows:

### *Cultural variables*

*H11a1,a2,a3,a4,a5,a6,a7: The relationship between (PEOU, PU, SN, QWL, SE, FC) and Behavioural Intention and actual usage of the e-learning system is moderated by the* **Power Distance** *value in the British context.*

*H11b1,b2,b3,b4,b5,b6,b7: The relationship between (PEOU, PU, SN, QWL, SE, FC) and Behavioural Intention and actual usage of the e-learning system is moderated by the* **Power Distance** *value in the Lebanese context.*

*H12a1,a2,a3,a4,a5,a6,a7: The relationship between (PEOU, PU, SN, QWL, SE, FC) and Behavioural Intention and actual usage of the e-learning system is moderated by the* **masculinity/femininity** *value in the British context.*

*H12b1,b2,b3,b4,b5,b6,b7: The relationship between (PEOU, PU, SN, QWL, SE, FC) and Behavioural Intention and actual usage of the e-learning system is moderated by the* **masculinity/femininity** *value in the Lebanese context.*

*H13a1,a2,a3,a4,a5,a6,a7: The relationship between (PEOU, PU, SN, QWL, SE, FC) and Behavioural Intention and actual usage of the e-learning system is moderated by the* **individualism /collectivism** *value in the British context.*

*H13b1,b2,b3,b4,b5,b6,b7: The relationship between (PEOU, PU, SN, QWL, SE, FC) and Behavioural Intention and actual usage of the e-learning system is moderated by the* **individualism /collectivism** *value in the Lebanese context.*

*H14a1,a2,a3,a4,a5,a6,a7: The relationship between (PEOU, PU, SN, QWL, SE, FC) and Behavioural Intention and actual usage of the e-learning system is moderated by the* **Uncertainty Avoidance** *value in the British context.*

*H14b1,b2,b3,b4,b5,b6,b7: The relationship between (PEOU, PU, SN, QWL, SE, FC) and Behavioural Intention and actual usage of the e-learning system is moderated by the* **Uncertainty Avoidance** *value in the Lebanese context.*

### Demographic variables:

*H15a1,a2,a3,a4,a5,a6,a7: The relationship between (PEOU, PU, SN, QWL, SE, FC) and Behavioural Intention and actual usage of the e-learning system will be moderated by the* **gender** *in the British context.*

*H15b1,b2,b3,b4,b5,b6,b7: The relationship between (PEOU, PU, SN, QWL, SE, FC) and Behavioural Intention and actual usage of the e-learning system will be moderated by the* **gender** *in the Lebanese context.*

*H16a1,a2,a3,a4,a5,a6,a7: The relationship between (PEOU, PU, SN, QWL, SE, FC) and Behavioural Intention and actual usage of the e-learning system will be moderated by the* **age** *in the British context.*

*H16b1,b2,b3,b4,b5,b6,b7: The relationship between (PEOU, PU, SN, QWL, SE, FC) and Behavioural Intention and actual usage of the e-learning system will be moderated by the* **age** *in the Lebanese context.*

*H17a1,a2,a3,a4,a5,a6,a7: The relationship between (PEOU, PU, SN, QWL, SE, FC) and Behavioural Intention and actual usage of the e-learning system will be moderated by* **Educational Level** *in the British context.*

*H17b1,b2,b3,b4,b5,b6,b7: The relationship between (PEOU, PU, SN, QWL, SE, FC) and Behavioural Intention and actual usage of the e-learning system will be moderated by* **Educational Level** *in the Lebanese context.*

*H18a1,a2,a3,a4,a5,a6,a7: The relationship between (PEOU, PU, SN, QWL, SE, FC) and Behavioural Intention and actual usage of the e-learning system will be moderated by* **Experience** *in the British context.*

*H18b1,b2,b3,b4,b5,b6,b7: The relationship between (PEOU, PU, SN, QWL, SE, FC) and Behavioural Intention and actual usage of the e-learning system will be moderated by **Experience** in the Lebanese context.*

## 6.5 Moderating effects within the British dataset

This section will discuss the moderating effect of Hofstede's four cultural dimensions at the individual level, as well as, four demographic variables namely gender, age, educational level and experience on the relationships between the exogenous (PEOU, PU, SN, QWL and FC) and the endogenous (BI and AU) latent constructs within the British model.

### 6.5.1 Hofstede's Cultural Dimensions

This section will report the results of the moderating impact of Hofstede's four cultural dimensions; namely power distance, masculinity\femininity, individualism\collectivism and uncertainty avoidance; on the relationships within the British model.

#### 6.5.1.1 Power distance

The construct PD was measured in the survey questionnaire using a five items using 7-point scales ranging from 1-strongly disagree to 7-strongly agree. The overall mean was 2.567/7 indicating a very low PD culture. Since the moderating construct PD is metric is nature, we used median-split method to transfer the metric scale into a nonmetric (categorical) scale (Hair et al., 2010). Out of the 602 respondents, there were 307 students within the low PD group (median<=2.4) and 295 within the high PD group (median>2.4).

Applying the measurement model for each group separately revealed the following: for the low-PD group [$\chi^2$ =658.380; df= 349; $\chi^2$/df= 1.964; GFI =.866; AGFI=.833; CFI=.926; RMSR=.061; RMSEA=.056; NFI=.862; PNFI=.741] and [$\chi^2$ =566.435; df= 349; $\chi^2$/df= 1.623; GFI =.883; AGFI=.854; CFI=.964;

RMSR=.078; RMSEA=.046, NFI=.911; PNFI=.783] for the high-PD group, indicating an acceptable fit of the data.

Table 6.11 shows that the CR and AVE values within both samples were all above 0.7 and above 0.5 respectively and therefore all factors have satisfactory reliability and convergent validity. Additionally, except for AU in high-PD group, the MSV for all other constructs is less than AVE and the square root of AVE is higher than their correlation value, which indicates that discriminant validity is also established for both samples.

| Factor Correlation Matrix with √AVE on the diagonal (Power Distance: low level group) | | | | | | | | | | | |
|---|---|---|---|---|---|---|---|---|---|---|---|
| Factor | CR | AVE | MSV | ASV | SE | PU | PEOU | SN | QWL | FC | AU | BI |
| SE | 0.869 | 0.689 | 0.441 | 0.238 | 0.830 | | | | | | | |
| PU | 0.883 | 0.603 | 0.335 | 0.200 | 0.387 | 0.776 | | | | | | |
| PEOU | 0.883 | 0.602 | 0.412 | 0.212 | 0.642 | 0.512 | 0.776 | | | | | |
| SN | 0.819 | 0.602 | 0.111 | 0.054 | 0.190 | 0.255 | 0.197 | 0.776 | | | | |
| QWL | 0.858 | 0.548 | 0.308 | 0.132 | 0.283 | 0.538 | 0.253 | 0.145 | 0.740 | | | |
| FC | 0.788 | 0.564 | 0.331 | 0.210 | 0.573 | 0.421 | 0.575 | 0.333 | 0.348 | 0.751 | | |
| AU | 0.711 | 0.514 | 0.441 | 0.149 | 0.664 | 0.346 | 0.366 | 0.165 | 0.193 | 0.341 | 0.666 | |
| BI | 0.806 | 0.581 | 0.335 | 0.233 | 0.468 | 0.579 | 0.491 | 0.281 | 0.555 | 0.534 | 0.407 | 0.762 |
| Factor Correlation Matrix with √AVE on the diagonal (Power Distance: high level group) | | | | | | | | | | | |
| Factor | CR | AVE | MSV | ASV | SE | PU | PEOU | SN | QWL | FC | AU | BI |
| SE | 0.856 | 0.666 | 0.542 | 0.229 | 0.816 | | | | | | | |
| PU | 0.918 | 0.691 | 0.480 | 0.340 | 0.430 | 0.831 | | | | | | |
| PEOU | 0.931 | 0.730 | 0.396 | 0.256 | 0.462 | 0.599 | 0.854 | | | | | |
| SN | 0.819 | 0.602 | 0.320 | 0.193 | 0.275 | 0.566 | 0.361 | 0.776 | | | | |
| QWL | 0.886 | 0.609 | 0.534 | 0.358 | 0.433 | 0.693 | 0.629 | 0.517 | 0.780 | | | |
| FC | 0.897 | 0.747 | 0.587 | 0.280 | 0.508 | 0.495 | 0.462 | 0.374 | 0.536 | 0.864 | | |
| AU | 0.705 | 0.543 | 0.587 | 0.346 | 0.736 | 0.602 | 0.367 | 0.421 | 0.503 | 0.766 | 0.607 | |
| BI | 0.893 | 0.735 | 0.634 | 0.340 | 0.373 | 0.655 | 0.585 | 0.492 | 0.796 | 0.476 | 0.608 | 0.858 |

**Table 6-11: Construct reliability, convergent validity and discriminant validity for moderator power distance**

The results presented in Table 6.12 show that the cultural variable power distance was found to moderate the relationship between PU_BI (supporting H11a2), SN_BI (supporting H11a3), and QWL_BI (supporting H11a4). The relationship was stronger for low-PD group in terms of PU_BI, QWL_BI, while the relationship was stronger for high-PD group in terms of SN_BI. Contrary to our expectation, PD did not moderate the relationship between PEOU_BI (rejecting H11a1), SE_AU (rejecting H11a5) and FC_AU (rejecting H11a6). These results indicate that hypothesis H11a was partially supported. Overall, the $R^2$ for BI was 49.7% and 48.2% for AU within the low-PD group, while within the high-PD group the variance explained ($R^2$) for BI was 57.9% and 55.6% for AU which indicates a moderate fit for the low-group model and good fit for the high-PD model.

| Hypothesis | Low PD | | High PD | | Z-score | Results |
|---|---|---|---|---|---|---|
| | $R^2$ | Estimate | $R^2$ | Estimate | | |
| PEOU →BI | 49.7% | 0. 184 | 57.9% | 0. 138 | -0.479 | Not supported |
| PU → BI | | 0. 241 | | 0.125 | -1.748* | Supported |
| SN→ BI | | 0.267 | | 0.344 | 1.684* | Supported |
| QWL→BI | | 0.458 | | 0.307 | -1.673* | Supported |
| SE→AU | 48.2% | 0.182 | 55.6% | 0.195 | 0.219 | Not supported |
| FC→AU | | 0.121 | | 0. 158 | 0.325 | Not supported |
| Notes: *** p-value < 0.01; ** p-value < 0.05; * p-value < 0.10 | | | | | | |

**Table 6-12: The summary of the moderating effect of power distance**

## 6.5.1.2 Masculinity\femininity

Six items were used to measure the moderating construct MF. The construct was measured using 7-point scales ranging from 1-strongly disagree to 7-strongly agree. The overall mean was 2.535/7 indicating a masculine culture. Since moderating construct MF is metric in nature, we used median-split method to transfer the metric scale into a nonmetric (categorical) scale. There were 290 within the masculine group (median<=2.833) and 312 within the feminine group (median>2.833).

The first run of the measurement model resulted the following fit for the low-MF sample [$\chi^2$ =637.930; df= 349; $\chi^2$/df= 1.828; GFI =.873; AGFI=.842; CFI=.950; RMSR=.066; RMSEA=.053, NFI=.893; PNFI=.771] and [$\chi^2$ =630.579; df= 349; $\chi^2$/df= 1.807; GFI =.866; AGFI=.834; CFI=.957; RMSR=.08; RMSEA=.052,

NFI=.91; PNFI=.782] for the high-MF group which indicates a good fit of the data.

The results presented in Table 6.13 shows that the CR value is higher than 0.7 and the AVE is higher than 0.5 for all the constructs within the two samples, which indicates an adequate reliability and convergent validity. Except for the AU construct for both samples, the MSV for all other constructs is less than AVE and the square root of AVE is higher than their correlation value, therefore discriminant validity is considered satisfactory for both samples.

| Factor Correlation Matrix with √AVE on the diagonal (Masculinity: low level group) | | | | | | | | | | | | |
|---|---|---|---|---|---|---|---|---|---|---|---|---|
| **Factor** | **CR** | **AVE** | **MSV** | **ASV** | **SE** | **PU** | **PEOU** | **SN** | **QWL** | **FC** | **AU** | **BI** |
| **SE** | 0.891 | 0.731 | 0.573 | 0.326 | 0.855 | | | | | | | |
| **PU** | 0.905 | 0.655 | 0.493 | 0.319 | 0.515 | 0.809 | | | | | | |
| **PEOU** | 0.918 | 0.692 | 0.432 | 0.281 | 0.657 | 0.644 | 0.832 | | | | | |
| **SN** | 0.850 | 0.654 | 0.166 | 0.114 | 0.317 | 0.337 | 0.325 | 0.809 | | | | |
| **QWL** | 0.886 | 0.609 | 0.461 | 0.229 | 0.385 | 0.649 | 0.411 | 0.278 | 0.780 | | | |
| **FC** | 0.861 | 0.679 | 0.489 | 0.325 | 0.699 | 0.545 | 0.596 | 0.407 | 0.447 | 0.824 | | |
| **AU** | 0.725 | 0.502 | 0.573 | 0.275 | 0.757 | 0.478 | 0.424 | 0.316 | 0.355 | 0.651 | 0.709 | |
| **BI** | 0.852 | 0.659 | 0.493 | 0.331 | 0.524 | 0.702 | 0.555 | 0.371 | 0.679 | 0.589 | 0.544 | 0.812 |
| **Factor Correlation Matrix with √AVE on the diagonal (Masculinity: high level group)** | | | | | | | | | | | | |
| **Factor** | **CR** | **AVE** | **MSV** | **ASV** | **SE** | **PU** | **PEOU** | **SN** | **QWL** | **FC** | **AU** | **BI** |
| **SE** | 0.896 | 0.742 | 0.711 | 0.325 | 0.861 | | | | | | | |
| **PU** | 0.926 | 0.714 | 0.471 | 0.396 | 0.529 | 0.845 | | | | | | |
| **PEOU** | 0.923 | 0.707 | 0.421 | 0.315 | 0.546 | 0.607 | 0.841 | | | | | |
| **SN** | 0.819 | 0.602 | 0.416 | 0.269 | 0.404 | 0.645 | 0.396 | 0.776 | | | | |
| **QWL** | 0.880 | 0.596 | 0.566 | 0.371 | 0.482 | 0.686 | 0.634 | 0.527 | 0.772 | | | |
| **FC** | 0.903 | 0.759 | 0.534 | 0.333 | 0.564 | 0.559 | 0.529 | 0.467 | 0.587 | 0.871 | | |
| **AU** | 0.715 | 0.511 | 0.711 | 0.441 | 0.843 | 0.686 | 0.528 | 0.566 | 0.518 | 0.731 | 0.715 | |
| **BI** | 0.905 | 0.760 | 0.599 | 0.416 | 0.522 | 0.675 | 0.649 | 0.576 | 0.774 | 0.569 | 0.712 | 0.872 |

**Table 6-13: Construct reliability, convergent validity and discriminant validity for moderator masculinity\femininity**

Table 6.14 shows that masculinity\femininity cultural variable moderates the relationship between PEOU_BI (supporting H12a1), PU_BI (supporting H12a2) and SN_BI (supporting H12a3) with the relationship found to be stronger for masculine group in terms of PU_BI while the relationship was stronger for feminine groups in terms of PEOU_BI and SN_BI. However, no differences were found on the relationship between QWL_BI (rejecting H12a4), SE_AU (rejecting H12a5) and FC_AU (rejecting H12a6). The results suggests that hypothesis H12a was partially supported. Overall, The $R^2$ for BI was 61.1% and 63.4% for AU in the low-MF group, whereas the $R^2$ for BI was 67.7% and 72.5% for AU in the high-group indicating an acceptable model fit of the data.

| Hypothesis | High Masculine | | Low Masculine | | Z-score | Results |
|---|---|---|---|---|---|---|
| | $R^2$ | Estimate | $R^2$ | Estimate | | |
| PEOU →BI | 61.1% | 0. 103 | 67.7% | 0.278 | 1.656* | Supported |
| PU → BI | | 0. 183 | | 0.068 | -1.889* | Supported |
| SN→ BI | | 0.189 | | 0. 293 | 1.940* | Supported |
| QWL→BI | | 0.346 | | 0.465 | 1.206 | Not supported |
| SE→AU | 63.4% | 0.127 | 72.5% | 0. 196 | 0.899 | Not supported |
| FC→AU | | 0.070 | | 0.128 | 1.310 | Not supported |
| Notes: *** p-value < 0.01; ** p-value < 0.05; * p-value < 0.10 | | | | | | |

**Table 6-14: The summary of the moderating effect of masculinity\femininity**

### 6.5.1.3 Individualism\collectivism

The moderating construct IC was measured by six items using 7-point scales ranging from 1-strongly disagree to 7-strongly agree. The overall mean was 3.82/7 indicating a moderate individualism\collectivism culture. Since the moderating construct MF is metric is nature, we used median-split method to transfer the metric scale into nonmetric (categorical) scale. Out of the 602 respondents in the study, the low- IC group (median<=3.82) were 286 students and the high-IC group (median>3.82) were 316.

Applying the measurement model for each group separately revealed the following: for the low-IC group [$\chi^2$ =674.289; df= 349; $\chi^2$/df= 1.932; GFI =.865; AGFI=.832; CFI=.953; RMSR=.076; RMSEA=.056, NFI=.909; PNFI=.781], and [$\chi^2$ =573.871; df= 349; $\chi^2$/df= 1.644; GFI =.885; AGFI=.857; CFI=.965;

RMSR=.068; RMSEA=.052, NFI=.914; PNFI=.786] for the high-PD group, indicating an acceptable fit of the data.

Table 6.15 shows that the CR and AVE values for all the constructs within both samples were all above 0.7 and above 0.5 respectively and therefore all factors have satisfactory reliability and convergent validity. Additionally, except for AU in both samples, the MSV for all other constructs is less than AVE and the square root of AVE is higher than their correlation value, which indicates that discriminant validity is also established for both samples.

| Factor Correlation Matrix with √AVE on the diagonal (Individualism: low level group) | | | | | | | | | | | | |
|---|---|---|---|---|---|---|---|---|---|---|---|---|
| **Factor** | **CR** | **AVE** | **MSV** | **ASV** | **SE** | **PU** | **PEOU** | **SN** | **QWL** | **FC** | **AU** | **BI** |
| **SE** | 0.902 | 0.754 | 0.719 | 0.408 | 0.869 | | | | | | | |
| **PU** | 0.932 | 0.732 | 0.539 | 0.407 | 0.591 | 0.856 | | | | | | |
| **PEOU** | 0.923 | 0.706 | 0.477 | 0.361 | 0.691 | 0.651 | 0.840 | | | | | |
| **SN** | 0.871 | 0.693 | 0.280 | 0.229 | 0.427 | 0.529 | 0.364 | 0.832 | | | | |
| **QWL** | 0.901 | 0.645 | 0.569 | 0.351 | 0.509 | 0.701 | 0.540 | 0.478 | 0.803 | | | |
| **FC** | 0.879 | 0.711 | 0.482 | 0.394 | 0.686 | 0.592 | 0.652 | 0.491 | 0.578 | 0.843 | | |
| **AU** | 0.746 | 0.599 | 0.719 | 0.434 | 0.848 | 0.646 | 0.587 | 0.527 | 0.530 | 0.694 | 0.774 | |
| **BI** | 0.899 | 0.748 | 0.569 | 0.453 | 0.631 | 0.734 | 0.659 | 0.509 | 0.754 | 0.675 | 0.721 | 0.865 |
| **Factor Correlation Matrix with √AVE on the diagonal (Individualism: high level group)** | | | | | | | | | | | | |
| **Factor** | **CR** | **AVE** | **MSV** | **ASV** | **SE** | **PU** | **PEOU** | **SN** | **QWL** | **FC** | **AU** | **BI** |
| **SE** | 0.905 | 0.760 | 0.667 | 0.316 | 0.872 | | | | | | | |
| **PU** | 0.910 | 0.669 | 0.465 | 0.368 | 0.529 | 0.818 | | | | | | |
| **PEOU** | 0.926 | 0.714 | 0.396 | 0.290 | 0.541 | 0.629 | 0.845 | | | | | |
| **SN** | 0.785 | 0.550 | 0.261 | 0.184 | 0.345 | 0.505 | 0.419 | 0.742 | | | | |
| **QWL** | 0.875 | 0.585 | 0.575 | 0.320 | 0.454 | 0.682 | 0.583 | 0.378 | 0.765 | | | |
| **FC** | 0.901 | 0.755 | 0.549 | 0.327 | 0.610 | 0.567 | 0.511 | 0.433 | 0.537 | 0.869 | | |
| **AU** | 0.724 | 0.533 | 0.667 | 0.377 | 0.817 | 0.629 | 0.447 | 0.384 | 0.475 | 0.741 | 0.730 | |
| **BI** | 0.893 | 0.737 | 0.575 | 0.386 | 0.521 | 0.682 | 0.606 | 0.511 | 0.758 | 0.556 | 0.675 | 0.859 |

**Table 6-15: Construct reliability, convergent validity and discriminant**

**validity for moderator individualism\collectivism**

The results presented in Table 6.16 show that individualism\collectivism cultural variable moderates the relationship between PEOU_BI (supporting H13a1), PU_BI (supporting H13a2), SN_BI (supporting H13a3) and FC_AU (supporting

H13a6). Specifically, the relationship was stronger for low-IC group in terms of PU_BI, while the relationship was stronger for high-IC group in terms of PEOU_BI, SN_BI and FC_AU. The results also show that no differences were detected on the relationship between QWL_BI (rejecting H13a4) and SE_AU (rejecting H13a5). This means that hypothesis H13a was partially supported. Overall, The $R^2$ for BI was 69.9% and for AU was 74.2% within the low-IC group, while within the high-IC group the variance explained ($R^2$) for BI was 66.3% and 71.4% for AU, indicating an acceptable model.

| Hypothesis | Individualism | | Collectivism | | Z-score | Results |
|---|---|---|---|---|---|---|
| | $R^2$ | Estimate | $R^2$ | Estimate | | |
| PEOU →BI | | 0.153 | | 0.259 | 1.672* | Supported |
| PU → BI | 69.9% | 0.265 | 66.3% | 0.103 | -2.136** | Supported |
| SN→ BI | | 0.095 | | 0.168 | 1.702* | Supported |
| QWL→BI | | 0.398 | | 0.289 | -1.002 | Not supported |
| SE→AU | 74.2% | 0.239 | 71.4% | 0.320 | 1.465 | Not supported |
| FC→AU | | 0.053 | | 0.146 | 1.672* | Supported |
| Notes: *** p-value < 0.01; ** p-value < 0.05; * p-value < 0.10 | | | | | | |

**Table 6-16: The summary of the moderating effect of individualism\collectivism**

### 6.5.1.4 Uncertainty Avoidance

Six items were used to measure the moderating construct MF. The construct was measured using 7-point scales ranging from 1-strongly disagree to 7-strongly agree. The overall mean was 2.535/7 indicating a moderate uncertainty avoidance culture. Since the moderating construct UA is metric in nature, we used median-split method to transfer the metric scale into nonmetric (categorical) scale. Out of the 602 respondents, the results of the descriptive statistics showed that there are 324 students within the low UA group (median<=4.4) and 278 within the high UA group (median>4.4).

The first run of the measurement model resulted the following fit for the low-UA sample [$\chi^2$ =689.260; df= 349; $\chi^2$/df= 2.001; GFI =.872; AGFI=.840; CFI=.955;

RMSR=.083; RMSEA=.056, NFI=.915; PNFI=.786] and [χ² =599.470; df= 349; χ²/df= 1.718; GFI =.871; AGFI=.839; CFI=.955; RMSR=.067; RMSEA=.051, NFI=.901; PNFI=.773] for the high-MF group, which also indicates a good fit of the data.

As can be shown in Table 6.17, the CR values for all the constructs within both samples were higher than 0.7 and therefore all factors have adequate reliability. The results also show that the AVE is higher than 0.5 and that CR is higher than AVE for all the constructs which establish a satisfactory convergent validity. Additionally, Except for AU in both groups, the MSV for all other constructs is less than AVE and the square root of AVE is higher than their correlation value, therefore discriminant validity was also established for both samples.

| Factor Correlation Matrix with √AVE on the diagonal (Uncertainty Avoidance: low level group) | | | | | | | | | | | |
|---|---|---|---|---|---|---|---|---|---|---|---|
| Factor | CR | AVE | MSV | ASV | SE | PU | PEOU | SN | QWL | FC | AU | BI |
| SE | 0.920 | 0.793 | 0.731 | 0.402 | 0.890 | | | | | | | |
| PU | 0.930 | 0.727 | 0.554 | 0.427 | 0.600 | 0.853 | | | | | | |
| PEOU | 0.932 | 0.733 | 0.472 | 0.385 | 0.666 | 0.687 | 0.856 | | | | | |
| SN | 0.849 | 0.654 | 0.287 | 0.226 | 0.424 | 0.491 | 0.438 | 0.808 | | | | |
| QWL | 0.897 | 0.635 | 0.552 | 0.361 | 0.517 | 0.722 | 0.612 | 0.435 | 0.797 | | | |
| FC | 0.887 | 0.728 | 0.465 | 0.388 | 0.644 | 0.638 | 0.667 | 0.500 | 0.599 | 0.853 | | |
| AU | 0.748 | 0.601 | 0.731 | 0.418 | 0.855 | 0.658 | 0.580 | 0.490 | 0.514 | 0.614 | 0.775 | |
| BI | 0.909 | 0.769 | 0.554 | 0.465 | 0.647 | 0.744 | 0.659 | 0.536 | 0.743 | 0.682 | 0.739 | 0.877 |
| Factor Correlation Matrix with √AVE on the diagonal (Uncertainty Avoidance: high level group) | | | | | | | | | | | |
| Factor | CR | AVE | MSV | ASV | SE | PU | PEOU | SN | QWL | FC | AU | BI |
| SE | 0.880 | 0.711 | 0.646 | 0.316 | 0.843 | | | | | | | |
| PU | 0.911 | 0.671 | 0.436 | 0.351 | 0.514 | 0.819 | | | | | | |
| PEOU | 0.913 | 0.678 | 0.364 | 0.254 | 0.562 | 0.572 | 0.823 | | | | | |
| SN | 0.820 | 0.603 | 0.333 | 0.183 | 0.333 | 0.577 | 0.303 | 0.776 | | | | |
| QWL | 0.879 | 0.594 | 0.544 | 0.323 | 0.464 | 0.658 | 0.485 | 0.430 | 0.789 | | | |
| FC | 0.889 | 0.730 | 0.681 | 0.342 | 0.643 | 0.536 | 0.491 | 0.409 | 0.545 | 0.855 | | |
| AU | 0.731 | 0.524 | 0.681 | 0.401 | 0.804 | 0.616 | 0.447 | 0.421 | 0.532 | 0.825 | 0.724 | |
| BI | 0.881 | 0.713 | 0.613 | 0.375 | 0.499 | 0.660 | 0.603 | 0.461 | 0.783 | 0.557 | 0.665 | 0.844 |

**Table 6-17: Construct reliability, convergent validity and discriminant validity for moderator uncertainty avoidance**

Table 6.18 shows that only three paths were moderated by UA cultural variable. These paths were PU_BI (supporting H14a2), SN_BI (supporting H140a3) and FC_AU (supporting H14a6). More specifically, the relationship was stronger for low-UA group in terms of PU_BI, while the relationship was stronger for high-UA in terms of SN_BI and FC_AU. The results also show that no differences were detected in the relationship between PEOU_BI (rejecting H14a1), QWL_BI (rejecting H14a4) and SE_AU (rejecting H14a5). Thus it can be concluded that hypothesis H14a was partially supported. It was also found that the $R^2$ for BI was 68.7% and 64.5% for AU within the low-UA sample, while within the high-UA sample the variance explained ($R^2$) for BI was 63.8% and 66.4% for AU, indicating an acceptable model fit.

| Hypothesis | Low UA | | High UA | | Z-score | Results |
|---|---|---|---|---|---|---|
| | $R^2$ | Estimate | $R^2$ | Estimate | | |
| PEOU➔BI | 68.7% | 0.163 | 63.8% | 0.208 | 0.720 | Not supported |
| PU➔ BI | | 0.289 | | 0.101 | -1.829* | Supported |
| SN➔ BI | | 0.132 | | 0.238 | 1.848* | Supported |
| QWL➔BI | | 0.369 | | 0.307 | -.506 | Not supported |
| SE➔AU | 64.5% | 0.078 | 66.4% | 0.145 | 1.019 | Not supported |
| FC➔AU | | 0.067 | | 0.132 | 1.692* | Supported |
| Notes: *** p-value < 0.01; ** p-value < 0.05; * p-value < 0.10 | | | | | | |

**Table 6-18: The summary of the moderating effect of Uncertainty Avoidance**

## 6.5.2 Demographic Characteristics of British sample

This section will report the results of the moderating impact of the demographic characteristics namely gender, age, educational level and experience; on the relationships within the British model.

### 6.5.2.1 Gender

Since gender was nonmetric (categorical) in nature, there was no need to refine the division of the groups within the sample (Hair et al., 2010). Out of the 602 respondents in the survey, there were 315 males and 287 females. It is essential to test whether each group achieve an adequate fit for the data separately before proceeding with testing the effect of moderators on the relationship between exogenous (independent) and endogenous (dependent) constructs (Hair et al., 2010).

The first run of the model revealed the following results for the male group [$\chi^2$ =658.905; df= 349; $\chi^2$/df= 1.965; GFI =.867; AGFI=.834; CFI=.945; RMSR=.074; RMSEA=.055, NFI=.895; PNFI=.769] and [$\chi^2$ =608.282; df= 349; $\chi^2$/df= 1.743; GFI =.871; AGFI=.839; CFI=.959; RMSR=.084; RMSEA=.051, NFI=.909; PNFI=.782] for the female group, which indicate a good fit of the data. As can be shown in Table 6.19, the CR for all the constructs within both sub-samples were all above 0.7 and therefore all factors have adequate reliability. The results also show that the AVE is higher than 0.5 and that CR is higher than AVE for all the constructs which establish a satisfactory convergent validity. Additionally, except for AU construct in both sub-samples, the MSV for all other constructs is less than AVE and the square root of AVE is higher than their correlation value, therefore discriminant validity was also established for both sub-samples.

| Factor Correlation Matrix with √AVE on the diagonal (Gender: Male group) | | | | | | | | | | | | |
|---|---|---|---|---|---|---|---|---|---|---|---|---|
| **Factor** | **CR** | **AVE** | **MSV** | **ASV** | **SE** | **PU** | **PEOU** | **SN** | **QWL** | **FC** | **AU** | **BI** |
| **SE** | 0.907 | 0.766 | 0.645 | 0.328 | 0.875 | | | | | | | |
| **PU** | 0.905 | 0.657 | 0.480 | 0.320 | 0.522 | 0.810 | | | | | | |
| **PEOU** | 0.912 | 0.675 | 0.450 | 0.305 | 0.624 | 0.628 | 0.821 | | | | | |
| **SN** | 0.790 | 0.557 | 0.222 | 0.129 | 0.303 | 0.384 | 0.314 | 0.746 | | | | |
| **QWL** | 0.879 | 0.593 | 0.445 | 0.249 | 0.423 | 0.609 | 0.447 | 0.344 | 0.770 | | | |
| **FC** | 0.864 | 0.686 | 0.393 | 0.291 | 0.627 | 0.512 | 0.582 | 0.353 | 0.497 | 0.828 | | |
| **AU** | 0.728 | 0.583 | 0.602 | 0.332 | 0.803 | 0.559 | 0.515 | 0.321 | 0.428 | 0.603 | 0.764 | |
| **BI** | 0.851 | 0.657 | 0.480 | 0.383 | 0.570 | 0.693 | 0.671 | 0.471 | 0.667 | 0.553 | 0.671 | 0.811 |
| Factor Correlation Matrix with √AVE on the diagonal (Gender: Female group) | | | | | | | | | | | | |
| **Factor** | **CR** | **AVE** | **MSV** | **ASV** | **SE** | **PU** | **PEOU** | **SN** | **QWL** | **FC** | **AU** | **BI** |
| **SE** | 0.884 | 0.718 | 0.669 | 0.318 | 0.847 | | | | | | | |
| **PU** | 0.923 | 0.705 | 0.545 | 0.374 | 0.527 | 0.840 | | | | | | |
| **PEOU** | 0.923 | 0.707 | 0.371 | 0.266 | 0.542 | 0.596 | 0.841 | | | | | |
| **SN** | 0.856 | 0.665 | 0.289 | 0.186 | 0.336 | 0.538 | 0.331 | 0.815 | | | | |
| **QWL** | 0.891 | 0.620 | 0.602 | 0.369 | 0.495 | 0.738 | 0.609 | 0.446 | 0.788 | | | |
| **FC** | 0.897 | 0.746 | 0.520 | 0.323 | 0.595 | 0.554 | 0.494 | 0.426 | 0.556 | 0.863 | | |
| **AU** | 0.702 | 0.531 | 0.669 | 0.379 | 0.818 | 0.621 | 0.414 | 0.462 | 0.513 | 0.721 | 0.729 | |
| **BI** | 0.914 | 0.779 | 0.654 | 0.382 | 0.520 | 0.678 | 0.561 | 0.443 | 0.809 | 0.589 | 0.657 | 0.883 |

**Table 6-19: Construct reliability, convergent validity and discriminant validity for moderator gender**

The results of the multi-group analysis (MGA) presented in Table 6.20 show that gender moderates the relationship between PU_BI (supporting H15a2), SN_BI (supporting H15a3), QWL_BI (supporting H15a4), and FC_AU(supporting H15a6). Specifically, the relationship was stronger for males in terms of PU_BI and QWL_BI, while the relationship was stronger for females in terms of SN_BI and FC_AU. However, no differences were detected on the relationship between PEOU_BI (rejecting H15a1) and SE_AU (rejecting H15a5). Therefore, the results indicate that hypothesis H15a was partially supported. Overall, The $R^2$ for BI was 66.7% and for AU was 69.1% within the male sample, while within the female group the variance explained ($R^2$) for BI was 67.8% and 73.3% for AU, which indicates a good fit of the data.

| Hypothesis | Male | | Female | | Z-score | Results |
|---|---|---|---|---|---|---|
| | $R^2$ | Estimate | $R^2$ | Estimate | | |
| PEOU →BI | | 0.214 | | 0.129 | -0.839 | Not supported |
| PU → BI | 66.7% | 0.106 | 67.8% | 0.271 | 2.219** | Supported |
| SN→ BI | | 0.155 | | 0.276 | 1.712* | Supported |
| QWL→BI | | 0.472 | | 0.269 | -2.169** | Supported |
| SE→AU | 69.1% | 0.242 | 73.3% | 0.283 | 0.527 | Not supported |
| FC→AU | | 0.091 | | 0.147 | 1.689* | Supported |
| Notes: *** p-value < 0.01; ** p-value < 0.05; * p-value < 0.10 | | | | | | |

**Table 6-20: The summary of the moderating effect of gender**

## 6.5.2.2 Age

Similar to gender, the age was also ordinal in nature and therefore there was no need to refine the division of the groups within the sample (Hair et al., 2010). In our research, age was split into two groups; namely younger (age<=22) and older age groups (age> 22). Out of the 602 respondents, there were 370 students within the younger-age group and 232 students within the older-age group.

The first run of the measurement model resulted the following fit for the younger-age sample [$\chi^2$ =681.412; df= 349; $\chi^2$/df= 1.952; GFI =.884; AGFI=.855;

CFI=.955; RMSR=.077; RMSEA=.055, NFI=.913; PNFI=.784] and [$\chi^2$ =615.858; df= 349; $\chi^2$/df= 1.765; GFI =.851; AGFI=.814; CFI=.932; RMSR=.065; RMSEA=.058, NFI=.859; PNFI=.738] for the older-age group, which indicates a good fit of the data. The results presented in Table 6.21 show that for all the constructs within the two age group sub-samples the CR values were all above 0.7 and the AVE is higher than 0.5, therefore establishing adequate reliability and convergent validity. Except for AU construct for younger-age group (age <=22), the MSV for all other constructs is less than AVE and the square root of AVE is higher than their correlation value, therefore discriminant validity was also established for both samples.

| Factor Correlation Matrix with √AVE on the diagonal (Age:  group <=22) | | | | | | | | | | | |
|---|---|---|---|---|---|---|---|---|---|---|---|
| Factor | CR | AVE | MSV | ASV | SE | PU | PEOU | SN | QWL | FC | AU | BI |
| SE | 0.872 | 0.695 | 0.682 | 0.285 | 0.834 | | | | | | | |
| PU | 0.912 | 0.676 | 0.444 | 0.321 | 0.467 | 0.822 | | | | | | |
| PEOU | 0.919 | 0.695 | 0.310 | 0.234 | 0.501 | 0.552 | 0.833 | | | | | |
| SN | 0.839 | 0.636 | 0.285 | 0.182 | 0.315 | 0.534 | 0.354 | 0.797 | | | | |
| QWL | 0.880 | 0.595 | 0.561 | 0.296 | 0.406 | 0.666 | 0.505 | 0.438 | 0.771 | | | |
| FC | 0.885 | 0.723 | 0.444 | 0.293 | 0.601 | 0.499 | 0.496 | 0.430 | 0.504 | 0.850 | | |
| AU | 0.717 | 0.508 | 0.516 | 0.340 | 0.826 | 0.565 | 0.385 | 0.414 | 0.452 | 0.666 | 0.713 | |
| BI | 0.894 | 0.738 | 0.561 | 0.351 | 0.469 | 0.655 | 0.557 | 0.468 | 0.749 | 0.556 | 0.639 | 0.859 |
| Factor Correlation Matrix with √AVE on the diagonal (Age: group >22) | | | | | | | | | | | |
| Factor | CR | AVE | MSV | ASV | SE | PU | PEOU | SN | QWL | FC | AU | BI |
| SE | 0.916 | 0.785 | 0.504 | 0.293 | 0.886 | | | | | | | |
| PU | 0.901 | 0.645 | 0.441 | 0.280 | 0.500 | 0.803 | | | | | | |
| PEOU | 0.892 | 0.625 | 0.429 | 0.264 | 0.655 | 0.648 | 0.791 | | | | | |
| SN | 0.792 | 0.561 | 0.132 | 0.064 | 0.267 | 0.259 | 0.168 | 0.749 | | | | |
| QWL | 0.855 | 0.543 | 0.333 | 0.174 | 0.379 | 0.564 | 0.416 | 0.170 | 0.737 | | | |
| FC | 0.848 | 0.656 | 0.306 | 0.214 | 0.536 | 0.498 | 0.502 | 0.230 | 0.401 | 0.810 | | |
| AU | 0.737 | 0.504 | 0.500 | 0.239 | 0.710 | 0.467 | 0.435 | 0.264 | 0.241 | 0.553 | 0.731 | |
| BI | 0.800 | 0.572 | 0.441 | 0.307 | 0.605 | 0.664 | 0.598 | 0.363 | 0.577 | 0.438 | 0.574 | 0.756 |

**Table 6-21: Construct reliability, convergent validity and discriminant validity for moderator age**

As can be shown in Table 6.22, the results of MGA show that age moderate the relationships between PU_BI (supporting H16a2), SN_BI (supporting H16a3), QWL_BI (supporting H16a4) and SE_AU (supporting H16a5), with the relationship stronger for the younger-age group in terms of PU_BI, while the relationship was stronger for older-age group in terms of SN_BI, QWL_BI and SE_AU. On the other hand, no moderating effect was found on the relationship between PEOU_BI (rejecting H16a1) and FC_AU (rejecting H16a6), which suggests that hypothesis H16a was partially supported. Overall, The $R^2$ for BI was 63.2% and for AU was 71.5% within the younger-age group, while within the older-age group the variance explained ($R^2$) for BI was 57.8% and 56.4% for AU, which indicates a good fit of the data.

| Hypothesis | Age group <=22 | | Age group >22 | | Z-score | Results |
|---|---|---|---|---|---|---|
| | $R^2$ | Estimate | $R^2$ | Estimate | | |
| PEOU →BI | | 0.153 | | 0.238 | 0.997 | Not Supported |
| PU → BI | 63.2% | 0.174 | 57.8% | 0.083 | -1.772* | supported |
| SN→ BI | | 0. 086 | | 0.179 | 1.651* | Supported |
| QWL→BI | | 0.257 | | 0. 365 | 1.717* | Supported |
| SE→AU | 71.5% | 0. 149 | 56.4% | 0.259 | 1.718* | supported |
| FC→AU | | 0.075 | | 0.081 | 0.134 | Not Supported |
| Notes: *** p-value < 0.01; ** p-value < 0.05; * p-value < 0.10 | | | | | | |

**Table 6-22: The summary of the moderating effect of age**

## 6.5.2.3 Educational Level

Similar to other demographic characteristics, educational level was categorical in nature and therefore it does not require any refinement. Out of the 602 respondents, the descriptive frequencies for the educational level showed that there are 347 undergraduate and 255 postgraduate students.

Applying the measurement model for each group separately revealed the following: for the undergraduate's group [$\chi^2$ =679.171; df= 349; $\chi^2$/df= 1.946; GFI =.878; AGFI=.848; CFI=.952; RMSR=.078; RMSEA=.052, NFI=.905; PNFI=.779] and [$\chi^2$ =602.727; df= 349; $\chi^2$/df= 1.727; GFI =.857; AGFI=.822; CFI=.942; RMSR=.064; RMSEA=.054, NFI=.873; PNFI=.751] for the postgraduate group, indicating a good fit of the data. The results presented in Table 6.23 show that for all the constructs within the two samples the CR values were all above 0.7 and the AVE is higher than 0.5, therefore establishing adequate reliability and convergent validity. Except for the QWL construct in undergraduate group and AU in postgraduate group, the MSV for all other constructs is less than AVE and the square root of AVE is higher than their correlation value, which indicates a satisfactory discriminant validity for both samples.

| Factor Correlation Matrix with √AVE on the diagonal (Educational level: Undergraduate group) | | | | | | | | | | | |
|---|---|---|---|---|---|---|---|---|---|---|---|
| Factor | CR | AVE | MSV | ASV | SE | PU | PEOU | SN | QWL | FC | AU | BI |
| SE | 0.866 | 0.684 | 0.613 | 0.263 | 0.827 | | | | | | | |
| PU | 0.912 | 0.674 | 0.458 | 0.328 | 0.435 | 0.821 | | | | | | |
| PEOU | 0.919 | 0.695 | 0.298 | 0.234 | 0.496 | 0.536 | 0.834 | | | | | |
| SN | 0.827 | 0.615 | 0.386 | 0.237 | 0.333 | 0.621 | 0.415 | 0.784 | | | | |
| QWL | 0.877 | 0.589 | 0.616 | 0.329 | 0.415 | 0.677 | 0.539 | 0.547 | 0.767 | | | |
| FC | 0.880 | 0.714 | 0.456 | 0.281 | 0.560 | 0.490 | 0.457 | 0.486 | 0.502 | 0.845 | | |
| AU | 0.725 | 0.533 | 0.487 | 0.333 | 0.783 | 0.556 | 0.370 | 0.445 | 0.466 | 0.675 | 0.733 | |
| BI | 0.898 | 0.747 | 0.616 | 0.351 | 0.440 | 0.655 | 0.546 | 0.505 | 0.785 | 0.512 | 0.634 | 0.864 |
| Factor Correlation Matrix with √AVE on the diagonal (Educational level: postgraduate group) | | | | | | | | | | | |
| Factor | CR | AVE | MSV | ASV | SE | PU | PEOU | SN | QWL | FC | AU | BI |
| SE | 0.911 | 0.773 | 0.581 | 0.295 | 0.879 | | | | | | | |
| PU | 0.895 | 0.630 | 0.406 | 0.259 | 0.505 | 0.794 | | | | | | |
| PEOU | 0.898 | 0.638 | 0.406 | 0.248 | 0.595 | 0.637 | 0.799 | | | | | |
| SN | 0.827 | 0.616 | 0.138 | 0.061 | 0.276 | 0.210 | 0.141 | 0.785 | | | | |
| QWL | 0.864 | 0.561 | 0.300 | 0.147 | 0.326 | 0.548 | 0.347 | 0.096 | 0.749 | | | |
| FC | 0.845 | 0.652 | 0.336 | 0.230 | 0.580 | 0.462 | 0.559 | 0.207 | 0.391 | 0.808 | | |
| AU | 0.733 | 0.516 | 0.581 | 0.241 | 0.762 | 0.451 | 0.404 | 0.308 | 0.236 | 0.494 | 0.718 | |
| BI | 0.791 | 0.558 | 0.393 | 0.314 | 0.599 | 0.627 | 0.602 | 0.372 | 0.534 | 0.558 | 0.588 | 0.747 |

**Table 6-23: Construct reliability, convergent validity and discriminant validity for moderator Education level**

Table 6.24 shows that four paths were moderated by educational level. These paths were PEOU_BI (supporting H17a1), SN_BI (supporting H17a3), QWL_BI (supporting H17a4) and FC_AU (supporting H17a6). Specifically, the relationship was stronger for undergraduate students in terms of PEOU_BI, SN_BI and FC_AU, while the relationship was stronger for postgraduates in terms of QWL_BI. The results also show that no differences were detected on the relationship between PU_BI (rejecting H17a2) and SE_AU (rejecting H17a5). Thus it can be concluded that hypothesis H17a was partially supported. It was also found that the $R^2$ for BI was 65.3% and 72.8% for AU in the undergraduate sample, while in the postgraduate sample the variance explained ($R^2$) for BI was 58.3% and 60.3% for AU; indicating an acceptable model fit.

| Hypothesis | Undergraduate | | Postgraduate | | Z-score | Results |
|---|---|---|---|---|---|---|
| | $R^2$ | Estimate | $R^2$ | Estimate | | |
| PEOU→BI | | 0. 288 | | 0.110 | -2.164** | Supported |
| PU → BI | | 0.184 | | 0.165 | -0.186 | Not supported |
| SN→ BI | 65.3% | 0. 154 | 58.3% | 0.031 | -1.963** | Supported |
| QWL→BI | | 0.237 | | 0. 447 | 2.318** | Supported |
| SE→AU | 72.8% | 0.267 | 60.5% | 0.234 | -0.573 | Not supported |
| FC→AU | | 0.125 | | 0.033 | -2.47** | Supported |
| Notes: *** p-value < 0.01; ** p-value < 0.05; * p-value < 0.10 | | | | | | |

**Table 6-24: The summary of the moderating effect of educational level**

## 6.5.2.4 Experience

The moderating variable experience was categorical in nature and therefore it does not require any refinement. Experience was split into two categories: low-experience, and experienced. The descriptive frequencies showed that there are 192 students having low level of experience, whereas the rest (N=410) were experienced in using the web-based learning system.

Applying the measurement model for each group separately revealed the following: for the some-experience group [$\chi^2$ =576.472; df= 349; $\chi^2$/df= 1.652; GFI =.827; AGFI=.784; CFI=.950; RMSR=.090; RMSEA=.058, NFI=.883; PNFI=.759] and [$\chi^2$ =760.074; df= 349; $\chi^2$/df= 2.178; GFI =.886; AGFI=.858; CFI=.943; RMSR=.064; RMSEA=.054, NFI=.90; PNFI=.774] for the experienced group, indicating an acceptable fit of the data.

As can be shown in Table 6.25, the CR for all the constructs within both samples was higher than 0.7 and therefore all factors have adequate reliability. Except for AU low-experience group, the AVE is higher than 0.5 and CR is higher than AVE for all other constructs which indicate a satisfactory convergent validity. Additionally, Except for AU construct in both samples, the MSV for all other constructs is less than AVE and the square root of AVE is higher than their correlation value, therefore discriminant validity is considered satisfactory for both samples.

| Factor Correlation Matrix with √AVE on the diagonal (Experience:  low-experience group) | | | | | | | | | | | |
|---|---|---|---|---|---|---|---|---|---|---|---|
| **Factor** | **CR** | **AVE** | **MSV** | **ASV** | **SE** | **PU** | **PEOU** | **SN** | **QWL** | **FC** | **AU** | **BI** |
| **SE** | 0.886 | 0.723 | 0.616 | 0.331 | 0.850 | | | | | | | |
| **PU** | 0.926 | 0.713 | 0.546 | 0.381 | 0.477 | 0.845 | | | | | | |
| **PEOU** | 0.930 | 0.725 | 0.452 | 0.320 | 0.580 | 0.649 | 0.852 | | | | | |
| **SN** | 0.858 | 0.668 | 0.325 | 0.217 | 0.412 | 0.570 | 0.328 | 0.818 | | | | |
| **QWL** | 0.913 | 0.677 | 0.623 | 0.382 | 0.504 | 0.739 | 0.604 | 0.437 | 0.823 | | | |
| **FC** | 0.919 | 0.792 | 0.672 | 0.385 | 0.639 | 0.561 | 0.540 | 0.473 | 0.622 | 0.890 | | |
| **AU** | 0.734 | 0.473 | 0.672 | 0.424 | 0.785 | 0.583 | 0.516 | 0.534 | 0.558 | 0.820 | 0.688 | |
| **BI** | 0.910 | 0.772 | 0.623 | 0.424 | 0.555 | 0.704 | 0.672 | 0.468 | 0.789 | 0.630 | 0.688 | 0.879 |
| Factor Correlation Matrix with √AVE on the diagonal (Experience: Experienced group) | | | | | | | | | | | |
| **Factor** | **CR** | **AVE** | **MSV** | **ASV** | **SE** | **PU** | **PEOU** | **SN** | **QWL** | **FC** | **AU** | **BI** |
| **SE** | 0.901 | 0.752 | 0.689 | 0.311 | 0.867 | | | | | | | |
| **PU** | 0.886 | 0.608 | 0.355 | 0.279 | 0.535 | 0.779 | | | | | | |
| **PEOU** | 0.903 | 0.652 | 0.333 | 0.222 | 0.577 | 0.522 | 0.807 | | | | | |
| **SN** | 0.807 | 0.583 | 0.180 | 0.120 | 0.279 | 0.370 | 0.308 | 0.763 | | | | |
| **QWL** | 0.848 | 0.529 | 0.464 | 0.222 | 0.399 | 0.561 | 0.417 | 0.333 | 0.728 | | | |
| **FC** | 0.862 | 0.682 | 0.378 | 0.260 | 0.610 | 0.494 | 0.518 | 0.354 | 0.414 | 0.826 | | |
| **AU** | 0.729 | 0.580 | 0.689 | 0.330 | 0.830 | 0.596 | 0.428 | 0.339 | 0.400 | 0.615 | 0.761 | |
| **BI** | 0.850 | 0.655 | 0.464 | 0.309 | 0.508 | 0.584 | 0.480 | 0.424 | 0.681 | 0.511 | 0.655 | 0.809 |

**Table 6-25: Construct reliability, convergent validity and discriminant validity for moderator experience**

As can be shown in Table 6.26, experience was found to moderate the relationship between PEOU→BI (supporting H18a1), PU→BI (supporting H18a2), SN→BI (supporting H18a3), SE→AU (supporting H18a5), and FC→AU (supporting H18a6). The result shows that the relationship between PEOU→BI, SN→BI, SE→AU and FC→AU was stronger for the low-experience group while the relationship between PU→BI was stronger for experienced group. On the other hand, no moderating effect of experience on the relationship between QWL→BI was found. This results indicate that hypothesis H18a was partially supported. Overall, the $R^2$ for BI was 70.4% and for AU was 74.9% within the low-experience group, whereas in the experienced group the $R^2$ for BI was 55.9% and 66.7% for AU, indicating an acceptable model fit of the data.

| Hypothesis | some-experience | | Experienced | | Z-score | Results |
| --- | --- | --- | --- | --- | --- | --- |
| | $R^2$ | Estimate | $R^2$ | Estimate | | |
| PEOU →BI | 70.4% | 0.252 | 55.9% | 0.112 | -1.746* | Supported |
| PU → BI | | 0.097 | | 0.262 | 1.687* | Supported |
| SN→ BI | | 0.227 | | 0.104 | -1.695* | Supported |
| QWL→BI | | 0.246 | | 0.308 | 0.577 | Not Supported |
| SE→AU | 74.9% | 0.309 | 66.7% | 0.188 | -2.083** | Supported |
| FC→AU | | 0.206 | | 0.105 | -2.466** | Supported |
| Notes: *** p-value < 0.01; ** p-value < 0.05; * p-value < 0.10 | | | | | | |

**Table 6-26: The summary of the moderating effect of experience**

# 6.6   Moderating effects within the Lebanese sample

This section will report the moderating effect of Hofstede's four cultural dimensions (MF, ID, PD, UA) at the individual level as well as four demographic variables namely gender, age, educational level and experience on the relationships between the exogenous (PEOU, PU, SN, QWL and FC) and the endogenous (BI and AU) latent constructs within the Lebanese dataset.

## 6.6.1  Hofstede's Cultural Dimensions Lebanese sample

This section will report the results of the moderating impact of Hofstede's four cultural dimensions on the relationships with the Lebanese model.

### 6.6.1.1 Power distance

Out of the 569 respondents, the descriptive statistics showed that the overall mean for PD was 3.1603/7 indicating a low PD culture. Using median-split method (median= 3.0), there were 324 (median<=3) students within the low-PD group and 278 within the high-PD group (median >3).

The first run of the measurement model resulted the following fit for the low-PD sample [$\chi^2$ =669.837; df= 349; $\chi^2$/df= 1.710; GFI =.878; AGFI=.848; CFI=.949; RMSR=.069; RMSEA=.049, NFI=.888; PNFI=.763] and [$\chi^2$ =647.993; df= 349;

$\chi^2/df=$ 1.857; GFI =.852; AGFI=.816; CFI=.957; RMSR=.098; RMSEA=.057, NFI=.882; PNFI=.758] for the high-PD group, which also indicates a good fit of the data. As can be shown in Table 6.27, the CR was higher than 0.7 and the AVE was higher than 0.5 which satisfied the criteria for reliability and convergent validity. Moreover, except for AU in the high-level PD group, the MSV for all other constructs is less than the AVE and the square root of AVE is higher than their correlation value, which indicates satisfactory discriminant validity for both samples.

| Factor Correlation Matrix with √AVE on the diagonal (Power Distance: low level group) | | | | | | | | | | | | |
|---|---|---|---|---|---|---|---|---|---|---|---|---|
| **Factor** | **CR** | **AVE** | **MSV** | **ASV** | **SE** | **PU** | **PEOU** | **SN** | **QWL** | **FC** | **AU** | **BI** |
| **SE** | 0.876 | 0.705 | 0.291 | 0.152 | 0.840 | | | | | | | |
| **PU** | 0.879 | 0.594 | 0.181 | 0.095 | 0.233 | 0.771 | | | | | | |
| **PEOU** | 0.924 | 0.708 | 0.199 | 0.121 | 0.446 | 0.426 | 0.842 | | | | | |
| **SN** | 0.813 | 0.527 | 0.135 | 0.092 | 0.327 | 0.316 | 0.263 | 0.726 | | | | |
| **QWL** | 0.824 | 0.541 | 0.342 | 0.160 | 0.379 | 0.402 | 0.365 | 0.348 | 0.736 | | | |
| **FC** | 0.903 | 0.757 | 0.291 | 0.117 | 0.539 | 0.169 | 0.336 | 0.247 | 0.394 | 0.870 | | |
| **AU** | 0.711 | 0.553 | 0.141 | 0.062 | 0.375 | 0.207 | 0.014 | 0.217 | 0.248 | 0.273 | 0.706 | |
| **BI** | 0.838 | 0.637 | 0.342 | 0.145 | 0.358 | 0.305 | 0.387 | 0.368 | 0.585 | 0.315 | 0.256 | 0.798 |
| Factor Correlation Matrix with √AVE on the diagonal (Power Distance: high level group) | | | | | | | | | | | | |
| **Factor** | **CR** | **AVE** | **MSV** | **ASV** | **SE** | **PU** | **PEOU** | **SN** | **QWL** | **FC** | **AU** | **BI** |
| **SE** | 0.897 | 0.745 | 0.430 | 0.268 | 0.863 | | | | | | | |
| **PU** | 0.919 | 0.694 | 0.397 | 0.235 | 0.331 | 0.833 | | | | | | |
| **PEOU** | 0.925 | 0.711 | 0.325 | 0.260 | 0.509 | 0.524 | 0.843 | | | | | |
| **SN** | 0.799 | 0.509 | 0.236 | 0.183 | 0.410 | 0.365 | 0.367 | 0.713 | | | | |
| **QWL** | 0.838 | 0.566 | 0.498 | 0.289 | 0.440 | 0.615 | 0.546 | 0.471 | 0.752 | | | |
| **FC** | 0.877 | 0.705 | 0.521 | 0.290 | 0.640 | 0.310 | 0.570 | 0.417 | 0.465 | 0.840 | | |
| **AU** | 0.712 | 0.514 | 0.590 | 0.353 | 0.656 | 0.507 | 0.495 | 0.459 | 0.469 | 0.722 | 0.628 | |
| **BI** | 0.872 | 0.696 | 0.590 | 0.372 | 0.551 | 0.630 | 0.535 | 0.486 | 0.706 | 0.539 | 0.768 | 0.835 |

**Table 6-27: Construct reliability, convergent validity and discriminant validity for moderator power distance**

The results presented in Table 6.28 show that the cultural variable power distance was found to moderate the relationship between PU_BI (supporting H11b2), SN_BI (supporting H11b3) and SE_AU (supporting H11b5). The relationship was stronger for the low-PD group in terms of PU_BI while the relationship was stronger for the high-PD group in terms of SN_BI and SE_AU. Contrary to our expectation, PD did not moderate the relationship between PEOU_BI (rejecting H11b1), QWL_BI (rejecting H11b4) and FC_AU (rejecting H11b6). The results indicate that hypothesis H11b was partially supported. Overall, the $R^2$ for BI was 40.3% and 51.9% for AU within the low-PD group, indicating a moderate model fit, while within the high-PD group the variance explained ($R^2$) for BI was 60.2% and 66.1% for AU, indicating a good fit for the high-PD model.

| Hypothesis | Low PD | | High PD | | Z-score | Results |
|---|---|---|---|---|---|---|
| | $R^2$ | Estimate | $R^2$ | Estimate | | |
| PEOU→BI | 40.3% | 0.166 | 60.2% | 0.202 | 0.606 | Not supported |
| PU → BI | | 0.256 | | 0.133 | -1.743* | Supported |
| SN→ BI | | 0.087 | | 0.153 | 1.677* | Supported |
| QWL→BI | | 0.341 | | 0.320 | -0.185 | Not supported |
| SE→AU | 51.9% | 0.108 | 66.1% | 0. 283 | 2.247** | Supported |
| FC→AU | | 0.111 | | 0.124 | 0.205 | Not supported |
| Notes: *** p-value < 0.01; ** p-value < 0.05; * p-value < 0.10 | | | | | | |

**Table 6-28: The summary of the moderating effect of power distance**

## 6.6.1.2 Masculinity\femininity

The overall mean was 3.22/7 indicating a moderately masculine culture. Using the median-split method, the descriptive statistics showed that there are 281 within the low-MF group (median<=3) and 288 within the high-MF group (median>3).

Applying the measurement model for each group separately revealed the following: for the low-MF group [$\chi^2$ =544.599; df= 349; $\chi^2$/df= 1.589; GFI =.881; AGFI=.852; CFI=.951; RMSR=.074; RMSEA=.054, NFI=.879; PNFI=.756] and [$\chi^2$ =667.890; df= 349; $\chi^2$/df= 1.914; GFI =.858; AGFI=.823; CFI=.942; RMSR=.078; RMSEA=.057, NFI=.886; PNFI=.762] for the high-MF group, indicating an acceptable fit of the data.

The results presented in Table 6.29 show that, except for AU in both samples, the CR was higher than 0.7, which indicates that the constructs have adequate reliability. Moreover, the results also show that the AVEs for all constructs were higher than 0.5 which satisfied the criterion of convergent validity. Additionally, except for AU in the high-level masculinity group, the MSV for all other constructs is less than the AVE and the square root of AVE is higher than their correlation value, which satisfied the criterion of discriminant validity for both samples.

| Factor Correlation Matrix with √AVE on the diagonal (Masculinity: low level group) | | | | | | | | | | | | |
|---|---|---|---|---|---|---|---|---|---|---|---|---|
| **Factor** | **CR** | **AVE** | **MSV** | **ASV** | **SE** | **PU** | **PEOU** | **SN** | **QWL** | **FC** | **AU** | **BI** |
| **SE** | 0.875 | 0.702 | 0.375 | 0.131 | 0.838 | | | | | | | |
| **PU** | 0.871 | 0.577 | 0.158 | 0.069 | 0.115 | 0.759 | | | | | | |
| **PEOU** | 0.914 | 0.681 | 0.114 | 0.073 | 0.300 | 0.337 | 0.825 | | | | | |
| **SN** | 0.749 | 0.542 | 0.196 | 0.162 | 0.250 | 0.216 | 0.171 | 0.665 | | | | |
| **QWL** | 0.810 | 0.518 | 0.336 | 0.150 | 0.324 | 0.398 | 0.330 | 0.310 | 0.720 | | | |
| **FC** | 0.894 | 0.739 | 0.375 | 0.128 | 0.612 | 0.130 | 0.307 | 0.255 | 0.395 | 0.859 | | |
| **AU** | 0.731 | 0.527 | 0.187 | 0.076 | 0.433 | 0.175 | -0.017 | 0.211 | 0.298 | 0.290 | 0.690 | |
| **BI** | 0.814 | 0.598 | 0.336 | 0.127 | 0.289 | 0.317 | 0.272 | 0.304 | 0.580 | 0.323 | 0.308 | 0.773 |
| **Factor Correlation Matrix with √AVE on the diagonal (Masculinity: high level group)** | | | | | | | | | | | | |
| **Factor** | **CR** | **AVE** | **MSV** | **ASV** | **SE** | **PU** | **PEOU** | **SN** | **QWL** | **FC** | **AU** | **BI** |
| **SE** | 0.893 | 0.738 | 0.349 | 0.261 | 0.859 | | | | | | | |
| **PU** | 0.912 | 0.675 | 0.375 | 0.212 | 0.821 | | | | | | | |
| **PEOU** | 0.929 | 0.723 | 0.349 | 0.267 | 0.591 | 0.538 | 0.850 | | | | | |
| **SN** | 0.843 | 0.576 | 0.236 | 0.172 | 0.434 | 0.396 | 0.385 | 0.759 | | | | |
| **QWL** | 0.845 | 0.578 | 0.494 | 0.271 | 0.467 | 0.579 | 0.530 | 0.458 | 0.761 | | | |
| **FC** | 0.887 | 0.725 | 0.423 | 0.240 | 0.551 | 0.282 | 0.542 | 0.370 | 0.447 | 0.851 | | |
| **AU** | 0.725 | 0.513 | 0.466 | 0.261 | 0.549 | 0.413 | 0.427 | 0.353 | 0.395 | 0.650 | 0.619 | |
| **BI** | 0.880 | 0.711 | 0.494 | 0.343 | 0.570 | 0.559 | 0.568 | 0.486 | 0.703 | 0.496 | 0.683 | 0.843 |

**Table 6-29: Construct reliability, convergent validity and discriminant validity for moderator masculinity\femininity**

Table 6.30 shows that masculinity\femininity cultural variable moderates the relationship between PEOU_BI (supporting H12b1), SN_BI (supporting H18b3) and SE_AU (supporting H12b5) with the relationship was stronger for low masculinity group. However, no differences were found on the relationship between PU_BI (rejecting H12b2), QWL_BI (rejecting H12b4), and FC_AU (rejecting H12b6). The result suggests that hypothesis H12b was partially supported. Overall, The $R^2$ for BI was 47.2% and 43.4% for AU within the low-MF group indicating a moderate model fit, while the $R^2$ for BI was 58.7% and 56.9% for AU within the high-group, indicating a good model fit of the data.

| Hypothesis | High Masculinity | | Low Masculinity | | Z-score | Results |
|---|---|---|---|---|---|---|
| | $R^2$ | Estimate | $R^2$ | Estimate | | |
| PEOU→BI | 47.2% | 0. 102 | 58.7% | 0. 223 | 1.813* | Supported |
| PU → BI | | 0.180 | | 0. 143 | -0.382 | Not supported |
| SN→ BI | | 0.086 | | 0. 197 | 1.679* | Supported |
| QWL→BI | | 0.345 | | 0.371 | 0.256 | Not supported |
| SE→AU | 43.4% | 0.139 | 56.9% | 0. 249 | 1.771* | Supported |
| FC→AU | | 0.209 | | 0.148 | -0.482 | Not supported |
| Notes: *** p-value < 0.01; ** p-value < 0.05; * p-value < 0.10 | | | | | | |

**Table 6-30: The summary of the moderating effect of masculinity\femininity**

### 6.6.1.3 Individualism\collectivism

The results of the descriptive statistics showed that the overall mean for the moderating construct IC was 5.011/7 indicating a collectivist cultural value. The data were split into two groups using the median-split method, there were 338 students within the low-IC group (median<=5) and 231 within the high-MF group (median>5).

The first run of the measurement model resulted in the following fit for the low-IC sample [$\chi^2$ =646.667; df= 349; $\chi^2$/df= 1.853; GFI =.883; AGFI=.854; CFI=.954; RMSR=.083; RMSEA=.050, NFI=.906; PNFI=.779] and [$\chi^2$ =586.034; df= 349; $\chi^2$/df= 1.679; GFI =.848; AGFI=.811; CFI=.945; RMSR=.081; RMSEA=.054, NFI=.875; PNFI=.752] for the high-IC group, which also indicates a good fit of the data. The results presented in Table 6.31 revealed that the CR and AVE were higher than 0.7 and 0.5 respectively, which indicates an adequate reliability and convergent validity. Furthermore, the AVE for all constructs is higher than the MSV and also the square root of AVE is higher than their correlation value, and therefore satisfactory discriminant validity was established for both samples.

| Factor Correlation Matrix with √AVE on the diagonal (Individualism: low level group) | | | | | | | | | | | |
|---|---|---|---|---|---|---|---|---|---|---|---|
| **Factor** | **CR** | **AVE** | **MSV** | **ASV** | **SE** | **PU** | **PEOU** | **SN** | **QWL** | **FC** | **AU** | **BI** |
| **SE** | 0.905 | 0.763 | 0.367 | 0.223 | 0.873 | | | | | | | |
| **PU** | 0.912 | 0.674 | 0.315 | 0.186 | 0.312 | 0.821 | | | | | | |
| **PEOU** | 0.932 | 0.734 | 0.267 | 0.203 | 0.449 | 0.511 | 0.857 | | | | | |
| **SN** | 0.785 | 0.548 | 0.260 | 0.158 | 0.394 | 0.300 | 0.365 | 0.699 | | | | |
| **QWL** | 0.834 | 0.560 | 0.533 | 0.258 | 0.417 | 0.533 | 0.454 | 0.434 | 0.748 | | | |
| **FC** | 0.897 | 0.744 | 0.367 | 0.218 | 0.606 | 0.260 | 0.463 | 0.380 | 0.454 | 0.863 | | |
| **AU** | 0.731 | 0.516 | 0.404 | 0.245 | 0.579 | 0.434 | 0.371 | 0.369 | 0.464 | 0.545 | 0.680 | |
| **BI** | 0.881 | 0.713 | 0.533 | 0.320 | 0.483 | 0.561 | 0.517 | 0.510 | 0.730 | 0.479 | 0.636 | 0.844 |
| Factor Correlation Matrix with √AVE on the diagonal (Individualism: high level group) | | | | | | | | | | | |
| **Factor** | **CR** | **AVE** | **MSV** | **ASV** | **SE** | **PU** | **PEOU** | **SN** | **QWL** | **FC** | **AU** | **BI** |
| **SE** | 0.865 | 0.683 | 0.426 | 0.269 | 0.826 | | | | | | | |
| **PU** | 0.900 | 0.645 | 0.336 | 0.219 | 0.803 | | | | | | | |
| **PEOU** | 0.927 | 0.719 | 0.426 | 0.274 | 0.653 | 0.530 | 0.848 | | | | | |
| **SN** | 0.832 | 0.555 | 0.203 | 0.141 | 0.397 | 0.451 | 0.306 | 0.745 | | | | |
| **QWL** | 0.846 | 0.580 | 0.362 | 0.259 | 0.478 | 0.580 | 0.564 | 0.423 | 0.762 | | | |
| **FC** | 0.898 | 0.747 | 0.397 | 0.253 | 0.630 | 0.348 | 0.555 | 0.323 | 0.482 | 0.864 | | |
| **AU** | 0.711 | 0.555 | 0.411 | 0.237 | 0.492 | 0.453 | 0.413 | 0.352 | 0.398 | 0.587 | 0.745 | |
| **BI** | 0.861 | 0.677 | 0.411 | 0.290 | 0.549 | 0.501 | 0.563 | 0.350 | 0.602 | 0.514 | 0.641 | 0.823 |

**Table 6-31: Construct reliability, convergent validity and discriminant validity for moderator individualism\collectivism**

The results presented in Table 6.32 show that individualism\collectivism cultural variable moderates the relationship between SN_BI (supporting H13b3), SE_AU (supporting H13b5) and FC_AU (supporting H13b6). Specifically, the relationship was stronger in the collectivistic group for all of them. The results also show that no differences were detected on the relationship between PEOU_BI (rejecting H13b1), PU_BI (rejecting H13b2) and QWL_BI (rejecting H13b4). This means that hypothesis H13b was partially supported. Overall, The $R^2$ for BI was 54.3% and for AU was 47.7% within the low-IC group, while within the high-IC group the variance explained ($R^2$) for BI was 46.1% and 50.5% for AU, indicating a moderate model fit.

| Hypothesis | Individualism | | Collectivism | | Z-score | Results |
|---|---|---|---|---|---|---|
| | $R^2$ | Estimate | $R^2$ | Estimate | | |
| PEOU →BI | 54.3% | 0. 113 | 46.1% | 0.179 | 0.656 | Not supported |
| PU → BI | | 0.168 | | 0.155 | -0.042 | Not supported |
| SN→ BI | | 0.128 | | 0.279 | 1.71* | Supported |
| QWL→BI | | 0.383 | | 0.231 | -1.392 | Not supported |
| SE→AU | 47.7% | 0.084 | 50.5% | 0.154 | 1.658* | Supported |
| FC→AU | | 0.125 | | 0.266 | 1.727* | Supported |
| Notes: *** p-value < 0.01; ** p-value < 0.05; * p-value < 0.10 | | | | | | |

**Table 6-32: The summary of the moderating effect of individualism\collectivism**

## 6.6.1.4 Uncertainty Avoidance

The overall mean for the moderating construct UA was 5.46/7, indicating a high UA cultural value. Based on median-split method, the data was split into two groups, there were 274 within the low-UA group (median <6) and 295 within the high-UA level group (median =>6).

Applying the measurement model for each group separately revealed the following: for the low-UA group [$\chi^2$ =642.541; df= 349; $\chi^2$/df= 1.841; GFI =.858; AGFI=.823; CFI=.946; RMSR=.078; RMSEA=.056, NFI=.889; PNFI=.764] and [$\chi^2$ =584.109; df= 349; $\chi^2$/df= 1.674; GFI =.886; AGFI=.850; CFI=.950; RMSR=.079; RMSEA=.048, NFI=.886; PNFI=.762] for the high-UA group, indicating an acceptable fit of the data. The results presented in Table 6.33revealed that the CR and AVE were higher than 0.7 and 0.5 respectively, which indicates an adequate reliability and convergent validity. Furthermore, the AVE for all constructs is higher than the MSV and also the square root of AVE is higher than their correlation value, and therefore satisfactory discriminant validity was established for both samples.

| Factor Correlation Matrix with √AVE on the diagonal (Uncertainty Avoidance: low level group) | | | | | | | | | | | |
|---|---|---|---|---|---|---|---|---|---|---|---|
| Factor | CR | AVE | MSV | ASV | SE | PU | PEOU | SN | QWL | FC | AU | BI |
| SE | 0.921 | 0.795 | 0.445 | 0.236 | 0.892 | | | | | | | |
| PU | 0.919 | 0.694 | 0.331 | 0.197 | 0.277 | 0.833 | | | | | | |
| PEOU | 0.935 | 0.742 | 0.331 | 0.243 | 0.522 | 0.575 | 0.862 | | | | | |
| SN | 0.827 | 0.547 | 0.213 | 0.169 | 0.434 | 0.409 | 0.396 | 0.739 | | | | |
| QWL | 0.796 | 0.536 | 0.483 | 0.251 | 0.434 | 0.505 | 0.528 | 0.449 | 0.704 | | | |
| FC | 0.893 | 0.737 | 0.445 | 0.223 | 0.667 | 0.288 | 0.507 | 0.362 | 0.469 | 0.859 | | |
| AU | 0.715 | 0.518 | 0.350 | 0.200 | 0.539 | 0.398 | 0.379 | 0.348 | 0.362 | 0.456 | 0.695 | |
| BI | 0.873 | 0.698 | 0.483 | 0.290 | 0.437 | 0.557 | 0.512 | 0.462 | 0.695 | 0.468 | 0.592 | 0.835 |
| Factor Correlation Matrix with √AVE on the diagonal (Uncertainty Avoidance: high level group) | | | | | | | | | | | |
| Factor | CR | AVE | MSV | ASV | SE | PU | PEOU | SN | QWL | FC | AU | BI |
| SE | 0.847 | 0.652 | 0.279 | 0.188 | 0.808 | | | | | | | |
| PU | 0.888 | 0.614 | 0.320 | 0.166 | 0.338 | 0.784 | | | | | | |
| PEOU | 0.913 | 0.679 | 0.228 | 0.150 | 0.444 | 0.395 | 0.824 | | | | | |
| SN | 0.773 | 0.539 | 0.148 | 0.100 | 0.297 | 0.290 | 0.226 | 0.687 | | | | |
| QWL | 0.838 | 0.567 | 0.329 | 0.194 | 0.367 | 0.566 | 0.364 | 0.353 | 0.753 | | | |
| FC | 0.894 | 0.738 | 0.416 | 0.192 | 0.502 | 0.244 | 0.424 | 0.312 | 0.380 | 0.859 | | |
| AU | 0.718 | 0.509 | 0.416 | 0.238 | 0.502 | 0.450 | 0.321 | 0.325 | 0.415 | 0.645 | 0.702 | |
| BI | 0.849 | 0.655 | 0.413 | 0.260 | 0.528 | 0.478 | 0.477 | 0.385 | 0.574 | 0.441 | 0.643 | 0.809 |

**Table 6-33: Construct reliability, convergent validity and discriminant validity for moderator uncertainty avoidance**

Table 6.34 shows that four paths were moderated by the UA cultural variable. These paths were PEOU_BI (supporting H14b1), SN_BI (supporting H14b3) and FC_AU (supporting H14b6). More specifically, the relationship was stronger for users with high UA cultural values. The results also show that no differences were detected in the relationship between PU_BI (rejecting H14b2), QWL_BI (rejecting H14b4) and SE_AU (rejecting H14b5). Thus, it can be concluded that hypothesis H14b was partially supported. It was also found that the $R^2$ for BI was 52.8% and 43.5% for AU within the low-UA sample, while within the high-UA sample the variance explained ($R^2$) for BI was 46.1% and 54.7% for AU, indicating a moderate model fit.

| Hypothesis | Low UA | | High UA | | Z-score | Results |
|---|---|---|---|---|---|---|
| | $R^2$ | Estimate | $R^2$ | Estimate | | |
| PEOU→BI | 52.8% | 0.129 | 46.1% | 0. 256 | 1.732* | Supported |
| PU → BI | | 0. 197 | | 0.134 | -0.662 | Not supported |
| SN→ BI | | 0.095 | | 0.213 | 1.926* | Supported |
| QWL→BI | | 0.370 | | 0.267 | -0.895 | Not supported |
| SE→AU | 43.5% | 0.109 | 54.7% | 0.134 | 0.383 | Not supported |
| FC→AU | | 0.128 | | 0.277 | 2.387** | Supported |
| Notes: *** p-value < 0.01; ** p-value < 0.05; * p-value < 0.10 | | | | | | |

**Table 6-34: The summary of the moderating effect of Uncertainty Avoidance**

## 6.6.2 Demographic Characteristics of Lebanese sample

This section will report the results of the moderating impact of four demographic variables (age, gender, educational level and experience) on the relationships between the main predictors and behavioural intention and usage of e-learning systems within the Lebanese model.

### 6.6.2.1 Gender

Out of the 569 respondents, the sample descriptive frequencies show that there are 306 males and 263 females. The first run of the model resulted in the following fit for the male group [$\chi^2$ =600.814; df= 349; $\chi^2$/df= 1.722; GFI =.877; AGFI=.847; CFI=.946; RMSR=.082; RMSEA=.049, NFI=.881; PNFI=.754] and [$\chi^2$ =606.026; df= 349; $\chi^2$/df= 1.736; GFI =.862; AGFI=.829; CFI=.955; RMSR=.084; RMSEA=.053, NFI=.901; PNFI=.755] for the female group, indicating an acceptable fit of the data.

As shown in Table 6.35, only AU in male group had lower than the recommended value (0.7), which indicates an adequate reliability for all the factors. The results also show that the AVE values were all above 0.5 and that CR is higher than AVE for all the constructs, which establish a satisfactory convergent validity. Additionally, except for AU construct in the male group, the MSV values for all other constructs is less than AVE values and the square root of AVE is higher than their correlation values, which indicates satisfactory discriminant validity for both samples.

| Factor Correlation Matrix with √AVE on the diagonal (Gender: Male group) | | | | | | | | | | | | |
|---|---|---|---|---|---|---|---|---|---|---|---|---|
| **Factor** | **CR** | **AVE** | **MSV** | **ASV** | **SE** | **PU** | **PEOU** | **SN** | **QWL** | **FC** | **AU** | **BI** |
| **SE** | 0.876 | 0.705 | 0.237 | 0.157 | 0.840 | | | | | | | |
| **PU** | 0.881 | 0.599 | 0.205 | 0.118 | 0.237 | 0.774 | | | | | | |
| **PEOU** | 0.916 | 0.685 | 0.234 | 0.160 | 0.484 | 0.410 | 0.828 | | | | | |
| **SN** | 0.781 | 0.583 | 0.218 | 0.185 | 0.292 | 0.262 | 0.277 | 0.695 | | | | |
| **QWL** | 0.819 | 0.532 | 0.312 | 0.159 | 0.367 | 0.453 | 0.373 | 0.344 | 0.729 | | | |
| **FC** | 0.887 | 0.724 | 0.237 | 0.157 | 0.487 | 0.192 | 0.460 | 0.331 | 0.365 | 0.851 | | |
| **AU** | 0.636 | 0.507 | 0.368 | 0.149 | 0.365 | 0.371 | 0.294 | 0.165 | 0.266 | 0.466 | 0.586 | |
| **BI** | 0.814 | 0.596 | 0.368 | 0.217 | 0.466 | 0.394 | 0.454 | 0.329 | 0.559 | 0.391 | 0.607 | 0.772 |
| **Factor Correlation Matrix with √AVE on the diagonal (Gender: Female group)** | | | | | | | | | | | | |
| **Factor** | **CR** | **AVE** | **MSV** | **ASV** | **SE** | **PU** | **PEOU** | **SN** | **QWL** | **FC** | **AU** | **BI** |
| **SE** | 0.893 | 0.737 | 0.480 | 0.282 | 0.858 | | | | | | | |
| **PU** | 0.928 | 0.721 | 0.401 | 0.270 | 0.379 | 0.849 | | | | | | |
| **PEOU** | 0.938 | 0.751 | 0.347 | 0.261 | 0.506 | 0.589 | 0.866 | | | | | |
| **SN** | 0.833 | 0.558 | 0.328 | 0.237 | 0.491 | 0.479 | 0.404 | 0.747 | | | | |
| **QWL** | 0.856 | 0.599 | 0.594 | 0.343 | 0.483 | 0.630 | 0.577 | 0.523 | 0.774 | | | |
| **FC** | 0.907 | 0.766 | 0.480 | 0.281 | 0.693 | 0.372 | 0.504 | 0.380 | 0.539 | 0.875 | | |
| **AU** | 0.728 | 0.575 | 0.382 | 0.295 | 0.608 | 0.484 | 0.401 | 0.526 | 0.528 | 0.602 | 0.758 | |
| **BI** | 0.902 | 0.756 | 0.594 | 0.367 | 0.502 | 0.633 | 0.559 | 0.573 | 0.771 | 0.545 | 0.618 | 0.869 |

**Table 6-35: Construct reliability, convergent validity and discriminant validity for moderator gender**

The results of the MGA presented in Table 6.36 show that gender moderates the relationship between PEOU_BI (supporting H15b1), SN_BI (supporting H15b3) and QWL_BI (supporting H15b4). Specifically, the relationship was stronger for males in terms of QWL_BI, while the relationship was stronger for female in terms of PEOU_BI and SN_BI. However no differences were detected in the relationship between PU_BI (rejecting H15b2), SE_AU (rejecting H15b5) and FC_AU (rejecting H15b6). This results indicates that hypothesis H15b was partially supported. Overall, the $R^2$ for BI was 40.2% and for AU was 44.5% within the male sample indicating a moderate model fit, while the variance explained ($R^2$) for BI was 57.2% and 51.3% for AU within the female group, which indicates a good model fit of the data.

| Hypothesis | Male | | Female | | Z-score | Results |
|---|---|---|---|---|---|---|
| | $R^2$ | Estimate | $R^2$ | Estimate | | |
| PEOU →BI | 40.2% | 0.096 | 57.2% | 0. 229 | 1.759* | Supported |
| PU → BI | | 0.176 | | 0. 109 | -0.610 | Not supported |
| SN→ BI | | 0.093 | | 0.186 | 1. 692* | Supported |
| QWL→BI | | 0.482 | | 0. 329 | -1.855* | Supported |
| SE→AU | 44.5% | 0.159 | 51.3% | 0.128 | -0.286 | Not supported |
| FC→AU | | 0.390 | | 0.340 | -0.492 | Not supported |
| Notes: *** p-value < 0.01; ** p-value < 0.05; * p-value < 0.10 | | | | | | |

**Table 6-36: The summary of the moderating effect of gender**

## 6.6.2.2 Age

Out of the 602 respondents, there were 412 students within the younger-age group and 157 students within the older-age group. The model fit indices for the younger-age sample are [$\chi^2$ =725.007; df= 349; $\chi^2$/df= 2.077; GFI =.889; AGFI=.861; CFI=.950; RMSR=.086; RMSEA=.051, NFI=.906; PNFI=.781] and [$\chi^2$ =537.8899; df= 349; $\chi^2$/df= 1.541; GFI =.814; AGFI=.768; CFI=.938; RMSR=.082; RMSEA=.059, NFI=.844; PNFI=.725] for the older-age group, which also indicate a good fit of the data.

The results presented in Table 6.37 show that for all the constructs within the two samples the CR were all above 0.7 and the AVE values were all above 0.5 and therefore established adequate reliability and convergent validity. Additionally, the MSV for all constructs is less than AVE and the square root of AVE is higher than their correlation value, therefore discriminant validity was also established for both samples.

| Factor Correlation Matrix with √AVE on the diagonal (Age: group <=22) | | | | | | | | | | | |
|---|---|---|---|---|---|---|---|---|---|---|---|
| **Factor** | **CR** | **AVE** | **MSV** | **ASV** | **SE** | **PU** | **PEOU** | **SN** | **QWL** | **FC** | **AU** | **BI** |
| **SE** | 0.891 | 0.734 | 0.318 | 0.222 | 0.856 | | | | | | | |
| **PU** | 0.912 | 0.675 | 0.308 | 0.198 | 0.320 | 0.822 | | | | | | |
| **PEOU** | 0.929 | 0.722 | 0.311 | 0.222 | 0.521 | 0.534 | 0.850 | | | | | |
| **SN** | 0.781 | 0.543 | 0.230 | 0.150 | 0.389 | 0.311 | 0.343 | 0.699 | | | | |
| **QWL** | 0.830 | 0.552 | 0.436 | 0.239 | 0.397 | 0.555 | 0.470 | 0.434 | 0.743 | | | |
| **FC** | 0.886 | 0.722 | 0.335 | 0.210 | 0.564 | 0.285 | 0.443 | 0.339 | 0.428 | 0.850 | | |
| **AU** | 0.731 | 0.529 | 0.375 | 0.245 | 0.551 | 0.472 | 0.388 | 0.392 | 0.422 | 0.579 | 0.655 | |
| **BI** | 0.867 | 0.688 | 0.436 | 0.304 | 0.502 | 0.537 | 0.558 | 0.480 | 0.660 | 0.489 | 0.612 | 0.829 |
| Factor Correlation Matrix with √AVE on the diagonal (Age: group >22) | | | | | | | | | | | |
| **Factor** | **CR** | **AVE** | **MSV** | **ASV** | **SE** | **PU** | **PEOU** | **SN** | **QWL** | **FC** | **AU** | **BI** |
| **SE** | 0.875 | 0.703 | 0.543 | 0.235 | 0.839 | | | | | | | |
| **PU** | 0.882 | 0.602 | 0.254 | 0.163 | 0.305 | 0.776 | | | | | | |
| **PEOU** | 0.932 | 0.734 | 0.398 | 0.211 | 0.474 | 0.430 | 0.857 | | | | | |
| **SN** | 0.844 | 0.577 | 0.211 | 0.119 | 0.363 | 0.459 | 0.301 | 0.760 | | | | |
| **QWL** | 0.843 | 0.574 | 0.497 | 0.257 | 0.515 | 0.483 | 0.536 | 0.358 | 0.758 | | | |
| **FC** | 0.918 | 0.789 | 0.543 | 0.266 | 0.737 | 0.272 | 0.631 | 0.370 | 0.529 | 0.888 | | |
| **AU** | 0.751 | 0.604 | 0.398 | 0.163 | 0.433 | 0.301 | 0.324 | 0.195 | 0.331 | 0.462 | 0.777 | |
| **BI** | 0.877 | 0.705 | 0.497 | 0.265 | 0.450 | 0.504 | 0.431 | 0.316 | 0.705 | 0.466 | 0.631 | 0.840 |

**Table 6-37: Construct reliability, convergent validity and discriminant validity for moderator age**

As shown in Table 6.38, age was found to moderate the relationship between PEOU_BI (supporting H16b1), QWL_BI (supporting H16b4) and FC_AU (supporting H16b6), with the relationship stronger for the younger-age group in terms of QWL_BI and FC_AU, while the relationship was stronger for the older-age group in terms of PEOU_BI. On the other hand, no moderating effect was found on the relationship between PU_BI (rejecting H16b2), SN_BI (rejecting H16b3) and SE_AU (rejecting H16b5), which suggests that hypothesis H16b was partially supported. Overall, the $R^2$ for BI was 56.5% and for AU was 48.2% within the younger-age group, while within the older-age group the variance explained ($R^2$) for BI was 53% and 42.4% for AU, indicating a moderate model fit.

| Hypothesis | Age group <=22 | | Age group >22 | | Z-score | Results |
|---|---|---|---|---|---|---|
| | $R^2$ | Estimate | $R^2$ | Estimate | | |
| PEOU →BI | | 0.108 | | 0.229 | 1.736* | Supported |
| PU → BI | 56.5% | 0.253 | 53% | 0.128 | -0.980 | Not supported |
| SN→ BI | | 0.105 | | 0.162 | 0.630 | Not supported |
| QWL→BI | | 0.378 | | 0.282 | -1.842* | Supported |
| SE→AU | 48.2% | 0.110 | 42.4 | 0.186 | 0.728 | Not supported |
| FC→AU | | 0.241 | | 0.106 | -1.673* | Supported |
| Notes: *** p-value < 0.01; ** p-value < 0.05; * p-value < 0.10 | | | | | | |

**Table 6-38: The summary of the moderating effect of age**

### 6.6.2.3 Educational Level

Out of the 602 respondents, the descriptive frequencies for the educational level showed that there are 365 undergraduate and 204 postgraduate students. Applying the measurement model for each group separately revealed the following: for the undergraduate's group are [$\chi^2$ =655.732; df= 349; $\chi^2$/df= 1.879; GFI =.886; AGFI=.858; CFI=.954; RMSR=.083; RMSEA=.049, NFI=.908; PNFI=.781] and [$\chi^2$ =592.409; df= 349; $\chi^2$/df= 1.697; GFI =.829; AGFI=.787; CFI=.932; RMSR=.077; RMSEA=.059, NFI=.852; PNFI=.732] for the postgraduate group, indicating an acceptable fit of the data.

The results presented in Table 6.39 revealed that the CR and AVE were higher than 0.7 and 0.5 respectively, which indicate an adequate reliability and convergent validity. Furthermore, the AVE for all constructs is higher than the MSV and also the square root of AVE is higher than their correlation value, and therefore satisfactory discriminant validity was established for both samples.

| Factor Correlation Matrix with √AVE on the diagonal (Educational level:  Undergraduate group) | | | | | | | | | | | |
|---|---|---|---|---|---|---|---|---|---|---|---|
| **Factor** | **CR** | **AVE** | **MSV** | **ASV** | **SE** | **PU** | **PEOU** | **SN** | **QWL** | **FC** | **AU** | **BI** |
| **SE** | 0.888 | 0.726 | 0.308 | 0.217 | 0.852 | | | | | | | |
| **PU** | 0.913 | 0.677 | 0.360 | 0.220 | 0.308 | 0.823 | | | | | | |
| **PEOU** | 0.931 | 0.731 | 0.333 | 0.230 | 0.499 | 0.541 | 0.855 | | | | | |
| **SN** | 0.765 | 0.554 | 0.250 | 0.160 | 0.376 | 0.347 | 0.339 | 0.680 | | | | |
| **QWL** | 0.830 | 0.552 | 0.464 | 0.258 | 0.399 | 0.600 | 0.501 | 0.468 | 0.743 | | | |
| **FC** | 0.895 | 0.741 | 0.347 | 0.211 | 0.554 | 0.303 | 0.455 | 0.305 | 0.433 | 0.861 | | |
| **AU** | 0.711 | 0.518 | 0.394 | 0.259 | 0.555 | 0.507 | 0.400 | 0.427 | 0.408 | 0.589 | 0.646 | |
| **BI** | 0.874 | 0.701 | 0.464 | 0.324 | 0.507 | 0.571 | 0.577 | 0.500 | 0.681 | 0.496 | 0.628 | 0.837 |
| Factor Correlation Matrix with √AVE on the diagonal (Educational level:  postgraduate group) | | | | | | | | | | | |
| **Factor** | **CR** | **AVE** | **MSV** | **ASV** | **SE** | **PU** | **PEOU** | **SN** | **QWL** | **FC** | **AU** | **BI** |
| **SE** | 0.879 | 0.710 | 0.491 | 0.214 | 0.842 | | | | | | | |
| **PU** | 0.887 | 0.612 | 0.175 | 0.119 | 0.310 | 0.782 | | | | | | |
| **PEOU** | 0.921 | 0.702 | 0.320 | 0.176 | 0.506 | 0.418 | 0.838 | | | | | |
| **SN** | 0.856 | 0.599 | 0.186 | 0.121 | 0.374 | 0.376 | 0.316 | 0.774 | | | | |
| **QWL** | 0.839 | 0.567 | 0.402 | 0.200 | 0.444 | 0.390 | 0.432 | 0.348 | 0.753 | | | |
| **FC** | 0.886 | 0.722 | 0.491 | 0.234 | 0.701 | 0.210 | 0.566 | 0.431 | 0.485 | 0.850 | | |
| **AU** | 0.715 | 0.535 | 0.307 | 0.130 | 0.375 | 0.245 | 0.260 | 0.208 | 0.330 | 0.431 | 0.731 | |
| **BI** | 0.850 | 0.657 | 0.402 | 0.209 | 0.422 | 0.403 | 0.359 | 0.336 | 0.634 | 0.417 | 0.554 | 0.811 |

**Table 6-39: Construct reliability, convergent validity and discriminant**

**validity for moderator education level**

Table 6.40 shows that three paths were moderated by educational level. These paths were PEOU_BI (supporting H17b1), PU_BI (supporting H17b2) and SN_BI (supporting H17b3). Specifically, the relationship was stronger for undergraduate students in terms of PEOU_BI and SN_AU, while the relationship was stronger for postgraduate in terms of PU_BI. The results also show that no differences were detected on the relationship between QWL_BI (rejecting H17b4), SE_AU (rejecting H17b5) and FC_AU (rejecting H17b6). Thus, it can be concluded that hypothesis H17b was partially supported. It was also found that the $R^2$ for BI was 59.1% and 49.5% for AU within the undergraduate sample, indicating a good model fit. Whereas, within the postgraduate sample, the variance explained ($R^2$) for BI was 44.1% and 42.9% for AU, indicating a moderate acceptable model fit.

| Hypothesis | Undergraduate | | Postgraduate | | Z-score | Results |
|---|---|---|---|---|---|---|
| | $R^2$ | Estimate | $R^2$ | Estimate | | |
| PEOU→BI | | 0.244 | | 0. 101 | -1.718* | Supported |
| PU → BI | 59.1% | 0.086 | 44.1% | 0. 231 | 1.936* | Supported |
| SN→ BI | | 0.254 | | 0. 156 | -1.664* | Supported |
| QWL→BI | | 0.396 | | 0.488 | 0.802 | Not supported |
| SE→AU | 49.5% | 0.186 | 42.9 | 0.158 | -0.290 | Not supported |
| FC→AU | | 0.228 | | 0.326 | 0.916 | Not supported |
| Notes: *** p-value < 0.01; ** p-value < 0.05; * p-value < 0.10 | | | | | | |

**Table 6-40: The summary of the moderating effect of educational level**

## 6.6.2.4 Experience

The descriptive frequencies for the category internet experience revealed that the majority of the students (N=344) were experienced in using the web-based learning system, whereas the rest (N= 225) have low experience.

Applying the measurement model for each group separately revealed the following: for the low-experience group [$\chi^2$ =576.385; df= 349; $\chi^2$/df= 1.652; GFI =.849; AGFI=.812; CFI=.949; RMSR=.093; RMSEA=.054, NFI=.883; PNFI=.759] and [$\chi^2$ =624.705; df= 349; $\chi^2$/df= 1.790; GFI =.887; AGFI=.859; CFI=.950; RMSR=.064; RMSEA=.048, NFI=.894; PNFI=.768] for the experienced group, indicating an acceptable fit of the data. The result in Table 6.41 show that the CR was higher than 0.7 and the AVE was higher than 0.5, which satisfied the criterion for reliability and convergent validity. Moreover, the MSV for all the constructs is less than AVE and the square root of AVE is higher than their correlation value, and therefore discriminant validity is also established for both samples.

| Factor Correlation Matrix with √AVE on the diagonal (Experience: low-experience group) | | | | | | | | | | | |
|---|---|---|---|---|---|---|---|---|---|---|---|
| Factor | CR | AVE | MSV | ASV | SE | PU | PEOU | SN | QWL | FC | AU | BI |
| SE | 0.909 | 0.771 | 0.419 | 0.266 | 0.878 | | | | | | | |
| PU | 0.910 | 0.669 | 0.321 | 0.218 | 0.370 | 0.818 | | | | | | |
| PEOU | 0.937 | 0.748 | 0.329 | 0.268 | 0.568 | 0.567 | 0.865 | | | | | |
| SN | 0.787 | 0.522 | 0.278 | 0.194 | 0.488 | 0.452 | 0.396 | 0.701 | | | | |
| QWL | 0.858 | 0.603 | 0.480 | 0.266 | 0.469 | 0.522 | 0.514 | 0.425 | 0.777 | | | |
| FC | 0.896 | 0.743 | 0.419 | 0.279 | 0.647 | 0.354 | 0.573 | 0.453 | 0.509 | 0.862 | | |
| AU | 0.717 | 0.526 | 0.441 | 0.232 | 0.517 | 0.387 | 0.390 | 0.307 | 0.427 | 0.583 | 0.703 | |
| BI | 0.877 | 0.706 | 0.480 | 0.340 | 0.510 | 0.562 | 0.574 | 0.527 | 0.693 | 0.526 | 0.664 | 0.840 |
| Factor Correlation Matrix with √AVE on the diagonal (Experience: Experienced group) | | | | | | | | | | | |
| Factor | CR | AVE | MSV | ASV | SE | PU | PEOU | SN | QWL | FC | AU | BI |
| SE | 0.877 | 0.705 | 0.340 | 0.218 | 0.840 | | | | | | | |
| PU | 0.857 | 0.550 | 0.230 | 0.129 | 0.300 | 0.742 | | | | | | |
| PEOU | 0.914 | 0.680 | 0.219 | 0.150 | 0.468 | 0.346 | 0.825 | | | | | |
| SN | 0.815 | 0.527 | 0.164 | 0.104 | 0.297 | 0.258 | 0.248 | 0.726 | | | | |
| QWL | 0.805 | 0.511 | 0.381 | 0.218 | 0.443 | 0.480 | 0.422 | 0.405 | 0.715 | | | |
| FC | 0.900 | 0.751 | 0.340 | 0.199 | 0.583 | 0.247 | 0.431 | 0.260 | 0.453 | 0.867 | | |
| AU | 0.713 | 0.519 | 0.354 | 0.230 | 0.567 | 0.429 | 0.335 | 0.394 | 0.410 | 0.559 | 0.699 | |
| BI | 0.857 | 0.669 | 0.381 | 0.240 | 0.521 | 0.389 | 0.416 | 0.354 | 0.617 | 0.470 | 0.595 | 0.818 |

**Table 6-41: : Construct reliability, convergent validity and discriminant validity for moderator experience**

As can be seen in Table 6.42, experience was found to moderate the relationship between PEOU_BI (supporting H18b1), SN_BI (supporting H18b3) and SE_AU (supporting H18b5). The result shows that the relationship between PEOU_BI, SN_BI and SE_AU was stronger for the low-experience group. On the other hand, no moderating effects of experience on the relationship between PU_BI (rejecting H18b2), QWL_BI (rejecting H18b4) and FC_AU (rejecting H18b6) were found. This result indicates that hypothesis H18b was partially supported. Overall, it was found that the $R^2$ for BI was 56.8% and 49.3% for AU within the some-experience group, while the $R^2$ for BI was 44.5% and 46.8% for AU, indicating that the model had a moderate acceptable fit of the data.

| Hypothesis | low-experience | | Experienced | | Z-score | Results |
|---|---|---|---|---|---|---|
| | $R^2$ | Estimate | $R^2$ | Estimate | | |
| PEOU → BI | 56.8% | 0.168 | 44.5% | 0.087 | -1.892* | Supported |
| PU → BI | | 0.089 | | 0.128 | 1.033 | Not supported |
| SN → BI | | 0.145 | | 0.082 | -1.694* | Supported |
| QWL → BI | | 0.350 | | 0.379 | 0.256 | Not supported |
| SE → AU | 49.3% | 0.165 | 46.8% | 0.057 | -1.827* | Supported |
| FC → AU | | 0.170 | | 0.191 | 0.250 | Not supported |
| Notes: *** p-value < 0.01; ** p-value < 0.05; * p-value < 0.10 | | | | | | |

**Table 6-42: The summary of the moderating effect of experience**

# 6.7   Summary and Conclusions

This chapter started by assessing the discriminant validity, convergent validity and reliability of all the constructs within the proposed research model for the two samples. The results of the confirmatory factor analysis revealed that four items (SE4, SE5, FC4, SN4) have to be deleted from the initial measurement model for the British sample. On the other hand, items (SE4, SE5, FC4 and QWL3) have to be deleted from the Lebanese sample. The criterion for deletion was based on the indicators that demonstrate high covariance and also had high regression weight. Having established validity and reliability of the constructs, the next step was to evaluate the structural model in order to test the hypothesised relationships among the constructs, within the proposed research model. The results of the squared multiple correlations ($R^2$), which provide information about the extent to which the model explains variance in the data set, suggested that the refined model exhibits a strong explanatory power for both samples. More specifically, the determinants PEOU, PU, SN, SE and QWL accounted for 68% and 57% of the variance of BI for the British and Lebanese sample respectively. Moreover, the determinants SE, FC and BI accounted for 79% of the variance of AU for the British sample and 52% for the Lebanese sample. The path model showed that nine out of ten of the direct hypotheses for both samples were accepted. Table 6.43summarise the results in terms of supported/ not supported hypotheses that are related to the direct determinates.

The final step of the analysis involves investigating the moderating impact of four cultural variables namely, power distance, individualism\collectivism, masculinity\femininity and uncertainty avoidance and four demographic variables namely, age, gender, experience and educational level and on the relationships between the exogenous (PEOU, PU, SN, QWL, SE and FC) and endogenous (BI and AU) latent constructs. A summary of the results related to the results in terms of supported/ not supported hypotheses that are related to the cultural dimensions and demographic characteristics are listed in Table 6.44 and 6.45 respectively. The results show that all the hypothesised relationships for the moderators within the two samples were partially supported.

This chapter presented the results of the main study; the next chapter will interpret the results and provide a discussion that links them to the literature.

| Summary of Results direct determinants | Study Results | |
|---|---|---|
| **Research Direct Hypotheses** | **England** | **Lebanon** |
| *H1a,b: Perceived Ease of Use will have a direct positive influence on the intention to use web-based learning System in the British and Lebanese context.* | **Supported** | **Supported** |
| *H2a,b: Perceived Usefulness will have a direct positive influence on the intention to use web-based learning system in the British and Lebanese context.* | **Supported** | **Supported** |
| *H3a,b: Students' BI will have a positive effect on his or her actual use of web-based learning system in the British and Lebanese context.* | **Supported** | **Supported** |
| *H4a,b: Social Norm will have a positive influence on student's behavioural intention to use and accept the e-learning technology in the British and Lebanese context.* | **Supported** | **Supported** |
| *H5a,b: QWL will have a positive influence on student's behavioural intention to use the web-based learning system in the British and Lebanese context.* | **Supported** | **Supported** |
| *H6a,b: Computer self-efficacy will have a positive influence on student's behavioural intention to use the web-based learning system in the British and Lebanese context.* | Not Supported | Not Supported |
| *H7a,b: Computer self-efficacy will have a positive influence on the actual usage of the web-based learning system in the British and Lebanese context.* | **Supported** | **Supported** |
| *H8a,b: Facilitating conditions will have a positive influence on actual usage of web-based learning system in the British and Lebanese context.* | **Supported** | **Supported** |
| *H9a,b: SN will have a positive influence on perceived usefulness of web-based learning system in the British and Lebanese context.* | **Supported** | **Supported** |
| *H10a,b: PEOU will have a positive influence on perceived usefulness of web-based learning in the British and Lebanese context.* | **Supported** | **Supported** |

**Table 6-43: Summary of Results related to the direct hypotheses**

| Summary of Results for moderating impact of the cultural dimensions | | *Study Results* | |
|---|---|---|---|
| **Research Moderators Hypotheses** | **Proposed Relationship** | *England* | *Lebanon* |
| *(H11a1,a2,a3,a4,a5,a6),* *(H11b1,b2,b3,b4,b5,b6)* The relationship between (PEOU, PU, SN, QWL, SE, FC) and Behavioural Intention and actual usage of the e-learning system is moderated by the **Power Distance** value in the British/Lebanese context. | PEOU →BI | Not Supported | Not Supported |
| | PU → BI | **Supported** | **Supported** |
| | SN→ BI | **Supported** | **Supported** |
| | QWL→BI | **Supported** | Not Supported |
| | SE→AU | Not Supported | **Supported** |
| | FC→AU | Not Supported | Not Supported |
| *(H12a1,a2,a3,a4,a5,a6),* *(H12b1,b2,b3,b4,b5,b6)* The relationship between (PEOU, PU, SN, QWL, SE, FC) and Behavioural Intention and actual usage of the e-learning system is moderated by the **masculinity/femininity** value in the British/Lebanese context. | PEOU →BI | **Supported** | **Supported** |
| | PU → BI | **Supported** | Not Supported |
| | SN→ BI | **Supported** | **Supported** |
| | QWL→BI | Not Supported | Not Supported |
| | SE→AU | Not Supported | **Supported** |
| | FC→AU | Not Supported | Not Supported |
| *(H13a1,a2,a3,a4,a5,a6)* *(H13b1,b2,b3,b4,b5,b6)* The relationship between (PEOU, PU, SN, QWL, SE, FC) and Behavioural Intention and actual usage of the e-learning system is moderated by the **individualism /collectivism** value in the British/ Lebanese context. | PEOU →BI | **Supported** | Not Supported |
| | PU → BI | **Supported** | Not Supported |
| | SN→ BI | **Supported** | **Supported** |
| | QWL→BI | Not Supported | Not Supported |
| | SE→AU | Not Supported | **Supported** |
| | FC→AU | **Supported** | **Supported** |
| *(H14a1,a2,a3,a4,a5,a6)* *(H14b1,b2,b3,b4,b5,b6)* The relationship between (PEOU, PU, SN, QWL, SE, FC) and Behavioural Intention and actual usage of the e-learning system is moderated by the **Uncertainty Avoidance** value in the British/ Lebanese context. | PEOU →BI | Not Supported | **Supported** |
| | PU → BI | **Supported** | Not Supported |
| | SN→ BI | **Supported** | **Supported** |
| | QWL→BI | Not Supported | **Supported** |
| | SE→AU | **Supported** | Not Supported |
| | FC→AU | **Supported** | **Supported** |

**Table 6-44: Summary of Results related to the moderating cultural dimensions hypotheses**

| Summary of Results for the moderating impact of the demographic characteristics | | *Study Results* | |
|---|---|---|---|
| **Research Moderators Hypotheses** | **Proposed Relationship** | *England* | *Lebanon* |
| *(H15a1,a2,a3,a4,a5,a6)* *(H15b1,b2,b3,b4,b5,b6)* *The relationship between (PEOU, PU, SN, QWL, SE, FC) and Behavioural Intention and actual usage of the e-learning system will be moderated by the* **gender** *in the British/Lebanese context.* | PEOU → BI | Not Supported | **Supported** |
| | PU → BI | **Supported** | Not Supported |
| | SN → BI | **Supported** | **Supported** |
| | QWL → BI | **Supported** | **Supported** |
| | SE → AU | Not Supported | Not Supported |
| | FC → AU | **Supported** | Not Supported |
| *(H16a1,a2,a3,a4,a5,a6),* *(H16b1,b2,b3,b4,b5,b6)* *The relationship between (PEOU, PU, SN, QWL, SE, FC) and Behavioural Intention and actual usage of the e-learning system will be moderated by the* **age** *in the British/Lebanese context.* | PEOU → BI | Not Supported | **Supported** |
| | PU → BI | **Supported** | Not Supported |
| | SN → BI | **Supported** | Not Supported |
| | QWL → BI | **Supported** | **Supported** |
| | SE → AU | **Supported** | Not Supported |
| | FC → AU | Not Supported | **Supported** |
| *(H17a1,a2,a3,a4,a5,a6),* *(H17b1,b2,b3,b4,b5,b6)* *The relationship between (PEOU, PU, SN, QWL, SE, FC) and Behavioural Intention and actual usage of the e-learning system will be moderated by* **Educational Level** *in the British/Lebanese context.* | PEOU → BI | **Supported** | **Supported** |
| | PU → BI | Not Supported | **Supported** |
| | SN → BI | **Supported** | **Supported** |
| | QWL → BI | **Supported** | Not Supported |
| | SE → AU | Not Supported | Not Supported |
| | FC → AU | **Supported** | Not Supported |
| *(H18a1,a2,a3,a4,a5,a6),* *(H18b1,b2,b3,b4,b5,b6)* *The relationship between (PEOU, PU, SN, QWL, SE, FC) and Behavioural Intention and actual usage of the e-learning system will be moderated by* **Experience** *in the British/Lebanese context.* | PEOU → BI | **Supported** | **Supported** |
| | PU → BI | **Supported** | Not Supported |
| | SN → BI | **Supported** | **Supported** |
| | QWL → BI | Not Supported | Not Supported |
| | SE → AU | **Supported** | **Supported** |
| | FC → AU | **Supported** | Not Supported |

**Table 6-45: Summary of Results related to the four demographic variables**

# Chapter 7:  Discussion

*"Unobstructed access to facts can produce unlimited good only if it is matched by the desire and ability to find out what they mean and where they lead." (Norman Cousins, Human Options: An Autobiographical Notebook, 1981)*

## 7.1   Introduction

Chapter 5 and 6 presented the results of the proposed research model to examine the potential factors that affect the adoption and usage of e-learning in Lebanon and England. Two-step approach was used during the data analysis process. In the first step, the Confirmatory Factor Analysis (CFA) was employed to assess the constructs' validity and test the model fit. The next step employed the Structural Equation Modeling (SEM) technique to test the hypothesised relationships among the independent and dependent variables and to examine the moderating impact of demographic characteristics (age, gender, educational level and experience) and cultural dimensions (masculinity\femininity, individualism\collectivism, power distance and Uncertainty Avoidance). In this chapter, the researcher seeks to discuss and provide an in-depth interpretation of the main findings presented in Chapter 5 and 6, and link them to the main aim and objectives of this research.

More specifically, this chapter is centred around 3 main parts. The first part provides a detailed discussion of the direct relationships in the research model. This part is dedicated to understanding the important roles that behavioural belief, social and organisational factors plays in affecting the student's beliefs towards adoption and acceptance of e-learning technology in Lebanon and England. The impact of two sets of moderators (demographic and cultural dimensions) on the relationships between the dependent and independent variables in the structural model are then discussed for Lebanon and England separately in section 2. It is followed in section 3 by a detailed discussion related to the similarities and dissimilarities between the two countries at the national level.

## 7.2   Validation of extended TAM across cultural settings (The impact of direct determinants)

This section is devoted to the discussion of the direct relationships between the 6 exogenous (perceived usefulness, perceived ease of use, social norm, quality of work life, self-efficacy and facilitating conditions) and the endogenous (behavioural intentions and actual usage) latent constructs. Figure 7.1 depicts the results of the direct hypothesised relationships in the proposed research model for both samples. The results revealed that 9 out of 10 paths for each sample were supported. The empirical results have shown that behavioural beliefs, social beliefs and organisational support have been found to affect the students' perceptions towards using the web-based learning system in Lebanon and England. This section will provide a brief discussion about the direct hypotheses testing in the British and Lebanese model, whereas a detailed discussion is provided in the sections that follow.

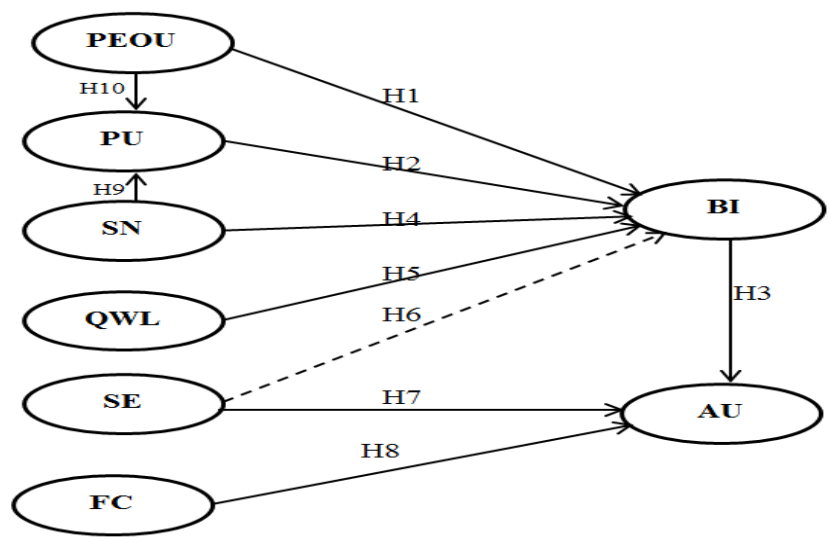

**Figure 7-1: Results of hypothesised direct relationships in the structural models in Lebanon and England.**

The standardised path coefficient of the factors affecting the Behavioural Intention within the proposed conceptual model for the British and Lebanese sample are shown in Table 6.7 and 6.9 respectively, whereas the results related to AU are shown in Table 6.8 and 6.10 respectively. The factor loadings (regression weights) output in AMOS indicates that 9 out of 10 hypothesised relationships for the two samples were supported. More specifically, PEOU ($\gamma$=0.12**, $\gamma$=0.106*) and PU ($\gamma$=0.201***, $\gamma$=0.195***) were found to have a significant positive influence on behavioural intention to use web-based learning system, supporting H1a,b and H2a,b respectively. Our result is in accordance with TAM which postulates that PU has more influence than PEOU on BI. Additionally, the influence of colleagues and instructors on students' behavioural intention to use the system was significant, SN ($\gamma$=0.082*,$\gamma$=0.095*) supporting H4a and H4b respectively. Moreover, Behavioural Intention were also influenced by the quality of work life ($\gamma$=0.430***, $\gamma$=0.408***) which supports H5a and H5b respectively. However, contrary to our expectation, our results fails to support the direct relationship between SE ($\gamma$=0.054, $\gamma$=0.041) and BI thus H6a and H6b were not supported.

The results of the squared multiple correlations (SMC) indicated that PEOU, PU, SN, SE and QWL account for 68% ($R^2 = 0.68$) of the variance of BI of the British sample and 57% of the Lebanese sample, with QWL contributing the most to behavioural intention compared to the other constructs. Additionally, SN ($\gamma$=0.299***, $\gamma$=0.167**) and PEOU ($\gamma$=0.398***, $\gamma$=0.318***) were found to have a significant positive influence on PU, supporting H9a,b and H10a,b respectively. Both SN and PEOU explained 54.1% of the variance of PU within the British sample and 32% within the Lebanese sample with PEOU contributing the most.

The results also shows that both SE ($\gamma$=0.284*** and $\gamma$=0.105*) and FC ($\gamma$=0.284*** and $\gamma$=0.176**) have a strong direct positive effect on AU which supports H7a,b and H8a,b respectively. Furthermore, AU is positively influenced by BI (ß=0.175***, ß=0.374***) which provide support for H3a,b. The results of the squared multiple correlation indicate that BI, SE and FC account for 79 % ($R^2$ = 0.79) of the variance of AU within the British sample, with SE contributing the most. While BI, SE and FC account for 52 % ($R^2$ = 0.52) of the variance of AU (indicating a reasonable explanation for AU) within the Lebanese sample, with BI having the highest impact on the actual usage compared to the other constructs. These results indicate that the higher the level of behavioural intention, the higher the actual usage will be.

In summary, the empirical results have shown that behavioural beliefs, social beliefs and organisational support have been found to affect the students' perceptions towards using the web-based learning system in the context of developing and developed countries, represented here in Lebanon and England. However, the level of significance varies among the two samples. Our results are in contrast with previous literature which suggests that Web-based learning system in Lebanon is still in its infancy (Nasser, 2000; UNDP, 2002; Baroud and Abouchedid, 2010). Our results support theoretically and empirically the ability of TAM to be a useful theoretical framework for better understanding the student's acceptance of e-learning technology in the context of developing countries where the generalizability of TAM is still questionable, and in the context of western/ developed countries where the majority of previous research has been conducted.

## 7.2.1  Behavioural beliefs and Behavioural Intention

The next two subsections provide empirical evidence of how behavioural beliefs (PEOU and PU) affect BI towards using the web-based learning system.

### 7.2.1.1 Perceived Ease of use and Behavioural Intention

As per the discussion provided in Chapter 3, perceived ease of use is defined as 'the degree to which a person believes that using a particular system would be free of effort' (Davis et al., 1989 p.320) and is similar to *effort expectancy* in UTAUT (Venkatesh et al., 2003). In the context of the research, previous TAM research suggests that if the web-based learning system is easy to use, then students are more likely to use it e.g. (Park, 2009; Chang and Tung, 2008; Zhang *et al.*, 2008). The results in Table 6.7 and 6.9 showed that PEOU has a significant positive influence on the behavioural intention to use the web-based learning system for both samples which provide support for hypothesis H1a and H1b. This finding is important as it validates and supports the relationship between PEOU and BI in the context of e-learning in Lebanon where such relationship is rarely explored, and also in the western/developed countries. Our results is in accordance with TAM model and previous literature e.g. (Saadé and Galloway, 2005; Liu *et al.*, 2010; Davis *et al.*, 1989), which posits that PEOU plays an important role in predicting the behavioural intention towards using technology. However, PEOU was found to be less important than other predictive variables with $\gamma=0.120*$ and $\gamma=0.106*$ in the UK and Lebanon sample respectively. This is maybe because students are experienced in using the computers in general and thus will use the system if it's useful in their education. This argument is supported by previous research (Davis et al., 1989; Venkatesh, 2002) which indicates that the direct influence of PEOU on BI will becomes non-significant over time due to the experience that individuals obtain when using the system.

In summary the path between PEOU and BI was significant for both samples indicating that H1a and H1b are supported.

### 7.2.1.2 Perceived usefulness and Behavioural Intention

Perceived usefulness (PU) is defined as "the degree to which a person believes that using a particular system would enhance his/her job performance" (Davis, 1989, p. 453). PU is similar to *relative advantage* in the model DOI and *performance expectancy* in UTAUT (Venkatesh et al., 2003). In the context of this research, if the students find the web-based learning system to be useful in their education, then they are more likely to use it (Park, 2009; Chang and Tung, 2008; Zhang *et al.*, 2008).

The parameter estimate results in Table 6.7 and 6.9 show that the hypothesised path between PU and BI was significant for both samples (UK and Lebanon) which provide support for hypothesis H2a and H2b. Our findings is in accordance with TAM and previous literature e.g. (Liu *et al.*, 2010; Chang and Tung, 2008) which found PU to have a strong positive effect on BI to use the technology. The strong significant influence of PU ($\gamma$=0.201*** in the UK and 0.195*** in Lebanon) on BI indicate that students think that the web-based learning system is useful in their education hence it is more likely to be accepted. Our results also suggest that students are more driven with the usefulness of the system rather than its ease of use.

In summary, the empirical results suggest that hypothesis H2a and H2b are supported. Therefore, our results support theoretically and empirically the relationship between PU and BI in the context of developing countries where such relationship has been rarely explored as well as in the context of western countries.

### 7.2.1.3 Perceived ease of use and perceived usefulness

The result of the structural model (see Table 6.7 and 6.9) show that perceived ease of use had a positive significant effect on perceived usefulness for both samples ($\gamma$=0.398*** in the UK and 0.318*** in Lebanon). Our results provide empirical evidence that hypothesis H10a and H10b were accepted. The results indicate that perceived ease of use influence the user's beliefs of usefulness towards the acceptance of the web-based learning system. In other words, PEOU was found to increase the perception of usefulness of the web-based learning system. Our finding is consistent with previous literature in IS e.g. (Davis *et al.*, 1989; Venkatesh and Davis, 2000; Igbaria *et al.*, 1997; Teo, 2009a; Teo, 2010) and in accordance with TAM model. Therefore, this finding contributes to the existing research on technology acceptance by validating this relationship in the Lebanese context among educational users where the majority of the previous studies were conducted in the context of western-countries. In summary, the results of this research suggest that hypotheses H10a and H10 were strongly supported.

## 7.2.2 Social Norm

Social Norm also known as Social Influence, is defined as 'the person's perception that most people who are important to him or her think he or she should or should not perform the behaviour in question' (Ajzen and Fishbein, 1990 p.188). In another words, SN refers to the social pressure coming from external environment which surrounds the individuals and may affect their perceptions and behaviours of engaging in a certain role. SN was included in many theories such as TRA, TPB, DTPB and TAM2 and is similar to social influence in UTAUT, social factors in MPCU and image in IDT (refer to Chapter 2).

The results in Table 6.7 and 6.9 show that the path between SN and BI was supported for the two samples, supporting hypotheses H4a and H4b. Our finding is consistent with previous research e.g. (Teo, 2010; Park, 2009; Van Raaij and Schepers, 2008) which indicates that others' opinions such as colleagues and superiors have an important influence on forming the perceptions towards using the technology. The significant impact of SN on BI maybe because the system is under mandatory usage conditions. This result provides more support that the impact of SN on BI will operate in mandatory setting (Venkatesh and Davis, 2000) even in a low power distance country such as England. However, it should be noted that SN was found to have less influence on behavioural intention compared to other factors (PEOU, PU, QWL, FC and SE), with $\gamma=0.082^*$ and $\gamma=0.095^{**}$ in the UK and Lebanon samples, respectively. This may be due to the fact that the impact of SN on BI will be reduced with the increase of experience and time (Venkatesh and Morris, 2000; Taylor and Todd, 1995b). Additionally, SN was found to have a direct positive effect on PU in both samples, providing support for H9a and H9b. Our result is accordance with the TAM2 (Venkatesh and Davis, 2000) and previous research (e.g., Teo, 2010).

In summary, the relationships between SN➔BI and SN➔PU were supported in both samples. Therefore, our results support theoretically and empirically the relationship between SN and BI and SN and PU for e-learning in the context of developing countries where such relationship has not been extensively explored as well as in the context of western countries.

## 7.2.3  Quality of work life and Behavioural Intention

In this research, QWL is defined in terms of students' perception and belief that using the e-learning system will improve their quality of work life such as saving expenses when downloading e-journals, or in communication when using email to communicate with their instructors and friends. Generally speaking, a mismatch between students and the impact of technology on their lives can be disadvantageous for both students and institutions and which in turn affect their behavioural intention to use the e-learning systems.

The results of parameter estimate presented in Table 6.7 and 6.9 show that the path between QWL and BI was highly significant for both samples. These results provide empirical evidence that highly support hypotheses H5a and H5b. It is noteworthy that perceived quality of work life had the strongest magnitude on behavioural intention compared with other predictors with $\gamma=0.430^{***}$ in the UK and $\gamma=0.408^{***}$ in Lebanon. Our findings is in line with previous studies in IS that suggest this extension may improve the TAM model e.g.(Zakour, 2004; Kripanont, 2007; Srite and karahanna, 2000), which posits that perceived quality of work life influence the users' intent to adopt and use the technology.

A significant contribution of this work is to demonstrate the relevance of quality of work life as an antecedent to behavioural intention within the context of e-learning adoption. This variable has previously been suggested as potentially important but had not been included in empirical work on TAM, nor had it been investigated in relation to e-learning acceptance or in an Lebanese or British cultural context. The results of our study confirm that quality of work life is an important consideration in the study of e-learning adoption.

In summary, hypotheses H5a and H5b were accepted.

## 7.2.4 Facilitating Technology and Actual Usage

As an organisational factor, facilitating conditions (FC) has been defined as "the degree to which an individual believes that an organizational and technical infrastructure exists to support use of the system" (Venkatesh *et al.*, 2003). More specifically, it refers to the availability of external resources (time, money and effort) and also the technological resources (PCs, broadband, etc...) needed to facilitate the performance of a particular behaviour.

The results in Table 6.7 and 6.9 suggest that the link between facilitating conditions and actual usage of the web-based learning system was significant for both samples. The results indicate that H8a and H8b were accepted with $\gamma=0.1**$ in the UK and $\gamma=0.176***$ in Lebanon. The result of this relationship suggests that providing the technological resources will motivate the students to use the e-learning system. Our results is consistent with previous research context (Teo, 2009a; Sivo and Brophy, 2003; Maldonado *et al.*, 2009; Chang *et al.*, 2007) and UTAUT model (Venkatesh *et al.*, 2003), which posits that FC has a direct positive influence on actual usage of technology. Another contribution of this study is the use of facilitating technology factors within an extended TAM to study the e-learning adoption within the context of developed countries and especially within the context of developing countries, where such countries typically support traditional styles of pedagogy in education, due to a lack of financial resources and appropriately trained staff.

## 7.2.5  Self-efficacy and usage behaviour

Self-efficacy (SE) - as an internal individual factor- has been defined as the belief "in one's capabilities to organise and execute the courses of action required to produce given attainments" (Bandoura, 1997, p.3).

The results of the parameter estimate in Table 6.7 and 6.9 indicate that the path between SE and BI was not significant which suggest that H6a and H6b were not supported, while the relationship between SE and AU were significant which support H7a and H7b. These results were not expected as the majority of the previous studies have found SE to be an important determinant that directly influences the user's behavioural intention towards using the e-learning acceptance (Chang and Tung, 2008; Yuen and Ma, 2008; Park, 2009; Vijayasarathy, 2004; Chatzoglou et al., 2009; Roca et al., 2006), however our results is similar to the findings of Venkatesh et al. (2003) who found that the a casual direct relationship between SE and BI was non-significant while the path between SE and AU was significant.

A plausible explanation for the failure to find a significant link between SE and BI could be that most of the students were experienced in using the computer and thus will form their perceptions on the bases of the usefulness of the system rather than self-efficacy or perceived ease of use. This study adds to the few studies which demonstrate the importance of SE as an antecedent to actual usage within the context of e-learning adoption in developing and developed countries. Our results support the importance of extending the TAM to include individual factors such as SE in the context of e-learning.

## 7.2.6  Behavioural Intention and Actual Usage

Table 6.7 and 6.9 presents the results of the direct hypothesised relationships in the structural model. The data analysis showed that the path between BI and AU was significant in both models; thus, supporting H3a and H3b. The results suggest that if students have a strong intention to engage in using the web-based learning systems, then they are more likely to use it. Our finding is consistent with previous research in IS e.g., (Davis *et al.*, 1989; Taylor and Todd, 1995b; Taylor and Todd, 1995c; Venkatesh and Davis, 2000; Venkatesh *et al.*, 2003) and in e-learning context (Zhang *et al.*, 2008; Yi-Cheng *et al.*, 2007; Chang and Tung, 2008; Park, 2009; Saeed and Abdinnour-Helm, 2008; McCarthy, 2006; Liu *et al.*, 2010; Walker and Johnson, 2008) and in accordance with TAM, DTPB, and TPB models.

In summary, the empirical results provide evidence to accept hypotheses H3a and H3b.

## 7.3   Discussion of results related to the impact of moderators

This section is devoted to discussing the moderating impact of the four cultural dimensions (MF, ID, PD, UA) in addition to four demographic variables namely gender, age, educational level and experience; on the relationships between the exogenous (PEOU, PU, SN, QWL and FC) and the endogenous (BI and AU) latent constructs. The impacts of these moderators were investigated through using multi-group analysis (as discussed in the previous chapter). In this approach, the data-sample is divided into subsamples and then the same structural model is run at the same time for both samples. It is then followed by pairwise comparison in path coefficients across the two groups and also considering the critical ratio for differences among the groups.

### 7.3.1  Discussion about the moderators related to cultural variables

This section will discuss the results of the moderating impact of the four cultural dimensions namely; power distance, masculinity\femininity, individualism\collectivism, uncertainty avoidance; on the direct relationships within the British and Lebanese model.

### 7.3.1.1 Power distance (PD)

In terms of the British sample, the result in Section 6.5.1.1 showed that the overall mean for the PD construct was 2.567/7 indicating a very low PD culture. The low level of PD in England is also consistent with the original findings of Hofstede's (1980) which indicate that the UK is low on power distance (scored 35/100). To analyse the moderating effects of PD the data were split into two groups using median split-method with 307 students included in the low PD group and 295 in the high PD group. Thus, the generalisability of the results of this variable should be treated with caution. The results revealed that both groups have an acceptable model fit of the data with the variance explained for BI was 49.7% and 48.2% for AU in the low PD group, whereas the $R^2$ for BI was 57.9% and 55.6% for AU in the high PD group. The highest path coefficient for both samples was between QWL→BI, while the lowest was between FC→AU in the low PD group and PEOU→BI in high level PD group. These results indicate that both groups perceived QWL as a major factor in forming their perceptions when using the technology. Furthermore, management support was found to be less important in the low level PD compared with social and individual factors. While in the high level PD group, although perceived usefulness was significant but its importance was limited compared to other factors in using and accepting the technology.

Regarding the *Lebanese sample*, the descriptive statistics in Section 6.6.1.1 showed that the overall mean for PD construct was 3.160/7 indicating a moderate PD culture. The low level of PD in Lebanon, as part of the Arab world, deviates from the original findings of Hofstede's (1980) which indicate that the Arab world are high on PD (scored 80/100). However, this result is acceptable as Hofstede's collected the data 30 years ago. The data were split into two groups using median split-method with 324 students included in the low PD group and 278 in the high PD group. Thus, the generalisability of the results of this variable should be treated with caution. Overall, the $R^2$ for BI was 40.3% and 51.9% for AU in the low-PD group indicating a moderate model fit, whereas in the high-PD group the $R^2$ for BI was 60.2% and 66.1% for AU indicating a good fit. The highest path coefficient for both samples was between QWL➔BI, while the lowest was between SN➔BI in the low PD group and FC➔BI in high level PD group. These results indicate that both groups perceived QWL as a major factor in forming their perceptions when using the technology. Furthermore, the influence of other colleagues and superiors were found to be less important in the low level PD compared with social and individual factors. While in the high level PD group, although management support was significant but its importance was limited compared to other factors in using and accepting the technology. Furthermore, the results of MGA showed that PD moderates the relationship between PU➔BI, SN➔BI and SE➔BI.

Table 7.1 presents the results of the moderating effect of power distance on the key determinants of behavioural intention and actual usage of an e-learning system for both samples.

| Research Moderator Hypotheses (Power Distance) | Proposed Relationship | England | Lebanon |
|---|---|---|---|
| *(H11a1,a2,a3,a4,a5,a6), (H11b1,b2,b3,b4,b5,b6) The relationship between (PEOU, PU, SN, QWL, SE, FC) and Behavioural Intention and actual usage of the e-learning system is moderated by the **Power Distance** value in the British/Lebanese context.* | PEOU ➔BI | Not Supported | Not Supported |
| | PU ➔ BI | **Supported** | **Supported** |
| | SN➔ BI | **Supported** | **Supported** |
| | QWL➔BI | **Supported** | Not Supported |
| | SE➔AU | Not Supported | **Supported** |
| | FC➔AU | Not Supported | Not Supported |

**Table 7-1: The summary of the moderating effect of power distance for both samples**

The results of the multi-group analysis showed that PD moderates the relationship between PU➔BI for the British and Lebanese sample. As hypothesised, the relationship was stronger for low PD. Thus, the findings show that users with lower level of PD will perceive higher importance to the usefulness of technology in the decision to adopt technology, which confirm the predictions of McCoy et al (2005a). Our findings is in line with previous research e.g. (Zmud, 1982) which suggests that individuals with high PD are characterised with lower rates of innovation and acceptance of technology and thus the effect of PD will be weaker on the relationship between PU and BI (Harris, 1997).

As hypothesised, PD was also found to moderate the relationship between SN and BI for the *UK and Lebanese* sample. The relationship between SN and BI was stronger for students with high Power Distance. Our results are consistent with the majority of previous research (Srite and Karahanna, 2006; Dinev *et al.*, 2009; McCoy *et al.*, 2005a). Thus, the results indicate that for users with higher power distance cultural values the opinions of others such as supervisors and peers play greater roles in the decision to adopt technology. It could be that users with higher level of PD were using the technology to comply with their superior opinions and suggestions and thus complete the required tasks given by supervisor and not because usefulness, but simply because she/he is the boss (Hofstede, 1980). Conversely, for people within low PD cultural values, the inequality is reduced to minimum and thus the impact of superiors are expected to be accessible to subordinates (Hoecklin et al., 1995).

The results of the MGA also revealed that PD moderates the relationship between QWL→BI for the UK sample. This means that users with lower level of PD will perceive higher importance on the impact of using the technology on their work life when forming their decision to adopt technology. However, contrary to our expectation, the relationship between QWL and BI was not found to be moderated by PD in the Lebanese sample, which means that the degree to which perceived QWL affects intention to use an e-learning system was not affected by power distance cultural variable.

Additionally, the empirical results showed that the relationship SE and BI was moderated by PD for the Lebanese sample. The relationship was stronger for users with high PD cultural values which confirm the results of the majority of previous research e.g. (Straub et al., 1997) which indicate that the freedom in working environment could increase the behavioural intention to adopt and use the new technological system  and thus individuals will use their own skills (i.e. SE) and accept the technology. On the other hand, PD did not moderate the relationship between SE and BI in the British sample, this maybe because students perceive the usefulness of the technology on their education. Therefore, the degree to which SE affect intention to use the e-learning system was not affected by PD in the British sample.

Finally, contrary to our expectations e.g. (McCoy *et al.*, 2005a; Li *et al.*, 2009), the power distance cultural variable did not moderate the relationship between PEOU→BI and FC→AU for the UK and Lebanese samples. This means that the degree to which perceived ease of use and facilitating conditions affects intention to use an e-learning system was not affected by power distance cultural variable.

Therefore, hypothesis H11a and H11b were partially supported.

## 7.3.1.2 Masculinity/Femininity (MF)

In terms of the *UK sample*, the results in Section 6.5.1.2 showed that the overall mean for MF construct was 2.535/7 indicating a very low feminine (high masculine) culture. The high level of masculinity in England is also consistent with the original findings of Hofstede's (1980) which indicate that the UK is high on masculinity cultural values (scored 66/100). Data were split into two groups using median split methods with 312 students included in the low masculine group and 290 in the high masculine group. Therefore, the generalizability of the results of this variable should be treated with caution since both groups were masculine but with different level. The results also showed that the variance explained for BI was 61.1% and 63.4% for AU in the masculine group, whereas the $R^2$ was 67.7% for BI and 72.5% for AU in the feminine group indicating an acceptable model fit of the data. The highest path coefficient for both samples was between QWL→BI, while the lowest was between FC→AU in the high masculine group and PU→BI in the low masculine group. These results indicate that both groups perceived the importance of technology on their quality of work life. Furthermore, management support was found to be less important in the high masculine group. Conversely, low masculine group give little attention to the usefulness of the systems rather than any other factors.

As for the Lebanese sample, the results in Section 6.6.1.3 showed that the overall mean for the MF construct was 3.22/7 indicating a moderately masculine culture. The moderate level of masculinity in Lebanon is very close to Hofstede's (1980) original findings, which indicates that the Arab world is moderate on masculinity cultural values (scored 53/100). The descriptive statistics showed that there are 281 respondents with high masculine cultural values and 288 with low masculine cultural values. The generalizability of the results of this variable should be treated with caution since both groups were masculine but with different levels. Overall, The $R^2$ for BI was 47.2% and 43.4% for AU in the high masculine group indicating a moderate model fit, while the $R^2$ for BI was 58.7% and 56.9% for AU in the low masculine group indicating a good model fit of the data. The highest path coefficient for both samples was between QWL➔BI, while the lowest was between SN➔BI in the high masculine group and PU➔BI in the low masculine group. These results indicate that both groups perceived QWL as a major factor in their decision to adopt the technology. Similar to the UK sample, social influence was found to be less important in the high masculine group compared to other factors. Conversely, low masculine group give little attention to the usefulness of the systems rather than social or individual factors.

Table 7.2 presents the results of the moderating effect of masculinity/femininity on the key determinants of behavioural intention and actual usage of an e-learning system for both samples.

| Research Moderator Hypotheses (masculinity/femininity) | Proposed Relationship | England | Lebanon |
|---|---|---|---|
| *(H12a1,a2,a3,a4,a5,a6)*, *(H12b1,b2,b3,b4,b5,b6)* *The relationship between (PEOU, PU, SN, QWL, SE, FC) and Behavioural Intention and actual usage of the e-learning system is moderated by the **masculinity/femininity** value in the British/Lebanese context.* | PEOU ➔BI | **Supported** | **Supported** |
| | PU ➔ BI | **Supported** | Not Supported |
| | SN➔ BI | **Supported** | **Supported** |
| | QWL➔BI | Not Supported | Not Supported |
| | SE➔AU | Not Supported | **Supported** |
| | FC➔AU | Not Supported | Not Supported |

**Table 7-2: The summary of the moderating effect of masculinity/femininity for both samples**

The results of the multi-group analysis showed that masculinity moderates the relationship between PEOU➔BI for the British and Lebanese sample, with the relationship stronger in the low masculine group. This result confirms that of previous work by (Li *et al.*, 2009; Dinev *et al.*, 2009; Srite and Karahanna, 2006; Zakour, 2004; McCoy *et al.*, 2005a) and demonstrates its applicability in the e-learning domain. This is maybe due to the fact that PEOU emphasise more on the creation of a pleasant and better work environment and emphasise less on work goals and achievements (McCoy *et al.*, 2007; Srite and Karahanna, 2006; Qingfei *et al.*, 2009) which is related to individuals with low masculine (feminine) cultural values.

The results also showed that masculinity moderates the relationship between PU➔BI for the British sample. In particular, the relationship was stronger for users with high masculine cultural values. This is maybe due to the fact that PU is closely related to achievements of work goals (Venkatesh and Davis, 2000). This finding is consistent with previous research e.g. (Venkatesh *et al.*, 2004) which found that instrumental beliefs influenced the technology acceptance intentions of individuals' with high masculine values. However, contrary to our expectation and hypotheses, the results have showed that the MF cultural variable did not moderate the relationship between PU and BI for the Lebanese *sample*. Therefore, our results suggest whether individual hold masculine or feminine cultural values, those who found the system useful will have a stronger intention to use the system than those with lower score of PU in the context of Lebanese culture.

Additionally, the results of the MGA have showed that PD moderates the relationship between SN➔BI for the UK and Lebanese sample, with the relationship was found to be stronger for low masculine (feminine) group. This result confirms that of previous work by (Srite and Karahanna, 2006; Dinev *et al.*, 2009; McCoy *et al.*, 2005a) and demonstrates its applicability in the e-learning domain in Lebanon and England. It is argued that individuals who hold low masculine cultural values are more influenced by interpersonal relationships and cooperation (Hofstede, 1991). Such individuals will tend to respond more to the opinions and suggestions of others (Li et al., 2009; Dinev et al., 2009; Srite and Karahanna, 2006; Venkatesh et al., 2004).

Furthermore, MF cultural variable was found to moderate the relationship between SE and AU for the Lebanese sample, with the relationship was stronger for individuals with low masculine values. Similar results were also reported by (Venkatesh *et al.*, 2004; Venkatesh and Morris, 2000) who found that individuals with high masculine values perceive analytical and competitive approaches to solving problems which will lead to higher score on *Self-efficacy*. However, MF did not moderate the relationship between SE and AU for the British sample. This is maybe due to the fact that the sample was relatively masculine. This means that the degree to which SE affects intention and acceptance of an e-learning system is not affected by masculinity\femininity cultural value an individual may hold.

Furthermore, masculinity\femininity cultural variable did not moderate the relationship between QWL→BI and FC→AU for the UK and Lebanese sample. This may be a result of the educational setting of the current study (as the quality of the experience may be more highly valued by all users in this context than it necessarily would in a purely work setting) or it may be that the measurement of QWL itself is ambiguous with respect to the relative emphasis on hedonic vs. instrumental features. Therefore, the degree to which QWL and FC affects intention to use an e-learning system is not affected by masculinity\femininity cultural variable.

Therefore, hypothesis H12a and H12b were partially supported.

### 7.3.1.3 Individualism/Collectivism (IC)

In terms of the *British sample*, the results of the moderating impact of IC cultural variable are presented in Section 6.5.1.3. The overall mean for IC construct was 3.82/7, indicating a moderate IC culture. This results deviates from Hofstede's (1980) findings which found that the UK scored 89 on the individualism scale. Thus, the generalisation of data should be treated with caution. As with other cultural variables, the data was split into two groups using median-split methods. There were 286 respondents in the individualistic group and 316 in the high collectivistic group. The results also showed that the $R^2$ for BI was 69.9% and 74.2% for AU in the individualistic group, whereas in the collectivistic group the variance explained ($R^2$) for BI was 66.3% and 71.4% for AU indicating a good model fit. In the individualistic group, the highest path coefficient was between QWL→BI, while the lowest was between FC→AU. This means that although organisational and technical infrastructure exists to support the use of the e-learning system, but their importance is limited compared to social and individual factors. It also means that individuals with individualistic cultural values perceive the importance of technology on their quality of work life. On the other hand, the highest path coefficient was between SE→AU in the collectivistic group, while the lowest was between PU→BI. Although perceived usefulness is important but it still get little attention compared to other factors. Conversely, such students perceive the highest importance to self-efficacy when forming their perceptions towards using the e-learning system. Furthermore, the test of the multi-group analysis showed that four paths were moderated by individualism\collectivism cultural variable. These paths were PEOU→BI, PU→ BI, SN→BI and FC→AU.

In terms of the *Lebanese sample*, the results of the descriptive statistics in Section 6.6.1.3 showed that the overall mean for the moderating construct IC was 5/7 indicating a collectivist cultural value. This results is consistent with Hofstede's (1980) findings which found that the Arab countries scored 38 on the individualism scale which means a moderate culture on I\C cultural values. As with other cultural variables, the data was split into two groups using median-split methods. There were 338 respondents in the low collectivistic group and 231 in the high collectivistic group. Overall, the $R^2$ for BI was 54.3% and for AU was 47.7% in the low collectivistic group, while in the high collectivistic group the variance explained ($R^2$) for BI was 46.1% and 50.5% for AU indicating a moderate model fit. In the low collectivistic group, the highest path coefficient was between QWL→BI, while the lowest was between SE→AU. This result suggests that for individuals with lower level of collectivism will perceive QWL as a major factor in forming their perceptions towards using the technology, whereas paid less attention to the importance of self-efficacy on their behaviours. On the other hand, in the high collectivistic group, the highest path was between SN→BI while the lowest was between PU→BI. The findings indicate that for users with high collectivistic cultural values, the influence of their colleagues and superiors plays an important role in their decision to adopt the web-based learning system, whereas such users paid less attention to the usefulness of the system compared with other factors.

The results of the moderating effect of individualism /collectivism on the key determinants of behavioural intention and actual usage of an e-learning system for both samples are presented in Table 7.3.

| Research Moderator Hypotheses (*individualism /collectivism*) | Proposed Relationship | England | Lebanon |
|---|---|---|---|
| (*H13a1,a2,a3,a4,a5,a6*), (*H13b1,b2,b3,b4,b5,b6*) *The relationship between (PEOU, PU, SN, QWL, SE, FC) and Behavioural Intention and actual usage of the e-learning system is moderated by the* **individualism /collectivism** *value in the British/Lebanese context.* | PEOU →BI | **Supported** | Not Supported |
| | PU → BI | **Supported** | Not Supported |
| | SN→ BI | **Supported** | **Supported** |
| | QWL→BI | Not Supported | Not Supported |
| | SE→AU | Not Supported | **Supported** |
| | FC→AU | **Supported** | **Supported** |

**Table 7-3: The summary of the moderating effect of masculinity/femininity for both samples**

The results of the MGA have showed that Individualism/Collectivism cultural values moderate the relationship between PEOU →BI for the UK sample. The relationship is stronger for users with high collectivistic cultural values. Thus, the findings show that for more collectivist users, the ease of use of the system plays greater roles in the decision to adopt technology. This result is in line with previous research e.g. (Lee *et al.*, 2007; McCoy *et al.*, 2005a; Sánchez-Franco *et al.*, 2009) which found that users with high individualistic cultural values will be more confident in using new technologies and will find the technology easier to use than users with high collectivistic cultural values. However, IC cultural variable did not moderate the relationship between PEOU →BI for the Lebanese sample. This result was unexpected. McCoy et al. (2007) found that the path from PEOU and BI was impaired in collectivist settings and speculated that people within these settings may be more willing to endure poor usability so long as they are achieving goals that are valued by the wider group. Therefore, the speculation that users with high individualistic cultural values will be more confident in using new technologies and will find the technology easier to use than users with high collectivistic cultural values was not supported. This means that the degree to which PEOU affects intention and acceptance of an e-learning system is not affected by IC cultural value an individual may hold.

The results of the multi-group analysis showed that an IC cultural variable also moderates the relationship between PU and BI to use the system for the British sample. The relationship is stronger for users with individualistic cultural values. Thus, the findings showed that users with individualistic cultural values may adopt and use the technology because of their potential usefulness on their education rather than the ease of use of the e-learning system. In other words, individualistic users will use the technology because it can enhance their productivity which is related to PU even if they do not have a positive attitude towards using the technology (Parboteeah et al., 2005). Our results is also in line with the social presence theory (Short et al., 1976), which suggests that people with collectivistic cultural values tend to underestimate the usefulness of a certain technology since they mute the group effects (Straub et al., 1997) as they prefer a more social presence such as face-to-face communication. However, our result fails to support this relationship within the Lebanese sample. This result was unexpected since individualistic cultures are characterised by an emphasis on the achievement of individual goals, so PU would appear to be a highly relevant factor for technology adoption in such settings, relating as it does to technology as a means for the achievement of specific goals. Therefore, the speculation that such users may adopt and use the technology because of their potential usefulness on their education rather than the ease of use of the e-learning system was not supported.

Additionally, the results revealed that IC moderates the relationship between SN and BI for the UK and Lebanese sample. The relationship was stronger for users with high collectivistic cultural values. Thus, the findings showed that for more collectivist users the opinions of others such as supervisors and peers play greater roles in the decision to adopt technology. This effect had been predicted by a number of authors, but previous support for it was limited to a few studies that compared samples across countries where there may have been other variables at work e.g. (Srite and Karahanna, 2006; Dinev *et al.*, 2009). The results also support Hofstede's (1980) findings which indicated that the relationship between the individual and the collectivity in human society goes beyond people living together; it is related and linked with social norms which mean that people with high collectivistic cultural values are more likely to comply with the opinions and suggestions of their direct referents.

Although no moderating effect of Individualism/Collectivism cultural variable was found on the relationship between QWL and BI for the UK and Lebanese sample, however QWL was found to be an important predictor in both samples. In other words, the degree to which QWL affect the intention to use the web-based learning is not affected by IC cultural value that an individual may hold.

Furthermore, the results of the MGA showed that IC cultural variable moderates the relationship between FC and AU for the Lebanese sample, with the relationship was stronger for users with high collectivistic cultural values. Our finding is in line with previous research e.g. (Csikszentmihalyi, 2000; Thatcher *et al.*, 2003) which show that for more collectivist users the management support such as training, technical infrastructure plays greater roles in the decision to adopt technology. However, this relationship was not supported in the British sample. This is maybe because technical infrastructures are related to organisational factors and not related directly to users' belief. This means that the degree to which FC affect the intention to use the web-based learning is not affected by IC cultural value that an individual may hold.

Thus, hypothesis H13a and H13b were partially supported.

### 7.3.1.4 Uncertainty avoidance (UA)

In terms of the UK sample, Section 6.5.1.4 presents the results of the moderating impact of uncertainty avoidance cultural variable. The overall mean for UA construct was 4.4/7, indicating a moderate uncertainty avoidance culture. Our results deviates from the original findings of Hofstede (1980) which indicate that the UK is low on uncertainty avoidance (scored 35/100). The results of the descriptive statistics showed that there are 324 students in the low UA group and 278 in the high UA group. The results indicated that both models had good model fit of the data. It was also found that the $R^2$ for BI was 68.7% and 64.5% for AU in the low-UA sample, while in the high-UA sample the $R^2$ for BI was 63.8% and 66.4% for AU. The highest path coefficient for both groups was between QWL→BI, while the lowest was between FC→AU for low UA group and PU→BI for high UA group. This means that both groups perceive QWL as a major factor in forming their perceptions towards using the e-learning technology. Additionally, for users with low UA cultural values, it was found that management support affects the user's decision to accept the e-learning technology however its importance was less than social and individual factors. While for users with high UA cultural values, the perceived usefulness was found to be important, however it gained little attention compared to other factors. Furthermore, the test of the multi-group analysis showed that three paths were moderated by the UA cultural variable. These paths were PU→BI, SN→BI SE→AU, and FC→AU.

In terms of the Lebanese sample, Section 6.6.1.4 presents the results of the moderating impact of the uncertainty avoidance cultural variable. The overall mean for the moderating construct UA was 5.46/7 indicating a high UA culture. Our results is consistent with the original findings of Hofstede's (1980) which reported that the Arab world scored 68 on uncertainty avoidance which means a high culture on UA. The results of the descriptive statistics showed that there are 274 students in the low UA group and 295 in the high UA group. The $R^2$ for BI was 52.8% and 43.5% for AU within the low-UA sample, while in the high-UA sample the variance explained ($R^2$) for BI was 46.1% and 54.7% for AU, indicating a moderate model fit. The highest path coefficient for both samples was between QWL→BI, while the lowest was between SN→BI for low UA group and PU→BI for high UA group. This means that both groups perceive QWL as a major factor in forming their perceptions towards using the e-learning technology. Additionally, for users with low UA cultural values the influence of colleagues and superiors affects their decision to accept the e-learning technology however it was less important than other factors. While users with high UA cultural values, perceived usefulness was found to be important, however less attention was paid to this factor compared to other social and individual factors. Furthermore, the test of the multi-group analysis showed that three paths were moderated by UA cultural variable. These paths were PEOU→BI, SN→BI, QWL→BI and FC→AU.

The results of the moderating effect of individualism /collectivism on the key determinants of behavioural intention and actual usage of an e-learning system for both samples are presented in Table 7.4.

| Research Moderator Hypotheses (*Uncertainty Avoidance*) | Proposed Relationship | England | Lebanon |
|---|---|---|---|
| (H14a1,a2,a3,a4,a5,a6), (H14b1,b2,b3,b4,b5,b6) *The relationship between (PEOU, PU, SN, QWL, SE, FC) and Behavioural Intention and actual usage of the e-learning system is moderated by the* **Uncertainty Avoidance** *value in the British/Lebanese context.* | PEOU →BI | Not Supported | **Supported** |
| | PU → BI | **Supported** | Not Supported |
| | SN→ BI | **Supported** | **Supported** |
| | QWL→BI | Not Supported | Not Supported |
| | SE→AU | **Supported** | Not Supported |
| | FC→AU | **Supported** | **Supported** |

**Table 7-4: The summary of the moderating effect of Uncertainty Avoidance for both samples**

The results of the MGA have showed that UA moderates the relationship between PEOU and BI for the Lebanese sample. The relationship was stronger for users with high UA cultural values. Thus, the findings suggest that for users high on UA, perceived ease of use plays greater role on their decision to adopt the technology. Our findings confirm the predictions of McCoy et al., (2007) and Straub et al. (1997) and Sanchez-Franco et al. (2009) which indicate that users with high UA tends to hold lower perceptions of using computers and is more concerned about ease of use of the technology in order to decrease their anxieties towards using the technology. However, our results fail to support the same relationship within the UK sample. This means the degree to which PEOU affect the intention to use the e-learning system is not affected by UA cultural values.

Furthermore, the relationship between PU and BI to use the e-learning system was moderated by UA for the UK sample. The relationship is stronger for users with low UA cultural values. Thus, the findings show that users with lower level of UA will perceive higher importance to the usefulness of technology in the decision to adopt technology. This supports the earlier findings McCoy *et al.* (2007) and the predictions of Sánchez-Franco *et al.* (2009). The fact that we examined culture at the individual level allowed our research to untangle the impact of several cultural variables and this may explain why we were able to demonstrate an impact of UA on the PU->BI relationship where Sánchez-Franco *et al.* (2009) failed. However, our results fail to support the same relationship within the Lebanese sample. Therefore, the speculation that users with low UA will accept the technology as they are less likely to be cautious towards technology and therefore perceive the system to be more useful than those with high UA cultural values was not supported in the context of e-learning in Lebanon.

Additionally, UA moderates the relationship between SN and BI for the UK and Lebanese sample. In particular, the relationship was stronger for users with high UA cultural values. In other words, students with high Uncertainty Avoidance will be highly influenced by their colleagues, peers and even their instructors to use the system as they are more likely to be cautious towards technology and the views of others provide useful information that reduces uncertainty. Again these results confirm some of the previous research (Srite and Karahanna, 2006; Dinev *et al.*, 2009).

Furthermore, the results showed that the relationship between FC and AU was stronger for users with high UA cultural values in the UK and Lebanese sample. Thus, the findings show that for high UA users management support, such as training and technical infrastructure, plays a greater role in the decision to adopt technology. In other words, such users will rely more on FC from the social environment (Venkatesh and Morris, 2000; Hwang, 2005) in order to reduce uncertainty and anxiety and improve performance.

Finally, contrary to our expectation, UA did not moderate the relationship between QWL and BI for the UK and Lebanese sample. This means that the degree to which QWL affects the users' decisions to use an e-learning system is not affected by Uncertainty Avoidance cultural variable.

Therefore, it can be concluded that hypothesis H14a and H14bwas partially supported.

## 7.3.2 Discussion about the moderators related to demographic variables

This section will discuss the results of the moderating impact of the demographic characteristics namely gender, age, educational level and experience; on the relationships within the British and Lebanese model.

### 7.3.2.1 Gender

*In terms of the UK sample*, the results in Section 6.5.2.1 showed that the variance explained ($R^2$) for BI was 66.7% and 69.1 % for AU in the male group, whereas in the female group the variance explained ($R^2$) for BI was 67.8% and 73.3% for AU, indicating a good fit of the data. In the male group, the highest path coefficient was between QWL→BI, while the lowest was between FC→AU. This means that although organisational and technical infrastructure exists to support use of the e-learning system, their importance is limited compared to social and individual factors. On the other hand, the highest path coefficient was between SE→AU in the female groups, while the lowest was between PU→BI. This result suggests that female students are more driven by self-efficacy in forming their perceptions towards using the e-learning system rather than its usefulness on their education. The test of the multi-group analysis showed that gender moderates the relationship between PU→BI, SN→BI, QWL→BI and FC→AU. In particular, the relationship was stronger for males in terms of PU→BI and QWL→BI, while the relationship was stronger for female in terms of SN→BI and FC→AU. However no differences were detected on the relationship between PEOU→BI, BI→AU and SE→AU.

*In the Lebanese sample*, the results in Section 6.6.2.1 revealed that the $R^2$ for BI was 40.2% and for AU was 44.5% within the male sample, indicating a moderate model fit, while the variance explained ($R^2$) for BI was 57.2% and 51.3% for AU within the female group, which indicates a good model fit of the data. The strongest magnitude for male and female was between QWL→BI while the lowest was between SN→BI in the male group and PU→BI in the female group. These results indicate that both groups perceived QWL as a major factor in forming their perceptions towards using the e-learning system. It also suggests that for male users the influence of other colleagues and instructors was important but less than other factors. The results also suggest that female students are more driven by social and individual factors rather than the usefulness of the e-learning system on their education. The results of the multi-group analysis showed that gender moderates the relationship between PEOU→BI, SN→BI and QWL→BI, while no moderating effects were detected in terms of PU→BI, SE→AU and FC→AU.

A summary of the moderating effect of gender for the British and Lebanese samples is presented in the Table 7.5:

| Research Moderator Hypotheses (Gender) | Proposed Relationship | England | Lebanon |
|---|---|---|---|
| *(H15a1,a2,a3,a4,a5,a6)* | PEOU →BI | Not Supported | **Supported** |
| *(H15b1,b2,b3,b4,b5,b6)* | PU → BI | **Supported** | Not Supported |
| *The relationship between (PEOU, PU,* | SN→ BI | **Supported** | **Supported** |
| *SN, QWL, SE, FC) and Behavioural* | QWL→BI | **Supported** | **Supported** |
| *Intention and actual usage of the e-* | SE→AU | Not Supported | Not Supported |
| *learning system will be moderated by the **gender** in the British/Lebanese context.* | FC→AU | **Supported** | Not Supported |

**Table 7-5: The summary of the moderating effect of gender for both samples**

Gender was found to moderate the relationship between PEOU and BI within the Lebanese sample, such that the relationship was stronger for the female group. This means that female students tend to place more emphasis on the ease of use of the system when deciding whether or not to adopt the system. Our finding is consistent with previous studies e.g. (Venkatesh *et al.*, 2003). This is may be due to the fact that men compared to women generally have lower computer anxiety and higher computer self-efficacy (Venkatesh and Morris, 2000). Our results also supports previous research in psychology e.g. (Cooper and Weaver, 2003; Roca *et al.*, 2006) which suggests that men perceive analytical and competitive approaches to solving problems (Venkatesh et al., 2004) which will lead to higher score on Self-efficacy which in turn lower the importance of perceived ease of use. On the contrary, gender did not moderate the relationship between PEOU →BI *within the British sample*, our results is inconsistent with previous research studies e.g. (Venkatesh and Morris, 2000; Venkatesh et al., 2003). In another words, no differences between male and female students were detected in terms of perceived ease of use. However, our results supports the findings of Wang et al (2009) who did not find any moderating impact of gender on the relationship between effort expectancy (similar to PEOU) and BI.

In terms of the moderating effect of gender on PU→BI, our results indicate that the relationship was stronger for male users within the *UK sample*. Our findings are consistent with literature in social psychology, which emphasises that men are more ''pragmatic'' compared to women and highly task-oriented (Minton et al., 1980) and usually have a greater emphasis on earnings and motivated by achievement needs (Hoffmann, 1980; Hofstede and Hofstede, 2005; Terzis and Economides, 2011) which is directly related to usefulness perceptions. However, contrary to our hypotheses and previous research e.g. (Venkatesh and Morris, 2000; Venkatesh *et al.*, 2003), no moderating effect of gender was found on the relationship between PU and BI within the *Lebanese sample*. Our results indicate that the strength of relationships among the PU and BI does not change with gender. This means that no matter what the gender an individual belonged to, those with higher score on perceived usefulness had a better intention to use e-learning system than those with lower perceived usefulness.

In addition, gender was also found to moderates the relationship between SN and BI *for both samples*, such that the relationship was stronger for female students. This result is consistent with the majority of previous literature (Venkatesh and Morris, 2000; Wang *et al.*, 2009; Hu *et al.*, 2010; He and Freeman, 2010) which report that men are less likely to accept behaviour even if it is confirmed by a majority of people. This might be because women are thought to rely more than men on others' opinion (Venkatesh and Morris, 2000; Hofstede and Hofstede, 2005) as they have a greater awareness of others' feelings compared to men and therefore are more easily motivated by social pressure and affiliation needs than men (Venkatesh and Morris, 2000). Another reason might be because  men are more confident in using computers and have higher self-efficacy compared with women (Venkatesh et al., 2003) and the impact of other people to adopt the technology will be weaker for men since they base their decision on their own personal knowledge and experience.

The test of the Multi-Group Analysis also showed that the relationship between QWL→BI was moderated by gender with the relationships found to be stronger for male *for both samples*. These results was expected, as QWL focuses on benefits of technology which are generally thought to be more relevant to males and focuses less on the issue of using technology to generate rapport (Srite and karahanna, 2000; Venkatesh *et al.*, 2004).

However, contrary to our hypotheses and previous research e.g. (Venkatesh and Morris, 2000; Venkatesh *et al.*, 2003), the results of the multi-group analysis revealed that gender did not moderate the relationship between SE and AU for the *British and Lebanese samples.* Our results indicate that the strength of relationships among SE and AU does not change with gender. This means that no matter what gender an individual belonged to, those with higher score on self-efficacy had a better intention to use e-learning system than those with lower perceived usefulness and self-efficacy.

Furthermore, gender was found to moderate the relationship between FC➔AU *within the British sample*, the relationship between was found to be stronger for female users, this results is consistent with Hofstede's cultural theory (Hofstede and Hofstede, 2005) proposition which is related to masculinity\femininity cultural dimensions, which indicates that women compared with men rated a higher importance towards FCs, with respect to the service aspects and the working environment. On the contrary, the results of MGA indicate that gender did not moderate the relationship between FC and AU for *the Lebanese sample.*

In summary, the results indicate that hypothesis H15a and H15b were partially supported.

## 7.3.2.2 Age

*In terms of the UK sample*, the results in Section 6.5.2.2 revealed that the variance explained for BI was 63.2% and 71.5% for AU in the younger-age group (age<=22), whereas in the older-age group (age>22) the $R^2$ for BI was 57.8% and 56.4% for AU, which indicates a good fit of the data. The highest path coefficient for both samples was between QWL➔BI, while the lowest was between FC➔AU. These results indicate that both groups perceived the importance of technology on their quality of work life. Furthermore, management support was found to be less important in both age groups compared to social and individual factors. the results revealed that age had a significant moderating impact on the relationship between PU➔BI, SN➔BI, QWL➔BI and SE➔AU with the relationship was stronger in younger-age group in terms of PU→BI, whereas the relationship was stronger for older-age group in terms of SN→BI, QWL→BI and SE→AU.

*In the Lebanese sample*, the results of moderating impact of age on the relationship between the exogenous and endogenous latent constructs presented in Section 6.6.2.2 indicated a good model fit of the data, with the $R^2$ was 56.5% for BI and 48.2% for AU in the younger-age group, whereas in the older-age group the $R^2$ for BI was 53% and 42.4% for AU. The highest path coefficient for both samples was between QWL→BI, while the lowest was between SN→BI in the younger-age group and between FC→AU in the older age group. These results indicate that both groups perceived QWL as a major factor in the decision to adopt the technology. Furthermore, for younger age group the social influence doesn't play an important role in determining their perceptions towards using the e-learning system. On the other hand, although management support was found to be an important factor for older age group, however it gained less attention compared to the social and other individual factors. The results of the Multi-group analysis showed that age moderates the relationship between PEOU→BI, QWL→BI and FC→AU.

As expected, age was found to moderate the relationship among most of the predictors and behavioural intention and actual usage (see Table 7.6). These results were expected as previous research showed that age plays an important role in the acceptance of technology e.g. (Venkatesh *et al.*, 2003; Morris *et al.*, 2005; Taylor and Todd, 1995b; Wang *et al.*, 2009).

| Research Moderator Hypotheses (Age) | Proposed Relationship | *England* | *Lebanon* |
|---|---|---|---|
| *(H16a1,a2,a3,a4,a5,a6)* *(H16b1,b2,b3,b4,b5,b6)* *The relationship between (PEOU, PU, SN, QWL, SE, FC) and Behavioural Intention and actual usage of the e-learning system will be moderated by the **age** in the British/Lebanese context.* | PEOU →BI | Not Supported | **Supported** |
| | PU → BI | **Supported** | Not Supported |
| | SN→ BI | **Supported** | Not Supported |
| | QWL→BI | **Supported** | **Supported** |
| | SE→AU | **Supported** | Not Supported |
| | FC→AU | Not Supported | **Supported** |

**Table 7-6: The summary of the moderating effect of age for both samples**

The results of the MGA showed that age moderates the relationship between PEOU_BI for the *Lebanese sample.* The relationship between PEOU and BI was stronger for older students which confirms the results of Wang et al (2009) and Venkatesh et al, (2003). These results indicate that older students are driven by the ease of use of the web-based learning system. The rationale could be that younger users usually tend to have higher self-efficacy, and their decision to accept the technology will be influenced by perceived usefulness rather than perceived ease of use (Wang *et al.*, 2009; Burton-Jones and Hubona, 2006). However, contrary to previous research and our hypotheses, no moderating role of age was found on the relationship between PEOU→BI within the *British sample*, which means that the degree to which PEOU affects the intention to use an e-learning system is not affected by the age group a user belongs to.

The results of the MGA also showed that age moderate the relationship between PU→BI within the *UK sample*, with the relationship being stronger for younger-age group. The rationale could be that younger users usually tend to have higher self-efficacy, and their decision to accept the technology will be influenced by perceived usefulness more than any other factors (Wang et al., 2009). On the other hand, age did not moderate the relationship between PU and BI within the *Lebanese sample.* Our result is inconsistent with the majority of previous research (Venkatesh *et al.*, 2003; Wang *et al.*, 2009; Czaja *et al.*, 2006), but confirm the results of Chung et al (2010). This means that no matter what age group an individual belonged to, those with higher score on perceived usefulness had a better intention to use e-learning system than those with lower perceived usefulness.

The results of the MGA also revealed that age moderates the relationship between SN and BI within the British sample, with the relationship was stronger for older users. Older users tends to be more influenced by other's opinions (Venkatesh *et al.*, 2003) compared to younger students. This may be due to the fact that affiliation needs increase with age (Morris and Venkatesh, 2000). However, contrary to our hypotheses and previous research e.g., (Wang *et al.*, 2009;

Venkatesh *et al.*, 2003; Jones *et al.*, 2009), no moderating role of age was found on the relationship between SN and BI within the Lebanese sample. The non-significant findings of the moderating effect of age may be because age differences were not large enough to demonstrate differences on the relationship between SN and BI. In this regards, the speculation that older users are affected more by others in terms of technology acceptance and adoption (Venkatesh et al., 2003) than younger users was not empirically supported.

Additionally, age moderate the relationship between QWL and BI for the UK and Lebanese samples. The relationship was found to be stronger for younger users. These results were expected since QWL focuses on benefits of technology which are generally more relevant to younger users (Srite and Karahanna, 2000).

Age was also found to moderate the relationship between FC and BI within the UK sample, with the relationship stronger for younger users which confirm the results of Venkatesh et al (2003) and Morris and Venkatesh (2000). Thus, the findings suggest that for younger students the management support plays an important role in the decision to adopt the web-based learning system. However, contrary to our hypotheses, our results fail to support the moderating impact of age on the relationship between FC and AU within the Lebanese sample. This means that degree to which FC affects the actual usage of an e-learning system is not affected by the age group a user belongs to.

Finally, age moderate the relationship between SE and AU for the Lebanese sample, with the relationship was found to be stronger for older users. The rationale could be that older adults often think that they are too old to learn a new technology (Turner et al., 2007). There is a clear evidence that younger adults have lower levels of computer anxiety than their older counterparts (Chaffin and Harlow, 2005; Saunders, 2004) and that lower levels of computer anxiety are associated with less reluctance to engage in opportunities to learn new Internet skills (Jung et al., 2010). However, contrary to our expectation, age did not moderate the relationship between SE and AU within the British sample. In another words, no matter what age group an individual belonged to, those with higher score on self-efficacy will use the e-learning system than those with lower self-efficacy.

Our results suggest that hypothesis H16a and H16b were partially supported.

### 7.3.2.3 Educational level

*In terms of the UK sample*, the results in Section 6.2.2.3 showed that the $R^2$ for BI was 65.3% and for AU was 72.8% in the undergraduate sample, whereas in the postgraduate group the variance explained ($R^2$) for BI was 58.3% and 60.5% for AU, indicating a good model fit. In the undergraduate group, the highest path coefficient was between PEOU→BI, while the lowest was between FC→AU. This means that although organisational and technical infrastructure exists to support use of the system, their importance is limited compared to other factors, such as social and individual factors. It also means that undergraduate students focus on the ease of use of the e-learning system to establish positive intentions towards the acceptance behaviour. On the other hand, the highest path coefficient was between QWL→BI in the postgraduate group, while the lowest was between SN→BI. This means that although postgraduate students are influenced by SN, but the influence of other's opinions (peers and superior) were very limited on their perceptions towards using the technology. Conversely, postgraduate students perceived the quality of work life as the major factor in forming their perceptions towards using the technology. Furthermore, the test of the MGA showed that four paths were moderated by educational level. These paths were PEOU→BI, SN→BI, QWL→BI and FC→AU.

Regarding the *Lebanese sample*, the results in Section 6.6.2.3 showed that the $R^2$ for BI was 59.1% and 49.5% for AU in the undergraduate group, whereas in the postgraduate sample the variance explained ($R^2$) for BI 44.1% and 42.9% for AU indicating a moderate acceptable model fit. The highest path coefficient for both samples was between QWL→BI, while the lowest was between PU→BI in the undergraduate group and between PEOU→BI in the postgraduate group. These results indicate that both groups perceived QWL as a major factor in the decision to adopt the technology. Furthermore, for undergraduate students, perceived usefulness was found to be less important than other factors in determining their perceptions towards using the web-based learning system. Conversely, postgraduate students emphasise more on the usefulness of the system rather than the ease of use. The test of the MGA showed that four paths were moderated by educational level. These paths were PEOU→BI, PU→BI and SN→BI.

The following table (7.7) provide a summary of the moderating effect of educational level.

| Research Moderator Hypotheses (Educational Level) | Proposed Relationship | England | Lebanon |
|---|---|---|---|
| *(H17a1,a2,a3,a4,a5,a6)* *(H17b1,b2,b3,b4,b5,b6)* *The relationship between (PEOU, PU, SN, QWL, SE, FC) and Behavioural Intention and actual usage of the e-learning system will be moderated by the **Educational level** in the British/Lebanese context.* | PEOU → BI | **Supported** | **Supported** |
| | PU → BI | Not Supported | **Supported** |
| | SN → BI | **Supported** | **Supported** |
| | QWL→BI | **Supported** | Not Supported |
| | SE→AU | Not Supported | Not Supported |
| | FC→AU | **Supported** | Not Supported |

**Table 7-7: The summary of the moderating effect of age for both samples**

Consistent with the previous research (Morris *et al.*, 2005; Burton-Jones and Hubona, 2006), educational level was found to have a significant influence on the relationship between SN→BI, PEOU→BI within the *UK and Lebanese* sample, where the relationship was stronger for users with lower educational level. These results were expected since less educated people would find the technology cumbersome and strenuous to learn and thus would rely on other's opinion regarding the adoption and usage of e-learning technology. Conversely, higher educational level will negatively affect the social influence on behaviour as both education and experience will empower the users (Burton-Jones and Hubona, 2006; Lymperopoulos and Chaniotakis, 2005). Furthermore, previous research have shown that when the education level of users increases, their intention to use e-learning systems increases (Calisir, 2009).

Furthermore, educational level was also found to moderate the relationship between PU and BI for the Lebanese sample where the relationship was stronger for postgraduate students, which confirm the majority of previous research e.g. (Porter and Donthu, 2006; Rogers, 2003). Thus, our findings suggest that postgraduate students place more emphasis on the usefulness of the system rather than the ease of use when forming their perceptions towards using the web-based learning system. However, contrary to our hypotheses; no moderating effect of educational level was found on the relationship between PU and BI. This means that the degree to which PU affects intention to use an e-learning system is not affected by students' educational level. This is maybe because both undergraduate and postgraduate students already familiar with the benefits of e-learning systems on their education.

Educational level was also found to moderate the relationship between QWL→BI and FC→AU within the UK sample, with the relationship was found to be stronger for postgraduate students. This result was expected as QWL focuses on benefits of technology which are generally thought to be more relevant to people with higher level of education and focuses less on the issue of using technology to generate rapport (Srite and karahanna, 2000; Venkatesh *et al.*, 2004). Additionally, the significant impact of educational level on the relationship between FC→AU indicates that management support has a strong effect on behavioural intention for undergraduate users but has very little effect on postgraduates. However, contrary to our expectation, no moderating effect of educational level was found on the relationship between FC→AU and QWL→BI within the Lebanese sample. This is maybe because both undergraduate and postgraduate students are already familiar with the benefits of e-learning systems on their education. Our result suggests that the degree to which FC and QWL affects intention to use an e-learning system is not affected by students' educational level.

Thus it can be concluded that hypothesis H17a and H17b were partially supported.

### 7.3.2.4 Experience

In terms of the British sample, the results in Section 6.4.1.4 showed that the variance explained ($R^2$) for BI was 70.4% and for AU was 74.9% within the some-experience group, whereas in the experienced group the $R^2$ for BI was 55.9% and 66.7% for AU, indicating an acceptable model fit of the data. The results show that the strongest path coefficient in the some-experience group was between SE→AU, while the lowest was between PU→BI. This result suggests that inexperienced users will form their perceptions based on their self-efficacy which is related to ease of use of the technology rather than its usefulness on their education. On the other hand, the results revealed that the highest path for the experienced group among the predictors of behavioural intention and usage was between QWL→BI, whereas the lowest was between SN→ BI. This result suggests that experienced users perceived the quality of work life as the major factor among other predictors in forming their perceptions towards using the e-learning system, while also found to give little attention to others' opinions (peers and superiors) towards using technology. Multi-group analysis showed that experience moderates the relationship between PEOU→BI, PU→BI, SN→BI, SE→AU, and FC→AU. The result shows that the relationship between PEOU→BI, SN→BI, SE→AU and FC→AU was stronger for the low-experience group while the relationship between PU→BI was stronger for the experienced group.

As for the Lebanese sample, the results in Section 6.6.2.4 showed that the variance explained ($R^2$) for BI was 56.8% and 49.3% for AU in the low-experience group, while in the experienced group, the $R^2$ for BI was 44.5% and 46.8% for AU, indicating that the model had a moderate acceptable fit of the data. The highest path coefficient for both groups was between QWL→BI, while the lowest was between PU→BI in the low-experience group and between SN→BI in the experienced group. This result suggests that both experienced and inexperienced users perceived the quality of work life as the major factor among other predictors in forming their perceptions towards using the e-learning system.

The result also suggests that individuals with low experience will give less attention to the usefulness of the system compared to other factors. On the other hand, experienced users paid very little attention to the opinions of others such as colleagues and superiors in their decision to adopt the technology. The results of the multi-group analysis showed that experience moderates the relationship between PEOU_BI, SN_BI and SE_AU.

Table 6.8 presents the results of the moderating effect of experience on the key determinants of behavioural intention and actual usage of an e-learning system.

| Research Moderator Hypotheses (Experience) | Proposed Relationship | England | Lebanon |
|---|---|---|---|
| *(H18a1,a2,a3,a4,a5,a6)* *(H18b1,b2,b3,b4,b5,b6)* *The relationship between (PEOU, PU, SN, QWL, SE, FC) and Behavioural Intention and actual usage of the e-learning system will be moderated by the* **Experience** *in the British/Lebanese context.* | PEOU →BI | Supported | Supported |
| | PU → BI | Supported | Not Supported |
| | SN→ BI | Supported | Supported |
| | QWL→BI | Not Supported | Not Supported |
| | SE→AU | Supported | Supported |
| | FC→AU | Supported | Not Supported |

**Table 7-8: The summary of the moderating effect of experience for both samples**

The results of the multi-group analysis showed that experience moderates the relationship between PEOU→BI, SE→AU for the UK and Lebanese sample, with the relationship stronger for the low-experience group. Our results supports the findings of the majority of previous research e.g. (Taylor and Todd, 1995b; Venkatesh *et al.*, 2000; Venkatesh *et al.*, 2003; Venkatesh and Morris, 2000; Choi and Han, 2009). These results suggest that experienced users may employ the knowledge that they have gained from their prior experience to form their intentions to use the technology (Morris and Venkatesh, 2000). The significant moderating effect of experience on the relationship between PEOU and BI is clear and stable in the literature (Venkatesh and Davis, 2000; Venkatesh and Morris, 2000). Generally speaking, when users have prior knowledge in using the technology, this will provide the users with a more robust base to learn as users will relate their incoming information with what their already know (Cohen and Levinthal, 1990). In other words, experienced users will perceive that PEOU and SE are not a big issue when learning a new technology (Taylor and Todd, 1995a; Venkatesh et al., 2003). In contrast, inexperienced users with no prior knowledge will prefer to use the technology which is easy to use (Vijayasarathy, 2004; Roca *et al.*, 2006).

The results also showed that experience moderate the relationship between SN and BI for the *UK and Lebanese* sample, the empirical result has demonstrated that the relationship was stronger for inexperienced users. Our findings is in line with previous research which suggest that the role of SN will be expected to be lower for experienced users as they more able to draw on their own past experiences to shape their perception rather than the opinions of others (Venkatesh *et al.*, 2003; Venkatesh and Davis, 2000; Venkatesh and Morris, 2000).

Experience also moderates the relationship between PU_BI for the British sample, with the relationship was stronger for the experienced group. These results is consistent with previous research (Venkatesh *et al.*, 2003; Maldonado *et al.*, 2009). Additionally, experience moderate the relationship between FC and AU for the British sample, with the relationship found to be stronger for low-experience group. On the other hand, contrary to our hypotheses, experience did not moderate the relationship between PU_BI, QWL_BI and FC_AU for the Lebanese sample. These results were not expected as it argued that the increase in users' experience will lead to user's wider options for help and support and this will lead to more usage of the system. It could be that both groups perceived the importance of the system on their education. In other words, the degree to which PU, QWL and FC affects intention to use an e-learning system is not affected by users' experience.

Thus, hypothesis H18a and H18b were partially supported.

# 7.4 Discussion related to the cross cultural differences between Lebanon and England

## 7.4.1 National culture differences

The last objective of this research was to investigate the differences between Lebanon and England on the individual- level cultural variables. The findings of this research (see Chapter 5, Table 5.24) revealed that there is a significant difference between the two countries on each of the four individual-level cultural variables.

In terms of PD, the mean score is 3.16 for Lebanon and 2.567 for England, indicating that both British and Lebanese students were low on PD, with the mean significantly higher in Lebanon (t=9.377, p<.001). The results deviate from the original findings of Hofstede's (1980) which indicate that Arab countries are high on power distance. Regarding masculinity/femininity (MF), the mean is 3.22 for Lebanon and 2.535 for England, which indicate a masculine cultural values in both samples, with the mean being significantly higher in Lebanon (t= 10.401, p<.001). This result is inconsistent with Hofstede's (1980) finding which indicated that Arab countries are high on femininity. In what concerns individualism/collectivism (IC), the mean score is 5.011 for Lebanon and 3.82 for England with the mean significantly higher in Lebanon (t=21.102, p<.001). The result reveals that British students have individualistic cultural values, while Lebanese students had collectivistic cultural values. The results is consistent with Hofstede's (1980) findings about Lebanon and England. Finally, concerning uncertainty avoidance (UA), the mean score is 5.46 for Lebanon, whereas for British students the mean is 4.4, revealing that the mean is significantly higher in Lebanon (t=20.171, p<.001). This means that Lebanese students find ambiguity more stressful and avoid unclear situation than British students. Our results is consistent with Hofstede's (1980) finding about the Arab countries and England.

It could be concluded that the mean for the individual-level cultural variables in the British sample were close to that of Hofstede's (1991). This result suggests that the sample characteristics satisfied the cultural criteria of the overall population. On the other hand, the findings from the Lebanese sample deviates on power distance and masculinity femininity. A plausible explanation could be that Arab countries were not as homogenous as Hofstede (1991) assumed, and also changes have occurred over time since Hofstede's collected his data 30 years ago. For example, the power distance is reduced to minimum especially in the Arab countries and more particularly in Lebanon while Hofstede's work was done 30 years ago and at that time a big gap between the people regarding the authority and power.

## 7.4.2 Differences on e-learning acceptance

The results of the group differences at the national level between Lebanon and the UK in terms of the predictors of the web-based learning system are presented in Chapter 5, Table 5.24. The results revealed that there are no significant differences in terms of perceived usefulness (t=-1.033, main difference=-.06750) and BI (t=-.563, mean difference= -.03795) between the two samples. This means that both Lebanese and British students found the system usefulness in their education and have good behavioural intention to use the system in the future.

In terms of PEOU, the mean score is 5.668 for Lebanon and 5.390 for England; unexpectedly the results revealed that PEOU is significantly higher in Lebanon compared to England (t=4.336, p<.001). In what concerns self-efficacy, the mean score is 5.191 for Lebanon and 4.977 for England, which is unexpectedly higher for Lebanon than England (t=3.414, p<.01).

Similarly, the results in Table 5.24 revealed that facilitating conditions (FC) was significantly higher in Lebanon (t=5.464, p<.001), with a mean of 5.442 for Lebanon and 5.045 for England. These results were not expected, however the acquisition of the data could explain the results because it was collected from two private universities in Lebanon who has been investing heavily in web-based learning systems to support their traditional teaching and to improve the learning experience as well as the performance of their students.

The results also showed that there is marginally significant difference for Lebanese and British students on social norm (SN), with a mean value 5.001 for Lebanon and 4.868 for England (t=2.010, p<.05), thus indicating that Lebanese students are more willing to accept the pressure from external environments (i.e., peers and superiors) to use the e-learning system.

This result was expected as there is a proportional relationship between SN and PD. Therefore; it could be that users with higher level of PD were using the technology simply to comply with their superior's opinions and suggestions in using the technology. Conversely, for people within low PD cultural values such as British, the inequality is reduced to minimum and thus the impacts of superiors are expected to be accessible to subordinates (Srite and Karahanna, 2006).

The results in Table 5.24 also showed that there is a significant difference for Lebanese and British students on QWL, with a mean score of 5.4 for Lebanon and 5.566 for England, indicating that a higher significant mean in England (t=-2.695, p<.01). This means that British students are more aware of the impact of technology on their quality of life. Similarly, with respect to AU, the mean score is 4.424 for England and 4.222 for Lebanon, thus revealing that British students are using the e-learning system more than the Lebanese students (t=-.20269, p<.01).

These results are important as it develops an inclusive categorisation of the similarities and differences between British and Lebanese students on the acceptance and usage of web-based learning systems. This will help in identifying and understanding any differences between the cultures of these two countries.

### 7.4.3  The Differences in the tasks the students perform using the web-based learning system

In terms of the British sample, the results in Chapter 5 (Table 5.25) showed that the majority of participants responded "to a great extent" on all other tasks except for "Announcement" which scored "to a moderate extent" and "Discussion Board" with the majority responded "to a small extent" where "Quizzes" scored the lowest mean (M=2.82) among other tasks with the majority of the participants answered "Not at all". In terms of the Lebanese sample, Table 5.26 showed that the participants responses was either 'to a moderate extent' or 'to a great extent', with the majority of the participants responded "to a moderate extent" for "Assessment", "Course Handbook", "Past Papers", "Websites" and "Quizzes".

Regarding the group differences on the tasks performed using e-learning system, the results in Table 5.27 revealed no significant differences in "lecture note", "course Handbook" and "Discussion Board" between Lebanon and England, significant difference were higher in Lebanon in terms of "Announcement" and "Quizzes" and lower in terms of "Email", "Past Papers", "Assessment" and "Websites". This indicates that Lebanese students use the e-learning system more for "Announcement" and "Quizzes" and lower for "Email", "Past Papers", "Assessment" and "Websites than British students.  It is worth noting that the highest mean for both samples was observed for "lecture note", with 318 (Mean=4.24) and 336 (mean=4.27) of respondents responded 'to a great extent' in Lebanon and England respectively.

These results are important since it will add to the few studies that determine the current usage of Web-based learning systems in the context of western/developed countries exemplified here in England, and in non-western/ developing countries exemplified here in Lebanon.

# 7.5   Conclusion

This chapter presented an in-depth interpretation of the key findings of the structural model. The hypotheses developed in the research study were discussed and linked to previous research literature. Specifically, this chapter was centred around 3 main parts. The first part provided a detailed discussion of hypotheses related to the direct relationships in the research model. This part helped the researcher to explain the overall relationships among the 6 exogenous (perceived ease of use, perceived usefulness, social norm, quality of work life, self-efficacy) and endogenous (behavioural intention and actual usage) latent constructs. In other words, this set of hypotheses helped to understand the important role the behavioural beliefs, social and organisational factors plays in affecting the student's beliefs towards adoption and acceptance of e-learning technology in Lebanon and England. The other part of this chapter was dedicated to discuss the results related to the impact of moderators namely; demographic characteristics and cultural dimensions; on the relationships in the structural model. Finally the last part of this chapter presented a detailed discussion related to the similarities and dissimilarities between the two countries at the national level.

In the next chapter (Chapter 8), a summary of the key findings of this research will be presented. Furthermore, Chapter 8 will highlight the methodological, theoretical and practical implications of the research study and discusses the potential limitations and directions for future research.

# Chapter 8:  Conclusion and Further Research

*"Unobstructed access to facts can produce unlimited good only if it is matched by the desire and ability to find out what they mean and where they lead." (Norman Cousins, Human Options: An Autobiographical Notebook, 1981)*

## 8.1   Introduction

This chapter provides a conclusion of the research findings. The chapter begins with an overall summary of this research in Section 8.2. It then presents discussion on the key methodological, theoretical and practical contributions to knowledge arising from this research in Section 8.3. This is followed by a brief discussion of the potential limitations and directions for future research in Sections 8.4 and 8.5 respectively. Finally, Section 8.6 concludes this chapter.

## 8.2   Research Overview

This section will provide a brief overview of the eight main chapters of this thesis and the steps undertaken to fulfil the research aim and objectives.

➢  **Defining the research problem and setting the research aims and objectives**

Chapter 1 introduces the reader to the research problem and the motivation behind conducting this research and its scope. It has been identified in literature that e-learning implementation is not simply a technological solution, but a process of many different factors such as social, behavioural, individual and cultural contexts. Such factors play an important role in how an information technology is developed and used. However, it has been argued that there is a lack of research covering the important role of such factors in technology adoption and use in the context of the developing countries such as Lebanon where universities and higher education institutions support traditional styles of pedagogy in education. Additionally, it has been identified in literature that although e-learning is considered as a global technology, the efficiency of such tools should also be measured locally since users usually work in local/national contexts. Therefore, this chapter highlights the importance of examining the effects of individual, social, organizational and individual level culture on the adoption and acceptance of e-learning tools by students in Lebanon and the UK.

➢ **Developing the Research Framework including the Direct and Moderators Hypothesised**

Chapter two provides a comprehensive literature review about the three main research areas that form the basis for this research: technology adoption, e-learning technologies, and individual-level culture. This chapter starts with a review of the nine most influential theories and models related to technology adoption including IDT, SCT, TRA, TPB, DTPB, TAM, TAM2, ATAM, and UTAUT, followed by the discussion of different e-learning tools being used by higher educational institutions. The importance of culture at the macro and micro level is also examined.

The literature review chapter discusses the various theories and models related to technology acceptance with its components and external factors which directly or indirectly are useful in developing the conceptual framework for this study. Chapter 3 explains and justifies the reasons behind choosing the Technology Acceptance Model (TAM) as a foundation for our conceptual model to study e-learning acceptance. This chapter also provides a further justification for including the personal, social, and situational factors as key determinants in addition to the integration of individual characteristics and Hofstede's cultural dimensions as moderators within the model to study e-learning adoption and acceptance. Moreover, research hypotheses are drawn and operational definitions are presented. The results of this chapter along with the detailed literature review in Chapter 2 helped achieve the first objective of this research, which is *"to develop a conceptual framework that captures the silent factors influencing the user adoption and acceptance of web-based learning system"*.

## ➢ Selecting the Method of Research and the Data Gathering Technique

Chapter 4 describes and justifies the philosophical approach, methods and techniques used in this research to achieve the main research objectives and to answer the research questions. Technically speaking, this research employs a quantitative method in order to understand and validate the conceptual framework. It also justifies the reasons behind employing the cross-sectional survey research approach which is based on positivism to guide the research. Additionally, a justification for the selection of Structural Equation Modelling (SEM) as a data analysis technique is provided. Furthermore, this chapter explained the sampling technique and the reasons behind choosing the convenience sampling technique.

## ➢ Preliminary data analysis

Chapter 5 describes the results of the pilot study to ensure the validity and reliability of the measuring instruments to be used in testing the hypotheses and presents the preliminary analysis of the data obtained from the respondents. The Statistical Package for the Social Science (SPSS) version 18.0 is employed for preliminary data analysis, including data screening, frequencies and percentages, reliability analysis, explanatory factor analysis and t-test. The results from the data analysis in this chapter focus on the cross-cultural differences between Lebanon and England (research objective number 2) and also investigate the different tasks that students perform using the e-learning systems (research objective number 3). This chapter helps achieve the second and third objectives of the research, which are *"to examine the similarities and differences between British and Lebanese students on the acceptance and usage of web-based learning systems"*.

➢ **Model Testing Phase**

Chapter 6 provides the results of the model testing phase where an in-depth analysis of the relationships among the constructs within the proposed research model is presented. Two steps have been used during the data analysis process. In the first step, the confirmatory factor analysis (CFA) is employed to assess the constructs' validity and test the model fit where both samples have been found to have satisfied reliability, discriminant validity and convergent validity. The next step employs the structural equation modeling (SEM) technique to test the hypothesised relationships among the independent and dependent variables. This step includes testing the direct relationships among the constructs of the revised measurement model in Lebanon and England respectively using the path analysis technique. The researcher then uses the multi-group analysis technique to investigate the moderating effect of demographic characteristics and individual-level culture on the relationships between the key factors in the generated model. The results of the structural model support the robustness of the model in Lebanon and England respectively. More specifically, this chapter helps achieve the main research question, which is *"to empirically identify the factors that affect the students' intention to adopt and use the web-based learning system, and to examine the moderating impact of individual-level culture and other demographic differences on the relationship between those factors"*

➢ **Discussion and Interpretation of the Main Research Findings**

Chapter 7 discusses the main findings from Chapters 5 and 6 through an in-depth interpretation of the demonstrated results and findings, and then links them to the main aim and objectives of this research. This chapter helps in understanding the important role the behavioural beliefs, social, organizational, individual and cultural factors play in affecting the student's beliefs towards adoption and acceptance of e-learning technology in Lebanon and England and also discusses the similarities and dissimilarities between the two countries at the national level.

> **Implications, Limitations and Future Research**

Having developed and tested the research hypothesis and thus answered the research questions, this chapter will highlight the methodological, theoretical and practical implications of the research study and discusses the potential limitations and directions for future research.

## 8.3   Contribution of the research

In this section, we summarise the contributions the work described in this thesis makes to theory, practice and methodology in the field.

### 8.3.1  Contribution to Theory

From the theoretical point of view, this study offers a number of significant contributions. The core outcomes of this research is to develop a conceptual research model that allow a better understanding of the factors that affect the acceptance of e-learning technology in UK and Lebanon, and to study the impact of two sets of moderators; namely cultural dimensions and individual characteristics; on the relationship between those factors and behavioural intention to use the technology. The literature review has revealed that there is a lack of research on the impact of social, individual, organizational and cultural factors especially in Lebanon and the UK. The results of this study contribute to the followings:

First, this thesis provides *a critical analysis of the adoption and acceptance models in the IS literature*. To answer the main research questions and thus achieve the objectives of this study, a thorough and critical literature review of the different constructs within each theoretical model is provided (Chapter 2). In addition, a comparative study between each of the models is conducted including the strength, weakness and explanatory power. Furthermore, individual-level culture is also investigated. This study employs the Technology Acceptance Model (Davis, 1989) due to its acceptable explanatory power and popularity in a number of application areas (Venkatesh and Bala, 2008). Although the TAM measures and predicts the acceptance and usage level of technology, there have been some criticisms concerning the theoretical contributions of the model, specifically its ability to fully explain technology adoption and usage e.g. see, (Bagozzi, 2007; Straub and Burton-Jones, 2007; Benbasat and Barki, 2007). Consequently, the existing parameters of TAM neglect the investigation of other essential predictors and factors that may affect the adoption and acceptance of technology such as social, organizational and individual factors. Additionally, the applicability of TAM is limited in the educational settings as much of the research has been carried out in non-educational contexts. Hence, this study extended TAM to include factors from different adoption models such as social norm, Quality of work life, self-efficacy and facilitating conditions towards understanding of acceptance intentions. Therefore, the detailed literature review and the parsimonious model used in this study make a contribution for the design of future technology acceptance models. In addition, this research added a further step to the studies that take into account the individual, social and organizational factors in technology acceptance and adoption.

The second contribution is to empirically confirm that TAM is applicable to e-learning acceptance within the Arab culture, exemplified by the Lebanon, and in the developed world exemplified by the UK. TAM has been criticised for showing bias in a cross-cultural context e.g. (McCoy *et al.*, 2005a; Straub *et al.*, 1997). Furthermore, many TAM studies focus on Western/developed countries, while TAM has not been widely tested within non-western/developing countries (Teo *et al.*, 2008). Consequently, Teo (2008) emphasizes on the importance of testing TAM in different cultures as it is argued that when Davis developed the TAM (Davis, 1989), he did not take into consideration the un-biased reliability of TAM in cross-cultural settings. Additionally, the applicability of TAM is limited in the educational settings as much of the research has been carried out in non-educational contexts. Abbad et al. (2009) had previously demonstrated support for an extended TAM in Jordan in the context of e-learning, but their study did not seek to characterise the Jordanian sample according to Hofstede's cultural dimensions. The current work therefore provides an additional contribution by clarifying the specific pattern of cultural responses for which TAM has been shown to apply. In addition, this is one of the few studies that validate the TAM model and other factors adopted from different theoretical models outside the context of North America. Our results indicate that TAM holds across cultures, in other words, it is applicable to both developing and developed worlds; therefore other Lebanese and British researchers may apply findings from previous research to local studies.

A third significant contribution of this work is to demonstrate the relevance of quality of work life as an antecedent to behavioural intention within the context of e-learning adoption. This variable has previously been suggested as potentially important but had not been included in empirical work on TAM, nor had it been investigated in relation to e-learning acceptance or in an Arab cultural context. The results of our study validate and confirm that quality of work life is an important consideration in the study of e-learning adoption. However, we have not been able to demonstrate the predicted cultural effect, so the applicability of this attribute may be relevant beyond the specific cultural focus of this study and further work on this issue is necessary.

The fourth contribution to knowledge is that, this is one of the few studies that combine technology acceptance theories and cultural theories at the micro-level within different cultural contexts who apparently exhibited unique psychological and personal characteristics. To our knowledge, no other research has measured cultural factors at the individual level in Lebanon and England. Therefore, this study is considered as a useful guide for other researchers to understand whether the acceptance of technology is mainly affected by individuals' cultural background (moderation effect) or whether the acceptance is mainly based on the key determinants of technology itself (without an indirect effect of moderation). Additionally, most of the literature about cultural studies in information systems research is based on the national or organizational level. Within the same country, individuals are usually influenced and motivated by professional, organizational, social groups they belong to. While national culture is a macro-level phenomenon, the acceptance of technology by end-users is at an individual level. Individual behaviour cannot be measured or predicted using the national measurement score since there are no means to generalise cultural characteristics about individuals within the same country, especially for measuring actual behaviour in the adoption and acceptance of technology (Ford *et al.*, 2003; McCoy *et al.*, 2005a; Straub *et al.*, 2002). There is more evidence about the misleading of using country score in explaining individual behaviour, for example Hofstede (1984) mentioned that his country-level analysis was not able to predict the individual behaviour. Therefore, this study differs from the majority of the research which uses the score of national culture to study the individual behaviour.

Therefore, the results of this research are beneficial in understanding the importance of individual level culture on the different factors that affect e-learning acceptance in Lebanon and England at individual and national level. For example, the moderating effect of culture on the relationship between social norm and behavioural intention are highly important in all of the four cultural dimensions in the Lebanese context, while less important in the context of the UK. In other words, the findings from technology acceptance models in one country may not be applicable to another country, especially in the developing world. Our work should therefore help towards building a clearer picture of precisely what moderating roles culture might play within the extended TAM models. Our work has particularly emphasised the role of multiple cultural variables in moderating the effect of social norms in technology acceptance, providing evidence for several relationships that have previously been hypothesised but for which supportive evidence was ambiguous at best.

The fifth contribution of this study is that, it adds to the few studies that take into account a set of individual factors and highlight their important role in user technology acceptance (Venkatesh et al., 2003). This study concludes that age, gender, experience and educational level play an important role between the key determinants and users' intentions and users' actual usage of e-learning technology. Previous research (Venkatesh *et al.*, 2003; Morris *et al.*, 2005; Wang *et al.*, 2009) has only investigated the impact of individual characteristics on many factors, but there is inconclusive evidence of whether it actually affects the relationship between the Quality of work life and behavioural intention.

The final theoretical contribution of this research is the development and validation of a survey instrument. It is essential to modify and validate the new measures in a situation where the theory is advanced, but no prior validation in the same context, and such efforts are considered an important contribution to scientific practice in the information system field (Straub et al, 2004). This study adopts the constructs' items from many different contexts and applies it to the context of e-learning, for example, the four individual-level cultural dimensions (Dorfman and Howell, 1988) and quality of work life constructs have never been used and validated in the context of this study (e-learning). Therefore, the modifications and validating measures of the cultural dimensions and quality of work life are considered as an important contribution to theory.

## 8.3.2  Implication to Practice

Our research question of this study has focused on identifying the factors that influence students' intention to adopt and use e-learning systems and explore whether there are differences among these factors between British and Lebanese students. This study identified many factors, which has led to a conceptual model that extends the TAM to include social norms, self-efficacy, quality of work life and facilitating conditions constructs as main determinants. This model also incorporates two sets of moderators namely; individual-level cultural dimensions (power distance, individualism\collectivism, masculinity\femininity, uncertainty avoidance) and individual characterises (age, gender, experience and educational level), in order to overcome the limitation of TAM. The empirical results have shown that behavioural beliefs, social beliefs, organizational support, cultural background and other individual characteristics have been found to affect the students' perceptions towards using the web-based learning system in Lebanon and England. These empirical findings have managerial implications which are relevant for system developers and university policy makers and administrators.

In terms of behavioural beliefs (perceived ease of use and perceived usefulness), the results shows that perceived usefulness (PU) contributed the most to behavioural intention compared to the perceived ease of use (PEOU). These findings are noticed more within respondents who are males, skilled in using technology, holds high masculine and individualistic values and low in power distance (PD) and uncertainty avoidance (UA) cultural values. In this context, it is therefore believed that students who find the system useful in their learning process and also find the system easy to use are more likely to adopt the system. The results also suggest that training is not necessary for individuals who belong to the first segment mentioned above; however it is crucial for the other one, since those users will form their perceptions about using the web-based learning system on the ease of use of the system no matter how useful the system is. Therefore, in order to attract more users of e-learning, instructors should improve the content quality of their e-learning systems by providing sufficient, up-to-date content that can fit the students' needs. In order to promote the ease of use of e-learning, system designers should provide a system which promotes ease of use.

Quality of work life (QWL) has been found to be the most important construct in explaining the causal process in the model for both samples. The demonstration that quality of work life is important in the e-learning context also suggests that system designers should pay attention to providing systems that address this concern and that educators should explain the benefits of e-learning in terms that relate to this construct. Additionally, this finding should inspire not only organizations but also the government in promoting the importance of introducing a new technology on the quality of work life.

We have also found that social norm is a significant determinant on behavioural intention to use e-learning especially in the Lebanese sample. The impact of this construct has been highly observed within users that are female, less experienced, with feminine, high on power distance and uncertainty values. It is therefore advisable for management and instructors in particular to target this segment of students. In this context, the instructor should announce to the students that using the system is mandatory and it is also advised that practitioners should persuade users who are familiar with the system to help in promoting it to other users. Thus, when the number of e-learning users reaches a critical mass point, the number of later e-learning adopters are likely to grow rapidly (Rogers, 2003). This emphasises the need to consider implementation strategies that develop buy-in from those within the wider social environment. The identified moderating effects of culture also suggest that educators may need to consider the balance of attributes (instrumental, hedonic or social) that they emphasise on encouraging technology up-take, depending on the dominant local cultural values.

As mentioned above, facilitating conditions has been found to be an important factor that positively influences actual usage of the web-based learning system. This construct is found to be less influential for males, older in age and less experienced users. Therefore, through categorization of the users into segments, the management may decrease the time and money constraints and thus provide efficient technical support in case they know the category the students belong to. In addition, university administrators can improve their strategic decision making about technology in the future.

Self-efficacy has been found to play an important role in predicting student's behavioural intention to use the e-learning. It is clear that individuals with higher self-efficacy induce a more active learning process (Chung *et al.*, 2010). Therefore, IT teams should provide both on- and off-line support in addition to training and this is necessary to increase e-learning self-efficacy. Training is very useful in boosting self-confidence in the use of technology and eventually individuals who demonstrate higher self-confidence in using technology are more likely to use the system.

This research sheds lights on the important role of the demographic characteristics and culture plays in the acceptance of technology especially when there is no homogeneity across the countries and even across the individuals within the same country. Therefore, e-learning designers and policy makers should advertise the benefits of e-learning tools and also provide training and user support programmes that consider the demographic differences. This is essential especially during the first stage of adoption as once users become familiar with an e-learning system they may persuade their friends to start using the system. Additionally, providing a user-friendly interface and standard way of course delivery is one possible solution to help students increase their perception towards acceptance and adoption of e-learning technologies.

The findings of this research also have practical implications to the higher educational institutions and universities in Lebanon and the UK. Although the government of these two countries are investing in e-learning technology, it should be noticed that students will not accept and use the technology only because it is useful. As previously mentioned, students' perceptions towards using the web-based learning system are formed through individual, social and organizational beliefs, in addition to cultural values and other demographic differences. In this context, all the major and different individual factors should be considered simultaneously; only then a more complete picture of the dynamic nature of individual technology may begin to emerge. In other words, it is futile to facilitate a technology which is implemented in a Western country or for specific group of users and then apply it in non-western countries that have substantial cultural differences without taking into consideration the cultural values. Therefore, policy makers should not consider the strategies related to content, design and structure in one country and simply apply it to another as it will be doomed to fail in other contexts. Additionally, it is recommended that educational authorities should decide on the best approach that fits their students before implementing any new technology.

Additionally, for the system developer of web-based learning systems and particularly the Blackboards Inc., this research provides the opinions of the British and Lebanese students on the important factors that affect the adoption and acceptance of such system. Similarly, the *users (students)* can understand what motivations and factors drive them into accepting the technology, and are aware of the impact of using technology on their working life and that using the technology is usually related to their social, attitudinal, cultural and individual differences.

In conclusion, the findings of this study provide a framework that allows the students, policy makers and system developers to understand the factors that affect students' intention to adopt and use e-learning systems in Lebanon and England.

### 8.3.3  Contribution to Research Methodology

From the methodological perceptive, although studies of user acceptance may need a methodological shift from quantitative research methodology that usually examines the phenomena under investigation from a positivist perspective to a mixed methods approach  in order to gain richer understanding of less studied factors. However, this research illustrates the power of quantitative methods in verifying and confirming the research models, thus achieving the research aims and objectives with several methodological contributions.

Firstly, this study contributes to the trends of IS research which uses the structural equation technique to test the measurement and structural models. Specifically, this research uses two-step approach (confirmatory factor analysis and structural equation modeling). Therefore, this research is one of the few studies to use SEM statistical methods in a cross-cultural investigation of the factors affecting the acceptance of e-learning environments. In addition, having conducted this research in Lebanon is another significant contribution. There is a lack of studies in the Arab world and specifically in Lebanon with applying SEM technique as a method of analysis. Therefore, this thesis provides a clear example to other researchers of how AMOS and SEM can be used in cultural research as a technique of analysis.

Secondly, the use of Multi-group Analysis (MGA) technique to examine the impact of moderators (individual characteristics and cultural dimensions) on the relationship between exogenous and endogenous latent contrasts in the proposed research model has been limited in previous literature; thus this research is among the few studies employing MGA to detect and analyse moderation effects. The results reveal that this method of testing the impact of moderators is successful in highlighting the effect of individual-level culture and individual characteristics on e-learning acceptance. Therefore, this technique is still new in AMOS and thus the inclusion of MGA to answer the questions related to group comparisons and differences could be useful for future cultural IS studies especially in the Lebanese context where researchers are not familiar at all with this technique.

Thirdly, the fact that the constructs used in the research have been drawn from previous literature and mostly designed and applied in American or European contexts highlights the importance of the validity and reliability of those predictors and their measurement scales in different cultural settings. For example, individual-level culture was tested in USA by Srite and Karahanna (2006) and McCoy et al., (2005). Our results have revealed that some indictors have to be deleted and thus are not the same as those of original scales to achieve convergent and discriminant validity in addition to reliability. For example, the items FC4 in facilitating conditions and SN4 from social norm have to be deleted from the Lebanese sample due to high covariance and loading on other variables. Therefore, this study contributes to examinations of the robustness of the constructs that has been used in the research in cross-cultural settings, particularly in Lebanon and in the UK. Nevertheless, this research also has shown some limitations.

## 8.4   Research Limitations

The findings of this research are encouraging and useful for academic institutions as they are based on a wide range of theoretical viewpoints and include a very large sample size (N=1168). However as with any research, our study has some potential limitations that need to be identified and discussed.

Firstly, our sample frame was based on convenience sampling technique and included participants studying at two private universities in Lebanon and one university in the UK. Although most of the cultural dimensions are by and large in line with Hofstede's (2001) findings and the sample characteristics satisfies the criteria for the target population, this sampling technique limits the ability to generalise the findings to the entire Lebanese and UK population.

Secondly, although our results find support for TAM in Lebanon and England respectively, however generalizability of the findings to other countries should be treated with caution. McCoy et al. (2007) found significant problems when applying TAM in a range of countries. He indicated that TAM may not produce satisfying results in countries with extreme cultural characteristics such as very high power distance or very low uncertainty avoidance.

Thirdly, we uses a self-report measure of the actual usage of the system rather than a log file or observation, which limits our findings in terms of capturing more thoroughly the students' way of using the web-based learning system. However, using self-report to measure the actual usage has been supported by previous researchers (Zhang *et al.*, 2008; Martinez-Torres *et al.*, 2008; Liaw and Huang, 2011) and it has been shown to be a strong predictor to actual usage of the system.

Another limitation is that we assume the moderators (demographic characteristics and individual cultural dimensions) to be statistically invariant. In other words, we did not investigate the measurement invariance which is related to the psychometric properties of the instrument (i.e., configural, metric, and measurement error) before applying the multi-group analysis to check for group differences among the groups to be compared. Thus, the findings of the moderators should be treated with cautions. Additionally, the two sets of moderators (age, gender, experience, educational level) and cultural dimensions (power distance, masculinity\femininity, individualism\collectivism, and uncertainty avoidance) were investigated in isolation. However, when we measure them all together only then may a more complete picture of the dynamic nature of individual technology begin to emerge.

Finally, our study has investigated the impact of moderating factors in a mandatory environment and specific user group (e-learning users) and thus our findings could not be generalised to a voluntary environment and other groups and other e-learning systems. Future research should investigate the impact of moderating factors in voluntary environment as it has previously been found that this variable can have a big influence on users' perception towards using technology (Venkatesh et al., 2003). Further work could also study different user groups (e.g., students with disability, children) and/or different organizational contexts (e.g., High schools or public institutions) to explore the validity of the model in different contexts.

Fourthly, the current work is only limited to one particular e-learning system (Blackboard). Thus, the students' perception might have been different when using other e-learning tools such as WebCT, Moodle etc due to its own features. Therefore, future research may replicate the study using different e-learning tools such as WebCT, mobile learning, IPAD and digital TV.

This research employs a cross-sectional method and quantitative survey to collect the data, in other words, the data were collected at one single time from Lebanon and the UK. Although the questionnaires have strong theoretical literature and were carefully distributed to students, using a purely quantitative analysis limits the ability to have an in-depth view of the phenomena being investigated which is mainly found in qualitative research. Hence, using this method only is justified due to time and resources taking into consideration that the research aim and objectives have been achieved. Therefore, future research could use a variety of methodologies (interviews, qualitative methods, longitudinal study, etc.) to understand culture-technology relationship.

Contrary to previous empirical studies in IS, the relationship between QWL and behavioural intention to use the technology did not mediate masculinity/femininity cultural value. Although masculinity/femininity is not referred to gender as defined by biological sex (male versus female), however, we recognize some of the answers are based on the biological sex and not on the work or study environment. Further study of the relationship between gender and masculinity/femininity is needed to check if those two variables are related to each other's.

## 8.5   Future research directions

This is the first study to focus on the examination of the potential factors that affect the adoption and usage of e-learning in Lebanon and England respectively, especially the factors related to the individual-level cultural dimensions. Therefore, there are a number of suggestions for future research arising from this work, especially that the Lebanese explanatory power of the Lebanese model is lower than the UK's one. In other words, there are other factors that might provide more power in explaining web-based learning behaviour in non-western countries. Therefore this study is only a beginning and suggests for several research directions.

First, although it was not possible to use triangulation method (both survey and interviews) due to time and resources constraints, the questionnaire findings would have been strengthened if it were supported by direct observation and interviews with the participants in an effort to produce more reliable data. A triangulation method would provide the researcher with an in-depth understanding about the students' opinions regarding the web-based learning systems and why they use or not use these tools. In other words, if adaptation of e-learning system is required, would the student choose to compromise their learning pedagogy or simply give up using e-learning technology?

Second, since user's behaviour may differ depending on culture, social, situational, beliefs and technology acceptance level and since the findings of our study are context-specific (Lebanon and the UK), it would be more typical to investigate if our developed model may hold for different nationalities and different geographical countries, like other Arab or Asian countries. This will be valuable in assessing the robustness and the validity of the research model across different cultural settings. Therefore, future research could replicate our study among mono-and-multi-cultural samples.

Third, our study has investigated the impact of moderating factors in a mandatory environment within one context. Future research should explore the impact of moderating factors in voluntary environment as it was found that this variable has a big influence on students' perception towards using technology (Venkatesh et al., 2003), or with different user groups (e.g., students with disability, children) and/or different organizational contexts (e.g., High schools or public institutions) to explore their validity in different contexts.

Fourth, the current work is only limited to web-based learning systems (such as Moodle and Blackboard). Therefore, future research may replicate the study using different e-learning tools or platforms (e.g. mobile learning, IPAD and digital TV) with a different pedagogical perceptive to find out how the results might differ from the current research. Furthermore, different technologies other than e-learning could be investigated, such as online banking, e-commerce, and telecommunication in order to generalise and prove the usefulness of the research model and to establish external validity of the model. Additionally, another area of possible research could be to investigate how cultural characteristics could be translated into e-learning interface guidelines. Such research would be beneficial to future developers of web-based learning systems targeting various cultures.

Fifth, future research may extend our study to integrate other potential constructs of interest to the education community and technology itself such as learning skills, perceived enjoyment, accessibility, flexibility, and system reliability and design characteristics, privacy and security of assessments in web-based educational technologies at higher educational institutions. In addition, further research could consider whether individual-level cultural variables also have a direct effect within the current research model or other competing models such as TRA and TPB; only then a more complete picture of the dynamic nature of individual technology may begin to emerge. Therefore, other research may be necessary to incorporate other 'new' factors related to web-based learning systems which are not included in this model.

Finally, the social norms construct that are included in this research combines the influence of both instructors and colleagues. Previous researchers have suggested that the different norm groups should be split and studied separately e.g. (Srite and Karahanna, 2006). While we note that in this study we are able to demonstrate significant interactions with cultural values and social norms without separating the different referent groups, future work could usefully examine the relative importance of these groups for the different cultural effects observed.

The findings of this research highlight the importance of each individual characteristic and cultural dimension, where one individual characteristic may override the others in web-based learning environments. Hence, future research is welcomed to examine the interaction between combinations of different dimensions to determine their relative importance in web-based learning environments. For example, research studies could be conducted to investigate the interaction effects between masculinity\femininity cultural variable and gender on student learning in web-based learning system by dividing participants into four groups (Masculinity and Female; Masculinity and Male; Femininity and Female, Femininity and Male). Another area of possible research could include the examination of the interaction effects between prior experience and gender (Expert and Female, Expert and Male; No-experience and Male; No-experience and female) on student learning in web-based learning systems.

## 8.6   Personal Reflection

The journey throughout my PhD is one of the most interesting and enjoyable experience in my whole life. I feel like research has become part of my life now, and I have been fortunate to be surrounded by inspirational researchers who are willing to help me grow in many areas. More importantly, this experience has not only taught me how to be a good researcher, but also changed the way I see life as many of the skills are transferable to my personal life.

# References

Aaker, D. A., Kumar, V., Day, G. S. & Leone, R. (2009) *Marketing Research*. John Wiley & Sons.

Abbasi, M. S., Tarhini, A., Elyas, T., & Shah, F. (2015). Impact of individualism and collectivism over the individual's technology acceptance behaviour: A multi-group analysis between Pakistan and Turkey. Journal of Enterprise Information Management, 28 (6), 747-768

Abbasi, M.S., Tarhini, A., Hassouna, M. & Shah, F. (2015). Social, Organizational, Demography and Individuals' Technology Acceptance Behaviour: A Conceptual Model. European Scientific Journal, 11 (9), 48-76.

Abbad, M. M., Morris, D. & De Nahlik, C. (2009). Looking under the bonnet: Factors affecting student adoption of e-learning systems in Jordan. *The International Review of Research in Open and Distance Learning* **10**.

Abu-Shanab, E. & Pearson, J. (2009). Internet banking in Jordan: an Arabic instrument validation process. *Int. Arab J. Inf. Technol.* **6,** 235-244.

Abu-Shanab, E. A. (2011) Education level as a technology adoption moderator. *Computer Research and Development (ICCRD), 2011 3rd International Conference on.* (pp. 324-328).

Adams, D. A., Nelson, R. R. & Todd, P. A. (1992). Perceived usefulness, ease of use, and usage of information technology: a replication. *MIS quarterly*, 227-247.

Agarwal, R. & Karahanna, E. (2000). Time flies when you're having fun: cognitive absorption and beliefs about information technology usage 1. *MIS quarterly* **24,** 665-694.

Agarwal, R. & Prasad, J. (1998). A conceptual and operational definition of personal innovativeness in the domain of information technology. *Information systems research* **9,** 204-215.

Agarwal, R. & Prasad, J. (1999). Are individual differences germane to the acceptance of new information technologies? *Decision Sciences* **30,** 361-391.

Ajjan, H. & Hartshorne, R. (2008). Investigating faculty decisions to adopt Web 2.0 technologies: Theory and empirical tests. *The Internet and Higher Education* **11,** 71-80.

Ajzen, I. (1985) *From intentions to actions: A theory of planned behavior*. Springer.

Ajzen, I. (1988) Attitudes, personality, and behavior. Dorsey Press (Chicago, IL).

Ajzen, I. (1991). The theory of planned behavior. *Organizational Behavior and Human Decision Processes* **50,** 179-211.

Ajzen, I. (2005) *Attitudes, personality, and behavior*. McGraw-Hill International.

Ajzen, I. & Fishbein, M. (1980) *Understanding attitudes and predicting social behavior (278)*. Prentice-Hall.

Akhter, S. H. (2003). Digital divide and purchase intention: Why demographic psychology matters. *Journal of Economic Psychology* **24,** 321-327.

Almajali, D. A., Masa'deh, R., & Tarhini, A. (2016). Antecedents of ERP Systems Implementation Success: A Study on Jordanian Healthcare Sector. Journal of Enterprise Information Management, 29 (4).

Al-Gahtani, S. S. (2008). Testing for the applicability of the TAM model in the Arabic context: exploring an extended TAM with three moderating factors. *Information Resources Management Journal (IRMJ)* **21,** 1-26.

Al-Jabri, I. M. & Al-Khaldi, M. A. (1997). Effects of user characteristics on computer attitudes among undergraduate business students. *Journal of Organizational and End User Computing (JOEUC)* **9,** 16-23.

Alenezi, H., Tarhini, A. & Sharma, S. K. (2015). Development of a Quantitative Model to Investigate the Strategic Relationship between Information Quality and e-Government Benefits. Transforming Government: People, Process and Policy, 9 (3), 324 - 351.

Alenezi, H., Tarhini, A. & Masa'deh, R. (2015). Investigating the Strategic Relationship between Information Quality and E-Government Benefits: A Literature Review. International Review of Social Sciences and Humanities, 9 (1), 33-50.

Alenezi, H., Tarhini, A. & Masa'deh, R. (2015). Exploring the Relationship between Information Quality and E-Government Benefits: A Literature Review. International Review of Social Sciences and Humanities, 9 (1), 33-50.

Al-Tamony, H., Tarhini, A., Al-Salti, Z., Gharaibeh, A. & Elyas, T. (Accepted). The relationship between Change Management Strategy and Successful Enterprise Resource Planning (ERP) Implementations: A Theoretical Perspective. International Journal of Business Management and Economic Research

Amoako-Gyampah, K. (2007). Perceived usefulness, user involvement and behavioral intention: an empirical study of ERP implementation. *Computers in Human Behavior* **23,** 1232-1248.

Arbaugh, J. & Duray, R. (2002). Technological and structural characteristics, student learning and satisfaction with web-based courses An exploratory study of two on-line MBA programs. *Management Learning* **33,** 331-347.

Arbuckle, J. (2009) *Amos 18 User's Guide*. SPSS Incorporated.

Arenas-Gaitán, J., Ramírez-Correa, P. E. & Javier Rondán-Cataluña, F. (2011). Cross cultural analysis of the use and perceptions of web Based learning systems. *Computers & Education* **57,** 1762-1774.

Armitage, C. J. & Conner, M. (2001). Efficacy of the theory of planned behaviour: A meta-analytic review. *British journal of social psychology* **40,** 471-499.

Arachchilage, N. A. G., Tarhini, A., & Love, S. (2015). Designing a mobile game to thwarts malicious IT threats: A phishing threat avoidance perspective. International Journal for Informatics, 8 (3), 1058-1065.

Bagozzi, R. P. (1981). An examination of the validity of two models of attitude. *Multivariate Behavioral Research* **16,** 323-359.

Bagozzi, R. P. (1984). Expectancy-value attitude models an analysis of critical measurement issues. *International Journal of Research in Marketing* **1,** 295-310.

Bagozzi, R. P. (2007). The Legacy of the Technology Acceptance Model and a Proposal for a Paradigm Shift. *Journal of the Association for Information Systems* **8,** 3.

Bandura, A. (1986) *Social foundations of thought and action: A social cognitive theory*. Prentice-Hall, Inc.

Bandura, A. (1995) *Self-efficacy in changing societies*. Cambridge University Press.

Bandura, A. (1997) *Self-Efficacy: The Exercise of Control* (604). New York: W.H. Freeman.

Bandura, A. & McClelland, D. C. (1977). Social learning theory.

Bandura, A. (2006). On integrating social cognitive and social diffusion theories. Communication of innovations: A journey with Ev Rogers, 111-135.

Barki, H. & Hartwick, J. (1994). Measuring user participation, user involvement, and user attitude. *MIS quarterly***,** 59-82.

Baroud, F. & Abouchedid, k. (2010) eLEARNING IN LEBANON: Patterns of E-learning Development in Lebanon's Mosaic Educational Context. *In Demiray, U. (Editor) e-learning practices: cases on challenges facing e-learning and national development, institutional studies and practices* (pp. 409-424). Eskisehir-Turkey, Anadolu University.

Bem, S. L. (1981). The BSRI and gender schema theory: A reply to Spence and Helmreich.

Benbasat, I. & Barki, H. (2007). Quo vadis, TAM. *Journal of the Association for Information Systems* **8,** 211-218.

Benbasat, I. & Weber, R. (1996). Research commentary: Rethinking "diversity" in information systems research. *Information systems research* **7,** 389-399.

Benbunan-Fich, R., Hiltz, S. R. & Harasim, L. (2005). The online interaction learning model: An integrated theoretical framework for learning networks. *Learning together online: Research on asynchronous learning networks***,** 19-37.

Berger, I. (1993). A framework for understanding the relationship between environmental attitudes and consumer behavior. *Marketing theory and application* **4,** 157-163.

Bhattacherjee, A. (2000). Acceptance of e-commerce services: the case of electronic brokerages. *Systems, Man and Cybernetics, Part A: Systems and Humans, IEEE Transactions on* **30,** 411-420.

Blaikie, N. (2000). Designing social research: the logic of anticipation. *Recherche* **67,** 02.

Blumberg, B., Cooper, D. R. & Schindler, P. S. (2008) *Business Research Methods*. McGraw-Hill Higher Education.

Blunch, N. J. (2008) *Introduction to structural equation modelling using SPSS and AMOS*. SAGE.

Blackboard Inc. (2012) Accelerate with Blackboard Learn. [online document] [Accessed 10.09.2013] Available at: http://blackboard.com/Platforms/Learn/Products/Blackboard-Learn.aspx

Bollen, K. A. (1989) *Structural equations with latent variables*. MA: Wiley.

Bouhnik, D. & Marcus, T. (2006). Interaction in distance-learning courses. *Journal of the American Society for Information Science and Technology* **57,** 299-305.

Bryman, A. (2008) *Social Research Methods*. Oxford University Press.

Bryman, A. & Bell, E. (2011) *Business research methods*. Cambridge; New York, NY: Oxford University Press.

Burton-Jones, A. & Hubona, G. S. (2006). The mediation of external variables in the technology acceptance model. *Information & Management* **43,** 706-717.

Byrne, B. M. (2006) *Structural Equation Modeling With Amos: Basic Concepts, Applications, and Programming*. Taylor & Francis Group.

Calisir, F., Gumussoy, A.C. & Bayram, A (2009). Predicting the behavioral intention to use enterprise resource planning systems: An exploratory extension of the technology acceptance model. *Management Research News* **32,** 597 - 613.

Callan, V. J., Bowman, K. & Framework, A. F. L. (2010). Sustaining e-learning innovations: a review of the evidence and future directions: final report, November 2010.

Carmines, E. G. & McIver, J. P. (1981). Analyzing models with unobserved variables: Analysis of covariance structures. *Social measurement: Current issues***,** 65-115.

Carter, L. & Weerakkody, V. (2008). E-government adoption: A cultural comparison. *Information Systems Frontiers* **10,** 473-482.

Cascio, W. F. & McEvoy, G. (2003) *Managing human resources: Productivity, quality of work life, profits*. McGraw-Hill.

Chaffin, A. J. & Harlow, S. D. (2005). Cognitive learning applied to older adult learners and technology. *Educational Gerontology* **31,** 301-329.

Chan, S.-c. & Lu, M.-t. (2004). Understanding internet banking adoption and use behavior: a Hong Kong perspective. *Journal of Global Information Management (JGIM)* **12,** 21-43.

Chang, I., Hwang, H.-G., Hung, W.-F. & Li, Y.-C. (2007). Physicians' acceptance of pharmacokinetics-based clinical decision support systems. *Expert Systems with Applications* **33,** 296-303.

Chang, S. C. & Tung, F. C. (2008). An empirical investigation of students' behavioural intentions to use the online learning course websites. *British Journal of Educational Technology* **39,** 71-83.

Chatzoglou, P. D., Sarigiannidis, L., Vraimaki, E. & Diamantidis, A. (2009). Investigating Greek employees' intention to use web-based training. *Computers & Education* **53,** 877-889.

Chau, P. Y. & Hu, P. J. H. (2001). Information Technology Acceptance by Individual Professionals: A Model Comparison Approach*. *Decision Sciences* **32,** 699-719.

Chau, P. Y. K. & Hu, P. J. (2002). Examining a model of information technology acceptance by individual professionals: An exploratory study. *Journal of management information systems* **18,** 191-230.

Chen, L.-d., Gillenson, M. L. & Sherrell, D. L. (2002). Enticing online consumers: an extended technology acceptance perspective. *Information & Management* **39,** 705-719.

Chen, W. S. & Hirschheim, R. (2004). A paradigmatic and methodological examination of information systems research from 1991 to 2001. *Information Systems Journal* **14,** 197-235.

Chesney, T. (2006). An acceptance model for useful and fun information systems.

Chin, W. W., Marcolin, B. L. & Newsted, P. R. (2003). A partial least squares latent variable modeling approach for measuring interaction effects: Results from a Monte Carlo simulation study and an electronic-mail emotion/adoption study. *Information systems research* **14,** 189-217.

Chin, W. W. & Todd, P. A. (1995). On the use, usefulness, and ease of use of structural equation modeling in MIS research: a note of caution. *MIS quarterly***,** 237-246.

Choi, B. & Han, G. (2009). Psychology of selfhood in china: Where is the collective ? *Culture & Psychology* **15,** 73-82.

Choudrie, J. & Dwivedi, Y., K. (2005). Investigating the Research Approaches for Examining Technology Adoption Issues. *Journal of Research Practice* **1**.

Christie, M. F. & Ferdos, F. (2004). The mutual impact of educational and information technologies: Building a pedagogy of e-learning. *Journal of Information Technology Impact* **4,** 15-26.

Chung, J. E., Park, N., Wang, H., Fulk, J. & McLaughlin, M. (2010). Age differences in perceptions of online community participation among non-users: An extension of the Technology Acceptance Model. *Computers in Human Behavior* **26,** 1674-1684.

Cohen, W. M. & Levinthal, D. A. (1990). Absorptive capacity: a new perspective on learning and innovation. *Administrative science quarterly***,** 128-152.

Compeau, D., Higgins, C. A., Huff, S., Redesign, B. P., Broadbent, M., Weill, P., Clair, D. S., Karahanna, E., Straub, D. W. & Chervany, N. L. (1999). Social Cognitive Theory and Individual Reactions to Computing Technology: A Longitudinal Study.

Compeau, D. R. & Higgins, C. A. (1991) A social cognitive theory perspective on individual reactions to computing technology. *Proceedings of the twelfth international conference on Information systems.* (pp. 187-198). University of Minnesota.

Compeau, D. R. & Higgins, C. A. (1995). Computer self-efficacy: Development of a measure and initial test. *MIS quarterly***,** 189-211.

Comrey, A. L. & Lee, H. B. (1992) *A First Course in Factor Analysis*. L. Erlbaum Associates.

Conner, M. & Armitage, C. J. (1998). Extending the theory of planned behavior: A review and avenues for further research. *Journal of Applied Social Psychology* **28,** 1429-1464.

Cooper, J. & Weaver, K. D. (2003) *Gender and computers: Understanding the digital divide*. Lawrence Erlbaum.

Creswell, J. W. (2008) *Research design : qualitative, quantitative, and mixed methods approaches*. Los Angeles: Sage.

Cronbach, L. J. (1951). Coefficient alpha and the internal structure of tests. *Psychometrika* **16,** 297-334.

Crotty, M. (1998) *The Foundations of Social Research: Meaning and Perspective in the Research Process*. Sage Publications.

Csikszentmihalyi, M. (2000) The contribution of flow to positive psychology: scientific essays in honour of Martin EP Seligman. The Science of Optimism and Hope. Philadelphia, PA: Templeton Foundation Press.

Czaja, S. J., Charness, N., Fisk, A. D., Hertzog, C., Nair, S. N., Rogers, W. A. & Sharit, J. (2006). Factors predicting the use of technology: findings from the Center for Research and Education on Aging and Technology Enhancement (CREATE). *Psychology and aging* **21,** 333.

Dadkhah, M & Tarhini, A. (2015). Phishers Attack on the Researchers for Financial Goals in Pharmaceutical and Medical Open Access Journals. International Journal of Pharmacy and Pharmaceutical Sciences, 7 (10), 1-2.

Dadkhah, M., Tarhini, A., Lyashenko, V. & Jazi, M. D. (2015). Hiring Editorial Member for Receiving Papers from Authors. Mediterranean Journal of Social Sciences, 6(4), 11-12.

David, M. & Sutton, C. D. (2004) *Social research: The basics*. Sage Publications Ltd.

Davis, F. D. (1989). Perceived usefulness, perceived ease of use, and user acceptance of information technology. *MIS quarterly,* 319-340.

Davis, F. D. (1993). User acceptance of information technology: system characteristics, user perceptions and behavioral impacts.

Davis, F. D., Bagozzi, R. P. & Warshaw, P. R. (1989). User acceptance of computer technology: a comparison of two theoretical models. *Management science,* 982-1003.

Denzin, N. K. & Lincoln, Y. S. (2005) *The Sage handbook of qualitative research*. Sage Publications, Inc.

DeVellis, R. F. (2003) *Scale development: theory and applications*. Newbury Park: CA: Sage.

Dinev, T., Goo, J., Hu, Q. & Nam, K. (2009). User behaviour towards protective information technologies: the role of national cultural differences. *Information Systems Journal* **19,** 391-412.

Dishaw, M. T. & Strong, D. M. (1999). Extending the technology acceptance model with task–technology fit constructs. *Information & Management* **36,** 9-21.

Dorfman, P. W. & Howell, J. (1988). Dimension of national culture and effective leadership patterns: Hofstede revisited. *Advances in International Comparative Management* **3,** 127-150.

Douglas, S. P. & Craig, C. S. (2006). On improving the conceptual foundations of international marketing research. *Journal of International Marketing,* 1-22.

Downey, J. (2006) Measuring general computer self-efficacy: The surprising comparison of three instruments in predicting performance, attitudes, and usage. *System Sciences, 2006. HICSS'06. Proceedings of the 39th Annual Hawaii International Conference on.* (pp. 210a-210a). IEEE.

Eagly, A. H. & Chaiken, S. (1993) *The psychology of attitudes*. Harcourt Brace Jovanovich College Publishers.

Eckhardt, G. M. & Houston, M. J. (2002). Cultural paradoxes reflected in brand meaning: McDonald's in Shanghai, China. *Journal of International Marketing*, 68-82.

Elenurm, T. (2008). Applying cross-cultural student teams for supporting international networking of Estonian enterprises. *Baltic Journal of Management* **3,** 145-158.

Elliott, M. A., Armitage, C. J. & Baughan, C. J. (2003). Drivers' compliance with speed limits: an application of the theory of planned behavior. *Journal of Applied Psychology* **88,** 964.

Evanschitzky, H. & Wunderlich, M. (2006). An examination of moderator effects in the four-stage loyalty model. *Journal of Service Research* **8,** 330-345.

Falconer, D. (2008). A demographic and content survey of critical research in information systems for the period 2001–2005. *Communications of the Association for Information Systems* **22,** 30.

Fishbein, M. & Ajzen, I. (1975) *Belief, attitude, intention and behavior: An introduction to theory and research*.

Fletcher, K. M. (2005). Self-efficacy as an evaluation measure for programs in support of online learning literacies for undergraduates. *The Internet and Higher Education* **8,** 307-322.

Ford, D. P., Connelly, C. E. & Meister, D. B. (2003). Information systems research and Hofstede's culture's consequences: an uneasy and incomplete partnership. *Engineering Management, IEEE Transactions on* **50,** 8-25.

Fornell, C. & Larcker, D. F. (1981). Structural equation models with unobservable variables and measurement error: Algebra and statistics. *Journal of marketing research*, 382-388.

Fowler, F. J. (2009) *Survey Research Methods*. Sage Publications.

Foxall, G. R. (1997) *Marketing psychology: The paradigm in the wings*. Palgrave Macmillan.

Freire, L. L., Arezes, P. M. & Campos, J. C. (2012). A literature review about usability evaluation methods for e-learning platforms. *Work: A Journal of Prevention, Assessment and Rehabilitation* **41,** 1038-1044.

Fu, J.-R., Farn, C.-K. & Chao, W.-P. (2006). Acceptance of electronic tax filing: A study of taxpayer intentions. *Information & Management* **43,** 109-126.

Galliers, R. (1992). Choosing Appropriate Information Systems Research Approaches: A Revised Taxonomy.

Garfield, M. J. & Watson, R. T. (1997). Differences in national information infrastructures: the reflection of national cultures. *The Journal of Strategic Information Systems* **6,** 313-337.

Garrison, D. R. (2011) *E-learning in the 21st century: A framework for research and practice*. Taylor & Francis.

Gefen, D. & Straub, D. W. (1997). Gender differences in the perception and use of e-mail: An extension to the technology acceptance model. *MIS quarterly*, 389-400.

Gefen, D. & Straub, D. W. (2000). The Relative Importance of Perceived Ease of Use in IS Adoption: A Study of E-Commerce Adoption. *J. AIS* **1,** 0-.

Gefen, D., Straub, D. W. & Boudreau, M. C. (2000). Structural equation modeling and regression: Guidelines for research practice.

Ghauri, P. N. & Grønhaug, K. (2005) *Research methods in business studies: a practical guide*. Prentice Hall.

Govindasamy, T. (2002). Successful implementation of e-Learning Pedagogical considerations. *Internet and Higher Education* **4,** 287-299.

Grandon, E. E., Alshare, K. & Kwun, O. (2005). Factors influencing student intention to adopt online classes: a cross-cultural study. *Journal of Computing Sciences in Colleges* **20,** 46-56.

Groves, R. M., Fowler, F. J., Couper, M. P., Lepkowski, J. M. & Singer, E. (2009) *Survey methodology*. John Wiley & Sons Inc.

Guba, E. G. & Lincoln, Y. S. (1994). Competing paradigms in qualitative research. *Handbook of qualitative research* **2,** 163-194.

Guo, Y. & Barnes, S. (2007). Why people buy virtual items in virtual worlds with real money. *ACM SIGMIS Database* **38,** 69-76.

Hair, J. F. J., Black, W. C., Babin, B. J., Anderson, R. E. & Tatham, R. L. (2010) *Multivariate data analysis*. New Jersey: Prentice-Hall.

Hair, J., Black W. C., Babin B. J., Anderson R. E. & Tatham R. L. (2006) Multivariate Data Analysis. Upper Saddle River, New Jersey: Pearson Prentice Hall, Pearson Education, Inc. .

Hall, B. & Howard, K. (2008). A Synergistic Approach: Conducting Mixed Methods Research With Typological and Systemic Design Considerations. *Journal of mixed methods research* **2,** 248-269.

Hall, E. T. (1973). The silent language.

Hannon, J. & D'Netto, B. (2007). Cultural diversity online: student engagement with learning technologies. *International Journal of Educational Management* **21,** 418-432.

Han, S. (2003) Individual adoption of information systems in organizations: A literature  review of technology acceptance model, TUCS.

Harris, R. W. (1997). Teaching, learning and information technology: Attitudes towards computers among Hong Kong's faculty. *Journal of computing in higher education* **9,** 89-114.

Hassouna, M., Tarhini, A. Elyas, T. & Abou Trab, M. S. (2015). Customer Churn in Mobile Markets: A Comparison of Techniques. International Business Research, 8(6), 224-237.

He, J. & Freeman, L. A. (2010). Are men more technology-oriented than women? The role of gender on the development of general computer self-efficacy of college students. *Journal of Information Systems Education* **21,** 203.

Hernandez, B., Jimenez, J. & Jose Martin, M. (2009). The impact of self-efficacy, ease of use and usefulness on e-purchasing: an analysis of experienced e-shoppers. *Interacting with Computers* **21,** 146-156.

Hesse-Biber, S. N. & Leavy, P. (2010) *The Practice of Qualitative Research* (400). Sage Publications.

Ho, C.-L. & Dzeng, R.-J. (2010). Construction safety training via e-Learning: Learning effectiveness and user satisfaction. *Computers & Education* **55,** 858-867.

Hoecklin, L. A., Unit, E. I. & Britain, G. (1995) *Managing cultural differences: Strategies for competitive advantage*. Addison-Wesley Wokingham,, England.

Hoffmann, L. W. (1980). Early childhood experiences and women's achievement motives. *Communication Research & Broadcasting*.

Hofstede, G. (1980) *Culture's consequences: comparing values, behaviours, institutions and organisations across nations*. Thousand Oaks, CA: Sage Publications.

Hofstede, G. (1991) *Cultures and Organizations: Software of the Mind: Intercultural Cooperation and its Importance for Survival*. McGraw-Hill International.

Hofstede, G. (1993). Cultural constraints in management theories. *The Academy of Management Executive* **7,** 81-94.

Hofstede, G. (2002). The Pitfalls of Cross-National Survey Research: A Reply to the Article by Spector et al. on the Psychometric Properties of the Hofstede Values Survey Module 1994. *Applied psychology* **51,** 170-173.

Hofstede, G. & Bond, M. H. (1988). The Confucius connection: From cultural roots to economic growth. *Organizational dynamics* **16,** 4-21.

Hofstede, G. & Hofstede, J. (2005) *Cultures and Organizations: Software of the Mind*. McGraw-Hill.

Hofstede, G. & Peterson, M. F. (2000). National values and organizational practices. *Handbook of organizational culture. and climate*, 401-414.

Hofstede, G. H. (1984) *Culture's consequences: International differences in work-related values*. Sage Publications, Inc.

Hoft, N. L. (1996) Developing a cultural model. *International users interface.* (pp. 41-73). John Wiley \& Sons, Inc.

Holmes-Smith, P. (2011). Structural Equation Modelling (Using Amos): From the Fundamentals to Advanced Topics. *Melbourne: SREAMS*.

Holmes, B. & Gardner, J. (2006) *E-learning: Concepts and practice*. Sage.

Horton, W. (2011) *E-learning by design*. Wiley. com.

Hsu, M.-H. & Chiu, C.-M. (2004). Predicting electronic service continuance with a decomposed theory of planned behaviour. *Behaviour & Information Technology* **23,** 359-373.

Hu, H., Al-Gahtani, S. S. & Hu, P. J. H. (2010). Examining Gender Effects in Technology Acceptance by Arabian Workers: A Survey Study.

Hu, L. & Bentler, P. M. (1999). Cutoff criteria for fit indexes in covariance structure analysis: Conventional criteria versus new alternatives. *Structural Equation Modeling: A Multidisciplinary Journal* **6,** 1-55.

Huang, W.-H. D., Hood, D. W. & Yoo, S. J. (2012). Gender divide and acceptance of collaborative Web 2.0 applications for learning in higher education. *The Internet and Higher Education*.

Hwang, Y. (2005). Investigating enterprise systems adoption: uncertainty avoidance, intrinsic motivation, and the technology acceptance model. *Eur. J. Inf. Syst.* **14,** 150-161.

Igbaria, M., Parasuraman, S. & Baroudi, J. J. (1996). A motivational model of microcomputer usage. *Journal of management information systems* **13,** 127-143.

Igbaria, M., Zinatelli, N., Cragg, P. & Cavaye, A. L. (1997). Personal computing acceptance factors in small firms: a structural equation model. *MIS quarterly***,** 279-305.

Iskander, M. (2008) *Innovative techniques in instruction technology, E-learning, E-assessment and education*. Springer.

Jackson, S. & Scott, S. (2001). Putting the body's feet on the ground: Towards a sociological reconceptualization of gendered and sexual embodiment. *EXPLORATIONS IN SOCIOLOGY* **59,** 9-24.

Janssens, W., de Pelsmacker, P., Wijnen, K. & Van Kenhove, P. (2008) *Marketing research with SPSS*. Prentice Hall.

Johns, S. K., Murphy Smith, L. & Strand, C. A. (2003) How culture affects the use of information technology. (pp. 84-109). Wiley Online Library.

Johnson, B. & Christensen, L. B. (2010) *Educational Research: Quantitative, Qualitative, and Mixed Approaches (p.34).* Thousand Oaks: Sage Publications.

Jones, G. H. & Jones, B. H. (2005). A comparison of teacher and student attitudes concerning use and effectiveness of web-based course management software. *Educational Technology & Society* **8,** 125-135.

Jones, S., Fox, S., Internet, P. & Project, A. L. (2009) *Generations online in 2009*. Pew Internet & American Life Project Washington, DC.

Jung, Y., Peng, W., Moran, M., Jin, S. A. A., McLaughlin, M., Cody, M., Jordan-Marsh, M., Albright, J. & Silverstein, M. (2010). Low-income minority seniors' enrollment in a cybercafé: Psychological barriers to crossing the digital divide. *Educational Gerontology* **36,** 193-212.

Kagitcibasi, C. (1997). Individualism and collectivism.

Kaplan, B. & Maxwell, J. (2005). Qualitative research methods for evaluating computer information systems. *Evaluating the Organizational Impact of Healthcare Information Systems***,** 30-55.

Karahanna, E., Straub, D. W. & Chervany, N. L. (1999). Information technology adoption across time: a cross-sectional comparison of pre-adoption and post-adoption beliefs. *MIS quarterly***,** 183-213.

Keller, C., Hrastinski, S. & Carlsson, S. A. (2007). Students' Acceptanc of E-learning Environments: A Comparative Study in Sweden and Lithuania. *International Business***,** 395-406.

Kijsanayotin, B., Pannarunothai, S. & Speedie, S. M. (2009). Factors influencing health information technology adoption in Thailand's community health centers: Applying the UTAUT model. *International journal of medical informatics* **78,** 404-416.

Kim, D. J. (2008). Self-perception-based versus transference-based trust determinants in computer-mediated transactions: A cross-cultural comparison study. *Journal of management information systems* **24,** 13-45.

Kim, K. & Moore, J. (2005). Web-based learning: Factors affecting student' satisfaction and learning experience. *First Monday* **10**.

King, W. R. & He, J. (2006). A meta-analysis of the technology acceptance model. *Information & Management* **43,** 740-755.

Kline, R. B. (2005) *Principles And Practice Of Structural Equation Modeling*. Guilford Press.

Kline, R. B. (2010) *Principles and practice of structural equation modeling*. The Guilford Press.

Koufaris, M. (2002). Applying the technology acceptance model and flow theory to online consumer behavior. *Information systems research* **13,** 205-223.

Krathwohl, D. R. (2004) *Methods of Educational and Social Science Research: An Integrated Approach*. Waveland Press.

Kripanont, N. (2007) Examining a technology acceptance model of internet usage by academics within Thai business schools. Victoria University Melbourne, Australia.

Kvasny, L. & Richardson, H. (2006). Critical research in information systems: looking forward, looking back. *Information Technology & People* **19,** 196-202.

Landry, B. J. L., Griffeth, R. & Hartman, S. (2006). Measuring student perceptions of blackboard using the technology acceptance model. *Decision Sciences Journal of Innovative Education* **4,** 87-99.

Lanseng, E. J. & Andreassen, T. W. (2007). Electronic healthcare: a study of people's readiness and attitude toward performing self-diagnosis. *International Journal of Service Industry Management* **18,** 394-417.

Last, J. & Abramson, J. (2001) International epidemiological association. A dictionary of epidemiology. New York: Oxford University Press.

Lederer, A. L., Maupin, D. J., Sena, M. P. & Zhuang, Y. (2000). The technology acceptance model and the World Wide Web. *Decision Support Systems* **29,** 269-282.

Lee, I., Choi, B., Kim, J. & Hong, S. J. (2007). Culture-technology fit: Effects of cultural characteristics on the post-adoption beliefs of mobile internet users. *International Journal of Electronic Commerce* **11,** 11-51.

Lewis, W., Agarwal, R. & Sambamurthy, V. (2003). Sources of influence on beliefs about information technology use: An empirical study of knowledge workers. *MIS quarterly*, 657-678.

Li, N. & Kirkup, G. (2007). Gender and cultural differences in Internet use: A study of China and the UK. *Computers & Education* **48,** 301-317.

Li, X., Hess, T. J., McNab, A. L. & Yu, Y. (2009). Culture and acceptance of global web sites: a cross-country study of the effects of national cultural values on acceptance of a personal web portal. *ACM SIGMIS Database* **40,** 49-74.

Liaw, S.-S. (2008). Investigating students' perceived satisfaction, behavioral intention, and effectiveness of e-learning: A case study of the Blackboard system. *Computers & Education* **51,** 864-873.

Liaw, S. & Huang, H. (2011). A study of investigating learners attitudes toward e-learning.

Lichtman, M. (2006) *Qualitative Research in Education: A User's Guide (pp. 7-8)*. Thousand Oaks: Sage Publications.

Light, R. J., Singer, J. D. & Willett, J. B. (1990) *By Design: Planning Research on Higher Education*. Harvard University Press.

Likert, R. (1932). A technique for the measurement of attitudes. *Archives of psychology* **22,** 1-55.

Limayem, M. & Hirt, S. G. (2000) Internet-based teaching: how to encourage university students to adopt advanced Internet-based technologies? , *System Sciences, 2000. Proceedings of the 33rd Annual Hawaii International Conference on.* (pp. 9 pp.). IEEE.

Lin, H.-F. (2006). Understanding behavioral intention to participate in virtual communities. *CyberPsychology & Behavior* **9,** 540-547.

Lincoln, Y. S., Lynham, S. A. & Guba, E. G. (2011). Paradigmatic controversies, contradictions, and emerging confluences, revisited. *The Sage handbook of qualitative research***,** 97-128.

Liu, I. F., Chen, M. C., Sun, Y. S., Wible, D. & Kuo, C. H. (2010). Extending the TAM model to explore the factors that affect Intention to Use an Online Learning Community. *Computers & Education* **54,** 600-610.

Liu, S. H., Liao, H. L. & Peng, C. J. (2005). Applying the Technology Acceptance Model And Flow Theory to Online E-learning  Users' Acectpance Behavior. *E-learning* **4,** H8.

Lymperopoulos, C. & Chaniotakis, I. E. (2005). Factors affecting acceptance of the internet as a marketing-intelligence tool among employees of Greek bank branches. *International Journal of Bank Marketing* **23,** 484-505.

Lynott, P. P. & McCandless, N. J. (2000). The impact of age vs. life experience on the gender role attitudes of women in different cohorts. *Journal of Women & Aging* **12,** 5-21.

Mahmood, M. A., Hall, L. & Swanberg, D. L. (2001). Factors affecting information technology usage: A meta-analysis of the empirical literature. *Journal of Organizational Computing and Electronic Commerce* **11,** 107-130.

Maldonado, U. P. T., Khan, G. F., Moon, J. & Rho, J. J. (2009) E-learning motivation, students' acceptance/use of educational portal in developing countries: a case study of Peru. *Computer Sciences and Convergence Information Technology, 2009. ICCIT'09. Fourth International Conference on.* (pp. 1431-1441). IEEE.

Mandell, S. L. (1987) *Computers and information processing: concepts and applications with BASIC*. West Pub. Co.

Manfreda, K. L., Batagelj, Z. & Vehovar, V. (2002). Design of web survey questionnaires: Three basic experiments. *Journal of Computer-Mediated Communication* **7,** 0-0.

Marakas, G. M., Mun, Y. Y. & Johnson, R. D. (1998). The multilevel and multifaceted character of computer self-efficacy: Toward clarification of the construct and an integrative framework for research. *Information systems research* **9,** 126-163.

Marcus, A. & Gould, E. W. (2000). Crosscurrents: cultural dimensions and global Web user-interface design. *interactions* **7,** 32-46.

Martinez-Torres, M., Marin, S. L. T., Garcia, F. B., Vazquez, S. G., Oliva, M. A. & Torres, T. (2008). A technological acceptance of e-learning tools used in practical and laboratory teaching, according to the European higher education area 1. *Behaviour & Information Technology* **27,** 495-505.

Mathieson, K. (1991). Predicting user intentions: comparing the technology acceptance model with the theory of planned behavior. *Information systems research* **2,** 173-191.

Masa'deh, R., Almajali, D., Obeidat, B.Y., Aqqad, N. & Tarhini, A. (2016). The Role of Knowledge Management Infrastructure in Enhancing Job Satisfaction. International Journal of Public Administration

Masa'deh, R., Gharaibeh, A., Tarhini, A. & Obeidat, B. (2015). Knowledge Sharing Capability: A Literature Review. Fourth Scientific & Research Conference on New Trends in Business, Management and Social Sciences, Istanbul, Turkey, 19-20 September 2015 (pp. 1-16).

Masa'deh, R., Tarhini, A., Mohammed, A.B., & Maqableh, M. (2016). Modeling Factors Affecting Student's Usage Behaviour of E-Learning Systems in Lebanon. International Journal of Business and Management, 11 (2), 299-312.

Masa'deh, R., Tayeh, M., Al-Jarrah, I. M., & Tarhini, A. (2015). Accounting vs. Market-based Measures of Firm Performance Related to Information Technology Investments. International Review of Social Sciences and Humanities, 9 (1), 129-145.

Masa'deh, R., Obeidat, B.Y., Al-Dmour, R.H. & Tarhini, A. (2015). Knowledge Management Strategies as Intermediary Variables between IT-Business Strategic Alignment and Firm Performance. European Scientific Journal, 11 (7), 344-368

Masa'deh, R., Tarhini, A., Al-Dmour, R. H. & Obeidat, B. Y. (2015). Strategic IT-Business Alignment as Managers' Explorative and Exploitative Strategies. European Scientific Journal, 11(7), 437-457.

Masa'deh, R., Shannak, R., Maqableh, M. & Tarhini, A. (2016). The Impact of Knowledge Management on Job Performance in Higher Education: The Case of the University of Jordan. Journal of Enterprise and Information Management.

McCarthy, R. (2006). MEASURING STUDENTS PERCEPTIONS OF BLACKBOARD USING THE TECHNOLOGY ACCEPTANCE MODEL: A PLS APPROACH. *E-learning* **26,** 18.

McCombs, B. (2011). Learner-centered practices: Providing the context for positive learner development, motivation, and achievement (Chapter 7). *Handbook of research on schools, schooling, and human development. Mahwah, NJ: Erlbaum*.

McCoy, S. (2002). The effect of national culture dimensions on the acceptance of information technology: A trait based approach. *Unpublished doctoral dissertation, University of Pittsburgh, Pittsburgh, PA. Diffusion of E-Medicine* **259**.

McCoy, S. (2003). Integrating national culture into individual IS adoption research: The need for individual level measures.

McCoy, S., Everard, A. & Jones, B. (2005a). An examination of the technology acceptance model in Uruguay and the U.S.: a focus on culture. *Journal of Global Information Technology* **8,** 27-45.

McCoy, S., Galletta, D. F. & King, W. R. (2005b). Integrating national culture into IS research: the need for current individual level measures. *Communications of the Association for Information Systems* **15,** 12.

McCoy, S., Galletta, D. F. & King, W. R. (2007). Applying TAM across cultures: the need for caution. *European Journal of Information Systems* **16,** 81-90.

McNeil, S. G., Robin, B. R. & Miller, R. M. (2000). Facilitating interaction, communication and collaboration in online courses. *Computers & Geosciences* **26,** 699-708.

Min, Q., Li, Y. & Ji, S. (2009) The Effects of Individual-Level Culture on Mobile Commerce Adoption: An Empirical Study. (pp. 305-312). IEEE.

Mingers, J. (2003). A classification of the philosophical assumptions of management science methods. *Journal of the Operational Research Society***,** 559-570.

Minton, H. L., Schneider, F. W. & Wrightsman, L. S. (1980) *Differential psychology*. Brooks/Cole Publishing Company.

Moon, J.-W. & Kim, Y.-G. (2001). Extending the TAM for a World-Wide-Web context. *Information & Management* **38,** 217-230.

Moore, G. C. & Benbasat, I. (1991). Development of an instrument to measure the perceptions of adopting an information technology innovation. *Information systems research* **2,** 192-222.

Morris, M. G. & Venkatesh, V. (2000). Age differences in technology adoption decisions: Implications for a changing work force. *Personnel psychology* **53,** 375-403.

Morris, M. G., Venkatesh, V. & Ackerman, P. L. (2005). Gender and age differences in employee decisions about new technology: An extension to the theory of planned behavior. *Engineering Management, IEEE Transactions on* **52,** 69-84.

Mumford, M. D. & Licuanan, B. (2004). Leading for innovation: Conclusions, issues, and directions. *The leadership quarterly* **15,** 163-171.

Myers, M. D. & Avison, D. (1997). Qualitative research in information systems. *Management Information Systems Quarterly* **21,** 241-242.

Nachmias, C. & Nachmias, D. (2008) *Research Methods in the Social Sciences*. WORTH PUBL Incorporated.

Nakata, C. (2009) *Beyond Hofstede: Culture frameworks for global marketing and management*. Palgrave Macmillan.

Nargundkar, R. (2003) *Marketing Research-Text & Cases 2E*. Tata McGraw-Hill Education.

Nassar, B., Arzoky, M., Tarhini, A. & Tarhini, T. (2015). An Empirical Analysis of the Seasonal Patterns in Aggregate Directors' Trades. International Journal of Economics and Finance, 7 (9), 59-84.

Nasser, R., Abouchedid, K. (2000). Attitudes and concerns towards distance education: the case of Lebanon. *Online Journal of Distance Learning Administration* **3,** 1-10.

Ndubisi, N. (2006). Factors of online learning adoption: a comparative juxtaposition of the theory of planned behaviour and the technology acceptance model. *International Journal on E-learning* **5,** 571-591.

Neuman, W. L. (2006) *Social research methods: qualitative and quantitative approaches*. Pearson/Allyn and Bacon.

Ngai, E. W. T., Poon, J. K. L. & Chan, Y. H. C. (2007). Empirical examination of the adoption of WebCT using TAM. *Computers & Education* **48,** 250-267.

Norman, P. & Conner, M. (1996). The role of social cognition models in predicting health behaviours: Future directions.

Nunnally, J. C. (1970) *Introduction to psychological measurement*. New York: McGraw-Hill.

Obeidat, B.Y., Al-Suradi, M., Masa'deh, R., & Tarhini, A. (2016). The Impact of Knowledge Management on Innovation: An Empirical Study on Jordanian Consultancy Firms. Management Research Review.

Obeidat, B.Y., Al-Suradi, M., Masa'deh, R., & Tarhini, A. (2016). The Impact of Knowledge Management on Innovation: An Empirical Study on Jordanian Consultancy Firms. Management Research Review, 39 (12).

Ong, C. S. & Lai, J. Y. (2006). Gender differences in perceptions and relationships among dominants of e-learning acceptance. *Computers in Human Behavior* **22,** 816-829.

Onwuegbuzie, A. J. (2002). Positivists, post-positivists, post-structuralists, and post-modernists: Why can't we all get along? Towards a framework for unifying research paradigms. *Education* **122,** 518–530.

Orlikowski, W. & Baroudi, J. J. (1991). Studying information technology in organizations: Research approaches and assumptions.

Orozco, J., Tarhini, A., Masa'deh, R. & Tarhini, T. (2015). A framework of IS/business alignment management practices to improve the design of IT Governance architectures. International Journal of Business and Management, 10(4), 1-12.

Ormrod, J. E. (2010) *Educational psychology : developing learners*. Boston: Pearson/Allyn & Bacon.

Pallant, J. (2010) *SPSS survival manual: A step by step guide to data analysis using SPSS*. Open University Press.

Parboteeah, K. P., Bronson, J. W. & Cullen, J. B. (2005). Does national culture affect willingness to justify ethically suspect behaviors? A focus on the GLOBE national culture scheme. *International Journal of Cross Cultural Management* **5,** 123-138.

Park, S. Y. (2009). An analysis of the technology acceptance model in understanding university students' behavioral intention to use e-learning. *Educational Technology & Society* **12,** 150-162.

Park, S. Y., Nam, M. W. & Cha, S. B. (2012). University students' behavioral intention to use mobile learning: Evaluating the technology acceptance model. *British Journal of Educational Technology*.

Peng, H., Su, Y. J., Chou, C. & Tsai, C. C. (2009). Ubiquitous knowledge construction: Mobile learning re-defined and a conceptual framework. *Innovations in Education and Teaching International* **46,** 171-183.

Pfeil, U., Zaphiris, P. & Ang, C. S. (2006). Cultural differences in collaborative authoring of Wikipedia. *Journal of Computer-Mediated Communication* **12,** 88-113.

Phang, C. W., Sutanto, J., Kankanhalli, A., Li, Y., Tan, B. C. & Teo, H.-H. (2006). Senior citizens' acceptance of information systems: A study in the context of e-government services. *Engineering Management, IEEE Transactions on* **53,** 555-569.

Pituch, K. A. & Lee, Y. (2006). The influence of system characteristics on e-learning use. *Computers & Education* **47,** 222-244.

Poon, W. C. (2007). Users' adoption of e-banking services: the Malaysian perspective. *Journal of Business & Industrial Marketing* **23,** 59-69.

Porter, C. E. & Donthu, N. (2006). Using the technology acceptance model to explain how attitudes determine Internet usage: The role of perceived access barriers and demographics. *Journal of Business Research* **59,** 999-1007.

Presser, S., Couper, M. P., Lessler, J. T., Martin, E., Martin, J., Rothgeb, J. M. & Singer, E. (2004). Methods for testing and evaluating survey questions. *Public opinion quarterly* **68,** 109-130.

Punch, K. F. & Punch, K. (2005) *Introduction to Social Research: Quantitative and Qualitative Approaches*. Sage.

Qingfei, M., Yuping, L. & Shaobo, J. (2009) The Effects of Individual-Level Culture on Mobile Commerce Adoption: An Empirical Study. *Mobile Business, 2009. ICMB 2009. Eighth International Conference on.* (pp. 305-312).

Reid, M. & Levy, Y. (2008). Integrating trust and computer self-efficacy with TAM: An empirical assessment of customers' acceptance of banking information systems (BIS) in Jamaica. *Journal of Internet Banking and Commerce* **12,** 2008-2012.

Richardson, H. & Robinson, B. (2007). The mysterious case of the missing paradigm: a review of critical information systems research 1991–2001. *Information Systems Journal* **17,** 251-270.

Robinson, J. P., Wrightsman, L. S. & Andrews, F. M. (1991) *Measures of personality and social psychological attitudes*. Academic Pr.

Roca, J. C., Chiu, C. M. & Martínez, F. J. (2006). Understanding e-learning continuance intention: An extension of the Technology Acceptance Model. *International Journal of Human-Computer Studies* **64,** 683-696.

Rodriguez, T. & Lozano, P. (2011). The acceptance of Moodle technology by business administration students. *Computers & Education*.

Rogers, E. M. (1995) *Diffusion of innovations*. New York: Simon and Schuster.

Rogers, E. M. (2003) *Diffusion of innovations*. New York: Free Press.

Roscoe, J. T. (1975) *Fundamental research statistics for the behavioral sciences*. Holt, Rinehart and Winston.

Rose, G. & Straub, D. (1998). Predicting general IT use: Applying TAM to the Arabic world. *Journal of Global Information Management (JGIM)* **6,** 39-46.

Russ-Eft, D. & Preskill, H. (2009) *Evaluation in Organizations: A Systematic Approach to Enhancing Learning, Performance, and Change*. Basic Books.

Saadé, R. G. & Galloway, I. (2005). Understanding intention to use multimedia information systems for learning. *Informing Science: International Journal of an Emerging Transdiscipline* **2,** 287-296.

Saeed, K. A. & Abdinnour-Helm, S. (2008). Examining the effects of information system characteristics and perceived usefulness on post adoption usage of information systems. *Information & Management* **45,** 376-386.

Sánchez-Franco, M. J., Martínez-López, F. J. & Martín-Velicia, F. A. (2009). Exploring the impact of individualism and uncertainty avoidance in Web-based electronic learning: An empirical analysis in European higher education. *Computers & Education* **52,** 588-598.

Sarantakos, S. (1993) *Social Research*. Basingstoke: Macmillan.

Saunders, E. J. (2004). Maximizing computer use among the elderly in rural senior centers. *Educational Gerontology* **30,** 573-585.

Saunders, M., Lewis, P. & Thornhill, A. (2009) *Research Methods for Business Students*. Financial Times/Prentice Hall.

Schepers, J. & Wetzels, M. (2007). A meta-analysis of the technology acceptance model: Investigating subjective norm and moderation effects. *Information & Management* **44,** 90-103.

Schreiber, J. B., Nora, A., Stage, F. K., Barlow, E. A. & King, J. (2006). Reporting structural equation modeling and confirmatory factor analysis results: A review. *The Journal of Educational Research* **99,** 323-338.

Schumacher, P. & Morahan-Martin, J. (2001). Gender, Internet and computer attitudes and experiences. *Computers in Human Behavior* **17,** 95-110.

Schumacker, R. E. & Lomax, R. G. (2004) *A beginner's guide to structural equation modeling*. Lawrence Erlbaum.

Schwartz, S. H. (1994). Are there universal aspects in the structure and contents of human values? *Journal of social issues* **50,** 19-45.

Schwartz, S. H. (1999). A theory of cultural values and some implications for work. *Applied psychology* **48,** 23-47.

Sekaran, U. & Bougie, R. (2011) *RESEARCH METHODS FOR BUSINESS : A SKILL BUILDING APPROACH, 5TH ED.* Wiley India Pvt. Ltd.

Sheppard, B. H., Hartwick, J. & Warshaw, P. R. (1988). The theory of reasoned action: A meta-analysis of past research with recommendations for modifications and future research. *Journal of consumer research*, 325-343.

Shih, Y.-Y. & Fang, K. (2004). The use of a decomposed theory of planned behavior to study Internet banking in Taiwan. *Internet Research* **14,** 213-223.

Short, J., Williams, E. & Christie, B. (1976). The social psychology of telecommunications.

Sivo, S. & Brophy, J. (2003). Students' Attitude in a Web-enhanced Hybrid Course: A Structural Equaadon Modeling Inquiry. *Journal of Educational Media & Library Sciences* **41,** 181-194.

Søndergaard, M. (1994). Research note: Hofstede's consequences: a study of reviews, citations and replications. *Organization studies* **15,** 447-456.

Sørnes, J.-O., Stephens, K. K., S_tre, A. S. & Browning, L. D. (2004). The reflexivity between ICTs and business culture: Applying Hofstede's theory to compare Norway and the United States. *Informing Science Journal* **7,** 1–30.

Srite, M. (2006). Culture as an explanation of technology acceptance differences: an empirical investigation of Chinese and US users. *Australasian Journal of Information Systems* **14**.

Srite, M. & karahanna, E. (2000) A Cross-Cultural Model of Technology Acceptance. *the Annual Diffusion of Innovations Group in Information Technology (DIGIT).* Charlotte, NC.

Srite, M. & Karahanna, E. (2006). The role of espoused national cultural values in technology acceptance. *MIS quarterly* **30,** 679-704.

Srite, M. & Karahnana, E. (1999). The influence of national culture on the acceptance and use of information technologies: An empirical study.

Stangor, C. (2010) *Research methods for the behavioral sciences*. Wadsworth Pub Co.

Steven, R. (2001). Is online learning right for you? *American Agent & Broker* **73,** 54.

Stoel, L. & Lee, K. H. (2003). Modeling the effect of experience on student acceptance of Web-based courseware. *Internet Research* **13,** 364-374.

Stoneman, P. (2002) *The economics of technological diffusion*. Blackwell Publishers Oxford,, UK.

Straub, D., Boudreau, M. C. & Gefen, D. (2004). Validation guidelines for IS positivist research. *Communications of the Association for Information Systems* **13,** 380-427.

Straub, D. & Burton-Jones, A. (2007). Veni, vidi, vici: Breaking the TAM logjam. *Journal of the Association for Information Systems* **8,** 223-229.

Straub, D., Keil, M. & Brenner, W. (1997). Testing the technology acceptance model across cultures: A three country study. *Information & Management* **33,** 1-11.

Straub, D., Loch, K., Evaristo, R., Karahanna, E. & Srite, M. (2002). Toward a theory-based measurement of culture. *Human factors in information systems***,** 61-82.

Struab, D., Gefen, D. & Boudreau, C. (2005) "Qunatiative Research" in Research In Information Systems: A Handbook For Research Supervisors And Their Students, , D. Avison and J. Pries-Heje, Elsevier/Butterworth-Heinemann, 221-238.

Šumak, B., Heričko, M. & Pušnik, M. (2011). A meta-analysis of e-learning technology acceptance: The role of user types and e-learning technology types. *Computers in Human Behavior* **27,** 2067-2077.

Sun, H. & Zhang, P. (2006). The role of moderating factors in user technology acceptance. *International Journal of Human-Computer Studies* **64,** 53-78.

Sun, P.-C., Tsai, R. J., Finger, G., Chen, Y.-Y. & Yeh, D. (2008). What drives a successful e-Learning? An empirical investigation of the critical factors influencing learner satisfaction. *Computers & Education* **50,** 1183-1202.

Szajna, B. (1994). Software evaluation and choice: predictive validation of the technology acceptance instrument. *MIS quarterly***,** 319-324.

Tabachnick, B. G. & Fidell, L. S. (2007) *Using Multivariate Statistics* (980). Pearson.

Tan, M. & Teo, T. S. (2000). Factors influencing the adoption of Internet banking. *Journal of the AIS* **1,** 5.

Tarde, G. (1903). The laws of imitation, trans. *EC Parsons, New York*.

Taylor, S. & Todd, P. (1995b). Assessing IT usage: The role of prior experience. *MIS quarterly***,** 561-570.

Taylor, S. & Todd, P. (1995c). Decomposition and crossover effects in the theory of planned behavior: A study of consumer adoption intentions. *International Journal of Research in Marketing* **12,** 137-155.

Taylor, S. & Todd, P. A. (1995a). Understanding information technology usage: A test of competing models. *Information systems research* **6,** 144-176.

Tella, A. (2012). System-related Factors that Predict Students' Satisfaction with the Blackboard Learning System at the University of Botswana. *AFRICAN JOURNAL OF LIBRARY ARCHIVES AND INFORMATION SCIENCE* **22,** 41-52.

Tarhini, A., Hone, K., & Liu, X. (2013). Extending the TAM to Empirically Investigate the Students' Behavioural Intention to Use E-Learning in Developing Countries. Science and Information Conference (IEEE), United Kingdom, 07-09 Oct 2013.

Tarhini, A., Hone, K., Liu, X., & Tarhini, T. (2016). Examining the Moderating Effect of Individual-level Cultural values on Users' Acceptance of E-learning in Developing Countries: A Structural Equation Modeling of an extended Technology Acceptance Model. Interactive Learning Environments, 1-23

Tarhini, A., Scott, M., Sharma, K.S., & Abbasi, M.S. (2015). Differences in Intention to Use Educational RSS Feeds Between Lebanese and British Students: A Multi‑Group Analysis Based on the Technology Acceptance Model. Electronic Journal of e-Learning, 13(1), 14-29.

Tarhini, A., Arachchilage, N. A. G., Masa'deh, R., & Abbasi, M.S. (2015). A Critical Review of Theories and Models of Technology Adoption and Acceptance in Information System Research. International Journal of Technology Diffusion (IJTD), 6 (4), 58-77.

Tarhini, A., Hone, K., & Liu, X. (2015). A cross-cultural examination of the impact of social, organisational and individual factors on educational technology acceptance between British and Lebanese university students. British Journal of Educational Technology. 46 (4), 739-755.

Tarhini, A., Teo, T. & Tarhini, T. (2015). A cross-cultural validity of the E-learning Acceptance Measure (ElAM) in Lebanon and England: A confirmatory factor analysis. Education and Information Technologies. DOI: 10.1007/s10639-015-9381-9

Tarhini, A., Mgbemena, C., AbouTrab, M.S., & Masa'deh, R. (2015). User Adoption of Online Banking in Nigeria: A Qualitative study. Journal of Internet Banking and Commerce, 20(3), 1-8.

Tarhini, A., Fakih, M., Arzoky, M. & Tarhini, T. (2015). Designing Guidelines to Discover Causes of Delays in Construction Projects: The case of Lebanon. International Business Research, 8(6), 73-88.

Tarhini, A., Ammar, H., Tarhini, T. & Masa'deh, R. (2015). Analysis of the Critical Success Factors for Enterprise Resource Planning Implementation from Stakeholders' Perspective: A Systematic Review. International Business Research, 8 (4), 25-40.

Tarhini, A., Hassouna, M., Abbasi, M.S., & Orozco, J. (2015). Towards the Acceptance of RSS to Support Learning: An empirical study to validate the Technology Acceptance Model in Lebanon. Electronic Journal of e-Learning, 13(1), 30-41

Tarhini, A., Hone, K., & Liu, X. (2014). The effects of individual differences on e-learning users' behaviour in developing countries: A structural equation model. Computers in Human Behavior, 41, 153-163

Tarhini, A., Hone, K., & Liu, X. (2014). Measuring the Moderating Effect of Gender and Age on E-Learning Acceptance in England: A Structural Equation Modelling Approach for an Extended Technology Acceptance Model. Journal of Educational Computing Research, 51(2), 163-184

Tarhini, A., Hone, K., & Liu, X. (2013). User Acceptance Towards Web-based Learning Systems: Investigating the role of Social, Organizational and Individual factors in European Higher Education. Procedia Computer Science 17, 189-197.

Tarhini, A., Hone, K., & Liu, X. (2013). Factors Affecting Students' Acceptance of E-learning Environments in Developing Countries: A Structural Equation Modeling Approach, International Journal of Information and Education Technology, 13(1), 54-59.

Terzis, V. & Economides, A. A. (2011). Computer based assessment: Gender differences in perceptions and acceptance. *Computers in Human Behavior* **27,** 2108-2122.

Thatcher, J. B., Srite, M., Stepina, L. P. & Liu, Y. (2003). Culture overload and personal innovativeness with information technology: Extending the nomological net. *Journal of Computer Information Systems* **44,** 74-81.

Trompenaars, F. (1993). Riding the Waves of Culture: Understanding Diversity in Global Business (Irwin, New York, NY).

Tsang, P., Kwan, R. & Fox, R. (2007) *Enhancing learning through technology*. World Scientific.

Turner, P., Turner, S. & Van de Walle, G. (2007). How older people account for their experiences with interactive technology. *Behaviour & Information Technology* **26,** 287-296.

UNDP (2002) United Nations Development Program. *Arab Human Development Report.* New York.

Van Raaij, E. M. & Schepers, J. J. L. (2008). The acceptance and use of a virtual learning environment in China. *Computers & Education* **50,** 838-852.

Venkatesh, V. (2000). Determinants of perceived ease of use: Integrating control, intrinsic motivation, and emotion into the technology acceptance model. *Information systems research* **11,** 342-365.

Venkatesh, V. & Bala, H. (2008). Technology Acceptance Model 3 and a Research Agenda on Interventions. *Decision Sciences* **39,** 273-315.

Venkatesh, V. & Davis, F. D. (1996). A model of the antecedents of perceived ease of use: Development and test*. *Decision Sciences* **27,** 451-481.

Venkatesh, V. & Davis, F. D. (2000). A theoretical extension of the technology acceptance model: Four longitudinal field studies. *Management science*, 186-204.

Venkatesh, V. & Morris, M. G. (2000). Why don't men ever stop to ask for directions? Gender, social influence, and their role in technology acceptance and usage behavior. *MIS quarterly*, 115-139.

Venkatesh, V., Morris, M. G. & Ackerman, P. L. (2000). A longitudinal field investigation of gender differences in individual technology adoption decision-making processes. *Organizational Behavior and Human Decision Processes* **83,** 33-60.

Venkatesh, V., Morris, M. G., Davis, G. B. & Davis, F. D. (2003). User acceptance of information technology: Toward a unified view. *MIS quarterly*, 425-478.

Venkatesh, V., Morris, M. G., Sykes, T. A. & Ackerman, P. L. (2004). Individual reactions to new technologies in the workplace: The role of gender as a psychological construct. *Journal of Applied Social Psychology* **34,** 445-467.

Vijayasarathy, L. R. (2004). Predicting consumer intentions to use on-line shopping: the case for an augmented technology acceptance model. *Information & Management* **41,** 747-762.

Vogt, W. P. (2005) *Dictionary of Statistics & Methodology: A Nontechnical Guide for the Social Sciences*. Sage Publications.

Wagner, E. D. (2008) Minding the Gap: Sustaining eLearning Innovation. *World Conference on E-Learning in Corporate, Government, Healthcare, and Higher Education 2008.* Las Vegas, Nevada, USA, AACE.

Walker, G. & Johnson, N. (2008). Faculty intentions to use components for Web-enhanced instruction. *INTERNATIONAL JOURNAL ON E LEARNING* **7,** 133.

Walsham, G. (1993) *Interpreting Information Systems in Organizations* (286). John Wiley; Sons, Inc.

Walsham, G. (1995). Interpretive case studies in IS research: nature and method. *European Journal of Information Systems* **4,** 74-81.

Wang, Y. S., Wu, M. C. & Wang, H. Y. (2009). Investigating the determinants and age and gender differences in the acceptance of mobile learning. *British Journal of Educational Technology* **40,** 92-118.

Weimer, M. (2013) *Learner-centered teaching: Five key changes to practice*. John Wiley & Sons.

Wilkins, L., Swatman, P. & Holt, D. (2009). Adding value to enterprisewide system integration: A new theoretical framework for assessing technology adoption outcomes. *e-COMMERCE 2009*, 53.

Williams, M. D., Rana, N. P., Dwivedi, Y. K. & Lal, B. (2011) Is UTAUT really used or just cited for the sake of it? a systematic review of citations of UTAUT's originating article. *ECIS.*

Williams, R. (1985) *Keywords: A vocabulary of culture and society*. Oxford University Press, USA.

Winfield, I. (1991) *Organisations and Information Technology: Systems, Power and Job Design*. Boston.

Wu, J., Tsai, R. J., Chen, C. C. & Wu, Y. (2006). An integrative model to predict the continuance use of electronic learning systems: hints for teaching. *International Journal on E-learning* **5**.

Wu, M.-Y., Chou, H.-P., Weng, Y.-C. & Huang, Y.-H. (2008) A Study of Web 2.0 Website Usage Behavior Using TAM 2. *Asia-Pacific Services Computing Conference, 2008. APSCC'08. IEEE.* (pp. 1477-1482). IEEE.

Yadav, R., Sharma, S.K. & Tarhini, A. (2016). A multi-analytical approach to understand and predict the mobile commerce adoption. Journal of Enterprise and Information Management, 29 (2).

Ya-Wen Teng, L. (2009). Collaborating and communicating online: A cross-bordered intercultural project between Taiwan and the US. *Journal of Intercultural Communication*.

YenYuen, Y. & Yeow, P. (2009) User acceptance of internet banking service in Malaysia. *Web information systems and technologies.* (pp. 295-306). Springer.

Yi-Cheng, C., Chun-Yu, C., Yi-Chen, L. & Ron-Chen, Y. (2007). Predicting College Student'Use of E-Learning Systems: an Attempt to Extend Technology Acceptance Model.

Yi, M. Y. & Hwang, Y. (2003). Predicting the use of web-based information systems: self-efficacy, enjoyment, learning goal orientation, and the technology acceptance model. *International Journal of Human-Computer Studies* **59,** 431-449.

Yi, M. Y., Jackson, J. D., Park, J. S. & Probst, J. C. (2006). Understanding information technology acceptance by individual professionals: Toward an integrative view. *Information & Management* **43,** 350-363.

Yin, R. K. (2009) *Case Study Research: Design and Methods*. Sage Publications.

Yousafzai, S. Y., Foxall, G. R. & Pallister, J. G. (2007a). Technology acceptance: a meta-analysis of the TAM: Part 1. *Journal of Modelling in Management* **2,** 251-280.

Yousafzai, S. Y., Foxall, G. R. & Pallister, J. G. (2007b). Technology acceptance: a meta-analysis of the TAM: Part 2. *Journal of Modelling in Management* **2,** 281-304.

Yousafzai, S. Y., Foxall, G. R. & Pallister, J. G. (2010). Explaining internet banking behavior: theory of reasoned action, theory of planned behavior, or technology acceptance model? *Journal of Applied Social Psychology* **40,** 1172-1202.

Yuan, K. H. (2005). Fit indices versus test statistics. *Multivariate Behavioral Research* **40,** 115-148.

Yuen, A. H. K. & Ma, W. W. K. (2008). Exploring teacher acceptance of e-learning technology. *Asia-Pacific Journal of Teacher Education* **36,** 229-243.

Zakaria, Z. (2001) Factors related to information technology implementation in the Malaysian Ministry of Education Polytechnics. Virginia Polytechnic Institute and State University.

Zakour, A. B. (2004) Cultural differences and information technology acceptance. *Proceedings of the 7th Annual Conference of the Southern Association for Information Systems.* (pp. 156-161).

Zhang, S., Zhao, J. & Tan, W. (2008). Extending TAM for online learning systems: An intrinsic motivation perspective. *Tsinghua Science & Technology* **13,** 312-317.

Zikmund, W. (2009) *Business Research Methods*. Texere Publishing Limited.

Zmud, R. W. (1982). Diffusion of modern software practices: influence of centralization and formalization. *Management science* **28,** 1421-1431.

# Appendix A

## Survey Questionnaire

**SECTION A: ABOUT YOURSELF (Please check √ only one option)**

| Gender | Male ☐ | Female ☐ | |
|---|---|---|---|
| Age | 17 – 22 years ☐ | 22>years ☐ | |
| Educational Level | Undergraduate ☐ | Postgraduate ☐ | |
| Web-based experience | Some Experience ☐ | Experienced ☐ | |
| Learner computer and Internet skills | Novice ☐ | Moderate ☐ | Expert ☐ |
| Number of Courses delivered using Web-based for the present year | 1-2 ☐ | 3-5 ☐ | > 5 ☐ |

**SECTION B: PERCEIVED USEFULNESS, PERCEIVED EASE OF USE AND VOLUNTARNESS ABOUT WEB-BASED SYSTEM USAGE**

*Please rate the extent to which you agree with each statement below.* (Please check √ the most appropriate option for each statement below)

| 1= Strongly Disagree 2= Disagree 3= Disagree somewhat 4= Undecided 5= Agree somewhat 6= Agree 7= Strongly Agree | | | | | | | |
|---|---|---|---|---|---|---|---|
| *B1. Perceived Usefulness about using Web-based learning systems* | | | | | | | |
| Using the Web-based learning system will | | | | | | | |
| 7. ...allow me to accomplish learning tasks more quickly | 1 | 2 | 3 | 4 | 5 | 6 | 7 |
| 8. ...improve my learning performance | 1 | 2 | 3 | 4 | 5 | 6 | 7 |
| 9. ...make it easier to learn course content | 1 | 2 | 3 | 4 | 5 | 6 | 7 |
| 10. ...increase my learning productivity | 1 | 2 | 3 | 4 | 5 | 6 | 7 |
| 11. ...enhance my effectiveness in learning | 1 | 2 | 3 | 4 | 5 | 6 | 7 |
| *B2. Perceived Ease Of Use about using the Web-based learning system* | | | | | | | |
| 12. Learning to operate the Web-based learning system is easy for me | 1 | 2 | 3 | 4 | 5 | 6 | 7 |
| 13. I find it easy to get the Web-based learning system to do what I want it to do | 1 | 2 | 3 | 4 | 5 | 6 | 7 |
| 14. My interaction with Web-based learning system is clear and understandable | 1 | 2 | 3 | 4 | 5 | 6 | 7 |
| 15. It is easy for me to become skillful at using the Web-based learning system | 1 | 2 | 3 | 4 | 5 | 6 | 7 |
| 16. I find the Web-based learning system easy to use | 1 | 2 | 3 | 4 | 5 | 6 | 7 |
| *B3. Voluntariness about using Web-based learning system* | | | | | | | |
| 17. My use of the Web-based system is voluntary | 1 | 2 | 3 | 4 | 5 | 6 | 7 |
| 18. My instructor does not require me to use the Web-based system | 1 | 2 | 3 | 4 | 5 | 6 | 7 |
| 19. Although it might be helpful, using Web-based learning system is certainly not compulsory in my learning | 1 | 2 | 3 | 4 | 5 | 6 | 7 |

**SECTION C: SELF-EFFICACY, FACILITATING CONDITIONS, SUBJECTIVE NORMS AND PERCEIVED QUALITY OF WORKING LIFE TOWARD WEB-BASED USAGE**

*Please rate the extent to which you agree with each statement below.* (Please check √ only one option for each statement below)

| 1= Strongly Disagree 2= Disagree 3= Disagree somewhat 4= Undecided 5= Agree somewhat 6= Agree 7= Strongly Agree | | | | | | | |
|---|---|---|---|---|---|---|---|
| **C1. Self-efficacy (SE) about the Web-based usage** | | | | | | | |
| I am confident in using the Web-based learning system: | | | | | | | |
| 20. Even if there is no one around to show me how to do it | 1 | 2 | 3 | 4 | 5 | 6 | 7 |
| 21. Even if I have only the online instructions for reference | 1 | 2 | 3 | 4 | 5 | 6 | 7 |
| 22. Even if I have never used such a system before | 1 | 2 | 3 | 4 | 5 | 6 | 7 |
| 23. As long as I have just seen someone using it before trying it myself | 1 | 2 | 3 | 4 | 5 | 6 | 7 |
| 24. As long as I have a lot of time to complete the job for which the software is provided | 1 | 2 | 3 | 4 | 5 | 6 | 7 |
| 25. As long as someone shows me how to do it | 1 | 2 | 3 | 4 | 5 | 6 | 7 |
| **C2. Facilitating Conditions (FC) about Using the Web-based learning system** | | | | | | | |
| 26. I have the knowledge necessary to use the Web-based system | 1 | 2 | 3 | 4 | 5 | 6 | 7 |
| 27. I have the resources necessary to use the Web-based leaning system | 1 | 2 | 3 | 4 | 5 | 6 | 7 |
| 28. When I needed help to use the Web-based system, Guidance was available to me | 1 | 2 | 3 | 4 | 5 | 6 | 7 |
| 29. A specific person was available to provide assistance | 1 | 2 | 3 | 4 | 5 | 6 | 7 |
| **C3. Subjective Norm (SN) about using Web-based learning systems** | | | | | | | |
| 30. My Instructors thinks that I should participate in the Web-based learning activities | 1 | 2 | 3 | 4 | 5 | 6 | 7 |
| 31. Other students think that I should participate in the Web-based learning activities | 1 | 2 | 3 | 4 | 5 | 6 | 7 |
| 32. Management of my university thinks that I should use the Web-based leaning activities | 1 | 2 | 3 | 4 | 5 | 6 | 7 |
| 33. Generally speaking, I would do what my instructor thinks I should do | 1 | 2 | 3 | 4 | 5 | 6 | 7 |
| **C4. Perceived Quality of work life** | | | | | | | |
| 34. The freedom to get the course information any time of the day will helps me to have more time for a creative thinking and leisure. | 1 | 2 | 3 | 4 | 5 | 6 | 7 |
| 35. Using the free resources such as web-based learning system and e-libraries helped me to save money and effort. | 1 | 2 | 3 | 4 | 5 | 6 | 7 |
| 36. Using the Web-based learning system provide more opportunities to participate in the class | 1 | 2 | 3 | 4 | 5 | 6 | 7 |
| 37. Using emails to communicate with other student groups help me to save my expense and effort. | 1 | 2 | 3 | 4 | 5 | 6 | 7 |
| 38. Overall, using the Web-based learning help improving my quality of working life. | 1 | 2 | 3 | 4 | 5 | 6 | 7 |

*Please rate the extent to which you agree with each statement below.* (Please check √ only one option for each statement below)

| 1= Strongly Disagree 2= Disagree 3= Disagree somewhat 4= Undecided 5= Agree somewhat 6= Agree 7= Strongly Agree | | | | | | | |
|---|---|---|---|---|---|---|---|
| *Behaviour Intention (BI) to use the Web-based learning system* | | | | | | | |
| 39. Given the chance, I intend to use the Web-based learning system to do different things, from downloading lecture notes and participating in chat rooms to learning on the Web | 1 | 2 | 3 | 4 | 5 | 6 | 7 |
| 40. I predict I would use web-based learning system in the next semester | 1 | 2 | 3 | 4 | 5 | 6 | 7 |
| 41. In general, I plan to use Web-based learning system frequently for my coursework and other activities in the next semester | 1 | 2 | 3 | 4 | 5 | 6 | 7 |

### SECTION E: ACTUAL USAGE of the Web-based learning system

42. On average, how **frequently** do you use the Web-based system?

| | | |
|---|---|---|
| (1) Less than once a month | (2) once a month | (3) a few times a month |
| (4) a few times a week | (5) about once a day | (6) several times a day |

43. On the average working **day**, how much time do you spend on the Web-based learning system?

| | | |
|---|---|---|
| (1) Almost never | (2) less than 30 min | (3) from 30 min to 1 h |
| (4) From 1 to 2 h | (5) from 2 to 3 h | (6) more than 3 h |

44. Please indicate the extent to which you use the Web-based learning system to perform the following tasks.

| | Not at all | To a small extent | To some extent | To a moderate extent | To a great extent |
|---|---|---|---|---|---|
| Lecture Notes | 1 | 2 | 3 | 4 | 5 |
| Announcements | 1 | 2 | 3 | 4 | 5 |
| Email | 1 | 2 | 3 | 4 | 5 |
| Assessments | 1 | 2 | 3 | 4 | 5 |
| Course Handbook | 1 | 2 | 3 | 4 | 5 |
| Discussion board | 1 | 2 | 3 | 4 | 5 |
| Websites | 1 | 2 | 3 | 4 | 5 |
| Take quizzes | 1 | 2 | 3 | 4 | 5 |
| Past Papers | 1 | 2 | 3 | 4 | 5 |

### SECTION F: POWER DISTANCE, MASCULINITY/FEMININITY, INDIVIDUALISM/ COLLECTIVISM AND UNCERTAINTY AVOIDANCE

*Please rate the extent to which you agree with each statement below.* (Please check √ only one option for each statement below)

### SECTION F: POWER DISTANCE, MASCULINITY/FEMININITY, INDIVIDUALISM/ COLLECTIVISM AND UNCERTAINTY AVOIDANCE

*Please rate the extent to which you agree with each statement below.* (Please check √ only one option for each statement below)

1= Strongly Disagree 2= Disagree 3= Disagree somewhat 4= Undecided 5= Agree somewhat 6= Agree 7= Strongly Agree

| | *F1. Power Distance (PD)* | | | | | | | |
|---|---|---|---|---|---|---|---|---|
| 45. | Instructors should make most decisions without consulting the students. | 1 | 2 | 3 | 4 | 5 | 6 | 7 |
| 46. | Instructors should not ask the opinions of people in lower of students too frequently | 1 | 2 | 3 | 4 | 5 | 6 | 7 |
| 47. | Instructors should avoid social interaction with students. | 1 | 2 | 3 | 4 | 5 | 6 | 7 |
| 48. | Instructors should not delegate important tasks students. | 1 | 2 | 3 | 4 | 5 | 6 | 7 |
| 49. | students should not disagree with decisions made by their Instructors and management | 1 | 2 | 3 | 4 | 5 | 6 | 7 |
| 50. | It is frequently necessary for a Instructors to use authority and power when dealing with students | 1 | 2 | 3 | 4 | 5 | 6 | 7 |
| | *F2. Masculinity/Femininity (MF)* | | | | | | | |
| 51. | It is preferable to have a male in high level position rather than a female | 1 | 2 | 3 | 4 | 5 | 6 | 7 |
| 52. | There are some majors in which a male student can always do better than a female student | 1 | 2 | 3 | 4 | 5 | 6 | 7 |
| 53. | It is more important for male student to have a professional career than it is for a female student to have a professional career | 1 | 2 | 3 | 4 | 5 | 6 | 7 |
| 54. | Female student do not value outstanding academic achievement in their studies as much as male student | 1 | 2 | 3 | 4 | 5 | 6 | 7 |
| 55. | Male students usually solve problems with logical analysis, female students usually solve problems with intuition | 1 | 2 | 3 | 4 | 5 | 6 | 7 |
| 56. | Male Students are more determined and competitive focusing on achievement and material success while female students are modest and humble focusing on relationships | 1 | 2 | 3 | 4 | 5 | 6 | 7 |
| | *F3. Individualism/Collectivism (IC)* | | | | | | | |
| 57. | Individuals should sacrifice self-interest for the group that they belong to | 1 | 2 | 3 | 4 | 5 | 6 | 7 |
| 58. | Individuals should stick with the group even through difficulties | 1 | 2 | 3 | 4 | 5 | 6 | 7 |
| 59. | Group welfare is more important than individual rewards | 1 | 2 | 3 | 4 | 5 | 6 | 7 |
| 60. | Group success is more important than individual success | 1 | 2 | 3 | 4 | 5 | 6 | 7 |
| 61. | Group loyalty should be encouraged even if individual goals suffer | 1 | 2 | 3 | 4 | 5 | 6 | 7 |
| 62. | Being accepted as a member of a group is more important than having autonomy and independence | 1 | 2 | 3 | 4 | 5 | 6 | 7 |
| | *F4. Uncertainty Avoidance (UA)* | | | | | | | |
| 63. | Rules and regulations are important because they inform students what the university is expected of them | 1 | 2 | 3 | 4 | 5 | 6 | 7 |
| 64. | It is important to have specific requirements and instructions spelled out in detail so that I always know what I am expected to do | 1 | 2 | 3 | 4 | 5 | 6 | 7 |
| 65. | It is important to closely follow instructions and procedures related to their Learning | 1 | 2 | 3 | 4 | 5 | 6 | 7 |
| 66. | Standardized work procedures are helpful for my learning | 1 | 2 | 3 | 4 | 5 | 6 | 7 |

# Appendix B

## Cover Letter

**Brunel University West London**
**Department of Information System and Computing**

Dear Participant,

I am a PhD research student at Brunel University West London, under the supervision of Professor Xiaohui Liu, Deputy Head of School (Research) for department of Information System, Computing and Mathematics. The research title is:
**The Effects of Individual-level Culture and Demographic Characteristics on E-learning Acceptance in Lebanon and England: A Structural Equation Modeling Approach**

The main aim of this study is to investigate and understand how cultural and individual differences affect students' perceptions and behaviours when using web-based learning systems in order to generate a model that determines the acceptance of technology in developing and developed worlds. This research will help to better understand the characteristics of students in the Arab worlds and England respectively. The outcome of the research should help policy makers, educators and experts to understand "what" the students expect from the learning management systems, and system for developers to understand "how" they could improve their learning management systems in the concerned cultural contexts and overcome problems that may occur during cross-cultural educational cooperation and e-learning implementation.

The questionnaire consists of five parts. The first part collects data about the participant's general information. The second and third parts assess the perceptions of students about web-based learning system. The fourth part measures the actual usage of the web-based learning system, while the last part measures the cultural factors about the participants. The questionnaire will take approximately 10 to 12 minutes of your time. Your participation is voluntary. If you do not wish to participate, simply discard the questionnaire at any time. All your information including your name will be kept completely anonymous and will be used for the purpose of this PhD research and destroyed after two years.

If you have any questions or concerns, please contact me Ali.Tarhini@brunel.ac.uk or my supervisor Xiaohui.Liu@brunel.ac.uk. If you have any concerns or complaints regarding the ethical elements of this project please contact siscm.srec@brunel.ac.uk or Professor Zidong Wang, Tel. No. 0044 (1) 895 266021.

# Appendix C

## Missing Data

| | N | Mean | Std. Deviation | Missing Data (England) | |
|---|---|---|---|---|---|
| | | | | Frequency | Percent |
| **Gender** | 600 | 1.48 | .500 | 4 | .5 |
| **Age** | 599 | 2.29 | 1.007 | 5 | .8 |
| **EdLevel** | 600 | 1.35 | .478 | 4 | .5 |
| **WebExp** | 601 | 2.23 | .707 | 2 | .3 |
| **CompExp** | 600 | 2.22 | .698 | 4 | .5 |
| **NOfCourses** | 600 | 1.78 | .759 | 3 | .5 |
| **PU1** | 599 | 5.23 | 1.354 | 5 | .8 |
| **PU2** | 601 | 5.16 | 1.305 | 3 | .5 |
| **PU3** | 600 | 5.35 | 1.313 | 4 | .5 |
| **PU4** | 598 | 5.23 | 1.333 | 6 | 1.0 |
| **PU5** | 595 | 5.24 | 1.342 | 9 | 1.5 |
| **PEU1** | 601 | 5.40 | 1.399 | 3 | .5 |
| **PEU2** | 599 | 5.32 | 1.289 | 5 | .8 |
| **PEU3** | 601 | 5.35 | 1.260 | 3 | .5 |
| **PEU4** | 602 | 5.39 | 1.279 | 2 | .3 |
| **PEU5** | 599 | 5.42 | 1.348 | 5 | .8 |
| **SE1** | 594 | 5.48 | 1.455 | 10 | 1.7 |
| **SE2** | 600 | 5.48 | 1.289 | 4 | .7 |
| **SE3** | 599 | 5.02 | 1.497 | 5 | .8 |
| **SE4** | 600 | 4.94 | 1.567 | 4 | .7 |
| **SE5** | 599 | 5.04 | 1.479 | 5 | .8 |
| **SE6** | 598 | 4.98 | 1.703 | 6 | 1.0 |
| **FC1** | 596 | 5.51 | 1.321 | 8 | 1.3 |
| **FC2** | 599 | 5.49 | 1.301 | 5 | .8 |
| **FC3** | 599 | 4.99 | 1.440 | 5 | .8 |
| **FC4** | 599 | 4.66 | 1.653 | 5 | .8 |
| **SN1** | 595 | 4.74 | 1.529 | 9 | 1.5 |
| **SN2** | 597 | 4.42 | 1.496 | 7 | 1.2 |
| **SN3** | 598 | 4.95 | 1.410 | 6 | 1.0 |
| **SN4** | 592 | 5314 | 1.387 | 12 | 2.0 |
| **QWL1** | 596 | 5.64 | 1.345 | 8 | 1.3 |
| **QWL2** | 595 | 5.68 | 1.343 | 9 | 1.5 |
| **QWL3** | 598 | 5.27 | 1.361 | 6 | 1.0 |
| **QWL4** | 598 | 5.55 | 1.329 | 6 | 1.0 |
| **QWL5** | 596 | 5.65 | 1.295 | 8 | 1.3 |
| **BI1** | 600 | 5.50 | 1.367 | 4 | .7 |
| **BI2** | 597 | 5.77 | 1.265 | 7 | 1.2 |
| **BI3** | 594 | 5.78 | 1.291 | 10 | 1.7 |
| **FreqUsage** | 595 | 4.85 | 1.128 | 9 | 1.5 |
| **DailyUsage** | 594 | 4.02 | 1.298 | 10 | 1.7 |
| **LecNotes** | 595 | 4.30 | .982 | 9 | 1.5 |
| **Announ** | 599 | 3.50 | 1.297 | 5 | .8 |
| **Email** | 596 | 4.00 | 1.199 | 8 | 1.3 |
| **Assess** | 593 | 4.27 | 1.011 | 11 | 1.8 |
| **Handbook** | 596 | 3.47 | 1.331 | 8 | 1.3 |
| **DiscBoard** | 601 | 2.76 | 1.397 | 3 | .5 |
| **Websites** | 597 | 3.36 | 1.361 | 7 | 1.2 |
| **Quizzes** | 599 | 2.83 | 1.505 | 5 | .8 |
| **PastPapers** | 597 | 3.66 | 1.362 | 7 | 1.2 |
| **PD1** | 599 | 2.77 | 1.541 | 5 | .8 |
| **PD2** | 600 | 2.58 | 1.420 | 4 | .7 |
| **PD3** | 594 | 2.23 | 1.357 | 10 | 1.7 |
| **PD4** | 595 | 2.77 | 1.443 | 9 | 1.5 |
| **PD5** | 599 | 2.65 | 1.411 | 5 | .8 |
| **PD6** | 597 | 2.83 | 1.296 | 7 | 1.2 |
| **MF1** | 593 | 2.30 | 1.481 | 11 | 1.8 |
| **MF2** | 596 | 2.52 | 1.589 | 8 | 1.3 |
| **MF3** | 597 | 2.17 | 1.493 | 7 | 1.2 |
| **MF4** | 593 | 2.14 | 1.547 | 11 | 1.8 |
| **MF5** | 594 | 2.57 | 1.600 | 10 | 1.7 |
| **MF6** | 596 | 2.49 | 1.633 | 8 | 1.3 |
| **IC1** | 600 | 3.59 | 1.538 | 4 | .7 |
| **IC2** | 593 | 4.24 | 1.559 | 11 | 1.8 |
| **IC3** | 599 | 4.04 | 1.398 | 5 | .8 |
| **IC4** | 593 | 3.99 | 1.496 | 11 | 1.8 |
| **IC5** | 598 | 3.84 | 1.540 | 6 | 1.0 |
| **IC6** | 598 | 3.66 | 1.540 | 6 | 1.0 |
| **UA1** | 597 | 4.35 | 1.302 | 7 | 1.2 |
| **UA2** | 597 | 4.18 | 1.218 | 7 | 1.2 |
| **UA3** | 596 | 4.30 | 1.348 | 8 | 1.3 |
| **UA4** | 589 | 4.29 | 1.381 | 15 | 2.5 |
| **UA5** | 593 | 4.27 | 1.459 | 11 | 1.8 |

**Table 1: British Sample Univariate missing data (individual-level)**

| | N | Mean | Std. Deviation | Missing Data (Lebanon) | |
|---|---|---|---|---|---|
| | | | | Frequency | Percent |
| **Gender** | 589 | 1.47 | .499 | 5 | .8 |
| **Age** | 589 | 2.06 | .904 | 5 | .8 |
| **EdLevel** | 587 | 1.36 | .480 | 7 | 1.2 |
| **WebExp** | 588 | 2.38 | .686 | 6 | 1.0 |
| **CompExp** | 589 | 2.23 | .704 | 5 | .8 |
| **NOfCourses** | 585 | 1.79 | .724 | 9 | 1.5 |
| **PU1** | 580 | 5.33 | 1.281 | 14 | 2.4 |
| **PU2** | 583 | 4.94 | 1.303 | 11 | 1.9 |
| **PU3** | 588 | 5.27 | 1.350 | 6 | 1.0 |
| **PU4** | 586 | 4.94 | 1.399 | 8 | 1.3 |
| **PU5** | 580 | 4.91 | 1.414 | 14 | 2.4 |
| **PEU1** | 583 | 5.80 | 1.261 | 11 | 1.9 |
| **PEU2** | 587 | 5.52 | 1.299 | 7 | 1.2 |
| **PEU3** | 583 | 5.57 | 1.261 | 11 | 1.9 |
| **PEU4** | 589 | 5.67 | 1.206 | 5 | .8 |
| **PEU5** | 588 | 5.76 | 1.172 | 6 | 1.0 |
| **SE1** | 586 | 5.52 | 1.390 | 8 | 1.3 |
| **SE2** | 587 | 5.48 | 1.307 | 7 | 1.2 |
| **SE3** | 588 | 5.07 | 1.420 | 6 | 1.0 |
| **SE4** | 589 | 5.71 | 1.034 | 5 | .8 |
| **SE5** | 588 | 4.97 | 1.467 | 6 | 1.0 |
| **SE6** | 588 | 4.82 | 1.750 | 6 | 1.0 |
| **FC1** | 588 | 5.62 | 1.246 | 6 | 1.0 |
| **FC2** | 587 | 5.49 | 1.255 | 7 | 1.2 |
| **FC3** | 583 | 5.69 | .974 | 11 | 1.9 |
| **FC4** | 584 | 5.61 | .994 | 10 | 1.7 |
| **SN1** | 585 | 4.83 | 1.608 | 9 | 1.5 |
| **SN2** | 587 | 4.28 | 1.507 | 7 | 1.2 |
| **SN3** | 585 | 5.16 | 1.363 | 9 | 1.5 |
| **SN4** | 585 | 5.49 | 1.121 | 9 | 1.5 |
| **QWL1** | 588 | 5.49 | 1.323 | 6 | 1.0 |
| **QWL2** | 588 | 5.51 | 1.295 | 6 | 1.0 |
| **QWL3** | 587 | 4.91 | 1.419 | 7 | 1.2 |
| **QWL4** | 584 | 5.48 | 1.333 | 10 | 1.7 |
| **QWL5** | 583 | 5.49 | 1.153 | 11 | 1.9 |
| **BI1** | 585 | 5.37 | 1.366 | 9 | 1.5 |
| **BI2** | 582 | 5.80 | 1.291 | 12 | 2.0 |
| **BI3** | 586 | 5.64 | 1.323 | 8 | 1.3 |
| **FreqUsage** | 583 | 4.90 | 1.104 | 11 | 1.9 |
| **DailyUsage** | 583 | 3.47 | 1.277 | 11 | 1.9 |
| **LecNotes** | 585 | 4.22 | 1.091 | 9 | 1.5 |
| **Announ** | 585 | 3.93 | 1.149 | 9 | 1.5 |
| **Email** | 580 | 3.47 | 1.335 | 14 | 2.4 |
| **Assess** | 580 | 3.47 | 1.274 | 14 | 2.4 |
| **Handbook** | 588 | 3.45 | 1.292 | 6 | 1.0 |
| **DiscBoard** | 586 | 2.67 | 1.393 | 8 | 1.3 |
| **Websites** | 579 | 3.09 | 1.397 | 15 | 2.5 |
| **Quizzes** | 585 | 3.30 | 1.296 | 9 | 1.5 |
| **PastPapers** | 585 | 3.44 | 1.366 | 9 | 1.5 |
| **PD1** | 585 | 3.16 | 1.740 | 9 | 1.5 |
| **PD2** | 588 | 3.26 | 1.696 | 6 | 1.0 |
| **PD3** | 587 | 3.12 | 1.603 | 7 | 1.2 |
| **PD4** | 586 | 3.33 | 1.642 | 8 | 1.3 |
| **PD5** | 582 | 3.32 | 1.661 | 12 | 2.0 |
| **PD6** | 580 | 3.86 | 1.577 | 14 | 2.4 |
| **MF1** | 586 | 3.28 | 1.759 | 8 | 1.3 |
| **MF2** | 584 | 3.71 | 1.819 | 10 | 1.7 |
| **MF3** | 585 | 3.38 | 1.779 | 9 | 1.5 |
| **MF4** | 585 | 3.19 | 1.700 | 13 | 2.2 |
| **MF5** | 585 | 3.38 | 1.698 | 9 | 1.5 |
| **MF6** | 586 | 3.32 | 1.663 | 8 | 1.3 |
| **IC1** | 587 | 4.73 | 1.440 | 7 | 1.2 |
| **IC2** | 583 | 5.21 | 1.245 | 11 | 1.9 |
| **IC3** | 584 | 4.99 | 1.300 | 10 | 1.7 |
| **IC4** | 585 | 5.00 | 1.325 | 9 | 1.5 |
| **IC5** | 586 | 5.06 | 1.369 | 8 | 1.3 |
| **IC6** | 588 | 4.61 | 1.498 | 6 | 1.0 |
| **UA1** | 589 | 5.44 | 1.200 | 5 | .8 |
| **UA2** | 588 | 5.47 | 1.259 | 6 | 1.0 |
| **UA3** | 584 | 5.31 | 1.220 | 10 | 1.7 |
| **UA4** | 582 | 5.31 | 1.225 | 12 | 2.0 |
| **UA5** | 584 | 5.46 | 1.198 | 10 | 1.7 |

**Table 2: Lebanese Sample Univariate missing data (individual-level)**

# Appendix D

## Outliers

### *Univariate outliers*

| Variable | Minimum | Maximum | Outlier Case Number |
|----------|---------|---------|---------------------|
| **PU1** | -3.13383 | 1.30944 | No case |
| **PU2** | -3.17424 | 1.41063 | No case |
| **PU3** | -3.32012 | 1.25831 | 838-1111 |
| **PU4** | -3.16435 | 1.33981 | No case |
| **PU5** | -3.11268 | 1.30579 | No case |
| **PEU1** | -3.14438 | 1.14761 | No case |
| **PEU2** | -3.15942 | 1.30516 | No case |
| **PEU3** | -3.45838 | 1.31434 | 803-1197 |
| **PEU4** | -3.42964 | 1.25878 | 803-1197 |
| **PEU5** | -3.29416 | 1.17367 | No case |
| **SE1** | -2.52576 | 1.21488 | No case |
| **SE2** | -2.76579 | 1.31533 | No case |
| **SE3** | -2.40962 | 1.43558 | No case |
| **SE4** | -2.50572 | 1.32905 | No case |
| **SE5** | -2.75217 | 1.33716 | No case |
| **SE6** | -2.35570 | 1.19401 | No case |
| **FC1** | -2.75016 | 1.10240 | No case |
| **FC2** | -2.79183 | 1.12062 | No case |
| **FC3** | -2.50757 | 1.32679 | No case |
| **FC4** | -2.10617 | 1.33977 | No case |
| **SN1** | -2.47619 | 1.48746 | No case |
| **SN2** | -2.35888 | 1.55531 | No case |
| **SN3** | -2.83278 | 1.46801 | No case |
| **SN4** | -2.99391 | 1.34780 | No case |
| **QWL1** | -3.21033 | 1.01309 | No case |
| **QWL2** | -3.50023 | .98521 | 616-751-803-1004 |
| **QWL3** | -3.14917 | 1.26576 | No case |
| **QWL4** | -3.19169 | 1.08771 | No case |
| **QWL5** | -3.55592 | 1.04153 | 751-757 |
| **BI1** | -3.29679 | 1.10379 | No case |
| **BI2** | -2.94906 | .97263 | No case |
| **BI3** | -2.84205 | .94568 | No case |
| **FreqUsage** | -3.36973 | 1.02620 | 637-1069-1197 |
| **DailyUsage** | -2.33257 | 1.54435 | No case |
| **PD1** | -1.24969 | 3.00019 | No case |
| **PD2** | -1.23364 | 2.69630 | No case |
| **PD3** | -1.13212 | 2.91883 | No case |
| **PD4** | -1.27889 | 2.71930 | No case |
| **PD5** | -1.27036 | 2.80234 | No case |
| **PD6** | -1.34659 | 2.76534 | No case |
| **MF1** | -1.15562 | 2.91669 | No case |
| **MF2** | -1.23394 | 2.83873 | No case |
| **MF3** | -1.18979 | 2.70736 | No case |
| **MF4** | -1.12426 | 2.81000 | No case |
| **MF5** | -1.28492 | 3.22287 | No case |
| **MF6** | -1.22312 | 3.29072 | 1183 |
| **IC1** | -1.54729 | 4.41913 | 1121 (Data Entry: 14 instead of 4) |
| **IC2** | -1.82962 | 1.97884 | No case |
| **IC3** | -1.91304 | 2.32206 | No case |
| **IC4** | -1.77593 | 2.24153 | No case |
| **IC5** | -1.76677 | 2.40714 | No case |
| **IC6** | -1.74761 | 2.44527 | No case |
| **UA1** | -2.30847 | 2.15334 | No case |
| **UA2** | -2.36260 | 2.24314 | No case |
| **UA3** | -2.27492 | 2.12439 | No case |
| **UA4** | -2.34087 | 2.15717 | No case |
| **UA5** | -2.19354 | 2.00138 | No case |

**Table 3: British sample Univariate outliers**

| Variable | Minimum | Maximum | Outlier Case Number |
|---|---|---|---|
| **PU1** | -3.55967 | 1.29947 | 57 |
| **PU2** | -3.23756 | 1.56185 | No case |
| **PU3** | -3.34091 | 1.27311 | 10 |
| **PU4** | -3.06897 | 1.46869 | No case |
| **PU5** | -2.99905 | 1.49163 | No case |
| **PEU1** | -3.12839 | .99211 | No case |
| **PEU2** | -2.84488 | 1.17782 | No case |
| **PEU3** | -2.97841 | 1.18142 | No case |
| **PEU4** | -3.10110 | 1.13444 | No case |
| **PEU5** | -3.16400 | 1.07603 | No case |
| **SE1** | -2.24074 | 1.13223 | No case |
| **SE2** | -2.33348 | 1.19402 | No case |
| **SE3** | -2.73085 | 1.39808 | No case |
| **SE4** | -4.53976 | 1.26936 | 169-222-313 |
| **SE5** | -2.73101 | 1.41284 | No case |
| **SE6** | -2.26729 | 1.26559 | No case |
| **FC1** | -2.55564 | 1.13325 | No case |
| **FC2** | -2.48722 | 1.21757 | No case |
| **FC3** | -3.93611 | 1.29875 | 107 |
| **FC4** | -2.96300 | 1.36216 | No case |
| **SN1** | -2.52498 | 1.39333 | No case |
| **SN2** | -2.36507 | 1.84251 | No case |
| **SN3** | -3.25863 | 1.38257 | No case |
| **SN4** | -4.25933 | 1.41867 | 213-195-143-222 |
| **QWL1** | -2.68959 | 1.14402 | No case |
| **QWL2** | -2.75537 | 1.15242 | No case |
| **QWL3** | -2.81427 | 1.47881 | No case |
| **QWL4** | -2.65317 | 1.15716 | No case |
| **QWL5** | -3.92402 | 1.29679 | 77 |
| **BI1** | -3.37989 | 1.22280 | 143 |
| **BI2** | -3.10777 | .94987 | No case |
| **BI3** | -3.63546 | 1.04811 | 143 |
| **FreqUsage** | -3.66245 | .98711 | 77-523 |
| **DailyUsage** | -1.95989 | 1.96403 | No case |
| **PD1** | -1.37616 | 2.52278 | No case |
| **PD2** | -1.50365 | 2.11656 | No case |
| **PD3** | -1.40661 | 2.42287 | No case |
| **PD4** | -1.35977 | 2.60458 | No case |
| **PD5** | -1.49777 | 2.61266 | No case |
| **PD6** | -1.49076 | 2.67752 | No case |
| **MF1** | -1.27915 | 2.30920 | No case |
| **MF2** | -1.32061 | 2.33301 | No case |
| **MF3** | -1.29828 | 2.47071 | No case |
| **MF4** | -1.38877 | 2.37116 | No case |
| **MF5** | -1.38342 | 2.37095 | No case |
| **MF6** | -1.69228 | 2.13384 | No case |
| **IC1** | -2.65210 | 1.60272 | No case |
| **IC2** | -3.45662 | 1.47274 | 497-541 |
| **IC3** | -3.09797 | 1.56127 | No case |
| **IC4** | -3.03288 | 1.51644 | No case |
| **IC5** | -2.94922 | 1.43799 | No case |
| **IC6** | -2.43872 | 1.61079 | No case |
| **UA1** | -3.77632 | 1.30742 | 63-225-455 |
| **UA2** | -3.57095 | 1.22308 | 225-248-352-455 |
| **UA3** | -3.00140 | 1.39712 | No case |
| **UA4** | -3.70107 | 1.42487 | 225-353 |
| **UA5** | -3.75738 | 1.30086 | 225 |

**Table 4: Lebanese sample Univariate outliers**

*Multivariate Outliers*

| Observation number | Mahalanobis d-squared | P1 | P2 |
|---|---|---|---|
| 544 | 111.728 | .000 | .000 |
| 17 | 105.275 | .000 | .000 |
| 156 | 103.695 | .003 | .000 |
| 108 | 103.470 | .006 | .000 |
| 188 | 97.280 | .008 | .000 |
| 122 | 96.959 | .011 | .000 |
| 406 | 90.396 | 0.16 | .000 |
| 157 | 86.905 | 0.26 | .000 |
| 405 | 86.005 | 0.37 | .000 |
| 327 | 84.484 | 0.41 | .000 |
| 80 | 80.853 | 0.41 | .000 |
| 268 | 78.774 | 0.48 | .000 |

**Table 5: British sample Multivariate outliers**

| Observation number | Mahalanobis d-squared | P1 | P2 |
|---|---|---|---|
| 390 | 89.484 | .000 | .000 |
| 557 | 81.739 | .000 | .000 |
| 204 | 80.930 | 0.001 | .000 |
| 371 | 76.709 | 0.002 | .000 |
| 333 | 71.080 | 0.004 | .000 |
| 339 | 69.625 | 0.004 | .000 |
| 263 | 69.176 | 0.007 | .000 |
| 114 | 69.122 | 0.009 | .000 |
| 285 | 68.797 | 0.013 | .000 |
| 106 | 67.810 | 0.017 | .000 |
| 167 | 67.525 | 0.019 | .000 |
| 2 | 66.647 | 0.022 | .000 |
| 207 | 66.601 | 0.026 | .000 |
| 127 | 65.179 | 0.290 | .000 |
| 500 | 65.036 | 0.030 | .000 |

**Table 6: Lebanese sample Multivariate outliers**

# Appendix E

## Normality

| Items | Mean | Std. Deviation | Skewness | Kurtosis |
|-------|------|----------------|----------|----------|
| PU1 | 5.25 | 1.330 | -.972 | .943 |
| PU2 | 5.17 | 1.289 | -.863 | .666 |
| PU3 | 5.37 | 1.288 | -.995 | .847 |
| PU4 | 5.23 | 1.312 | -.842 | .522 |
| PU5 | 5.24 | 1.338 | -.760 | .481 |
| PEU1 | 5.41 | 1.377 | -.930 | .564 |
| PEU2 | 5.34 | 1.264 | -.750 | .465 |
| PEU3 | 5.36 | 1.234 | -.742 | .549 |
| PEU4 | 5.40 | 1.257 | -.887 | .887 |
| PEU5 | 5.44 | 1.321 | -.880 | .617 |
| SE1 | 5.06 | 1.596 | -.562 | -.559 |
| SE2 | 5.07 | 1.470 | -.503 | -.551 |
| SE3 | 4.76 | 1.561 | -.299 | -.745 |
| SE4 | 4.93 | 1.563 | -.534 | -.506 |
| SE5 | 5.05 | 1.459 | -.687 | -.075 |
| SE6 | 4.99 | 1.683 | -.646 | -.496 |
| FC1 | 5.29 | 1.553 | -1.032 | .463 |
| FC2 | 5.29 | 1.525 | -.939 | .222 |
| FC3 | 4.93 | 1.565 | -.684 | -.175 |
| FC4 | 4.67 | 1.743 | -.474 | -.761 |
| SN1 | 4.75 | 1.514 | -.472 | -.217 |
| SN2 | 4.62 | 1.535 | -.427 | -.294 |
| SN3 | 4.96 | 1.392 | -.725 | .442 |
| SN4 | 5.14 | 1.381 | -.760 | .436 |
| QWL1 | 5.64 | 1.342 | -1.116 | 1.020 |
| QWL2 | 5.70 | 1.318 | -1.082 | .857 |
| QWL3 | 5.29 | 1.347 | -.700 | .031 |
| QWL4 | 5.56 | 1.321 | -.848 | .066 |
| QWL5 | 5.65 | 1.301 | -.913 | .439 |
| BI1 | 5.51 | 1.346 | -.924 | .479 |
| BI2 | 5.77 | 1.258 | -1.007 | .386 |
| BI3 | 5.76 | 1.309 | -1.012 | .383 |
| FreqUsage | 4.84 | 1.136 | -.995 | 1.016 |
| DailyUsage | 4.01 | 1.291 | -.162 | -.876 |
| PD1 | 2.65 | 1.326 | .588 | -.217 |
| PD2 | 2.57 | 1.273 | .538 | -.432 |
| PD3 | 2.39 | 1.233 | .514 | -.597 |
| PD4 | 2.60 | 1.252 | .294 | -.855 |
| PD5 | 2.56 | 1.228 | .494 | -.380 |
| PD6 | 2.64 | 1.218 | .296 | -.861 |
| MF1 | 2.42 | 1.228 | .359 | -.909 |
| MF2 | 2.51 | 1.229 | .372 | -.730 |
| MF3 | 2.52 | 1.285 | .268 | -.977 |
| MF4 | 2.43 | 1.273 | .546 | -.500 |
| MF5 | 2.71 | 1.332 | .437 | -.342 |
| MF6 | 2.62 | 1.328 | .441 | -.527 |
| IC1 | 3.36 | 1.471 | .088 | -.783 |
| IC2 | 3.88 | 1.576 | -.122 | -.775 |
| IC3 | 3.71 | 1.418 | -.197 | -.640 |
| IC4 | 3.65 | 1.492 | -.071 | -.712 |
| IC5 | 3.53 | 1.435 | -.097 | -.804 |
| IC6 | 3.50 | 1.432 | -.032 | -.747 |
| UA1 | 4.10 | 1.346 | -.348 | -.795 |
| UA2 | 4.08 | 1.304 | -.304 | -.696 |
| UA3 | 4.10 | 1.366 | -.290 | -.707 |
| UA4 | 4.12 | 1.335 | -.277 | -.614 |
| UA5 | 4.14 | 1.432 | -.165 | -.870 |
| | | | | |

**Table 7: British sample Normality (Skewness and Kurtosis values)**

| Items | Mean | Std. Deviation | Skewness | Kurtosis |
|---|---|---|---|---|
| PU1 | 5.41 | 1.225 | -.633 | .040 |
| PU2 | 5.06 | 1.244 | -.451 | -.193 |
| PU3 | 5.36 | 1.288 | -.683 | -.115 |
| PU4 | 5.07 | 1.310 | -.494 | -.106 |
| PU5 | 5.02 | 1.323 | -.492 | -.119 |
| PEU1 | 5.80 | 1.206 | -1.017 | .568 |
| PEU2 | 5.54 | 1.237 | -.768 | .107 |
| PEU3 | 5.59 | 1.195 | -.721 | -.056 |
| PEU4 | 5.67 | 1.173 | -.738 | .059 |
| PEU5 | 5.74 | 1.171 | -.906 | .480 |
| SE1 | 5.33 | 1.476 | -.728 | -.443 |
| SE2 | 5.32 | 1.407 | -.665 | -.385 |
| SE3 | 4.98 | 1.446 | -.369 | -.690 |
| SE4 | 5.70 | 1.014 | -.813 | .762 |
| SE5 | 4.96 | 1.440 | -.463 | -.446 |
| SE6 | 4.86 | 1.696 | -.629 | -.442 |
| FC1 | 5.47 | 1.355 | -.828 | -.024 |
| FC2 | 5.36 | 1.352 | -.628 | -.319 |
| FC3 | 5.51 | 1.149 | -.942 | .986 |
| FC4 | 5.43 | 1.159 | -.915 | 1.061 |
| SN1 | 4.88 | 1.521 | -.447 | -.445 |
| SN2 | 4.38 | 1.419 | -.135 | -.381 |
| SN3 | 5.23 | 1.280 | -.500 | .077 |
| SN4 | 5.51 | 1.039 | -1.094 | .993 |
| QWL1 | 5.52 | 1.291 | -.642 | -.317 |
| QWL2 | 5.54 | 1.270 | -.650 | -.244 |
| QWL3 | 4.94 | 1.397 | -.413 | -.328 |
| QWL4 | 5.49 | 1.309 | -.685 | -.242 |
| QWL5 | 5.52 | 1.142 | -.444 | -.304 |
| BI1 | 5.42 | 1.290 | -.797 | .214 |
| BI2 | 5.84 | 1.219 | -.958 | .305 |
| BI3 | 5.67 | 1.260 | -.822 | -.005 |
| FreqUsage | 4.94 | 1.074 | -.961 | .686 |
| DailyUsage | 3.50 | 1.267 | .149 | -.898 |
| MF1 | 3.12 | 1.541 | .435 | -.439 |
| MF2 | 3.49 | 1.658 | .185 | -.924 |
| MF3 | 3.20 | 1.568 | .349 | -.699 |
| MF4 | 3.05 | 1.514 | .285 | -.774 |
| MF5 | 3.18 | 1.460 | .288 | -.568 |
| MF6 | 3.15 | 1.442 | .440 | -.588 |
| PD1 | 3.13 | 1.671 | .520 | -.599 |
| PD2 | 3.17 | 1.644 | .340 | -.988 |
| PD3 | 3.06 | 1.593 | .444 | -.691 |
| PD4 | 3.21 | 1.597 | .398 | -.703 |
| PD5 | 3.21 | 1.600 | .368 | -.843 |
| PD6 | 3.66 | 1.568 | .248 | -.849 |
| IC1 | 4.74 | 1.412 | -.490 | -.291 |
| IC2 | 5.21 | 1.220 | -.588 | .177 |
| IC3 | 4.99 | 1.290 | -.512 | -.060 |
| IC4 | 5.00 | 1.319 | -.449 | -.305 |
| IC5 | 5.04 | 1.368 | -.480 | -.272 |
| IC6 | 4.62 | 1.477 | -.457 | -.491 |
| UA1 | 5.47 | 1.168 | -.725 | .495 |
| UA2 | 5.48 | 1.240 | -.845 | .720 |
| UA3 | 5.32 | 1.203 | -.561 | -.048 |
| UA4 | 5.33 | 1.173 | -.547 | .115 |
| UA5 | 5.47 | 1.179 | -.674 | .274 |

**Table 8: Lebanese sample Normality (Skewness and Kurtosis values)**

# Appendix F

## Multicollinearity

## *England*

### *Perceived Usefulness*

**Variables Entered/Removed[b]**

| Model | Variables Entered | Variables Removed | Method |
|---|---|---|---|
| 1 | QWL, SN, PEOU, FC, SE[a] | . | Enter |

a. All requested variables entered.
b. Dependent Variable: PU

**Coefficients[a]**

| Model | | Collinearity Statistics | |
|---|---|---|---|
| | | Tolerance | VIF |
| 1 | PEOU | .559 | 1.790 |
| | SE | .460 | 2.173 |
| | FC | .512 | 1.954 |
| | SN | .726 | 1.377 |
| | QWL | .619 | 1.615 |

a. Dependent Variable: PU

**Collinearity Diagnostics[a]**

| Model | Dimension | Eigenvalue | Condition Index | Variance Proportions | | | | | |
|---|---|---|---|---|---|---|---|---|---|
| | | | | (Constant) | PEOU | SE | FC | SN | QWL |
| 1 | 1 | 5.869 | 1.000 | .00 | .00 | .00 | .00 | .00 | .00 |
| | 2 | .041 | 12.027 | .11 | .00 | .08 | .37 | .24 | .02 |
| | 3 | .034 | 13.107 | .11 | .09 | .01 | .14 | .64 | .03 |
| | 4 | .021 | 16.740 | .23 | .30 | .22 | .33 | .12 | .12 |
| | 5 | .018 | 17.941 | .52 | .00 | .00 | .03 | .00 | .82 |
| | 6 | .017 | 18.728 | .02 | .61 | .69 | .13 | .00 | .01 |

a. Dependent Variable: PU

## *Perceived Ease Of Use*

**Variables Entered/Removed[b]**

| Model | Variables Entered | Variables Removed | Method |
|---|---|---|---|
| 1 | PU, SN, FC, QWL, SE[a] | . | Enter |

a. All requested variables entered.
b. Dependent Variable: PEOU

**Coefficients[a]**

| Model | | Collinearity Statistics | |
|---|---|---|---|
| | | Tolerance | VIF |
| 1 | SE | .490 | 2.039 |
| | FC | .511 | 1.956 |
| | SN | .691 | 1.448 |
| | QWL | .549 | 1.822 |
| | PU | .460 | 2.176 |

a. Dependent Variable: PEOU

**Collinearity Diagnostics[a]**

| Model | Dimension | Eigenvalue | Condition Index | Variance Proportions | | | | | |
|---|---|---|---|---|---|---|---|---|---|
| | | | | (Constant) | SE | FC | SN | QWL | PU |
| 1 | 1 | 5.872 | 1.000 | .00 | .00 | .00 | .00 | .00 | .00 |
| | 2 | .041 | 12.021 | .12 | .09 | .41 | .18 | .01 | .00 |
| | 3 | .032 | 13.530 | .15 | .03 | .08 | .74 | .07 | .01 |
| | 4 | .022 | 16.284 | .60 | .00 | .12 | .00 | .09 | .36 |
| | 5 | .019 | 17.488 | .00 | .82 | .37 | .05 | .16 | .01 |
| | 6 | .014 | 20.342 | .13 | .07 | .01 | .03 | .67 | .62 |

a. Dependent Variable: PEOU

## *Self-Efficacy*

**Variables Entered/Removed[b]**

| Model | Variables Entered | Variables Removed | Method |
|---|---|---|---|
| 1 | PEOU, SN, QWL, FC, PU[a] | . | Enter |

a. All requested variables entered.
b. Dependent Variable: SE

**Coefficients**[a]

| Model | | Collinearity Statistics | |
|---|---|---|---|
| | | Tolerance | VIF |
| 1 | FC | .602 | 1.662 |
| | SN | .690 | 1.448 |
| | QWL | .545 | 1.836 |
| | PU | .444 | 2.253 |
| | PEOU | .575 | 1.740 |

a. Dependent Variable: SE

**Collinearity Diagnostics**[a]

| Model | Dimension | Eigenvalue | Condition Index | Variance Proportions | | | | | |
|---|---|---|---|---|---|---|---|---|---|
| | | | | (Constant) | FC | SN | QWL | PU | PEOU |
| 1 | 1 | 5.874 | 1.000 | .00 | .00 | .00 | .00 | .00 | .00 |
| | 2 | .037 | 12.547 | .14 | .82 | .04 | .01 | .00 | .00 |
| | 3 | .034 | 13.216 | .04 | .00 | .87 | .03 | .01 | .10 |
| | 4 | .022 | 16.250 | .63 | .15 | .00 | .04 | .31 | .02 |
| | 5 | .019 | 17.405 | .01 | .02 | .05 | .37 | .02 | .74 |
| | 6 | .014 | 20.589 | .17 | .00 | .03 | .55 | .66 | .14 |

a. Dependent Variable: SE

## *Facilitating Conditions*

**Variables Entered/Removed**[b]

| Model | Variables Entered | Variables Removed | Method |
|---|---|---|---|
| 1 | SE, SN, QWL, PEOU, PU[a] | . | Enter |

a. All requested variables entered.
b. Dependent Variable: FC

**Coefficients**[a]

| Model | | Collinearity Statistics | |
|---|---|---|---|
| | | Tolerance | VIF |
| 1 | SN | .713 | 1.403 |
| | QWL | .545 | 1.834 |
| | PU | .437 | 2.291 |
| | PEOU | .530 | 1.888 |
| | SE | .532 | 1.880 |

a. Dependent Variable: FC

**Collinearity Diagnostics**[a]

| Model | Dimension | Eigenvalue | Condition Index | Variance Proportions | | | | | |
|---|---|---|---|---|---|---|---|---|---|
| | | | | (Constant) | SN | QWL | PU | PEOU | SE |
| 1 | 1 | 5.885 | 1.000 | .00 | .00 | .00 | .00 | .00 | .00 |
| | 2 | .036 | 12.779 | .00 | .81 | .00 | .00 | .07 | .12 |
| | 3 | .026 | 14.912 | .61 | .11 | .05 | .02 | .00 | .27 |
| | 4 | .021 | 16.608 | .20 | .05 | .30 | .30 | .04 | .21 |
| | 5 | .018 | 18.285 | .03 | .01 | .11 | .02 | .77 | .41 |
| | 6 | .014 | 20.597 | .17 | .03 | .54 | .66 | .12 | .00 |

a. Dependent Variable: FC

## *Social Norms*

**Variables Entered/Removed**[b]

| Model | Variables Entered | Variables Removed | Method |
|---|---|---|---|
| 1 | FC, QWL, PEOU, PU, SE[a] | . | Enter |

a. All requested variables entered.
b. Dependent Variable: SN

**Coefficients**[a]

| Model | | Collinearity Statistics | |
|---|---|---|---|
| | | Tolerance | VIF |
| 1 | QWL | .548 | 1.826 |
| | PU | .455 | 2.200 |
| | PEOU | .526 | 1.903 |
| | SE | .448 | 2.231 |
| | FC | .523 | 1.911 |

a. Dependent Variable: SN

**Collinearity Diagnostics**[a]

| Model | Dimension | Eigenvalue | Condition Index | Variance Proportions | | | | | |
|---|---|---|---|---|---|---|---|---|---|
| | | | | (Constant) | QWL | PU | PEOU | SE | FC |
| 1 | 1 | 5.886 | 1.000 | .00 | .00 | .00 | .00 | .00 | .00 |
| | 2 | .039 | 12.291 | .20 | .04 | .01 | .01 | .05 | .51 |
| | 3 | .022 | 16.204 | .58 | .00 | .18 | .12 | .06 | .28 |
| | 4 | .022 | 16.524 | .07 | .27 | .18 | .22 | .24 | .07 |
| | 5 | .017 | 18.734 | .00 | .06 | .00 | .54 | .65 | .14 |
| | 6 | .014 | 20.426 | .15 | .64 | .62 | .11 | .00 | .00 |

a. Dependent Variable: SN

## Quality of Work Life

**Variables Entered/Removed[b]**

| Model | Variables Entered | Variables Removed | Method |
|-------|-------------------|-------------------|--------|
| 1 | SN, PEOU, FC, PU, SE[a] | . | Enter |

a. All requested variables entered.
b. Dependent Variable: QWL

**Coefficients[a]**

| Model | | Collinearity Statistics | |
|-------|------|-----------|-------|
| | | Tolerance | VIF |
| 1 | PU | .495 | 2.020 |
| | PEOU | .534 | 1.874 |
| | SE | .452 | 2.214 |
| | FC | .511 | 1.956 |
| | SN | .699 | 1.430 |

a. Dependent Variable: QWL

**Collinearity Diagnostics[a]**

| Model | Dimension | Eigenvalue | Condition Index | Variance Proportions | | | | | |
|-------|-----------|-----------|-----------------|------------|-----|------|-----|-----|-----|
| | | | | (Constant) | PU | PEOU | SE | FC | SN |
| 1 | 1 | 5.871 | 1.000 | .00 | .00 | .00 | .00 | .00 | .00 |
| | 2 | .040 | 12.114 | .12 | .00 | .00 | .08 | .33 | .33 |
| | 3 | .033 | 13.270 | .17 | .01 | .11 | .01 | .21 | .49 |
| | 4 | .022 | 16.282 | .64 | .28 | .08 | .03 | .24 | .02 |
| | 5 | .018 | 18.189 | .01 | .44 | .00 | .67 | .18 | .12 |
| | 6 | .016 | 19.014 | .06 | .28 | .80 | .22 | .04 | .03 |

a. Dependent Variable: QWL

# *Lebanon*

## Perceived Usefulness

**Variables Entered/Removed[b]**

| Model | Variables Entered | Variables Removed | Method |
|---|---|---|---|
| 1 | QWL, SN, FC, PEOU, SE[a] | . | Enter |

a. All requested variables entered.
b. Dependent Variable: PU

**Coefficients[a]**

| Model | | Collinearity Statistics | |
|---|---|---|---|
| | | Tolerance | VIF |
| 1 | PEOU | .673 | 1.487 |
| | SE | .560 | 1.786 |
| | SN | .782 | 1.279 |
| | FC | .661 | 1.513 |
| | QWL | .716 | 1.396 |

a. Dependent Variable: PU

**Collinearity Diagnostics[a]**

| Model | Dimension | Eigenvalue | Condition Index | Variance Proportions | | | | | |
|---|---|---|---|---|---|---|---|---|---|
| | | | | (Constant) | PEOU | SE | SN | FC | QWL |
| 1 | 1 | 5.897 | 1.000 | .00 | .00 | .00 | .00 | .00 | .00 |
| | 2 | .031 | 13.849 | .00 | .04 | .02 | .75 | .19 | .00 |
| | 3 | .022 | 16.249 | .06 | .13 | .03 | .17 | .44 | .28 |
| | 4 | .020 | 17.270 | .01 | .50 | .06 | .02 | .13 | .46 |
| | 5 | .016 | 19.296 | .71 | .00 | .27 | .00 | .04 | .23 |
| | 6 | .014 | 20.220 | .22 | .33 | .61 | .06 | .20 | .03 |

a. Dependent Variable: PU

## Perceived Ease Of Use

**Variables Entered/Removed[b]**

| Model | Variables Entered | Variables Removed | Method |
|---|---|---|---|
| 1 | PU, FC, SN, QWL, SE[a] | . | Enter |

a. All requested variables entered.
b. Dependent Variable: PEOU

**Coefficients**[a]

| Model | | Collinearity Statistics | |
|---|---|---|---|
| | | Tolerance | VIF |
| 1 | SE | .607 | 1.647 |
| | SN | .770 | 1.298 |
| | FC | .676 | 1.479 |
| | QWL | .631 | 1.585 |
| | PU | .711 | 1.405 |

a. Dependent Variable: PEOU

**Collinearity Diagnostics**[a]

| Model | Dimension | Eigenvalue | Condition Index | Variance Proportions | | | | | |
|---|---|---|---|---|---|---|---|---|---|
| | | | | (Constant) | SE | SN | FC | QWL | PU |
| 1 | 1 | 5.889 | 1.000 | .00 | .00 | .00 | .00 | .00 | .00 |
| | 2 | .034 | 13.257 | .00 | .07 | .02 | .32 | .02 | .39 |
| | 3 | .029 | 14.303 | .00 | .00 | .89 | .05 | .03 | .13 |
| | 4 | .017 | 18.567 | .31 | .12 | .02 | .06 | .41 | .41 |
| | 5 | .016 | 19.191 | .49 | .22 | .06 | .26 | .31 | .00 |
| | 6 | .016 | 19.407 | .20 | .59 | .01 | .30 | .23 | .07 |

a. Dependent Variable: PEOU

## Self-Efficacy

**Variables Entered/Removed**[b]

| Model | Variables Entered | Variables Removed | Method |
|---|---|---|---|
| 1 | PEOU, SN, FC, QWL, PU[a] | . | Enter |

a. All requested variables entered.
b. Dependent Variable: SE

**Coefficients**[a]

| Model | | Collinearity Statistics | |
|---|---|---|---|
| | | Tolerance | VIF |
| 1 | SN | .800 | 1.250 |
| | FC | .752 | 1.329 |
| | QWL | .641 | 1.561 |
| | PU | .640 | 1.562 |
| | PEOU | .657 | 1.523 |

a. Dependent Variable: SE

**Collinearity Diagnostics^a**

| Model | Dimension | Eigenvalue | Condition Index | Variance Proportions | | | | | |
|-------|-----------|-----------|-----------------|-------------|-----|-----|-----|-----|------|
| | | | | (Constant) | SN | FC | QWL | PU | PEOU |
| 1 | 1 | 5.888 | 1.000 | .00 | .00 | .00 | .00 | .00 | .00 |
| | 2 | .031 | 13.722 | .00 | .01 | .47 | .01 | .40 | .00 |
| | 3 | .030 | 14.008 | .00 | .89 | .09 | .00 | .03 | .07 |
| | 4 | .019 | 17.440 | .00 | .03 | .02 | .51 | .00 | .56 |
| | 5 | .017 | 18.775 | .60 | .06 | .36 | .04 | .31 | .03 |
| | 6 | .015 | 20.121 | .39 | .01 | .05 | .43 | .26 | .35 |

a. Dependent Variable: SE

## Social Norms

**Variables Entered/Removed^b**

| Model | Variables Entered | Variables Removed | Method |
|-------|-------------------|-------------------|--------|
| 1 | SE, PU, FC, QWL, PEOU^a | . | Enter |

a. All requested variables entered.
b. Dependent Variable: SN

**Coefficients^a**

| Model | | Collinearity Statistics | |
|-------|------|-----------|-------|
| | | Tolerance | VIF |
| 1 | FC | .659 | 1.517 |
| | QWL | .645 | 1.551 |
| | PU | .650 | 1.537 |
| | PEOU | .606 | 1.651 |
| | SE | .582 | 1.719 |

a. Dependent Variable: SN

**Collinearity Diagnostics^a**

| Model | Dimension | Eigenvalue | Condition Index | Variance Proportions | | | | | |
|-------|-----------|-----------|-----------------|-------------|-----|-----|-----|------|-----|
| | | | | (Constant) | FC | QWL | PU | PEOU | SE |
| 1 | 1 | 5.900 | 1.000 | .00 | .00 | .00 | .00 | .00 | .00 |
| | 2 | .033 | 13.281 | .00 | .29 | .03 | .39 | .00 | .06 |
| | 3 | .020 | 17.182 | .03 | .04 | .44 | .01 | .47 | .05 |
| | 4 | .017 | 18.545 | .53 | .45 | .01 | .32 | .02 | .01 |
| | 5 | .016 | 19.393 | .35 | .21 | .22 | .03 | .00 | .52 |
| | 6 | .014 | 20.550 | .09 | .01 | .31 | .24 | .51 | .35 |

a. Dependent Variable: SN

## Facilitating Condition

**Variables Entered/Removed**[b]

| Model | Variables Entered | Variables Removed | Method |
|---|---|---|---|
| 1 | SN, PEOU, QWL, PU, SE[a] | . | Enter |

a. All requested variables entered.
b. Dependent Variable: FC

**Coefficients**[a]

| Model | | Collinearity Statistics | |
|---|---|---|---|
| | | Tolerance | VIF |
| 1 | QWL | .647 | 1.546 |
| | PU | .644 | 1.554 |
| | PEOU | .622 | 1.607 |
| | SE | .640 | 1.561 |
| | SN | .772 | 1.296 |

a. Dependent Variable: FC

**Collinearity Diagnostics**[a]

| Model | Dimension | Eigenvalue | Condition Index | Variance Proportions | | | | | |
|---|---|---|---|---|---|---|---|---|---|
| | | | | (Constant) | QWL | PU | PEOU | SE | SN |
| 1 | 1 | 5.896 | 1.000 | .00 | .00 | .00 | .00 | .00 | .00 |
| | 2 | .030 | 13.940 | .00 | .02 | .24 | .05 | .01 | .68 |
| | 3 | .025 | 15.310 | .02 | .03 | .35 | .12 | .29 | .19 |
| | 4 | .019 | 17.605 | .10 | .57 | .13 | .27 | .00 | .10 |
| | 5 | .016 | 19.149 | .75 | .11 | .02 | .03 | .36 | .00 |
| | 6 | .014 | 20.673 | .13 | .28 | .26 | .54 | .34 | .03 |

a. Dependent Variable: FC

## Quality of Work Life

**Variables Entered/Removed**[b]

| Model | Variables Entered | Variables Removed | Method |
|---|---|---|---|
| 1 | FC, PU, SN, PEOU, SE[a] | . | Enter |

a. All requested variables entered.
b. Dependent Variable: QWL

**Coefficients**[a]

| Model | | Collinearity Statistics | |
|---|---|---|---|
| | | Tolerance | VIF |
| 1 | PU | .729 | 1.372 |
| | PEOU | .607 | 1.648 |
| | SE | .570 | 1.755 |
| | SN | .788 | 1.269 |
| | FC | .676 | 1.480 |

a. Dependent Variable: QWL

**Collinearity Diagnostics**[a]

| Model | Dimension | Eigenvalue | Condition Index | Variance Proportions | | | | | |
|---|---|---|---|---|---|---|---|---|---|
| | | | | (Constant) | PU | PEOU | SE | SN | FC |
| 1 | 1 | 5.889 | 1.000 | .00 | .00 | .00 | .00 | .00 | .00 |
| | 2 | .033 | 13.384 | .00 | .42 | .00 | .05 | .06 | .33 |
| | 3 | .030 | 14.027 | .00 | .15 | .08 | .00 | .82 | .01 |
| | 4 | .017 | 18.346 | .00 | .28 | .44 | .16 | .00 | .48 |
| | 5 | .017 | 18.876 | .96 | .10 | .00 | .10 | .05 | .04 |
| | 6 | .014 | 20.299 | .03 | .05 | .48 | .68 | .07 | .14 |

a. Dependent Variable: QWL

# Appendix G

## Measurement Model (Model Fit Summary)

### *Measurement model (British sample first run)*

CMIN

| Model | NPAR | CMIN | DF | P | CMIN/DF |
|---|---|---|---|---|---|
| Default model | 94 | 1640.667 | 467 | .000 | 3.513 |
| Saturated model | 561 | .000 | 0 | | |
| Independence model | 33 | 15295.398 | 528 | .000 | 28.969 |

RMR, GFI

| Model | RMR | GFI | AGFI | PGFI |
|---|---|---|---|---|
| Default model | .125 | .853 | .823 | .710 |
| Saturated model | .000 | 1.000 | | |
| Independence model | .807 | .143 | .089 | .134 |

Baseline Comparisons

| Model | NFI Delta1 | RFI rho1 | IFI Delta2 | TLI rho2 | CFI |
|---|---|---|---|---|---|
| Default model | .893 | .879 | .921 | .910 | .921 |
| Saturated model | 1.000 | | 1.000 | | 1.000 |
| Independence model | .000 | .000 | .000 | .000 | .000 |

Parsimony-Adjusted Measures

| Model | PRATIO | PNFI | PCFI |
|---|---|---|---|
| Default model | .884 | .790 | .814 |
| Saturated model | .000 | .000 | .000 |
| Independence model | 1.000 | .000 | .000 |

NCP

| Model | NCP | LO 90 | HI 90 |
|---|---|---|---|
| Default model | 1173.667 | 1054.003 | 1300.888 |
| Saturated model | .000 | .000 | .000 |
| Independence model | 14767.398 | 14367.255 | 15173.894 |

FMIN

| Model | FMIN | F0 | LO 90 | HI 90 |
|---|---|---|---|---|
| Default model | 2.730 | 1.953 | 1.754 | 2.165 |
| Saturated model | .000 | .000 | .000 | .000 |
| Independence model | 25.450 | 24.571 | 23.906 | 25.248 |

RMSEA

| Model | RMSEA | LO 90 | HI 90 | PCLOSE |
|---|---|---|---|---|
| Default model | .065 | .061 | .068 | .000 |
| Independence model | .216 | .213 | .219 | .000 |

AIC

| Model | AIC | BCC | BIC | CAIC |
|---|---|---|---|---|
| Default model | 1828.667 | 1839.941 | 2242.292 | 2336.292 |
| Saturated model | 1122.000 | 1189.280 | 3590.544 | 4151.544 |
| Independence model | 15361.398 | 15365.356 | 15506.607 | 15539.607 |

### ***Measurement model (Lebanese sample first run)***

CMIN

| Model | NPAR | CMIN | DF | P | CMIN/DF |
|---|---|---|---|---|---|
| Default model | 94 | 1377.704 | 467 | .000 | 2.950 |
| Saturated model | 561 | .000 | 0 | | |
| Independence model | 33 | 12300.361 | 528 | .000 | 23.296 |

RMR, GFI

| Model | RMR | GFI | AGFI | PGFI |
|---|---|---|---|---|
| Default model | .084 | .860 | .832 | .716 |
| Saturated model | .000 | 1.000 | | |
| Independence model | .544 | .207 | .158 | .195 |

Baseline Comparisons

| Model | NFI Delta1 | RFI rho1 | IFI Delta2 | TLI rho2 | CFI |
|---|---|---|---|---|---|
| Default model | .888 | .873 | .923 | .913 | .923 |
| Saturated model | 1.000 | | 1.000 | | 1.000 |
| Independence model | .000 | .000 | .000 | .000 | .000 |

Parsimony-Adjusted Measures

| Model | PRATIO | PNFI | PCFI |
|---|---|---|---|
| Default model | .884 | .785 | .816 |
| Saturated model | .000 | .000 | .000 |
| Independence model | 1.000 | .000 | .000 |

NCP

| Model | NCP | LO 90 | HI 90 |
|---|---|---|---|
| Default model | 910.704 | 803.192 | 1025.822 |
| Saturated model | .000 | .000 | .000 |
| Independence model | 11772.361 | 11414.640 | 12136.454 |

FMIN

| Model | FMIN | F0 | LO 90 | HI 90 |
|---|---|---|---|---|
| Default model | 2.438 | 1.612 | 1.422 | 1.816 |
| Saturated model | .000 | .000 | .000 | .000 |
| Independence model | 21.771 | 20.836 | 20.203 | 21.480 |

RMSEA

| Model | RMSEA | LO 90 | HI 90 | PCLOSE |
|---|---|---|---|---|
| Default model | .059 | .055 | .062 | .000 |
| Independence model | .199 | .196 | .202 | .000 |

AIC

| Model | AIC | BCC | BIC | CAIC |
|---|---|---|---|---|
| Default model | 1565.704 | 1577.742 | 1973.532 | 2067.532 |
| Saturated model | 1122.000 | 1193.842 | 3555.951 | 4116.951 |
| Independence model | 12366.361 | 12370.587 | 12509.535 | 12542.535 |

# Appendix H

## Standardised regression weights

| Lebanon (Standardised regression weight) | | | Estimates | England (Standardised regression weight) | | | Estimates |
|---|---|---|---|---|---|---|---|
| PU1 | <--- | PU | .727 | PU1 | <--- | PUU | .824 |
| PU2 | <--- | PU | .836 | PU2 | <--- | PUU | .861 |
| PU3 | <--- | PU | .777 | PU3 | <--- | PUU | .818 |
| PU4 | <--- | PU | .858 | PU4 | <--- | PUU | .850 |
| PU5 | <--- | PU | .840 | PU5 | <--- | PUU | .839 |
| PEOU1 | <--- | PEOU | .807 | PEOU1 | <--- | PEU | .852 |
| PEOU2 | <--- | PEOU | .866 | PEOU2 | <--- | PEU | .879 |
| PEOU3 | <--- | PEOU | .880 | PEOU3 | <--- | PEU | .863 |
| PEOU4 | <--- | PEOU | .855 | PEOU4 | <--- | PEU | .850 |
| PEOU5 | <--- | PEOU | .843 | PEOU5 | <--- | PEU | .767 |
| SN1 | <--- | SN | .713 | SN1 | <--- | SNN | .829 |
| SN2 | <--- | SN | .641 | SN2 | <--- | SN | .765 |
| SN3 | <--- | SN | .862 | SN3 | <--- | SN | .771 |
| SN4 | <--- | SN | .712 | SN4 | <--- | SN | .632 |
| QWL1 | <--- | QWL | .731 | QWL1 | <--- | QWL | .802 |
| QWL2 | <--- | QWL | .797 | QWL2 | <--- | QWL | .820 |
| QWL3 | <--- | QWL | .575 | QWL3 | <--- | QWL | .753 |
| QWL4 | <--- | QWL | .664 | QWL4 | <--- | QWL | .740 |
| QWL5 | <--- | QWL | .797 | QWL5 | <--- | QWL | .808 |
| SE1 | <--- | SE | .887 | SE1 | <--- | SE | .903 |
| SE2 | <--- | SE | .913 | SE2 | <--- | SE | .888 |
| SE3 | <--- | SE | .768 | SE3 | <--- | SE | .818 |
| SE4 | <--- | SE | .517 | SE4 | <--- | SE | .490 |
| SE5 | <--- | SE | .341 | SE5 | <--- | SE | .390 |
| FC1 | <--- | FC | .880 | FC1 | <--- | FC | .920 |
| FC2 | <--- | FC | .899 | FC2 | <--- | FC | .920 |
| FC3 | <--- | FC | .821 | FC3 | <--- | FC | .729 |
| FC4 | <--- | FC | .719 | FC4 | <--- | FC | .627 |
| BI1 | <--- | BI | .722 | BI1 | <--- | BI | .807 |
| BI2 | <--- | BI | .852 | BI2 | <--- | BI | .874 |
| BI3 | <--- | BI | .914 | BI3 | <--- | BI | .907 |
| FreqUsage | <--- | AU | .777 | FreqUsage | <--- | AU | .634 |
| DailyUsage | <--- | AU | .628 | DailyUsage | <--- | AU | .867 |

**Table 10: Standardized Regression Weights for both samples: the red items are below the cut off 0.5**